\documentclass[11pt]{article}
\pdfoutput=1 

\usepackage{jheppub} 

\usepackage[T1]{fontenc} 

\usepackage[export]{adjustbox}
\usepackage{color}
\usepackage{xcolor}
\usepackage{autobreak}
\usepackage{multirow}
\usepackage{subfig}
\usepackage{amsmath}
\usepackage[vcentermath]{youngtab}
\usepackage{ytableau}
\usepackage{pifont}
\usepackage{extarrows}
\usepackage{makecell}
\usepackage{diagbox}
\usepackage{lscape}
\usepackage{ulem} 
\usepackage{bbm}
\usepackage{diagbox}
\usepackage{mathtools}

\usepackage{tikz}
\usepackage[compat=1.1.0]{tikz-feynhand}
\usepackage{tikz-cd}

\tikzset{thick bos/.style={very thick,decorate,decoration={snake,amplitude=1pt,segment length=4pt}}}
\tikzset{thick sca/.style={very thick,dash=on 4pt off 2pt phase 0pt}}

\newcommand{\sca}[1] {\draw[brown,thick sca] (#1)--(v1)} 
\newcommand{\fer}[3]{\propag[#1,fer] (#2)--(#3);
\draw[#1,style=very thick] (#2)--(#3);} 
\newcommand{\antfer}[3]{\propag[#1,antfer] (#2)--(#3);
\draw[#1,style=very thick] (#2)--(#3);} 
\newcommand{\ferflip}[4]{
\draw[#3,very thick,rotate=#2] (0.7*#1,0)--(0.46*#1,0);
\draw[#4,very thick,decoration={markings,mark=at position 0.56 with {\arrow{Triangle[length=4pt,width=4pt]}}},postaction={decorate},rotate=#2] (0.46*#1,0)--(v1);
\draw[very thick,rotate=#2] plot[mark=x,mark size=2.5] coordinates {(0.46*#1,0)};
\draw[rotate=#2] (0.1*#1,-0.08) -- (0.16*#1,0.08);} 
\newcommand{\antferflip}[4]{
\draw[#3,very thick,rotate=#2] (0.7*#1,0)--(0.46*#1,0);
\draw[#4,very thick,decoration={markings,mark=at position 0.82 with {\arrow{Triangle[length=4pt,width=4pt]}}},postaction={decorate},rotate=#2] (v1)--(0.46*#1,0);
\draw[very thick,rotate=#2] plot[mark=x,mark size=2.5] coordinates {(0.46*#1,0)};
\draw[rotate=#2] (0.1*#1,-0.08) -- (0.16*#1,+0.08);
} 
\newcommand{\bos}[2] {\draw[#2,thick bos] (#1)--(v1)} 
\newcommand{\bosflip}[4] {
\begin{scope}[rotate=#2]
\clip (0,-0.1) rectangle (0.44*#1,0.1); 
\draw[#3,thick bos] (0,0)--(#1,0);
\end{scope}
\begin{scope}[rotate=#2]
\clip (0.44*#1,-0.1) rectangle (0.7*#1,0.1); 
\draw[#4,thick bos] (0,0)--(#1,0);
\end{scope}
\draw[very thick,rotate=#2] plot[mark=x,mark size=2.5] coordinates {(0.43*#1,0)};
\draw[rotate=#2] (0.17*#1,-0.08) -- (0.23*#1,+0.08);
} 
\newcommand{\bosflipflip}[5] {
\begin{scope}[rotate=#2]
\clip (0,-0.1) rectangle (0.35*#1,0.1); 
\draw[#3,thick bos] (0,0)--(#1,0);
\end{scope}
\begin{scope}[rotate=#2]
\clip (0.35*#1,-0.1) rectangle (0.55*#1,0.1); 
\draw[#4,thick bos] (0,0)--(#1,0);
\end{scope}
\begin{scope}[rotate=#2]
\clip (0.55*#1,-0.1) rectangle (0.7*#1,0.1); 
\draw[#5,thick bos] (0,0)--(#1,0);
\end{scope}
\draw[very thick,rotate=#2] plot[mark=x,mark size=2.5] coordinates {(0.55*#1,0)};
\draw[very thick,rotate=#2] plot[mark=x,mark size=2.5] coordinates {(0.35*#1,0)};
\draw[rotate=#2] (0.17*#1,-0.08) -- (0.23*#1,+0.08);
\draw[rotate=#2] (0.07*#1,-0.08) -- (0.13*#1,+0.08);
} 

\newcommand{\Ampone}[3] {
\begin{tikzpicture}[baseline=-0.1cm] \begin{feynhand}
\setlength{\feynhandarrowsize}{4pt}
\vertex [particle] (i1) at (#1,0) {$#2$}; 
\vertex (v1) at (0,0);
#3;
\vertex[dot] (v1) at (0,0) {};
\end{feynhand} \end{tikzpicture}
}

\newcommand{\Ampthree}[6] {
\begin{tikzpicture}[baseline=-0.1cm] \begin{feynhand}
\setlength{\feynhandarrowsize}{4pt}
\vertex [particle] (i1) at (-1.01,0) {$#1$}; 
\vertex [particle] (i2) at (0.579,0.827) {$#2$}; 
\vertex [particle] (i3) at (0.579,-0.827) {$#3$}; 
\vertex (v1) at (0,0);
#4;
#5;
#6;
\end{feynhand} \end{tikzpicture}
}

\usepackage{eucal}

\usepackage[utf8]{inputenc}
\usepackage{times} 
\usepackage{mathptmx} 

\newcommand{\eq}[1]{\begin{equation}\begin{split} #1 \end{split}\end{equation}}

\allowdisplaybreaks

\title{Massless-Massive Amplitude Correspondence I: Helicity-chirality Matching and On-shell Higgsing}

\author[a,b,e]{Yu-Han Ni, }
\author[a]{Chao Wu, }
\author[a, b, c, d]{Jiang-Hao Yu, }

\affiliation[a]{Institute of Theoretical Physics, Chinese Academy of Sciences, Beijing 100190, China}
\affiliation[b]{School of Physical Sciences, University of Chinese Academy of Sciences, Beijing 100049, China}
\affiliation[c]{School of Fundamental Physics and Mathematical Sciences, Hangzhou Institute for Advanced
Study, UCAS, Hangzhou 310024, China}
\affiliation[d]{International Center for Theoretical Physics Asia-Pacific, Beijing/Hangzhou, China}
\affiliation[e]{School of Science and Engineering, Chinese University of Hong Kong, Shenzhen, Shenzhen 518172, China}

\emailAdd{niyuhan@cuhk.edu.cn}
\emailAdd{wuch7@itp.ac.cn}
\emailAdd{jhyu@itp.ac.cn}

\abstract{

In this work, the massless-massive correspondence for the on-shell scattering amplitudes is constructed so the massive amplitudes could inherit advantageous techniques developed in the massless calculation. This correspondence is established by matching massless amplitudes to Minimal Helicity-Chirality (MHC) amplitudes, which arise from an expansion of massive spin-spinor amplitudes in terms of the chirality-flip $m\eta$ order by order. The primary MHC amplitude deforms into a massless amplitude of the same helicity; if a vector boson is involved, it may instead vanish due to the associated conserved current. In cases where the primary amplitude vanishes, the leading contributions originate from descendant MHC amplitudes, each corresponding to a distinct massless amplitude in the ultraviolet theory containing either a transverse gauge boson or a Goldstone boson. We propose a systematic amplitude deformation procedure for three-point massless-massive matching based on helicity-chirality unification and the scaling properties of $m\eta$. Sub-leading MHC amplitudes are matched to massless amplitudes with additional on-shell Higgs splitting, a process known as on-shell Higgsing. In this work, we extend and reinterpret on-shell Higgsing as a transversality flip between different MHC states, and obtain all the 3-point massless-massive matching results in the spontaneous broken standard model.

}

\begin{document}

\maketitle

\flushbottom

\section{Introduction}

In recent years, significant progress has been made in understanding and computing scattering amplitudes for massless particles. The introduction of spinor-helicity variables has been central to this development, as they trivialize on-shell constraints and eliminate gauge redundancies~\cite{Parke:1986gb,Bern:1996je,Dixon:1996wi,Elvang:2013cua,Cheung:2017pzi,Travaglini:2022uwo,Badger:2023eqz}. For massive particles, however, the on-shell description appears to lose this advantage, since gauge redundancy is no longer present. Unlike in the massless case, there are multiple possible choices for massive spinors, which correspond to different choices of spin quantization axis~\cite{Kleiss:1985yh,Hagiwara:1985yu,Kleiss:1988xr,Dittmaier:1998nn,Schwinn:2005pi,Schwinn:2006ca,Badger:2005zh,Badger:2005jv,Conde:2016vxs,Conde:2016izb,Basile:2024ydc}. A common approach introduces a light-like reference vector to define the spin axis; a particularly important special case is the helicity axis, aligned with the particle's momentum, which permits a smooth massless limit. To systematize the construction, Arkani-Hamed, Huang, and Huang (AHH) introduced a spin-spinor formalism~\cite{Arkani-Hamed:2017jhn} that carries an additional $SU(2)$ little-group index, thereby making little-group covariance manifest. This framework allows all possible three-particle massive amplitudes to be enumerated directly from Poincar\'e symmetry. The AHH formalism has since been applied to a wide range of areas, including electroweak theory~\cite{Franken:2019wqr, Bachu:2019ehv, Ballav:2020ese, Wu:2021nmq, Ballav:2021ahg, Liu:2022alx, Bachu:2023fjn, Ema:2024rss}, effective field theories~\cite{Shadmi:2018xan, Aoude:2019tzn, Durieux:2019eor, Ma:2019gtx, Durieux:2019siw, Li:2020gnx, Li:2020xlh, Li:2020tsi, Li:2020zfq, Durieux:2020gip, Li:2021tsq, AccettulliHuber:2021uoa, Dong:2021vxo, Li:2022tec, Balkin:2021dko, DeAngelis:2022qco, Dong:2022mcv, Ren:2022tvi, Liu:2023jbq, Goldberg:2024eot}, and gravitational wave physics~\cite{Guevara:2018wpp, Chung:2018kqs, Guevara:2019fsj, Bern:2019nnu, Maybee:2019jus, Arkani-Hamed:2019ymq}.

In massless theories, three-particle amplitudes are completely determined by consistency conditions derived from little-group scaling~\cite{Benincasa:2007xk,Witten:2003nn}. Four-point and higher-point amplitudes can then be constructed systematically using bootstrap or BCFW techniques that enforce consistent factorization properties~\cite{Britto:2004ap,Britto:2005fq}. 
Similarly, the bootstrap method has been utilized to construct the higher-point AHH massive amplitudes from the three-point massive ones~\cite{Arkani-Hamed:2017jhn}. However, there are special cases, when all the external particles are massive and at least one particle has spin-one and higher, the massive amplitudes cannot proceed by a direct application of the massless bootstrap or BCFW methods. In the bootstrap approach, additional four-point contact terms must be included to satisfy unitarity constraints~\cite{Liu:2022alx}. Similarly, in a BCFW recursive construction, different momentum shifts must be chosen for different polarizations when external massive vector bosons are present~\cite{Lai:2023upa,Ema:2024vww}. Therefore, it is not so obvious to directly apply various higher-point construction techniques developed for massless theories to the on-shell formulation of massive amplitudes.


Spontaneous symmetry breaking offers a pathway for massive on-shell amplitudes to inherit the computational advantages developed for massless scattering amplitudes, thereby establishing a direct correspondence between massless amplitudes in the unbroken phase and massive amplitudes in the broken phase. In the Standard Model, the Higgs mechanism endows fermions and gauge bosons with mass, transforming initially massless states into massive ones. Consequently, one expects that multiple massless helicity amplitudes should be unified into their corresponding massive counterparts. Conversely, in the high-energy limit where the electroweak gauge symmetry is restored, it becomes possible to probe the theory in its unbroken phase. A massive amplitude should then decompose into a sum of massless helicity amplitudes. This decomposition is formalized in the AHH spinor formalism~\cite{Arkani-Hamed:2017jhn}, which exploits the fact that a massive spin-spinor can be decomposed into two massless helicity spinors. Thus, the AHH framework provides a powerful and systematic on-shell description of electroweak symmetry restoration.

In the high-energy limit $m \to 0$, a massive amplitude must smoothly transition to a massless one. To connect these two regimes, one performs an amplitude deformation, identifying the mass term with the $\eta$ spinor~\cite{Arkani-Hamed:2017jhn}. In practice, such deformations have often been applied empirically, underscoring the need for a more systematic procedure. Furthermore, a massive amplitude contains not only the leading massless contribution but also sub-leading terms that correspond to massless amplitudes with additional Higgs bosons. When the Higgs field acquires a vacuum expectation value, these multi-Higgs massless amplitudes must be matched to the single massive amplitude. This implies that a given massive amplitude corresponds to an entire family of massless amplitudes with increasing numbers of Higgses. This observation motivates the development of a systematic massless-massive correspondence that encompasses both leading and sub-leading orders.


To address the above issues, we propose to extend the massive AHH formalism to the one with complex masses, which extends the $SU(2)$ little group to an enlarged $SU(2) \times ISO(2)$ structure. In this extended framework, an additional quantum number, transversality, is introduced to characterize particle states. In the high-energy expansion, massive amplitudes are expanded in terms of Minimal Helicity-Chirality (MHC) amplitudes, organized order-by-order in powers of $m\eta$, representing the chirality flip. This organization differs from the standard $\eta$-expansion in the AHH high-energy limit. The ladder operators of the extended little group connect different MHC amplitudes, implying that a single MHC amplitude can, in principle, determine all others at different orders. In the high-energy limit, each MHC amplitude must correspond to a specific massless amplitude or vanishing one due to conserved current, where the helicity and transversality quantum numbers coalesce into a single helicity. The precise nature of this massless-massive correspondence depends on the order in the $m\eta$ expansion.

The primary MHC amplitudes, constructed solely from $\lambda$ spinors, exhibit a one-to-one correspondence with massless amplitudes. For descendant MHC amplitudes, which contain both $\lambda$ and $m\eta$ factors, the matching is more involved. In processes that do not involve massive gauge bosons, a descendant amplitude with a specific number of chirality flips (i.e., $m\eta$ insertions) should be matched to a massless amplitude with the same helicity but with an additional Higgs boson for each flip. This matching is implemented through a leading-order deformation that incorporates on-shell Higgs splitting, a procedure known as {\it on-shell Higgsing}~\cite{Balkin:2021dko}. In this work the on-shell Higgsing is extended and explained as the tranversality flip between different MHC states. Once the Higgs-splitting function is known, the leading-order correspondence fully determines the structure, and all sub-leading matching conditions can be derived systematically from this leading result.

Special care is required when establishing the massless-massive correspondence for amplitudes involving massive vector bosons. For a spin-1 MHC particle state, spontaneous symmetry breaking implies that the massive vector originates from a massless gauge boson and a Goldstone boson. The massless limit of an amplitude containing a massive vector boson is singular unless the vector couples to a conserved current. Consequently, the three-point primary MHC amplitude for such a process vanishes, consistent with this current conservation. Instead, the leading contributions arise from several descendant amplitudes, related by the ladder operators, in which their helicity satisfy the massless helicity condition, each corresponding to a massless amplitude in the ultraviolet (UV) theory involving either a transverse gauge boson or a Goldstone boson. Although these descendant amplitudes contain chirality-flip factors proportional to $m\eta$, amplitude deformation converts each $m\eta$ factor into a reference momentum $\lambda$ in the light-cone gauge. This implies that a single massive amplitude receives UV contributions from two distinct classes of massless amplitudes: those with a transverse gauge boson and those with a Goldstone boson. Conversely, both the massless gauge-boson amplitude and the massless scalar amplitude unify into a single massive AHH amplitude in the infrared (IR), recovering the {\it IR unification}~\cite{Arkani-Hamed:2017jhn}. For instance, the massless $FFS$ and $FFV$ amplitudes unify into the massive $FFV$ amplitude, illustrating the on-shell realization of gauge-Yukawa unification through the Higgs mechanism.

Following the logic of massless-massive correspondence, we propose a systematic deformation procedure based on the scaling behavior associated with chirality flips. For matching amplitudes containing a gauge boson, a chirality flip must be applied to that boson. Conversely, for Goldstone-boson amplitudes, the chirality flip is applied to the conserved current, ensuring consistency with the Goldstone equivalence theorem. Implementing this procedure yields the complete set of three-point massless--massive matchings for the fundamental Standard Model sectors: $VVV$, $FFS$, $FFV$, and $VVS$. Once the leading-order matching is established, the sub-leading matching can be systematically derived by augmenting the leading-order result with additional on-shell Higgs splitting. We find that this on-shell Higgs splitting operation is equivalent to a helicity flip between distinct MHC states. Consequently, once the leading-order matching is known, all sub-leading matchings can be obtained straightforwardly, without introducing any new UV physics.

This paper is organized as follows. In section 2, all the 3-point massless and massive amplitudes are written and related by the spacetime symmetry, and the MHC states and MHC amplitudes are introduced for the high energy expansion. In section 3, the Massless-massive matching is obtained for the $FFS$ amplitudes, and the on-shell Higgsed insertion is introduced for the systematic sub-leading matching. In section 4, the primary MHC amplitude is vanishing due to the conserved current. We perform the leading FFV massless-massive matching by amplitude deformation, and the sub-leading matching is performed by the Higgs insertion. In section 5, all the 3-point amplitudes are decomposed into the MHC conserved current and the gauge boson or Goldstone boson, and then a systematic deformation procedure is proposed for all the 3-point amplitudes. In section 6, all the SM 3-point amplitudes are matched from the massless SM amplitudes, with the gauge structure identified. We conclude in section 7.

\section{Massless and Massive Amplitudes}

\subsection{Spinor under extended little group}

The spacetime symmetry of Minkowski space has the following group structure:
\begin{equation}
\text{Poincaré}=\text{Lorentz}\rtimes \text{Translation}.
\end{equation}
In quantum field theory, there are two distinct approaches to utilizing the representations of the spacetime symmetry group:
\begin{itemize}
\item \textit{Field perspective}: Fields transform as finite dimension representations of the Lorentz group. Conventionally, one employs the representations of  $SL(2,\mathbbm{C}) \sim SU(2)_l\times SU(2)_r$, which constitutes the universal cover of the Lorentz group $SO(3,1) \sim SL(2,\mathbbm{C})/\mathbb{Z}_2$. Since the Lorentz algebra possesses two Casimir operators, its irreducible representations are characterized by the quantum numbers $j_1$ and $j_2$, corresponding to the $SU(2)_l$ and $SU(2)_l$ subgroups, respectively. The representations of local fields are commonly denoted as $|j_1,j_2\rangle$, with the left-handed spinor field $\psi_L \equiv \psi_\alpha$ and right-handed spinor field $\psi_R \equiv \psi^{\dot\alpha}$ as the building blocks to construct the local fields with general spin.

\item \textit{Particle perspective}: Particles are identified with irreducible representations of the Poincaré group. The Poincaré algebra also features two Casimir operators. Following Wigner's classification, one distinguishes between massless representations $|p, h\rangle$ and massive representations $|\mathbf{p}, s\rangle$. Here, $p$ (for massless particles) and $\mathbf{p}$ (for massive particles) denote the eigenvalues of the translation generators. Meanwhile, helicity $h$ and spin $s$ represent the quantum numbers associated with the little groups $U(1)_{\text{LG}}$ and $SU(2)_{\text{LG}}$, respectively, which leave the corresponding momentum invariant.
\end{itemize}

Let us first take the first perspective, using a chiral fermion field as example. The Lagrangian can be written in most general way as
\begin{eqnarray} \label{eq:chiralfermion-Lag}
\mathcal{L} = i \psi^\dagger \bar{\sigma}^\mu \partial_\mu \psi - \frac{1}{2} m \psi \psi - \frac{1}{2} m^* \psi^\dagger \psi^\dagger,
\end{eqnarray}
where $m$ is a complex parameter with dimensions of mass.
Usually the phase of $m$ is irrelevant: if $m = |m| e^{i\alpha}$, we can set $\psi = e^{-i\alpha/2} \tilde{\psi}$ in Lagrangian. Then we get a typical Lagrangian for $\tilde{\psi}$ that is identical to \eqref{eq:chiralfermion-Lag}, but with $m$ replaced by $|m|$. Therefore, the $U(1)$ phase can be rotated away. In this description, chirality is not a good quantum number because the mass parameter would relate particle and anti-particle with different chirality.

For our purpose, let us promote the complex mass parameter as the {\it complex mass spurion}, which carries the chirality quantum number. In this way, the chirality, described by the phase of $m$, would become a good quantum number in the Lagrangian. When all the masses in the Lagrangian become dynamical spurion field, the Lagrangian would effectively describe a massless theory, although essentially it is a massive one after the spurion field becomes parameter. There are advantages for this spurion description: every field is in gauge eigenstate, instead of mass eigenstate, and the mass mixing becomes interaction, and the mass effect is described gy the chirality flip. Therefore, its connection to massless UV theory becomes transparent. All of these effects has been investigated in Ref.~\cite{Dreiner:2008tw}, although the chirality quantum number is not specified. In this work, we would extend this $U(1)$ phase further to the $ISO(2)$ to describe the chirality, and chirality flip. This can be equivalently described by the particle states in the second perspective.

In the second perspective, the on-shell method is utilized to describe particle. Let us consider here for the spinors the spinor wave functions that appear in the Fourier decomposition of the Weyl spinor fields
\begin{eqnarray}
\psi_\alpha &=& \int \frac{d^3 \mathbf{p}}{(2\pi)^4 2E_{\mathbf p}}  \left( 
\lambda_{\alpha}^{I} \, a_{I}\, e^{-i \mathbf{p} \cdot x} + {\lambda}_{\alpha I} \, a^{I\dagger}\, e^{i \mathbf{p} \cdot x}
\right),  \quad \ \ \text{left-handed field}\nonumber \\ 
\psi^{\dot\alpha} &=&  
\int \frac{d^3 \mathbf{p}}{(2\pi)^4 2E_{\mathbf p}}  \left( 
\tilde{\lambda}^{\dot \alpha I} \, a_{I}\, e^{-i \mathbf{p} \cdot x} + {\tilde{\lambda}}^{\dot\alpha}_{I} \, a^{I\dagger}\, e^{i \mathbf{p} \cdot x}
\right),   \quad \text{right-handed field}
\end{eqnarray}
summing over the spin index $I$, in which $I = 1, 2$ for massive particles, $I =1$ for massless particles. 
For massless particles, the index $I$ is usually neglected because the a phase can be absorbed due to the complex nature of spinor. 
By introducing the Levi-Civita tensor $\varepsilon_{IK}$ and $\varepsilon^{IK}$, we can raise and lower the little group indices of massive spinors as follows
\begin{equation}
\lambda_{\alpha I}=\varepsilon_{IK}\lambda^K_{\alpha},\quad
\tilde\lambda_{\dot\alpha}^I=\varepsilon^{IK}\tilde\lambda_{\dot\alpha K}.
\end{equation}
Note that the basic building blocks, $\lambda$ and $\tilde{\lambda}$, are the $c$-numbers and are not Grassmann numbers
\begin{eqnarray}
    \lambda^I_\alpha &:& \quad \text{left-handed spinor}, \\
    \tilde{\lambda}^{\dot\alpha I} &:& \quad \text{right-handed spinor}. 
\end{eqnarray}
Since they carry both the Lorentz indices $\alpha$ and $\dot{\alpha}$ and the spin indices $I$, they transform under both the Lorentz group
\begin{eqnarray}
   \lambda_{\alpha} \rightarrow \Lambda_\alpha{}^\beta \lambda_{\beta}, \quad \tilde{\lambda}_{\dot{\alpha}} \rightarrow \widetilde{\Lambda}_{\dot{\alpha}} {}^{\dot{\beta}}\tilde{\lambda}_{\dot{\beta}}, 
\end{eqnarray}
and the little group 
\begin{eqnarray}
\lambda^{I}_\alpha \rightarrow W_{J}^{I} \lambda^{J}_\alpha, \quad \tilde{\lambda}_{I} \rightarrow\left(W^{-1}\right)_{I}^{J} \tilde{\lambda}_{J}, 
\end{eqnarray} 
where $W$ is the little group $SU(2)$ matrix for massive particles, while $W$ is reduced to a phase for massless particles, and the index could be neglected due to the complex nature of spinor.

Using these basic spinors $\lambda$ and $\tilde{\lambda}$, let us define the scalar product in the spinor space. The momentum in the spinor space is defined as 
\begin{eqnarray}
    p_{\alpha \dot{\beta}} &\equiv& p_\mu (\sigma^\mu)_{\alpha \dot{\beta}} =  \begin{pmatrix}
p_0 + p_3 & p_1 - i p_2 \\
p_1 + i p_2 & p_0 - p_3
\end{pmatrix},  \\
\bar{p}^{\dot{\alpha} \beta} &\equiv& p_\mu (\bar{\sigma}^\mu)^{\dot{\alpha} \beta} 
= \begin{pmatrix}
p_0 + p_3 & p_1 - i p_2 \\
p_1 + i p_2 & p_0 - p_3
\end{pmatrix},
\end{eqnarray}
so that $\text{det} p= \text{det} \bar{p}= p^2$ is invariant under Lorentz transformations. For real momenta, $p$ and $\bar{p}$ are hermitian. In general, the momentum can be expressed as an inner product of two massive spinors
\begin{eqnarray}
\mathbf{p}_{\alpha\dot{\alpha}} \equiv
\lambda_{\alpha}^I \tilde{\lambda}_{\dot{\alpha} I}, \quad I = 1, 2,
\end{eqnarray}
where $I$ is the $SU(2)_{\text{LG}}$ index. 
While for massless particles, $\text{det} p= \text{det} \bar{p}= 0$, the $\mathbf{p}_{\alpha\dot{\alpha}} $ matrix is not full rank. Since every $2\times 2$ matrix of rank 1, can be written as the outer product of two vectors, the massless momentum can be written as a product of two Weyl spinors
\begin{eqnarray}
p_{\alpha \dot{\alpha}} = p_\mu \sigma^{\mu} _{\alpha \dot{\alpha}}=\lambda_{\alpha} \tilde{\lambda}_{\dot{\alpha}}. 
\end{eqnarray}

For massive particles, these basic spinors could also form another two scalar products, defined as
\begin{equation}
m \equiv \frac{1}{2}\lambda_{\alpha I}\lambda^{\alpha I},\quad \tilde{m} \equiv \frac{1}{2} \tilde{\lambda}_{\dot{\alpha}I}\tilde{\lambda}^{\dot{\alpha}I}.
\end{equation}
For real momentum, $p$ and $\bar{p}$ are hermitian, and thus although the scalar product is real, it has
\begin{eqnarray}
\mathbf{p}_{\alpha\dot{\alpha}} \equiv
\lambda_{\alpha}^I \tilde{\lambda}_{\dot{\alpha} I} 
\quad \rightarrow \quad  \tilde\lambda_{\dot \alpha} = \left(\lambda_{\alpha}\right)^{\dagger} . 
\end{eqnarray}
Thus in general both $m$ and $\tilde{m}$ are {\it complex} despite the real momentum.  
The three scalar products should satisfy the following on-shell condition for an on-shell particle
\begin{eqnarray}
    \mathbf{p}^2 = \mathbf{m}^2 \rightarrow \operatorname{det} \lambda \times \operatorname{det} \tilde{\lambda}= \frac{1}{2}\lambda^{I\alpha} \lambda^J_{\alpha} \tilde{\lambda}_{J\dot{\alpha}} \tilde{\lambda}_I^{\dot{\alpha}} = m\tilde{m}  = \mathbf{m}^2,
\end{eqnarray}
where $m$ and $\tilde{m}$ could be complex but $\mathbf{m}^2$ is real~\footnote{In literature~\cite{Arkani-Hamed:2017jhn}, usually $m$ and $\tilde{m}$ are taken to be real. In this work, although $m$ and $\tilde{m}$ are complex, the $\textbf{m}^2$ is real and thus the pole structure in the propagator does not modified. }. Note that the on-shell condition tells the spinors $\lambda$ and $\tilde\lambda$ automatically satisfy the equation of motion (EOM):
\begin{eqnarray}
\mathbf{p}_{\alpha \dot{\alpha}} \tilde{\lambda}^{\dot{\alpha} I} = \tilde{m} \lambda_{\alpha}^{I}, \quad \mathbf{p}_{\alpha \dot{\alpha}} \lambda^{\alpha I} = -m \tilde{\lambda}_{\dot{\alpha}}^{I}.
\end{eqnarray}

Now it is ready to discuss the one-particle state for massless and massive particles. For a massless particle, the two massless spinors $\lambda$ and $\tilde\lambda$ carry different helicity $-\frac12$ and $+\frac12$, which correspond to the quantum number of massless little group $U(1)_{\text{LG}}$.
Thus, the massless representation can be written in terms of these spinors
\begin{equation}
|p,h\rangle=
\begin{cases}
\lambda_{\alpha_1}...\lambda_{\alpha_{2h}},\;\;\;\quad h\ge 0, \\
\tilde{\lambda}_{\dot{\alpha}_1}... \tilde{\lambda}_{\dot{\alpha}_{(-2h)}},\quad h<0.
\end{cases}
\end{equation}
For a massive particle, the generator of the massive little group $SU(2)_{\text{LG}}$ can be defined as
\begin{equation}
J_{IK} \equiv i\left(\lambda_{\alpha (I}\frac{\partial}{\partial \lambda^{K)}_{\alpha}} + \tilde{\lambda}_{\dot{\alpha} (I}\frac{\partial}{\partial \tilde{\lambda}^{K)}_{\dot{\alpha}}}\right), 
\end{equation}
where the indices $(I K)$ are symmetrized. The quadratic Casimir $J^{IK}J_{IK}$ yields the spin $s$ of the representation. For notational simplicity, we restrict our consideration to massive spinor representations with upper $SU(2)_{\text{LG}}$ indices (e.g. $\lambda^I$, $\tilde{\lambda}^I$). Unlike the massless case, both $\lambda^I$ and $\tilde{\lambda}^I$ transform identically under the little group with quantum number $1/2$. Therefore, either massive spinor can be used to define the Poincaré representation, leading to multiple equivalent forms,
\begin{equation}
|\mathbf{p},s\rangle=
\begin{cases}
\lambda^{(I_1}_{\alpha_1}\lambda^{I_2}_{\alpha_2}\dots\lambda^{I_{2s})}_{\alpha_{2s}},\\
\tilde\lambda^{(I_1}_{\dot\alpha_1}\lambda^{I_2}_{\alpha_2}\dots\lambda^{I_{2s})}_{\alpha_{2s}},\\
\cdots,\\
\tilde\lambda^{(I_1}_{\dot\alpha_1}\tilde\lambda^{I_2}_{\dot{\alpha}_2}\dots\tilde\lambda^{I_{2s})}_{\dot{\alpha}_{2s}},
\end{cases}
\end{equation}
where the indices $(I_1 I_2 \dots I_{2s})$ are fully symmetrized. In Ref.~\cite{Arkani-Hamed:2017jhn}, the authors adopt the representation involving only $\lambda^I$, corresponding to the first case in the equation above.

Let us look at how a spin-$s$ spinor transform under the Poincar\'e and Lorentz representations. Define the chirality as $j_2-j_1$, in which the Lorentz representation is $|j_1,j_2\rangle$. 
We classify these chirality representations as follows,
\begin{equation} \begin{aligned}
&c = j_2-j_1<0:\quad \text{chiral representation},\\
&c = j_2-j_1=0:\quad \text{symmetric representation},\\
&c = j_2-j_1>0:\quad \text{anti-chiral representation}.\\
\end{aligned} \end{equation}
For spin-$\frac{1}{2}$ particles, there are chiral $c= -\frac12$ and anti-chiral $c = +\frac12$ representations. 
For massless particle, the Lorentz representation with given $j_1$ and $j_2$ has a unique chirality $j_2-j_1=h$, which one-to-one corresponds to the Poincaré representation. For example, the massless fermion has two states
\begin{equation}
\begin{tabular}{c|c|c}
\hline
Poincaré: $|p,h\rangle$ & Lorentz: $|j_1,j_2\rangle$ & massless fermion \\
\hline
$|p,-\frac{1}{2}\rangle$ & $|\frac{1}{2},0\rangle$ & $\lambda_{\alpha}$ \\
$|p,+\frac{1}{2}\rangle$ & $|0,\frac{1}{2}\rangle$ & $\tilde\lambda_{\dot\alpha}$ \\
\hline
\end{tabular}
\end{equation}
However, for massive particle, the Lorentz representation is not in a one-to-one correspondence to Poincaré representation,
\begin{equation}
\begin{tabular}{c|c|c}
\hline
Poincaré: $|\mathbf{p},s\rangle$ & Lorentz: $|j_1,j_2\rangle$ & massive fermion \\
\hline
\multirow{2}{*}{$|\mathbf{p},\frac{1}{2}\rangle$} & $|\frac{1}{2},0\rangle$ & $\lambda_{\alpha}^I$ \\
& $|0,\frac{1}{2}\rangle$ & $\tilde\lambda_{\dot\alpha}^I$ \\
\hline
\end{tabular}
\end{equation}
The different Lorentz representation are related by the equation of motion $\mathbf{p}_{\alpha \dot{\alpha}} \tilde{\lambda}^{\dot{\alpha} I} = \mathbf{m} \lambda_{\alpha}^{I}$, where $\mathbf m$ denotes the particle mass.

In this work, we study the correspondence between massless and massive amplitudes. To establish this correspondence, it is crucial to maintain a one-to-one mapping between the Lorentz and Poincaré representations for massive particles. This can be achieved by introducing an additional $U(1)$ transformation, whose generator is given by
\begin{equation} \begin{aligned}
D_- \equiv \frac{1}{2} \left( \tilde{\lambda}_{\dot{\alpha}}^I \frac{\partial}{\partial \tilde{\lambda}_{\dot{\alpha}}^I} - \lambda_{\alpha}^I \frac{\partial}{\partial \lambda_{\alpha}^I} \right).
\end{aligned} \end{equation}
Under this $U(1)$ transformation, the massive spinors transform as
\begin{equation}
\lambda_{\alpha}^{I}\rightarrow e^{-\frac{i}{2}\phi}\lambda_{\alpha}^{J}\;,\quad\tilde{\lambda}_{\dot{\alpha}I}\rightarrow e^{\frac{i}{2}\phi}\tilde{\lambda}_{\dot{\alpha}J}. 
\end{equation}
This $U(1)$ symmetry identify the quantum number {\it transversality} $t$. 
In this case, the mass parameters transform under this additional $U(1)$ symmetry as
\begin{equation}
m \equiv \frac{1}{2}\lambda_{\alpha I}\lambda^{\alpha I} \rightarrow e^{-i\phi}m,\quad \tilde{m} \equiv \frac{1}{2} \tilde{\lambda}_{\dot{\alpha}I}\tilde{\lambda}^{\dot{\alpha}I}\rightarrow e^{i\phi}\tilde{m},
\end{equation}
and thus they {\it must be complex}. Note that the massive momentum $\mathbf{p}$ remains invariant under this $U(1)$ transformation, so the massive little group extends to $U(2)_{\text{LG}}=SU(2)_{\text{LG}}\times U(1)$~ \footnote{There is another justification of this $U(2)$ from the conformal-helicity duality~\cite{Conde:2016izb}, in which the massive little group $U(2)_{\text{LG}}$ and the conformal group $SU(2,2)$ together form a dual pair group. }. 
This leads to an extended Poincaré representation $|\mathbf{p},s,t\rangle$, where the transversality quantum number $t$  corresponds to the $U(1)$ charge.
Similarly, the Lorentz group extends to $U(1) \times SO(3,1)$, with the associated representation now characterized by $[t,j_1,j_2]$, where $|t| \leq s$. When $t=j_2-j_1$, this establishes a one-to-one correspondence between the extended Poincaré and Lorentz representations
\begin{equation}
[t,j_1,j_2]\sim |\mathbf{p},s,t\rangle.
\end{equation}
Take the spin-$\frac12$ fermion as example. The two massive spinors are now distinguishable in both extended Poincaré and Lorentz representations
\begin{equation}
{\textrm{spin-}}\frac{1}{2}: 
\begin{cases}
    \tilde{\lambda}^I \ \sim \ [+\frac{1}{2}, 0, \frac{1}{2}] \ \sim \ |\mathbf p, \frac{1}{2}, +\frac{1}{2}\rangle,\\
    \lambda^I \ \sim \  [-\frac{1}{2}, \frac{1}{2}, 0] \ \sim \  |\mathbf p, \frac{1}{2}, -\frac{1}{2}\rangle.
\end{cases}
\label{eq:primaryhalf-transversality}
\end{equation}

There would be an additional benefit for such extension in the spinor space. 
The $U(1)$ generator $D_-$, together with $m$ and $\tilde{m}$, composes the Lie algebra of a $ISO(2)$ group:
\begin{equation}
    [D_-,m]=-m\;,\quad [D_-,\tilde{m}]=+\tilde{m}\;,\quad [m,\tilde{m}]=0.
\end{equation}
This extends the Poincar\'e symmetry to be $ISO(3,1) \times ISO(2)$, and thus the particle states are also extended to be $|\mathbf{p}, s, m,\tilde{m}\rangle$ with
\begin{eqnarray}
    P^2 |\mathbf{p}, s, m,\tilde{m}\rangle &=& \mathbf{p}^2 |\mathbf{p}, s, m,\tilde{m}\rangle, \\
W^2 |\mathbf{p}, s, m,\tilde{m}\rangle &=& -s(s+1) \mathbf{p}^2 |\mathbf{p}, s, m,\tilde{m}\rangle.
\end{eqnarray}
Here $m$ and $\tilde{m}$ are the eigenvalue of two generators in $ISO(2)$. 
The two generators $m$ and $\tilde{m}$ generate the wavefunctions of a state $|\mathbf{p},s,m,\tilde{m}\rangle$ excited by spin half field $\psi_\alpha$ and $\psi_{\dot\alpha}^\dagger$ with transversality $t$ as
\begin{eqnarray}
\label{eq:TableofWavefunctions}
\begin{matrix}
&&\lambda_{\alpha}^I&& \\
&m\lambda_{\alpha}^I& & \tilde{m}\lambda_{\alpha}^I& \\ m^2\lambda_{\alpha}^I && m\tilde{m}\lambda_{\alpha}^I && \tilde{m}^2\lambda_{\alpha}^I \\ \vdots && \vdots &&\vdots \end{matrix} \qquad,  \qquad \begin{matrix} && \tilde{\lambda}_{\dot\alpha}^I && \\
&m\tilde{\lambda}_{\dot\alpha}^I& & \tilde{m}\tilde{\lambda}_{\dot\alpha}^I &\\ m^2\tilde{\lambda}_{\dot\alpha}^I && m\tilde{m}\tilde{\lambda}_{\dot\alpha}^I && \tilde{m}^2\tilde{\lambda}_{\dot\alpha}^I \\ \vdots && \vdots&&\vdots
\end{matrix}
\end{eqnarray}
in which we define the primary and descendant representations of a particle state.
These are the full spaces of wavefunctions with all possible field configurations, generated by $m$ and $\tilde{m}$. The wavefunctions are categorized into one formula,
\begin{eqnarray}
\label{eq:allmstates2}
\begin{pmatrix}
    \dots &m&1&\tilde{m}&\dots
\end{pmatrix}\times
\begin{pmatrix}
    1\\
    m\tilde{m}\\
    \vdots
\end{pmatrix}\times
\begin{bmatrix}
    \begin{pmatrix}
    \lambda_{\alpha}^I\\
     \tilde{m}\lambda_{\alpha}^I
\end{pmatrix}&\begin{pmatrix}
    \tilde{\lambda}_{\dot\alpha}^I\\
    m\tilde{\lambda}_{\dot\alpha}^I
\end{pmatrix}
\end{bmatrix},
\end{eqnarray}
where the first matrix $(\cdots)$ accounts for the contributions from $m,\tilde{m}$ to traverse all the possible $t$, while the second matrix contains only the function of $(m\tilde{m})$, and the last matrix $[\cdots]$ contains non-trivial wave functions.
Note that the $ISO(2)$ symmetry is restricted due to the on-shell condition $\mathbf{p}^2 = \mathbf{m}^2 = m \tilde{m}$. 
The rotation element $D_-$ of $ISO(2)$ reduce to the $SO(2)\simeq U(1)$ with $m\to e^{-i\phi}m$, $\tilde{m}\to e^{-i\phi}\tilde{m}$. It would be clearer to define 
\begin{eqnarray}
|\mathbf{p},s,t\rangle=\int\frac{\text{d}\phi}{2\pi}|\mathbf{p},s,\mathbf{m}e^{-it\phi},\mathbf{m}e^{it\phi}\rangle,
\end{eqnarray}
which recovers the particle state with the $U(1)$ extension, but with more representations than just $t = j_2-j_1$:
\begin{eqnarray}
D_-|\mathbf{p},s, t\rangle= t|\mathbf{p},s, t\rangle.
\end{eqnarray}
From Eq.~\eqref{eq:allmstates2}, we note that the first and second matrices are trivial structures, and the third matrix
gives the particle state running over $t \le s$. 
For the chiral representation, the particle state takes the form
\begin{eqnarray}
\text{left-handed chiral $c = -\frac12$: } \quad \begin{cases}
    \lambda_{\alpha}^I & \text{for } \ \  t = -\frac12\,,\\
     \tilde{m}\lambda_{\alpha}^I & \text{for } \ \  t = +\frac12\,,
\end{cases}
\end{eqnarray}
containing both primary representation $\lambda_{\alpha}^I$ and descendant representation $\tilde{m}\lambda_{\alpha}^I$, and similarly the anti-chiral 
\begin{eqnarray}
\text{right-handed chiral $c = +\frac12$: } \quad \begin{cases}
    \tilde{\lambda}_{\dot\alpha}^I & \text{for } \ \  t = +\frac12\,,\\
     m\tilde{\lambda}_{\dot\alpha}^I & \text{for } \ \  t = +\frac12\,.
\end{cases}
\end{eqnarray}
They are related by the EOMs
\begin{equation} \begin{aligned} \label{eq:EOM}
t&=+\frac{1}{2}:& \mathbf{p}_{\alpha \dot{\alpha}} \tilde{\lambda}^{\dot{\alpha} I} &= \tilde{m} \lambda_{\alpha}^{I},\\
t&=-\frac{1}{2}:& \mathbf{p}_{\alpha \dot{\alpha}} \lambda^{\alpha I} &= -m \tilde{\lambda}_{\dot{\alpha}}^{I}.    
\end{aligned} \end{equation}
This shows that the particle states $\tilde\lambda^I$ and $\lambda^I$ on the left side have chirality opposite to $\tilde m\lambda^I$ and $m \tilde\lambda^I$ on the right side. Thus, while the chirality $j_2-j_1$ flips between the two sides of these equations due to the EOM, the transversality $t$ remains unchanged. The reason is that the spurion masses $m$ and $\tilde m$ carry transversality but do not affect chirality.

Finally let us summarize the particle states containing both the primary and descendant spinors
\begin{equation}
\begin{tabular}{c|c|c}
\hline
& $c = -\frac12 $ & $c = +\frac12 $ \\
\hline
primary & $\lambda^I_\alpha$ ($t = -\frac12$)& $\tilde\lambda^I_{\dot\alpha}$ ($t = \frac12$) \\
descendant & $\tilde{m}\lambda^I_\alpha$ ($t = +\frac12$) &  $m \tilde\lambda^I_{\dot\alpha}$ ($t = -\frac12$) \\
\hline
\end{tabular}
\end{equation}
which belongs to different transversality and chirality. Given $t \le s$, there is no 2nd descendant representation for the spin-$\frac{1}{2}$ particle. 
For spin-$1$ particle, the particle states are
\begin{eqnarray}
\begin{tabular}{c|ccc}
\hline
\diagbox{$\Delta$}{$c$} & $-1$ & $0$ & $1$ \\
\hline
primary & $\lambda^{(I}_{\alpha} \lambda^{J)}_{\beta}$ ($t= -1$) & $\lambda^{(I}_{\alpha} \tilde{\lambda}^{J)}_{\dot{\beta}}$ ($t= 0$) & $\tilde{\lambda}^{(I}_{\dot{\alpha}} \tilde{\lambda}^{J)}_{\dot{\beta}}$ ($t= 1$) \\
1st descendant & $\tilde{m} \lambda^{(I}_{\alpha} \lambda^{J)}_{\beta}$ ($t= 0$)  & \makecell{$m \lambda^{(I}_{\alpha} \tilde{\lambda}^{J)}_{\dot{\beta}}$ ($t=-1$), \\ $\tilde{m} \lambda^{(I}_{\alpha} \tilde{\lambda}^{J)}_{\dot{\beta}}$ ($t=+1$)} & $ \ m \tilde{\lambda}^{(I}_{\dot{\alpha}} \tilde{\lambda}^{J)}_{\dot{\beta}}$  ($t= 0$) \\
2nd descendant & $\tilde{m}^2 \lambda^{(I}_{\alpha} \lambda^{J)}_{\beta}$ ($t= 1$) & ${\color{gray} \mathbf{m}^2 \lambda^{(I}_{\alpha} \tilde{\lambda}^{J)}_{\dot{\beta}}}$ & $m^2 \tilde{\lambda}^{(I}_{\dot{\alpha}} \tilde{\lambda}^{J)}_{\dot{\beta}}$  ($t= -1$)\\
\hline
\end{tabular}
\end{eqnarray}
The one colored in gray, $\mathbf{m}^2 \lambda^{(I}_{\alpha} \tilde{\lambda}^{J)}_{\dot{\beta}}$, is considered equivalent to $\lambda^{(I}_{\alpha} \tilde{\lambda}^{J)}_{\dot{\beta}}$, because we mode out the mass shell $\mathbf{m}^2$ for massive representations. For spin-$1$ particle, since $t \le s$, there is no 3rd descendant representation.

\subsection{Massive helicity-chirality formalism}
\label{sec:MHC_state}

Before investigating scattering amplitudes, we should first analyze the high-energy (H.E.) behavior of massive particles in extended spacetime representations. 

It is useful to decompose the massive momenta into two light-like vectors as defined in Ref.~\cite{Dittmaier:1998nn},
\begin{eqnarray}
    k^{\mu} = k^{\flat;\mu} + \frac{m^2}{2\,(q \cdot k)} q^{\mu},
\end{eqnarray}
where the reference vector $q$ is used to define the spin axis $n$, orthogonal to $k$, in a covariant way
\begin{eqnarray}
    n_q^\mu = k^\mu- \frac{m^2 \, q^\mu}{q \cdot k} = k^{\flat;\mu} - \frac{m^2}{2\,(q \cdot k)} q^{\mu}.
\end{eqnarray}

In the spinor-helicity formalism, taking $n$ along the direction of momentum $\hat{n}$, we obtain  $n^\mu = (|k| , E\hat{n})$, orthogonal to $k^\mu = (E, |k| \hat{n})$. 
Further identifying two massless momenta $p^\mu = k^{\flat;\mu}$ and $\eta^\mu  = \frac{m^2}{2\,(q \cdot k)} q^{\mu}$, the massive momentum $\mathbf{p}$ can be decomposed into two massless ones
\begin{equation} \begin{aligned}
\mathbf{p}_{\alpha\dot\alpha}=p_{\alpha\dot\alpha}+\eta_{\alpha\dot\alpha}=\lambda_{\alpha}\tilde\lambda_{\dot\alpha}+\eta_{\alpha}\tilde\eta_{\dot\alpha},
\end{aligned} \end{equation}
with the quantization axis is along the direction of momentum $\hat{n}$ and 
\begin{equation} \begin{aligned}
\mathbf{n}_{\alpha\dot\alpha}=p_{\alpha\dot\alpha}-\eta_{\alpha\dot\alpha}=\lambda_{\alpha}\tilde\lambda_{\dot\alpha}-\eta_{\alpha}\tilde\eta_{\dot\alpha}.
\end{aligned} \end{equation}
In the $SU(2)_{\text{LG}}$ representation space, the massive spinor can be decomposed into two massless helicity spinors:
\begin{equation}
\begin{cases}
\lambda_{\alpha}^{I} = -\lambda_{\alpha} \zeta^{-I} +\eta_{\alpha} \zeta^{+I}, \\
\tilde{\lambda}_{\dot{\alpha}}^I = \tilde{\lambda}_{\dot{\alpha}} \zeta^{+I} +\tilde{\eta}_{\dot{\alpha}} \zeta^{-I},
\end{cases}
\label{eq:massless-decompose}    
\end{equation}
where $\zeta^{\pm I}$ are the basis vectors of the $SU(2)_{\text{LG}}$ representation space, satisfying $\epsilon_{IJ} \zeta^{+I} \zeta^{-J} = 1$. The additional $U(1)$ phase is hidden in the massless spinors. 
The EOMs are than reduced to the following form,
\begin{align}
p_{\alpha \dot{\alpha}} \tilde{\eta}^{\dot{\alpha}} &= \tilde{m} \lambda_{\alpha},\qquad
p_{\alpha \dot{\alpha}} \eta^{\alpha} = -m \tilde{\lambda}_{\dot{\alpha}}.\label{eq:EOM_LO}\\
\eta_{\alpha \dot{\alpha}} \tilde{\lambda}^{\dot{\alpha}} &= \tilde{m} \eta_{\alpha},\qquad
\eta_{\alpha \dot{\alpha}} \lambda^{\alpha} = -m \tilde{\eta}_{\dot{\alpha}}.\label{eq:EOM_NLO}
\end{align}

Let us look at the scaling behavior of the momenta and spinors. 
The two massless light-like vectors are 
\begin{eqnarray}
p^{\mu} =\frac{E+|p|}{2}(1, \hat{n}), \qquad \eta^{\mu} =\frac{E-|p|}{2}(1, -\hat{n}),
\end{eqnarray}
with $E$ and $|p|$, the energy and momentum values of the massive vector $\mathbf{p}$. 
The large-component spinors ($\lambda,\tilde\lambda$) and small-component spinors ($\eta,\tilde\eta$) exhibit different scaling behaviors
\begin{equation} \begin{aligned} \label{eq:spinor_scaling}
\lambda_{\alpha} &\sim \sqrt{E},& \tilde{\lambda}_{\dot{\alpha}} &\sim \sqrt{E}, \\
\eta_{\alpha} &\sim \frac{\mathbf{m}}{\sqrt{E}},& \tilde{\eta}_{\dot{\alpha}} &\sim \frac{\mathbf{m}}{\sqrt{E}},
\end{aligned} \end{equation}
where $\mathbf{m}$ is the absolute value of mass spurions $m$ and $\tilde m$.

This decomposition induces a reduction of the extended little group $U(2)_{\text{LG}}$ to its subgroup $U(1)_w\times U(1)_z$. In this formalism, the $U(2)_{\text{LG}}$ generators $J_{IK}$ and $D_-$ can be expressed as:
\begin{eqnarray}
D_- &=& \frac{1}{2} \left[ \Big( \tilde{\lambda}_{\dot{\alpha}} \frac{\partial}{\partial \tilde{\lambda}_{\dot{\alpha}}} + \tilde{\eta}_{\dot{\alpha}} \frac{\partial}{\partial \tilde{\eta}_{\dot{\alpha}}} \Big) - \Big( \lambda_{\alpha} \frac{\partial}{\partial \lambda_{\alpha}} + \eta_{\alpha} \frac{\partial}{\partial \eta_{\alpha}} \Big) \right], \\
J^{3} &=& \frac{i}{2}J_{IK}(\sigma_3)^K_L\varepsilon^{LI}= \frac{1}{2} \left[ \Big( \eta_{\alpha} \frac{\partial}{\partial\eta_{\alpha}} +\tilde{\lambda}_{\dot{\alpha}} \frac{\partial}{\partial \tilde{\lambda}_{\dot{\alpha}}} \Big) -\Big( \lambda_{\alpha} \frac{\partial}{\partial\lambda_{\alpha}} +\tilde{\eta}_{\dot{\alpha}} \frac{\partial}{\partial \tilde{\eta}_{\dot{\alpha}}} \Big) \right], \\
J^{+} &=& \frac{i}{2} J_{IK}(\sigma_1+i\sigma_2)^K_L\varepsilon^{LI}= -\eta_{\alpha} \frac{\partial}{\partial \lambda_{\alpha}} +\tilde{\lambda}_{\dot{\alpha}} \frac{\partial}{\partial \tilde{\eta}_{\dot{\alpha}}}, \label{eq:J+}\\
J^{-} &=& \frac{i}{2} J_{IK}(\sigma_1-i\sigma_2)^K_L\varepsilon^{LI}= -\lambda_{\alpha} \frac{\partial}{\partial\eta_{\alpha}} +\tilde{\eta}_{\dot{\alpha}} \frac{\partial}{\partial\tilde{\lambda}_{\dot{\alpha}}}, \label{eq:J-}
\end{eqnarray}
where $\sigma_1$, $\sigma_2$ and $\sigma_3$ are the Pauli matrices. Here, $U(1)_w$ is associated with the generator $J^3$ and the helicity quantum number $h$~\footnote{Although $U(1)_w$ and the massless little group $U(1)_\text{LG}$ both relate to helicity, they are different groups. Specifically, $\eta$ and $\tilde\eta$ are charged under $U(1)_w$, whereas they are irrelevant to $U(1)_{\text{LG}}$. This distinction leads us to use different notation.}, while $U(1)_z$ corresponds to the generator $D_-$ and the transversality $t$. Therefore, a spin-$s$ particle state is characterized by the triplet $(s,h,t)$.

This construction establishes the \textit{helicity-chirality formalism}, which enables simultaneous tracking of both helicity and chirality information in the H.E. expansion of massive particles. For the primary particle states, the chirality $c$ is equal to the transversality $t$. Let us first consider decomposition of the primary states. 
For spin-$\frac{1}{2}$ particles, both chiral and anti-chiral representations exist. Combined with the two possible helicity states $h=\pm\frac{1}{2}$, this yields four distinct states:
\begin{eqnarray}
\begin{array}{c|cc}
\hline
 & h=-\frac12 & h=+\frac12 \\
\hline
t=-\frac12 & -\lambda_{\alpha}  &  \eta_{\alpha}  \\
t=+\frac12 & \tilde{\eta}_{\dot{\alpha}}  & \tilde{\lambda}_{\dot{\alpha}} \\
\hline
\end{array}.
\label{eq:spin-half-states}
\end{eqnarray}
In contrast, massless fermions possess only a single helicity state: $\lambda$ for the chiral representation and $\tilde{\lambda}$ for the anti-chiral representation. For spin-1 particles, three chirality states exist. When combined with helicity, this results in nine possible states:
\begin{eqnarray}
\begin{array}{l|ccc}
\hline
& h=-1 & h=0 & h=+1 \\
\hline
t=-1 & \lambda_{\alpha_1}\lambda_{\alpha_2}   &  -\lambda_{\alpha_1}\eta_{\alpha_2} -\eta_{\alpha_1}\lambda_{\alpha_2} &  \eta_{\alpha_1}\eta_{\alpha_2}  \\
t=0 & -\lambda_{\alpha_1}\tilde{\eta}_{\dot{\alpha}_2}  & \eta_{\alpha_1}\tilde{\eta}_{\dot{\alpha}_2}-\lambda_{\alpha_1}\tilde{\lambda}_{\dot{\alpha}_2}  & \eta_{\alpha_1}\tilde{\lambda}_{\dot{\alpha}_2} \\
t=+1 & \tilde{\eta}_{\alpha_1} \tilde{\eta}_{\alpha_2} & \tilde{\lambda}_{\alpha_1}\tilde{\eta}_{\dot{\alpha}_2}+\tilde{\eta}_{\alpha_1}\tilde{\lambda}_{\dot{\alpha}_2} & \tilde{\lambda}_{\alpha_1} \tilde{\lambda}_{\alpha_2} \\
\hline
\end{array}. 
\end{eqnarray}

Among these states, some are composed exclusively of the large-component spinors $\lambda$ and $\tilde \lambda$, and thus provide the leading contributions. These particle states can be expressed diagrammatically as:
\begin{equation} \label{eq:primary_state}
\includegraphics[width=0.5\linewidth]{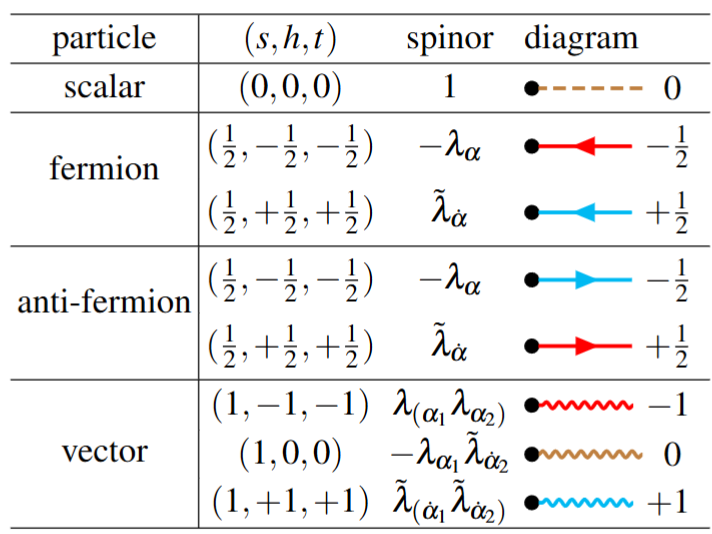}
\end{equation}
where the red, brown and blue colors label the three chirality states: chiral, symmetric and anti-chiral. The number on the right of a diagram labels the helicity. Unlike transversality, chirality is additionally tied to the distinction between particles and anti-particles.

The remaining particle states can be obtained by the $h$-flip operation, which transform a state $(s,h,t)$ into $(s,h \pm 1,t)$.  Due to the constraint $|h|\ge s$, this flip  can be applied at most $2s$ times. The operation is mediated by the ladder operators $J^{\pm}$, as defined in Eqs.~\eqref{eq:J-} and \eqref{eq:J+}. Diagrammatically, it is represented by adding a "$/$" over the diagram in Eq.~\eqref{eq:primary_state}. For example, acting $J^{\pm}$ on the primary fermionic state yields
\begin{equation} \begin{aligned} \label{eq:flipF1}
\begin{tikzpicture}[baseline=-0.1cm] \begin{feynhand}
\setlength{\feynhandarrowsize}{4pt}
\vertex [particle] (i1) at (1.5,0) {$+\frac12$}; 
\vertex (v1) at (0,0);
\draw[red,very thick] (0.7*1.5,0)--(0.6*1.5,0);
\draw[red,very thick,decoration={markings,mark=at position 0.56 with {\arrow{Triangle[length=4pt,width=4pt]}}},postaction={decorate}] (0.6*1.5,0)--(v1);
\draw (0.2*1.5,-0.08) -- (0.26*1.5,0.08);
\vertex[dot] (v1) at (0,0) {};
\end{feynhand} \end{tikzpicture}
&=J^+\circ 
\begin{tikzpicture}[baseline=-0.1cm] \begin{feynhand}
\setlength{\feynhandarrowsize}{4pt}
\vertex [particle] (i1) at (1.5,0) {$-\frac12$}; 
\vertex (v1) at (0,0);
\draw[red,very thick,decoration={markings,mark=at position 0.56 with {\arrow{Triangle[length=4pt,width=4pt]}}},postaction={decorate}] (0.7*1.5,0)--(v1);
\vertex[dot] (v1) at (0,0) {};
\end{feynhand} \end{tikzpicture}
=J^+\circ \lambda_\alpha=-\eta_\alpha,\\
\quad
\begin{tikzpicture}[baseline=-0.1cm] \begin{feynhand}
\setlength{\feynhandarrowsize}{4pt}
\vertex [particle] (i1) at (1.5,0) {$-\frac12$}; 
\vertex (v1) at (0,0);
\draw[cyan,very thick,decoration={markings,mark=at position 0.56 with {\arrow{Triangle[length=4pt,width=4pt]}}},postaction={decorate}] (0.7*1.5,0)--(v1);
\draw (0.2*1.5,-0.08) -- (0.26*1.5,0.08);
\vertex[dot] (v1) at (0,0) {};
\end{feynhand} \end{tikzpicture}
&=J^-\circ
\begin{tikzpicture}[baseline=-0.1cm] \begin{feynhand}
\setlength{\feynhandarrowsize}{4pt}
\vertex [particle] (i1) at (1.5,0) {$+\frac12$}; 
\vertex (v1) at (0,0);
\draw[cyan,very thick,decoration={markings,mark=at position 0.56 with {\arrow{Triangle[length=4pt,width=4pt]}}},postaction={decorate}] (0.7*1.5,0)--(v1);
\vertex[dot] (v1) at (0,0) {};
\end{feynhand} \end{tikzpicture}
=J^-\circ\tilde\lambda_{\dot\alpha}=\tilde\eta_{\dot\alpha}.
\end{aligned} \end{equation}

Then we consider the descendant particle states, in which transversality and chirality differ, $t \neq c$. 
After decomposition, these representations are equivalent to multiplying the massless spinor structures above by mass spurion $m$ or $\tilde{m}$. We refer to this operation as the $m$-flip, which not only increases the mass dimension of the state, but also transforms the primary state $(s,h,t)$ to a descendant one $(s,h,t\pm 1)$. In the spin-$\frac12$ case, we use the following diagrammatic representation:
\begin{equation} \begin{aligned}  \label{eq:flipF2}
\begin{tikzpicture}[baseline=-0.1cm] \begin{feynhand}
\setlength{\feynhandarrowsize}{4pt}
\vertex [particle] (i1) at (1.5,0) {$-\frac12$}; 
\vertex (v1) at (0,0);
\draw[cyan,very thick] (0.7*1.5,0)--(0.46*1.5,0);
\draw[red,very thick,decoration={markings,mark=at position 0.56 with {\arrow{Triangle[length=4pt,width=4pt]}}},postaction={decorate}] (0.46*1.5,0)--(v1);
\draw[very thick] plot[mark=x,mark size=2.5] coordinates {(0.46*1.5,0)};
\vertex[dot] (v1) at (0,0) {};
\end{feynhand} \end{tikzpicture}
&=\tilde m\circ 
\begin{tikzpicture}[baseline=-0.1cm] \begin{feynhand}
\setlength{\feynhandarrowsize}{4pt}
\vertex [particle] (i1) at (1.5,0) {$-\frac12$}; 
\vertex (v1) at (0,0);
\draw[red,very thick,decoration={markings,mark=at position 0.56 with {\arrow{Triangle[length=4pt,width=4pt]}}},postaction={decorate}] (0.7*1.5,0)--(v1);
\vertex[dot] (v1) at (0,0) {};
\end{feynhand} \end{tikzpicture}
=-\tilde{m} \lambda_\alpha,\\
\begin{tikzpicture}[baseline=-0.1cm] \begin{feynhand}
\setlength{\feynhandarrowsize}{4pt}
\vertex [particle] (i1) at (1.5,0) {$+\frac12$}; 
\vertex (v1) at (0,0);
\draw[red,very thick] (0.7*1.5,0)--(0.46*1.5,0);
\draw[cyan,very thick,decoration={markings,mark=at position 0.56 with {\arrow{Triangle[length=4pt,width=4pt]}}},postaction={decorate}] (0.46*1.5,0)--(v1);
\draw[very thick] plot[mark=x,mark size=2.5] coordinates {(0.46*1.5,0)};
\vertex[dot] (v1) at (0,0) {};
\end{feynhand} \end{tikzpicture}
&=m\circ
\begin{tikzpicture}[baseline=-0.1cm] \begin{feynhand}
\setlength{\feynhandarrowsize}{4pt}
\vertex [particle] (i1) at (1.5,0) {$+\frac12$}; 
\vertex (v1) at (0,0);
\draw[cyan,very thick,decoration={markings,mark=at position 0.56 with {\arrow{Triangle[length=4pt,width=4pt]}}},postaction={decorate}] (0.7*1.5,0)--(v1);
\vertex[dot] (v1) at (0,0) {};
\end{feynhand} \end{tikzpicture}
=m \tilde \lambda_{\dot\alpha},
\end{aligned} \end{equation}
where "$\boldsymbol{\times}$" denotes the action of $m$ or $\tilde m$. The colors on the two sides of "$\boldsymbol{\times}$" differ because these representations imply a chirality flip resulting from the application of the EOM. Using Eq.~\eqref{eq:EOM_LO}, we track the chirality information as follows,
\begin{equation}
\begin{tabular}{c|c|c}
\hline
$(s,h,t)$ & EOM & chirality flip \\
\hline
$(\frac12,-\frac12,+\frac12)$ & $\tilde{m} \lambda_{\alpha}\rightarrow p_{\alpha \dot{\alpha}} \tilde{\eta}^{\dot{\alpha}}$ & $-\frac12\rightarrow+\frac12$ \\
\hline
$(\frac12,+\frac12,-\frac12)$ & $m \tilde{\lambda}_{\dot{\alpha}}\rightarrow p_{\alpha \dot{\alpha}} \eta^{\alpha}$ & $-\frac12\rightarrow+\frac12$  \\
\hline
\end{tabular}
\end{equation}
We see that applying the EOM changes $\lambda$ ($\tilde\lambda$) to $\tilde\eta$ ($\eta$),  resulting in a chirality flip. Note that the quantum numbers $(s,h,t)$ remain unchanged under the EOM.

In this work, we focus on the massive representations that correspond directly to massless states. This would select some of the helicity-chirality states, satisfying the {\it helicity-transversality unification}, namely $h=t$.   
This can be understood in the H.E. limit. The $U(1)_w\times U(1)_z$ are unified into one $U(1)_w$, indicating that the massive spinors with both the transversality and the helicity quantum numbers should be unified into the massless helicity quantum number. We refer to these as the \textit{minimal helicity-chirality} (MHC) representations. The $h=t$ condition indicates a feature of the MHC states: every $\eta$ ($\tilde\eta$) is associated with an $\tilde{m}$ ($m$) to make sure that the transversality of the state matches the helicity of $\eta$ ($\tilde\eta$). Thus the basic building blocks of the MHC states are 
\begin{equation} \begin{aligned} \label{eq:spinor_scaling}
\lambda_{\alpha} &\sim \sqrt{E},& \tilde{\lambda}_{\dot{\alpha}} &\sim \sqrt{E}, \\
\tilde{m}\eta_{\alpha} &\sim \frac{\mathbf{m}^2}{\sqrt{E}},& m\tilde{\eta}_{\dot{\alpha}} &\sim \frac{\mathbf{m}^2}{\sqrt{E}},
\end{aligned} \end{equation}
with the scaling behaviors. 
Thus for the spin-$\frac12$ fermion, the MHC states are selected to be 
\begin{equation}
\begin{tabular}{c|cc}
\hline
& $c=-\frac12$ & $c=+\frac12$  \\
\hline
primary  & $\lambda_\alpha$ ($h=t=-\frac12$) &   $\tilde\lambda_{\dot\alpha}$ ($h=t=+\frac12$) \\
descendant &  $\tilde m\eta_\alpha$ ($h=t=+\frac12$)  &   $m\tilde\eta_{\dot\alpha}$ ($h=t=-\frac12$) \\
\hline
\end{tabular}
\end{equation}
Reorganized in terms of chirality, the MHC states for the fermion can be written as 
\begin{equation}
\begin{tabular}{c|cc}
\hline
& $h=t=-\frac12$ & $h=t=+\frac12$ \\
\hline
$c=-\frac12$ & $\lambda_\alpha$ & $\tilde m\eta_\alpha$  \\
$c=+\frac12$ & $m\tilde\eta_{\dot\alpha}$ & $\tilde\lambda_{\dot\alpha}$  \\
\hline
\end{tabular}
\end{equation}
where $c$ denotes the chirality of each state.
Similarly, the MHC particle states for the vector boson read
\begin{equation}\label{eq:vector_table_all}
\begin{tabular}{c|c|c|c} 
\hline
& $h=t=-1$ & $h=t=0$ & $h=t=+1$ \\
\hline
$c=-1$ & $\lambda_{\alpha_1}\lambda_{\alpha_2}$ & $\tilde m\lambda_{(\alpha_1}\eta_{\alpha_2)}$ & $\tilde m^2\eta_{\alpha_1}\eta_{\alpha_2}$ \\
\hline
$c=0$ & $m\tilde\eta_{\dot\alpha}\lambda_{\alpha}$ & \makecell{$\tilde\lambda_{\dot\alpha}\lambda_{\alpha}$\\$m\tilde m\tilde\eta_{\dot\alpha}\eta_{\alpha}$} & $\tilde m\tilde\lambda_{\dot\alpha}\eta_{\alpha}$ \\
\hline
$c=+1$ & $m^2\tilde\eta_{\dot\alpha_1}\tilde{\eta}_{\dot\alpha_2}$ & $m\tilde\lambda_{(\dot\alpha_1}\tilde\eta_{\dot\alpha_2)}$ & $\tilde\lambda_{\dot\alpha_1}\tilde\lambda_{\dot\alpha_2}$ \\
\hline
\end{tabular}
\end{equation}
It can be reorganized as primary, 1st descendant, 2nd descendant MHC states. In particular, let us look at the vector boson with chirality $c=0$. Note that the vector boson of helicity-0 are written as the combination of the primary and 2nd descendant states, which is identified as the Goldstone boson. For the vector boson with the $\pm1$ helicity, it should be the transverse gauge boson for $h=t = \pm 1$. In later section, we will discuss this in detail.

For a spin-$s$ particle, the particle states can be related by applying $J^{-}$ and $m$, or $J^+$ and $\tilde m$. In such representations, both $h$-flip and $m$-flip occur when we  apply $J^{-}$ and $m$, or $J^+$ and $\tilde m$, to the primary states. 
Therefore, the following composite ladder operator is utilized
\begin{equation}
m J^-,\quad \tilde m J^+
\end{equation}
to change the helicity $h$ and transversality $t$ simultaneously. 
These operators acting on the spinors gives the chirality flips
\begin{eqnarray}
\begin{array}{c|cc}
\hline
 &  c=-\frac12  &  c=+\frac12  \\
\hline
h=-\frac{1}{2} & \multirow{3}{*}{\begin{tikzpicture}
\node (A) at (0,1) {$\lambda_{\alpha}$};
\node (B) at (0,0) {$\tilde{m}\eta_{\alpha}$};
\draw [->] (-0.1,0.7) -- (-0.1,0.3);
\node at (-0.6,0.5) {$\tilde mJ^+$};
\draw [->] (0.1,0.3) -- (0.1,0.7);
\node at (0.6,0.5) {$m J^-$};
\end{tikzpicture}} & \multirow{3}{*}{\begin{tikzpicture}
\node (A) at (0,1) {$m\tilde{\eta}_{\dot{\alpha}}$};
\node (B) at (0,0) {$\tilde{\lambda}_{\dot{\alpha}}$};
\draw [->] (-0.1,0.7) -- (-0.1,0.3);
\node at (-0.6,0.5) {$\tilde m J^+$};
\draw [->] (0.1,0.3) -- (0.1,0.7);
\node at (0.6,0.5) {$m J^-$};
\end{tikzpicture}} \\
&& \\
h = +\frac{1}{2} && \\
\hline
\end{array} \label{eq:helicityflip} 
\end{eqnarray}
We have introduced a slash and a cross to denote respectively the action of spurion masses $m,\tilde m$ and ladder operators $J^\pm$: 
\begin{equation}
\left\{\begin{aligned}
\begin{tikzpicture}[baseline=-0.1cm] \begin{feynhand}
\setlength{\feynhandarrowsize}{4pt}
\vertex [particle] (i1) at (1.5,0) {$+$}; 
\vertex (v1) at (0,0);
\draw[cyan,very thick] (0.7*1.5,0)--(0.6*1.5,0);
\draw[cyan,very thick,decoration={markings,mark=at position 0.56 with {\arrow{Triangle[length=4pt,width=4pt]}}},postaction={decorate}] (0.6*1.5,0)--(v1);
\draw (0.2*1.5,-0.08) -- (0.26*1.5,0.08);
\vertex[dot] (v1) at (0,0) {};
\end{feynhand} \end{tikzpicture}&=J^-\circ\tilde\lambda_{\dot\alpha}\\
\begin{tikzpicture}[baseline=-0.1cm] \begin{feynhand}
\setlength{\feynhandarrowsize}{4pt}
\vertex [particle] (i1) at (1.5,0) {$-$}; 
\vertex (v1) at (0,0);
\draw[red,very thick] (0.7*1.5,0)--(0.46*1.5,0);
\draw[cyan,very thick,decoration={markings,mark=at position 0.56 with {\arrow{Triangle[length=4pt,width=4pt]}}},postaction={decorate}] (0.46*1.5,0)--(v1);
\draw[very thick] plot[mark=x,mark size=2.5] coordinates {(0.46*1.5,0)};
\vertex[dot] (v1) at (0,0) {};
\end{feynhand} \end{tikzpicture}&=m\circ\tilde\lambda_{\dot\alpha}
\end{aligned}\right.
\quad\Rightarrow\quad
\begin{tikzpicture}[baseline=-0.1cm] \begin{feynhand}
\setlength{\feynhandarrowsize}{4pt}
\vertex [particle] (i1) at (1.5,0) {$-$}; 
\vertex (v1) at (0,0);
\draw[red,very thick] (0.7*1.5,0)--(0.46*1.5,0);
\draw[cyan,very thick,decoration={markings,mark=at position 0.45 with {\arrow{Triangle[length=4pt,width=4pt]}}},postaction={decorate}] (0.46*1.5,0)--(v1);
\draw[very thick] plot[mark=x,mark size=2.5] coordinates {(0.46*1.5,0)};
\draw (0.1*1.5,-0.08) -- (0.16*1.5,+0.08);
\vertex[dot] (v1) at (0,0) {};
\end{feynhand} \end{tikzpicture}=mJ^-\circ\tilde\lambda_{\dot\alpha}
\end{equation}

These two types of operations are interchangeable because the commutators satisfy $[m,J^{\pm}]=[\tilde{m},J^{\pm}]=0$. 
For example, applying either $mJ^-$ or $J^- m$ to the state $(s,h,t)=(\frac12,+\frac12,+\frac12)$ yields the same result
\begin{equation}  \label{eq:flipF3}
\begin{tikzpicture}[baseline=-0.1cm] \begin{feynhand}
\setlength{\feynhandarrowsize}{4pt}
\vertex [particle] (i1) at (1.5,0) {$-\frac12$}; 
\vertex (v1) at (0,0);
\draw[red,very thick] (0.7*1.5,0)--(0.46*1.5,0);
\draw[cyan,very thick,decoration={markings,mark=at position 0.45 with {\arrow{Triangle[length=4pt,width=4pt]}}},postaction={decorate}] (0.46*1.5,0)--(v1);
\draw[very thick] plot[mark=x,mark size=2.5] coordinates {(0.46*1.5,0)};
\draw (0.1*1.5,-0.08) -- (0.16*1.5,+0.08);
\vertex[dot] (v1) at (0,0) {};
\end{feynhand} \end{tikzpicture}=
\begin{tikzpicture}[baseline=-0.1cm] \begin{feynhand}
\setlength{\feynhandarrowsize}{4pt}
\vertex [particle] (i1) at (1.5,0) {$-\frac12$}; 
\vertex (v1) at (0,0);
\draw[cyan,very thick] (v1)--(0.2*1.5,0);
\draw[red,very thick,decoration={markings,mark=at position 0.8 with {\arrow{Triangle[length=4pt,width=4pt]}}},postaction={decorate}] (0.7*1.5,0)--(0.2*1.5,0);
\draw[very thick] plot[mark=x,mark size=2.5] coordinates {(0.2*1.5,0)};
\draw (0.5*1.5,-0.08) -- (0.56*1.5,+0.08);
\vertex[dot] (v1) at (0,0) {};
\end{feynhand} \end{tikzpicture}=m \tilde\eta_{\dot\alpha},
\end{equation}
Although both operations lead to the same physical state, we adopt the convention of using the first type of operation (i.e. $mJ^-$ or $\tilde m J^+$)  to represent states with both helicity and chirality flips.

To obtain the MHC states for a spin-$s$ particle, we could begin with a primary MHC state composed solely of $\lambda$ and $\tilde\lambda$, and then apply $m J^-$ or $\tilde m J^+$ to generate descendant MHC states
\begin{equation}
\text{primary}\xrightarrow{mJ^-,\tilde mJ^+}
\text{1st descendant}\xrightarrow{mJ^-,\tilde mJ^+}
\cdots
\end{equation}
After applying these operations $2s$ times, we obtain all MHC states. 
In the following we will use the following diagram to denote the chirality flip. For the primary state with $c=h=+\tfrac12$, acting with ${m} J^-$ yields the corresponding descendant state with $h=-\tfrac12$,
\begin{equation} \begin{aligned}
\Ampone{1.5}{+}{\fer{cyan}{i1}{v1}}=\tilde\lambda_{\dot\alpha}
\quad&\xrightarrow{ m J^-}\quad
\Ampone{1.5}{-}{\ferflip{1.5}{0}{red}{cyan}}= m \tilde\eta_{\dot\alpha} \,,
\end{aligned} \end{equation}
Further action of $ m J^-$ annihilates this state, consistent with the condition $|h|\le s$. For the primary state with $c=h=-\tfrac12$, we can act with $ {\tilde m}  J^+$  and obtain a similar descendant state with $h=+\tfrac12$
\begin{equation} \begin{aligned}
\Ampone{1.5}{+}{\fer{red}{i1}{v1}}= \lambda_{ \alpha}
\quad&\xrightarrow{ {\tilde m}  J^+}\quad
\Ampone{1.5}{-}{\ferflip{1.5}{0}{cyan}{red}}= {\tilde m}  \eta_{ \alpha} \,.
\end{aligned} \end{equation}

For spins $s=0,\frac{1}{2},1$, the resulting states can be drawn as follow
\begin{equation}  \label{eq:MHC_particle}
\includegraphics[width=\linewidth]{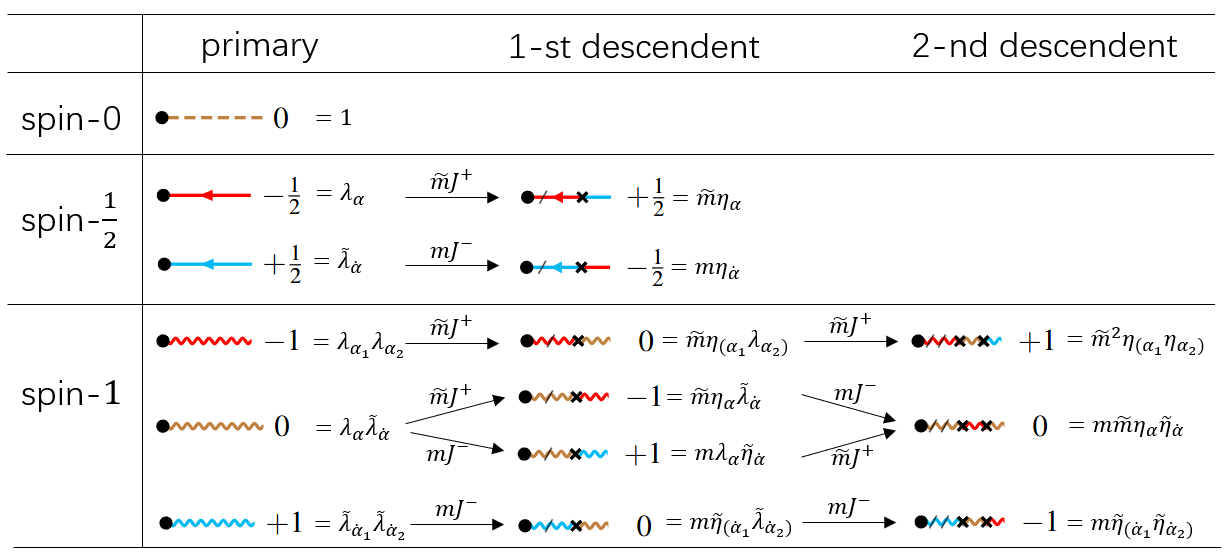}
\end{equation}
Note that $m\tilde{m} = \tilde{m}m$, so the diagram for the 2nd descendant representations with $t=h=0$ is not unique
\begin{equation}
\Ampone{1.5}{0}{\bosflipflip{1.5}{0}{brown}{red}{brown}}=
\Ampone{1.5}{0}{\bosflipflip{1.5}{0}{brown}{cyan}{brown}}
\end{equation}
In the following, we will choose the first diagram as the standard representation of the state $\tilde m m\eta_{\alpha}\tilde\eta_{\dot\alpha}$.

\subsection{Three-point massless amplitudes}
\label{sec:3pt_massless}


For massless particle, the 3-pt amplitude can be fully determined by the spacetime symmetry (i.e. the representation of $U(1)_{\text{LG}}$) and the locality. 

We first consider the little group representations. A 3-pt massless amplitude is characterized by the helicity of all particles, $(h_1,h_2,h_3)$. Note that the 3-particle kinematics shows that $\lambda_1\propto \lambda_2\propto\lambda_3$ or $\tilde\lambda_1\propto \tilde\lambda_2\propto\tilde\lambda_3$, so 3-pt amplitudes are constructed exclusively from either $\lambda$s or $\tilde{\lambda}$s. 
The little group scaling determines the form of the amplitude,
\begin{eqnarray} \label{eq:scaling3pt}
\mathcal{A}(1^{h_1},2^{h_2},3^{h_3}) = \mathcal{G}\times
\begin{cases}
\langle12\rangle^{h_1+h_2-h_3} \langle23\rangle^{h_2+h_3-h_1} \langle31\rangle^{h_3+h_1-h_2}, & h_1+h_2+h_3<0, \\
[12]^{-h_1-h_2+h_3} [23]^{-h_2-h_3+h_1} [31]^{-h_3-h_1+h_2}, & h_1+h_2+h_3\ge 0,
\end{cases} \label{eq:3-ptmassless}
\end{eqnarray}
where $\mathcal{G}$ is a coupling constant, $|i\rangle_{\alpha} \equiv {\lambda_i}_{\alpha}$ and $|i]^{\dot{\alpha}} \equiv {\tilde{\lambda}_i}^{\dot{\alpha}} $. 

The scattering amplitude must also satisfy the locality. To examine this, we perform a dimensional analysis on the massless amplitude. The massless spinors $\lambda$ and $\tilde{\lambda}$ have mass dimension $\frac{1}{2}$, so the Lorentz structure contributes a total mass dimension of $h_1 + h_2 + h_3$, which equals to the total helicity of the amplitude. On the other hand, the mass dimension of a 3-point amplitude must be $1$. This imposes a constraint on the coupling constant $\mathcal{G}$,
\begin{equation}
\text{dim }\mathcal{G}=1-|h_1+h_2+h_3|.
\end{equation}
Depending on the value of the mass dimension, there are three cases:
\begin{itemize}
\item $\text{dim }\mathcal{G}>0$: A positive mass dimension implies a non-local interaction involving $1/\partial^2$ in the Lagrangian, violating locality.
\item $\text{dim }\mathcal{G}=0$: A dimensionless coupling constant corresponds to a fundamental interaction in renormalizable theories.
\item $\text{dim }\mathcal{G}<0$: A coupling constant with negative mass dimension can be expressed as a power of $1/\Lambda$, indicating an effective interaction with a cutoff scale $\Lambda$.
\end{itemize}
Only the second and third cases satisfy locality. In this work, we focus on the second case. Therefore, the total helicity of the 3-pt massless amplitude is constrained to
\begin{equation}
h_1+h_2+h_3=\pm 1.
\end{equation}
This determines the total helicity for the 3-pt amplitude in the Standard Model. Therefore, the amplitudes such as $(h_1,h_2,h_3)=(+\frac12,-\frac12,0)$, $(+1,-1,0)$ or $(0,0,0)$ are forbidden.

Under this constraint, there are four types of 3-pt amplitudes. For each type, we list two typical structures corresponding to total helicity $h_1+h_2+h_3=+1$ and $-1$, 
\begin{eqnarray}
\begin{array}{c|cc|cc}
\hline
\multirow{2}{*}{\text{External particles}} & \multicolumn{2}{c|}{h_1+h_2+h_3=1} & \multicolumn{2}{c}{h_1+h_2+h_3=-1} \\
\cline{2-5}
& (h_1,h_2,h_3) & \text{Lorentz structures} & (h_1,h_2,h_3) & \text{Lorentz structures} \\
\hline
FFV & (-\frac12,+\frac12,+1) & \frac{[23]^2}{[12]} & (-\frac12,+\frac12,-1) & \frac{\langle13\rangle^2}{\langle12\rangle} \\
\hline
FFS & (+\frac12,+\frac12,0) & [12] & (-\frac12,-\frac12,0) & \langle12\rangle \\
\hline
VVV & (+1,+1,-1) & \frac{[12]^3}{[23][31]} & (-1,-1,+1) & \frac{\langle12\rangle^3}{\langle23\rangle\langle31\rangle} \\
\hline
VSS & (+1,0,0) & \frac{[12][31]}{[23]} & (-1,0,0) & \frac{\langle12\rangle\langle31\rangle}{\langle23\rangle} \\
\hline
\end{array}
\end{eqnarray}
where $S$, $F$ and $V$ represent scalar boson, fermion and vector boson.

\subsection{Three-point massive amplitudes}

In the $U(1)$ extended spacetime symmetry, massive amplitudes are characterized by the quantum numbers of $U(2)_{\text{LG}}$, namely the spin $s$ and transversality $t$. We therefore refer to such amplitudes as spin-transversality (ST) amplitudes. These amplitudes can be used to systematically construct all the possible 3-point massive Lorentz structures and inherently incorporate the effects of the EOM.

Various transversality categories $[t_1,t_2,t_3]$ exist under a given spin category $(s_1,s_2,s_3)$, because the only constraint on the transversalities is $|t_i|\leq s_i$. To relate these different representations, we introduce the following operators,
\begin{equation} \label{eq:GeneratorT}
T^{+}_{\alpha \dot{\alpha}}\equiv \sum_{i}\tilde{\lambda}_{\dot{\alpha},i}^{I}\frac{\partial}{\partial \lambda_i^{\alpha I}},\quad
T^{-}_{\alpha \dot{\alpha}}\equiv \lambda_{\alpha,i}^{I}\frac{\partial}{\partial \tilde{\lambda}_i^{\dot{\alpha}I}}. 
\end{equation}
where $i$ labels the particle. Although these operators are not directly associated with the extended Lorentz group $U(1)\times SO(3,1)$, they can be interpreted as generators of a larger spacetime group $SO(5,1)$~\cite{Ni:2024yrr, Ni:2025xkg}. When acting on the representation $[t_1,t_2,t_3]$, these generators yield
\begin{equation} \begin{aligned}
T^{+}\circ [t_1,t_2,t_3]&=[t_1+1,t_2,t_3]+[t_1,t_2+1,t_3]+[t_1,t_2,t_3+1], \\
T^{-}\circ [t_1,t_2,t_3]&=[t_1-1,t_2,t_3]+[t_1,t_2-1,t_3]+[t_1,t_2,t_3-1]. 
\end{aligned} \end{equation}
The resulting primary ST representations still satisfy the constraint $|t_i|\le s_i$ for every particle-$i$. For example, the action on a spin-$\frac{1}{2}$ representation is given by
\begin{equation} \begin{aligned}
T^+_{\alpha\dot{\alpha}}\circ \lambda_{\beta,i}^I &= \epsilon_{\alpha\beta}\tilde{\lambda}_{\dot{\alpha},i}^I,& T^-_{\alpha\dot{\alpha}}\circ \lambda_{\beta,i}^I&=0\,, \\
T^-_{\alpha\dot{\alpha}}\circ \tilde{\lambda}_{\dot{\beta},i}^I &= \epsilon_{\dot{\alpha}\dot{\beta}}\lambda_{\alpha,i}^I,& T^+_{\alpha\dot{\alpha}}\circ \tilde{\lambda}_{\dot{\beta},i}^I &= 0\,, 
\end{aligned} \end{equation}
where $\delta^{\alpha}_\beta$ and $\delta^{\dot \alpha}_{\dot \beta}$ correspond to trivial representations with $t_i = 0$.

This formalism provides a systematic method for generating all the massive 3-pt structures. The total transversality $t\equiv t_1+t_2+t_3$ serves as a weight for the amplitude. One can begin with the highest-weight amplitude, which has maximal transversality $t=s_1+s_2+s_3$, and then successively apply the $T^-$ generators to derive representations with lower weights, until we find the lowest-weight represesentation with transversality $t=-s_1-s_2-s_3$. Finally, the effects of the EOM are incorporated through appropriate factors of $m$ and $\tilde{m}$.

In the Standard Model, the highest-weight amplitude takes the form
\begin{equation} \label{eq:HW_3pt}
[\mathbf{12}]^{s_1+s_2-s_3}[\mathbf{23}]^{s_2+s_3-s_1}[\mathbf{31}]^{s_3+s_1-s_2},
\end{equation}
which is originated from the fact that the unbolded highest weight amplitude, which is also a primary MHC amplitude, has the form $
[12]^{y_{12}}[23]^{y_{23}}[31]^{y_{31}}$, and thus has the same scaling behavior as the the massless little group scaling
\begin{equation} \begin{aligned}
y_{12}=s_1+s_2-s_3=\Delta s_3,\\ 
y_{23}=s_2+s_3-s_1=\Delta s_1,\\
y_{31}=s_3+s_1-s_2=\Delta s_2. 
\end{aligned} \end{equation} 
Since the scattering amplitude must be a Lorentz scalar, we require not the vector-valued operator $T^-$, but its scalar product:
\begin{equation}
T^-\cdot T^-=\sum_i (T^-_i)^2+2(T^-_1\cdot T^-_2+T^-_2\cdot T^-_3+T^-_1\cdot T^-_3),
\end{equation}
where $T_i^-$ acts only on the spinor structure of particle $i$. Although the full $SO(5,1)$ spacetime symmetry is not present, each term in this expansion can individually generate lower-weight amplitudes. Applying $(T^-_i)^2$ to Eq.~\eqref{eq:HW_3pt} yields zero, so only the cross terms $T^-_i \cdot T^-_j$ (with $i \ne j$) contribute.

For the massive $FFS$ amplitude, the highest-weight amplitude is $[\mathbf{12}]$, which has total transversality $+1$. Only $T_1^-\cdot T_2^-$ gives a non-vanishing result, 
\begin{equation} \begin{aligned}
T_1^-\cdot T_2^- \circ[\mathbf{12}]=\langle\mathbf{12}\rangle,
\end{aligned} \end{equation}
producing the lowest-weight amplitude with total transversality $-1$. Therefore, the $FFS$ amplitude has two Lorentz structures $[\mathbf{12}]$ and $\langle\mathbf{12}\rangle$. 

We then study the $FFV$ amplitude to involve the nontrivial operation of $T^-_1\cdot T^-_3$ and $T^-_2\cdot T^-_3$. Its Lorentz structures are organized as follows,
\begin{equation}
\begin{tikzpicture}[baseline=-0.1cm]
\path
(0,0) node(C1) [rectangle] {$[\mathbf{23}][\mathbf{31}]$}
(3,1) node(C2) [rectangle] {$[\mathbf{23}]\langle\mathbf{31}\rangle$}
(3,-1) node(C3) [rectangle] {$\langle\mathbf{23}\rangle[\mathbf{31}]$}
(6,0) node(C4) [rectangle] {$\langle\mathbf{23}\rangle\langle\mathbf{31}\rangle$};
\draw [thick,->] (C1)--node[above]{\small $T_1^-\cdot T_3^-$} (C2);
\draw [thick,->] (C1)--node[above]{\small $T_2^-\cdot T_3^-$} (C3);
\draw [thick,->] (C2)--node[above]{\small $T_2^-\cdot T_3^-$} (C4);
\draw [thick,->] (C3)--node[above]{\small $T_1^-\cdot T_3^-$} (C4);
\end{tikzpicture}
\end{equation}
In the $FFV$ case, the highest-weight amplitude is $[\mathbf{23}][\mathbf{31}]$, and the lowest-weight amplitude is $\langle\mathbf{23}\rangle\langle\mathbf{31}\rangle$. The two intermediate structures, $[\mathbf{23}]\langle\mathbf{31}\rangle$ and $\langle\mathbf{23}\rangle[\mathbf{31}]$, both have total transversality $0$. 

In the Standard Model, the most complicated amplitude is $VVV$, whose highest-weight amplitude is $[\mathbf{12}][\mathbf{23}][\mathbf{31}]$. In this case, all three products $T^-_1\cdot T^-_2$, $T^-_2\cdot T^-_3$ and $T^-_3\cdot T^-_1$ will generate massive structures with lower weight:
\begin{equation}
\begin{tikzpicture}[baseline=-0.1cm]
\path
(0,0) node(C1) [rectangle] {$[\mathbf{12}][\mathbf{23}][\mathbf{31}]$}
(3.5,2) node(C2) [rectangle] {$[\mathbf{12}][\mathbf{23}]\langle\mathbf{31}\rangle$}
(3.5,0) node(C3) [rectangle] {$[\mathbf{12}]\langle\mathbf{23}\rangle[\mathbf{31}]$}
(3.5,-2) node(C4) [rectangle] {$\langle\mathbf{12}\rangle[\mathbf{23}][\mathbf{31}]$}
(7,2) node(C5) [rectangle] {$[\mathbf{12}]\langle\mathbf{23}\rangle\langle\mathbf{31}\rangle$}
(7,0) node(C6) [rectangle] {$\langle\mathbf{12}\rangle[\mathbf{23}]\langle\mathbf{31}\rangle$}
(7,-2) node(C7) [rectangle] {$\langle\mathbf{12}\rangle\langle\mathbf{23}\rangle[\mathbf{31}]$}
(10.5,0) node(C8) [rectangle] {$\langle\mathbf{12}\rangle\langle\mathbf{23}\rangle\langle\mathbf{31}\rangle$};
\draw [thick,->] (C1)--node[left, xshift=-10pt, yshift=-4pt]{\small $T_1^-\cdot T_3^-$} (C2);
\draw [thick,->] (C1)--node[above]{\small $T_2^-\cdot T_3^-$} (C3);
\draw [thick,->] (C1)--node[above, xshift=-6pt, yshift=8pt]{\small $T_1^-\cdot T_2^-$} (C4);
\draw [thick,->] (C2)--node[above]{\small $T_2^-\cdot T_3^-$} (C5);
\draw [thick,->] (C2)--node[above, xshift=-6pt, yshift=8pt]{\small $T_1^-\cdot T_2^-$} (C6);
\draw [thick,->] (C3)--node[left, xshift=-10pt, yshift=-4pt]{\small $T_1^-\cdot T_3^-$} (C5);
\draw [thick,->] (C3)--node[above, xshift=-6pt, yshift=8pt]{\small $T_1^-\cdot T_2^-$} (C7);
\draw [thick,->] (C4)--node[left, xshift=-10pt, yshift=-4pt]{\small $T_1^-\cdot T_3^-$} (C6);
\draw [thick,->] (C4)--node[above]{\small $T_2^-\cdot T_3^-$} (C7);
\draw [thick,->] (C5)--node[above, xshift=-6pt, yshift=8pt]{\small $T_1^-\cdot T_2^-$} (C8);
\draw [thick,->] (C6)--node[above]{\small $T_2^-\cdot T_3^-$} (C8);
\draw [thick,->] (C7)--node[below right]{\small $T_1^-\cdot T_3^-$} (C8);
\end{tikzpicture}
\end{equation}
This method allows us to systematically enumerate all the independent three-point Lorentz structures for particles of spin $0$, $\frac{1}{2}$, and $1$:
\begin{eqnarray}
\begin{array}{c|c|c}
\hline
\mbox{external particles} & \mbox{SM structures} & \mbox{EFT structures} \\
\hline
(\mathbf{1}^{1/2}, \mathbf{2}^{1/2}, \mathbf{3}^{1}) &  \langle\mathbf{23}\rangle [\mathbf{31}], [\mathbf{23}] \langle\mathbf{31}\rangle & [\mathbf{23}][\mathbf{31}],\langle\mathbf{23}\rangle \langle\mathbf{31}\rangle \\
\hline
(\mathbf{1}^{1/2}, \mathbf{2}^{1/2}, \mathbf{3}^{0}) & [\mathbf{12}], \langle\mathbf{12}\rangle &  \\
\hline
(\mathbf{1}^{1}, \mathbf{2}^{1}, \mathbf{3}^{1}) & \makecell{ [\mathbf{12}] [\mathbf{23}] \langle\mathbf{31}\rangle, [\mathbf{12}] \langle\mathbf{23}\rangle [\mathbf{31}], \langle\mathbf{12}\rangle [\mathbf{23}] [\mathbf{31}],  \\  \langle\mathbf{12}\rangle [\mathbf{23}] \langle\mathbf{31}\rangle,[\mathbf{12}]\langle\mathbf{23}\rangle \langle\mathbf{31}\rangle,\langle\mathbf{12}\rangle \langle\mathbf{23}\rangle [\mathbf{31}] }
& \makecell{[\mathbf{12}] [\mathbf{23}] [\mathbf{31}],\\\langle\mathbf{12}\rangle \langle\mathbf{23}\rangle \langle\mathbf{31}\rangle} \\
\hline
(\mathbf{1}^{1}, \mathbf{2}^{1}, \mathbf{3}^{0}) &  \langle\mathbf{12}\rangle [\mathbf{12}] & \langle\mathbf{12}\rangle^2,[\mathbf{12}]^2 \\
\hline
(\mathbf{1}^{0}, \mathbf{2}^{0}, \mathbf{3}^{0}) & 1 & \\
\hline
\end{array}\label{eq:3-ptmassive}
\end{eqnarray}
Here, we distinguish between structures that appear in the Standard Model and those that only arise in effective field theory.

Finally, each Lorentz structure is multiplied by spurion masses $m_i$ or $\tilde{m}_i$ to form the full ST amplitude, subject to the constraint $|t_i| \le s_i$.
For the $FFS$ amplitude, we have
\begin{equation} \begin{aligned}
\mathbf{M}(\mathbf{1}^{\frac12},\mathbf{2}^{\frac12},\mathbf{1}^0)
=&(\mathbf{c}_1+\mathbf{c}_2 \tilde{m}_1+\mathbf{c}_3 \tilde{m}_2+\mathbf{c}_4 \tilde{m}_1 \tilde{m}_2)\langle \mathbf{12}\rangle+(\mathbf{c}_5+\mathbf{c}_6 m_1+\mathbf{c}_7 m_2+\mathbf{c}_8 m_1 m_2)[\mathbf{12}].
\end{aligned} \end{equation}
where $\mathbf{c}_i$ are the coefficients. For the $FFV$ amplitude, we consider only the structures $\langle\mathbf{23}\rangle [\mathbf{31}]$ and $[\mathbf{23}] \langle\mathbf{31}\rangle$, yielding
\begin{eqnarray}
\mathbf{M}(\mathbf{1}^{\frac12},\mathbf{2}^{\frac12},\mathbf{1}^1)
&=&
(\mathbf c_1+ \mathbf c_2 \tilde{m}_1  + \mathbf c_3 \tilde{m}_2 + \mathbf c_4 \tilde{m}_3 + \mathbf c_5 \tilde{m}_1 \tilde{m}_2 + \mathbf c_6 \tilde{m}_1 \tilde{m}_3 + \mathbf c_7 \tilde{m}_2 \tilde{m}_3 \nonumber \\
&&+ \mathbf c_8 \tilde{m}_3^2 + \mathbf c_9 \tilde{m}_1 \tilde{m}_2 \tilde{m}_3+ \mathbf c_{10} \tilde{m}_1 \tilde{m}_3^2+ \mathbf c_{11} \tilde{m}_2 \tilde{m}_3^2+ \mathbf c_{12} \tilde{m}_1 \tilde{m}_2 \tilde{m}_3^2 ) \langle\mathbf{23}\rangle\langle\mathbf{31}\rangle \\
&+& (\mathbf c_{13}+ \mathbf c_{14} \tilde{m}_1  + \mathbf c_{15} m_2 + \mathbf c_{16} \tilde{m}_3 + \mathbf c_{17} m_3 + \mathbf c_{18} \tilde{m}_1 m_2 + \mathbf c_{19} \tilde{m}_1 \tilde{m}_3 \nonumber \\
&&+ \mathbf c_{20} \tilde{m}_1 m_3 + \mathbf c_{21} m_2 \tilde{m}_3 + \mathbf c_{22} m_2 m_3 + \mathbf c_{23} \tilde{m}_1 m_2 \tilde{m}_3 + \mathbf c_{24} \tilde{m}_1 m_2 m_3) [\mathbf{23}]\langle\mathbf{31}\rangle \nonumber\\
&+& (\mathbf c_{25}+  \mathbf c_{26} m_1 + \mathbf c_{27} \tilde{m}_2 + \mathbf c_{28} \tilde{m}_3 + \mathbf c_{29} m_3 + \mathbf c_{30} m_1 \tilde{m}_2 + \mathbf c_{31} m_1 \tilde{m}_3 \nonumber \\
&&+ \mathbf c_{32} m_1 m_3 + \mathbf c_{33} \tilde{m}_2 \tilde{m}_3 + \mathbf c_{34} \tilde{m}_2 m_3 + \mathbf c_{35} m_1 \tilde{m}_2 \tilde{m}_3 + \mathbf c_{36} m_1 \tilde{m}_2 m_3) \langle\mathbf{23}\rangle[\mathbf{31}] \nonumber\\
&+& (\mathbf c_{37} + \mathbf c_{38} m_1 + \mathbf c_{39} m_2 + \mathbf c_{40} m_3 + \mathbf c_{41} m_1 m_2 + \mathbf c_{42} m_1 m_3 + \mathbf c_{43} m_2 m_3 \nonumber\\
&&+ \mathbf c_{44} m_3^2 + \mathbf c_{45} m_1 m_2 m_3+ \mathbf c_{46} m_1 m_3^2+ \mathbf c_{47} m_2 m_3^2+ \mathbf c_{48} m_1 m_2 m_3^2 )[\mathbf{23}][\mathbf{31}]. \nonumber
\end{eqnarray}
where the mass dependence on $m_3^2$ or $\tilde m_3^2$ due to double chirality flip for massive vector boson.

\subsection{High-energy expansion and power counting analysis}
\label{sec:expansion}

We now consider the H.E. expansion of the 3-pt massive amplitude. In this expansion, the ST amplitude $\mathbf{M}$ with spin $(s_1,s_2,s_3)$ can be decomposed into a sum of helicity-chirality amplitude $\mathcal{M}^{\mathcal{H},\mathcal{T}}$, where $\mathcal{H}=(h_1,h_2,h_3)$ denotes the helicity and $\mathcal{T}=(t_1,t_2,t_3)$ the transversality. Using Eq.~\eqref{eq:massless-decompose}, this decomposition takes the form
\begin{equation}
\mathbf{M}(\mathbf{1}^{s_1}, \mathbf{2}^{s_2}, \mathbf{3}^{s_3})
\to\sum_{\mathcal{H}} \prod_{i=1}^3 \left((\zeta^+_i)^{s_i+h_i}(\zeta^-_i)^{s_i-h_i}\right) \mathcal{M}^{\mathcal{H},\mathcal{T}},
\end{equation}
To illustrate this expansion, let us consdier the $FFS$ amplitude as an example. The massive structure $\langle\mathbf{12}\rangle$ expands as
\begin{equation}
\langle\mathbf{12}\rangle\to 
\xi_1^- \xi_2^- \langle12\rangle 
- \xi_1^+ \xi_2^- \langle\eta_12\rangle 
- \xi_1^- \xi_2^+ \langle1\eta_2\rangle 
+ \xi_1^+ \xi_2^+ \langle\eta_1\eta_2\rangle.
\end{equation}
Each term corresponds to $\mathcal{M}^{\mathcal{H},\mathcal{T}}$ with the same transversality but different helicity,
\begin{equation}
\begin{tabular}{c|c|c|c|c}
\hline
$\mathcal{T}$ & \multicolumn{4}{c}{$(-\frac12,-\frac12,0)$} \\
\hline
$\mathcal{H}$ & $(-\frac12,-\frac12,0)$ & $(+\frac12,-\frac12,0)$ & $(-\frac12,+\frac12,0)$ & $(+\frac12,+\frac12,0)$ \\
\hline
$\mathcal{M}^{\mathcal{H},\mathcal{T}}$ & $\langle12\rangle $  & $-\langle\eta_12\rangle$ & $-\langle1\eta_2\rangle $ & $\langle\eta_1\eta_2\rangle $ \\
\hline
\end{tabular}
\end{equation}
Similarly, we examine the $FFV$ massive amplitude $\langle\mathbf{13}\rangle [\mathbf{32}]$, which can be decomposed into $2\times 2 \times 3 = 12$ helicity categories. The expansion takes the form
\begin{equation} \begin{aligned}
\mathcal{M}^{(\frac{1}{2},\frac{1}{2},1)}
=&\zeta^+_1\zeta^+_2(\zeta^+_3)^2\mathcal{M}^{(+\frac{1}{2},+\frac{1}{2},+1)}+\zeta^+_1\zeta^-_2(\zeta^+_3)^2\mathcal{M}^{(+\frac{1}{2},-\frac{1}{2},+1)}\\
&+\zeta^-_1\zeta^+_2(\zeta^+_3)^2\mathcal{M}^{(-\frac{1}{2},+\frac{1}{2},+1)}+\zeta^-_1\zeta^-_2(\zeta^+_3)^2\mathcal{M}^{(-\frac{1}{2},-\frac{1}{2},+1)}\\
&+\zeta^+_1\zeta^+_2(\zeta^+_3\zeta^-_3)\mathcal{M}^{(+\frac{1}{2},+\frac{1}{2},0)}+\zeta^+_1\zeta^-_2(\zeta^+_3\zeta^-_3)\mathcal{M}^{(+\frac{1}{2},-\frac{1}{2},0)}\\
&+\zeta^-_1\zeta^+_2(\zeta^+_3\zeta^-_3)\mathcal{M}^{(-\frac{1}{2},+\frac{1}{2},0)}+\zeta^-_1\zeta^-_2(\zeta^+_3\zeta^-_3)\mathcal{M}^{(-\frac{1}{2},-\frac{1}{2},0)}\\
&+\zeta^+_1\zeta^+_2(\zeta^-_3)^2\mathcal{M}^{(+\frac{1}{2},+\frac{1}{2},-1)}+\zeta^+_1\zeta^-_2(\zeta^-_3)^2\mathcal{M}^{(+\frac{1}{2},-\frac{1}{2},-1)}\\
&+\zeta^-_1\zeta^+_2(\zeta^-_3)^2\mathcal{M}^{(-\frac{1}{2},+\frac{1}{2},-1)}+\zeta^-_1\zeta^-_2(\zeta^-_3)^2\mathcal{M}^{(-\frac{1}{2},-\frac{1}{2},-1)}.
\end{aligned} \end{equation}

Terms with different transversality are associated with the mass spurions $m$ or $\tilde m$. The contributions containing the structure $\langle\mathbf{12}\rangle$ and mass spurions to the amplitude expansion are given by
\begin{eqnarray}
\mathbf{M}(\mathbf{1}^{\frac12},\mathbf{2}^{\frac12},\mathbf{1}^0)\rightarrow &&
\mathbf{c}_1 \left(\langle12\rangle - \langle\eta_12\rangle - \langle1\eta_2\rangle + \langle\eta_1\eta_2\rangle \right) \nonumber \\
&& + \mathbf{c}_2 m_1\left(\langle12\rangle - \langle\eta_12\rangle - \langle1\eta_2\rangle + \langle\eta_1\eta_2\rangle \right) \nonumber \\
&& + \mathbf{c}_3 m_2\left(\langle12\rangle - \langle\eta_12\rangle - \langle1\eta_2\rangle + \langle\eta_1\eta_2\rangle \right) \nonumber \\
&& + \mathbf{c}_4 m_1 m_2\left(\langle12\rangle - \langle\eta_12\rangle - \langle1\eta_2\rangle + \langle\eta_1\eta_2\rangle \right). \label{eq:FFS_expand}
\end{eqnarray}
These amplitudes can be represented diagrammatically. For the scalar and fermion states, we employ eqs.~\eqref{eq:primary_state}, \eqref{eq:flipF1}, \eqref{eq:flipF2}, and \eqref{eq:flipF3} to represent individual particles. The diagram for an anti-fermion is assigned the opposite color. This yields the following representation
\begin{align}
\includegraphics[width=0.7\linewidth]{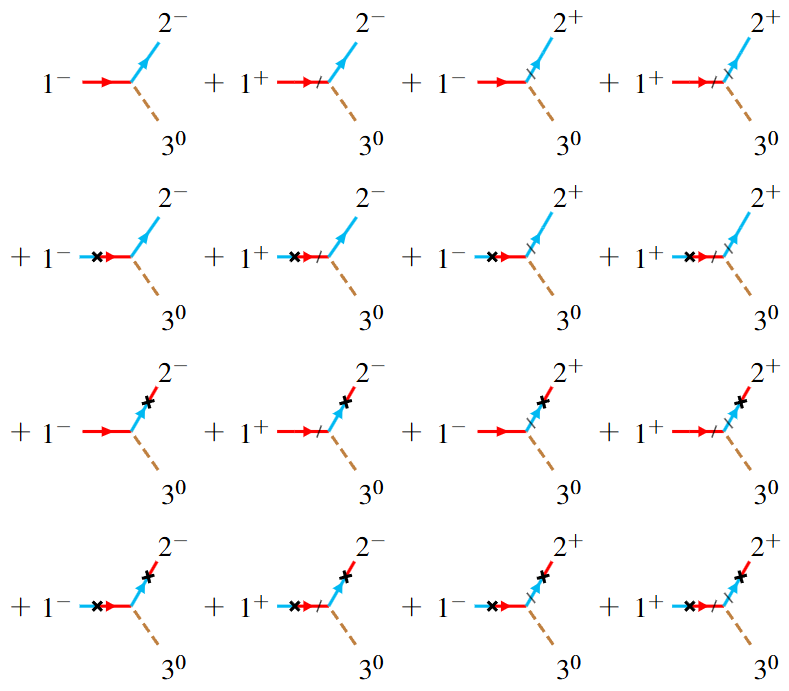}
\label{eq:FFS_expand_diagram2}
\end{align}

To connect with the massless amplitudes, we select terms that satisfy chirality-helicity unification to define the MHC amplitude:
\begin{equation}
\mathcal{M}=\sum_{\mathcal{T}=\mathcal{H}}\mathcal{M}^{\mathcal{H},\mathcal{T}},
\quad 
\end{equation}

The H.E. behavior of these massive amplitudes can be analyzed through the power counting in the energy scale $E$ and mass $\mathbf m$. In the MHC amplitude, the building blocks scale as:
\begin{equation} \begin{aligned}
\lambda_{\alpha},\tilde{\lambda}_{\dot{\alpha}} &\sim \sqrt{E},&  & \tilde m\eta_{\alpha},m\tilde{\eta}_{\dot{\alpha}} \sim \frac{\mathbf{m}^2}{\sqrt{E}}. \\
\end{aligned} \end{equation}
Replacing $\lambda$ with $\tilde{m}\eta$ therefore changes the power counting as
\begin{equation} \begin{aligned}
\sqrt{E}\to\sqrt{E}\times\frac{\mathbf m^2}{E}\,.
\end{aligned} \end{equation}
Thus, in the H.E. regime $E\gg \mathbf{m}$, a 3-pt MHC amplitude exhibits the following expansion,
\begin{equation} \label{eq:expand3pt}
\mathcal{M}=\sum_{l=0}^{2s} [\mathcal{M}]_l\sim \sum_{l=0}^{2s} \underbrace{\frac{1}{\mathbf m^{s-1+l}}}_{\text{coefficient}}\times \underbrace{E^s\left(\frac{\mathbf m^2}{E}\right)^l}_{\text{spinor structure}},
\end{equation}
where $[\mathcal{M}]_l$ denotes the component of order $l$ in the H.E. expansion, and $s = s_1 + s_2 + s_3$ is the total spin. The number of terms is determined by $s$. By analogy with the classification of particle states, we refer to these structures as primary and descendant MHC amplitudes.

To ensure consistent power counting, the MHC amplitude must satisfy the following momentum conservation conditions
\begin{equation}
\begin{cases}
\sum_i p_{i,\alpha\dot{\alpha}}=0, \\
\sum_i \eta_{i,\alpha\dot{\alpha}}=0,
\end{cases}
\label{eq:momenta-conserv}
\end{equation}
For a 3-pt MHC amplitude, this implies $\lambda_1\propto \lambda_2\propto\lambda_3$ or $\tilde\lambda_1\propto \tilde\lambda_2\propto\tilde\lambda_3$, mirroring the kinematics of massless 3-point amplitudes. Note that the primary amplitude involving a symmetric vector boson vanishes due to this condition. In the $FFV$ case, the structures $[23]\langle 31\rangle$ and $\langle 23\rangle[31]$ vanish, while $[23][31]$ and $\langle 23\rangle\langle 31\rangle$ remain. Therefore, the former correspond to the vanishing amplitudes with conserved currents, satisfying the Ward identity.

Take the $FFS$ as example. We take the diagonal elements in Eq.~\eqref{eq:FFS_expand}, corresponding to the diagonal diagrams in Eq.~\eqref{eq:FFS_expand_diagram2}. Incorporating contributions from the Lorentz structure $[\mathbf{12}]$, we obtain the full MHC amplitude
\begin{equation} \begin{aligned}
\mathcal{M}(FFS)= 
&\mathbf{c}_1 \langle12\rangle  - \mathbf{c}_2 \tilde{m}_1 \langle\eta_12\rangle - \mathbf{c}_3 \tilde{m}_2 \langle1\eta_2\rangle + \mathbf{c}_4 \tilde{m}_1 \tilde{m}_2 \langle\eta_1\eta_2\rangle\\
&+\mathbf{c}_5 [12] + \mathbf{c}_6 {m}_1[\eta_12] + \mathbf{c}_7 {m}_2 [1\eta_2] + \mathbf{c}_8 {m}_1 {m}_2 [\eta_1\eta_2].
\end{aligned} \end{equation}
Let us reorganize the MHC amplitudes as the primary and descendant ones. Using Eq.~\eqref{eq:spinor_scaling}, the $FFS$ amplitude ($s=1$) consists of three terms
\begin{equation}
\mathcal{M}(FFS)=[\mathcal{M}]_{0}+[\mathcal{M}]_{1}+[\mathcal{M}]_{2},
\end{equation}
which corresponds to 
\begin{equation} \begin{aligned}
&\text{primary}:& [\mathcal{M}]_{0}&=\mathbf{c}_1\langle12\rangle+\mathbf{c}_5[12],\\
&\text{1st descendant}:& [\mathcal{M}]_{1}&=- \mathbf{c}_2 \tilde{m}_1 \langle\eta_12\rangle - \mathbf{c}_3 \tilde{m}_2 \langle1\eta_2\rangle + \mathbf{c}_6 {m}_1[\eta_12] + \mathbf{c}_7 {m}_2 [1\eta_2],\\
&\text{2nd descendant}:& [\mathcal{M}]_{2}&=\mathbf{c}_4 \tilde{m}_1 \tilde{m}_2 \langle\eta_1\eta_2\rangle+\mathbf{c}_8 {m}_1 {m}_2 [\eta_1\eta_2].
\end{aligned} \end{equation}

We next consider the $FFV$ amplitude. The Lorentz structures $\langle\mathbf{23}\rangle\langle\mathbf{31}\rangle$ and $[\mathbf{23}][\mathbf{31}]$ correspond to non-renormalizable dipole interactions. In this work, we focus on the two structures $[\mathbf{23}]\langle\mathbf{31}\rangle$ and $\langle\mathbf{23}\rangle[\mathbf{31}]$. The resulting $FFV$ MHC amplitudes are given by
\begin{align}
&\mathcal{M}(FFV) \nonumber \\
=& \mathbf c_{13} [23]\langle31\rangle - \frac{\mathbf c_{13}}{\mathbf {m}_3^2}{m}_3\tilde {m}_3 [2\eta_3]\langle\eta_31\rangle - \mathbf c_{14} \tilde {m}_1 [23]\langle3 \eta_1\rangle + \frac{\mathbf c_{14}}{\mathbf {m}_3^2}\tilde {m}_1 {m}_3\tilde {m}_3  [2\eta_3]\langle\eta_3\eta_1\rangle  \nonumber \\
&+ \mathbf c_{15} {m}_2 [\eta_23]\langle31\rangle - \frac{\mathbf c_{15}}{\mathbf {m}_3^2}{m}_2 {m}_3\tilde {m}_3  [\eta_2\eta_3]\langle\eta_31\rangle - \mathbf c_{16}\tilde {m}_3 [23]\langle\eta_31\rangle + \mathbf c_{17} {m}_3 [2\eta_3]\langle31\rangle \nonumber \\
&- \mathbf c_{18} \tilde {m}_1 {m}_2 [\eta_23]\langle3\eta_1\rangle + \frac{\mathbf c_{18}}{\mathbf {m}_3^2}\tilde {m}_1 {m}_2 {m}_3\tilde {m}_3 [\eta_2\eta_3]\langle\eta_3\eta_1\rangle + \mathbf c_{19} \tilde {m}_1 \tilde {m}_3 [23]\langle\eta_3\eta_1\rangle \nonumber \\
&- \mathbf c_{20} \tilde {m}_1 {m}_3 [2\eta_3]\langle3\eta_1\rangle
- \mathbf c_{21} {m}_2 \tilde {m}_3 [\eta_23]\langle\eta_31\rangle + \mathbf c_{22} {m}_2 {m}_3 [\eta_2\eta_3]\langle31\rangle \nonumber \\
&+ \mathbf c_{23} \tilde {m}_1 {m}_2\tilde {m}_3 [\eta_23]\langle\eta_3\eta_1\rangle - \mathbf c_{24} \tilde {m}_1 {m}_2 {m}_3[\eta_2\eta_3]\langle3\eta_1\rangle 
+ \mathbf c_{25} \langle23\rangle[31] \nonumber \\
&- \frac{\mathbf c_{25}}{\mathbf {m}_3^2}{m}_3\tilde {m}_3 \langle2\eta_3\rangle[\eta_31]  + \mathbf c_{26} {m}_1 \langle23\rangle[3 \eta_1] - \frac{\mathbf c_{26}}{\mathbf {m}_3^2}{m}_1 {m}_3\tilde {m}_3  \langle2\eta_3\rangle[\eta_3\eta_1] \nonumber \\
&- \mathbf c_{27} \tilde {m}_2 \langle\eta_23\rangle[31] + \frac{\mathbf c_{27}}{\mathbf {m}_3^2}\tilde {m}_2 {m}_3\tilde {m}_3  \langle\eta_2\eta_3\rangle[\eta_31] - \mathbf c_{28} \tilde {m}_3 \langle2\eta_3\rangle[31] + \mathbf c_{29} {m}_3 \langle23\rangle[\eta_31] \nonumber \\
&- \mathbf c_{30} {m}_1 \tilde {m}_2 \langle\eta_23\rangle[3\eta_1] + \frac{\mathbf c_{30}}{\mathbf {m}_3^2}{m}_1\tilde {m}_2 {m}_3\tilde {m}_3  \langle\eta_2\eta_3\rangle[\eta_3\eta_1] - \mathbf c_{31} {m}_1\tilde {m}_3 \langle2\eta_3\rangle[3\eta_1] \nonumber \\
&+ \mathbf c_{32} {m}_1 {m}_3 \langle23\rangle[\eta_3\eta_1] 
+ \mathbf c_{33} \tilde {m}_2 \tilde {m}_3 \langle\eta_2\eta_3\rangle[31] - \mathbf c_{34} \tilde {m}_2 {m}_3 \langle\eta_23\rangle[\eta_31] \nonumber \\
&+ \mathbf c_{35} {m}_1\tilde {m}_2 \tilde {m}_3\langle\eta_2\eta_3\rangle[3\eta_1] - \mathbf c_{36} {m}_1\tilde {m}_2 {m}_3 \langle\eta_23\rangle[\eta_3\eta_1]. \label{eq:FFV_expand}
\end{align}
Reorganizing in terms of the power counting, we obtain the MHC amplitudes
\begin{eqnarray}
    \mbox{primary: }& \langle13\rangle [32];  \nonumber\\
\mbox{1st descendant: }& -\tilde{m}_1 \langle\eta_13\rangle [32], {m}_2 \langle13\rangle [3\eta_2], -{m}_3 \langle1\eta_3\rangle [32], \tilde{m}_3 \langle13\rangle [\eta_32];  \nonumber\\
\mbox{2nd descendant: }& -\tilde{m}_1 {m}_2 \langle\eta_13\rangle [3\eta_2], -\tilde{m}_1 {m}_3 \langle\eta_13\rangle [\eta_32], \tilde{m}_1 \tilde{m}_3 \langle\eta_1\eta_3\rangle [32], {m}_2 {m}_3 \langle13\rangle [\eta_3\eta_2], -{m}_2 \tilde{m}_3 \langle1\eta_3\rangle [3\eta_2],  \nonumber\\
& {\color{gray} -{m}_3\tilde{m}_3 \langle1\eta_3\rangle [\eta_32]};  \nonumber\\
\mbox{3rd descendant: }& -\tilde{m}_1 {m}_2 {m}_3 \langle\eta_13\rangle [\eta_3\eta_2], \tilde{m}_1 {m}_2 \tilde{m}_3 \langle\eta_1\eta_3\rangle [3\eta_2],  \nonumber\\
&{\color{gray} \tilde{m}_1 {m}_3 \tilde{m}_3 \langle\eta_1\eta_3\rangle [\eta_32], -{m}_2 {m}_3 \tilde{m}_3 \langle1\eta_3\rangle [\eta_3\eta_2]};  \nonumber\\
{\color{gray} \mbox{4th descendant: }}&{\color{gray} \tilde{m}_1 {m}_2 {m}_3 \tilde{m}_3 \langle\eta_1\eta_3\rangle [\eta_3\eta_2]}. 
\end{eqnarray}
From above, we note that the primary MHC amplitudes are obtained by just simply unbolding the ST amplitudes.  In the $FFV$ case, this term vanishes as a consequence of momentum conservation. Terms containing one factor of $m\eta$ constitute the first descendant amplitudes, those with two $m\eta$ factors form the second descendant amplitudes, and the term containing only $m\eta$ factors corresponds to the fourth descendant amplitude. Once the primary MHC amplitudes are determined, all sub-leading amplitudes can be systematically generated through the application of ladder operators. 
Several comments are in order:

To identify the UV origins of these massive amplitudes, we match their scaling behavior to massless structures. An
$n$-point massless structure scales as $E^{4-n}$, which aligns with Eq.~\eqref{eq:expand3pt} when $n = 4-s+l$. Thus, for a given order $l$, we establish the correspondence
\begin{equation}
[\mathcal{M}]_l\sim (4-s+l)\text{-pt massless structures}.
\end{equation}
This mapping allows us to systematically derive the UV origins for all 3-point massive amplitudes in the Standard Model. The results are summarized in table~\ref{tab:PC3pt}.
\begin{table}[htbp]
\centering
\caption{The correspondence between MHC amplitudes and massless structures. Here, "pri." and "des." denote primary and descendant MHC amplitude.}
\begin{tabular}{c|c|ccccccc}
\hline
total spin & massive & \multicolumn{7}{c}{$n$-pt massless structure} \\
\cline{3-9}
$s$ & amplitude & 1 & 2 & 3 & 4 & 5 & 6 & 7 \\
\hline
$0$ & $SSS$ & \mbox{-} & \mbox{-} & \mbox{-} & pri. & \mbox{-} & \mbox{-} & \mbox{-} \\
$1$ & $FFS$ & \mbox{-} & \mbox{-} & pri. & 1st des. & 2nd des. & \mbox{-} & \mbox{-} \\
$2$ & $FFV$ & \mbox{-} & pri. & 1st des. & 2nd des. & 3rd des. & 4th des. & \mbox{-} \\
$2$ & $VVS$ & \mbox{-} & pri. & 1st des. & 2nd des. & 3rd des. & 4th des. & \mbox{-} \\
$3$ & $VVV$ & pri. & 1st des. & 2nd des. & 3rd des. & 4th des. & 5th des. & 6th des. \\
\hline
\end{tabular}  \label{tab:PC3pt}
\end{table}

This analysis shows that primary representations with different total spins correspond to distinct massless structures. In particular, the UV origins of primary MHC amplitudes fall into two categories:
\begin{itemize}
\item The primary $SSS$ and $FFS$ amplitudes correspond to 4-pt and 3-pt massless amplitudes, i.e. $\lambda \phi^4$ and Yukawa interaction in the Standard Model.

\item Other primary MHC amplitudes including massive vector (i.e. $FFV$, $VVS$, $VVV$) are associated with 2-pt or 1-pt massless structures. These do not correspond to physical massless amplitudes and will be further discussed in section~\ref{sec:vector}.
\end{itemize}

\section{Massless-massive Matching via On-shell Higgsed Insertion}
\label{sec:FFS}

In previous section, we write the 3-point massless and massive amplitudes based on the Extended Poincar/'e symmetry. In this section we aim to build the relationship between the massless and massive ones via the on-shell matching procedure. Note that although the massive amplitudes are expanded by the MHC amplitudes with the massless spinors order by order in the large energy effective theory framework~\cite{Ni:2025xkg}, still it is not obvious how the massless amplitudes are matched to the MHC amplitudes. Thus we will perform the matching for massless amplitudes and MHC ones.

\subsection{Matching from massless to MHC Amplitudes}

Let us first investigate 3-point amplitudes involving in only fermion and scalar. For scalar particles, the massless and massive scalar can be directly related. For fermion particles, the massive spinor can be decomposed into two massless spinors $\lambda$ and $\eta$, while the massless spinor only has $\lambda$ component.  
The 3-point massive amplitude can be expanded into MHC amplitude order by order. For the $FFS$ amplitude it gives 
\begin{equation} \begin{aligned}
\mathcal{M}(FFS)= 
&\mathbf{c}_1 \langle12\rangle  - \mathbf{c}_2 m_1 \langle\eta_12\rangle - \mathbf{c}_3 m_2 \langle1\eta_2\rangle + \mathbf{c}_4 m_1 m_2 \langle\eta_1\eta_2\rangle\\
&+\mathbf{c}_5 [12] + \mathbf{c}_6 \tilde{m}_1[\eta_12] + \mathbf{c}_7 \tilde{m}_2 [1\eta_2] + \mathbf{c}_8 \tilde{m}_1 \tilde{m}_2 [\eta_1\eta_2].
\end{aligned} \end{equation}
where $\mathbf{c}_i$ are massive coefficients to be determined. For each term, following eq.~\eqref{eq:MHC_particle}, we can associate a diagrammatic representation for each particle. Combining these yields the MHC diagrams for the three-particle amplitude,
\begin{equation} \begin{aligned} \label{eq:FFS_expand_diagram}
\includegraphics[width=0.7\linewidth]{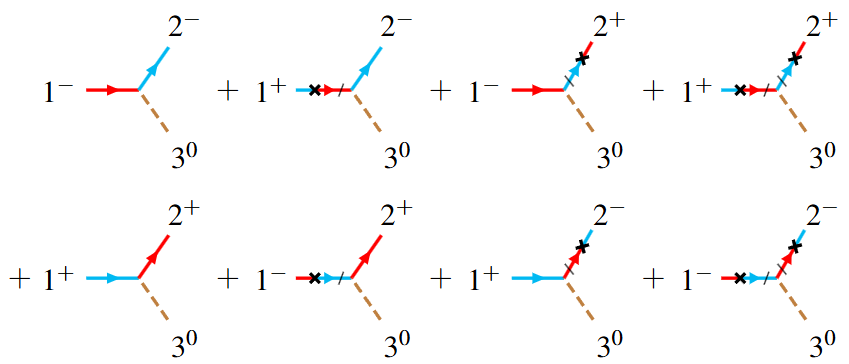}
\end{aligned} \end{equation}

According to the principle of chirality-helicity unification, the 3-pt massless amplitude should exactly match to the 3-pt primary MHC amplitude,
\begin{equation}
\mathcal{A}(1^{h_1}, 2^{h_2}, 3^{h_3})\rightarrow
[\mathcal{M}(\mathbf{1}^{h_1=t_1}, \mathbf{2}^{h_2=t_2}, \mathbf{3}^{h_3=t_3})]_0.
\end{equation}
This provides a clear correspondence between massless and massive particles. Specifically, a massless fermion with helicity $\pm\frac12$ matches to the leading component of a massive fermion with the same helicity and chirality:
\begin{eqnarray}
\begin{tikzpicture}[baseline=0.7cm]
\begin{feynhand}
\setlength{\feynhandblobsize}{6mm}
\setlength{\feynhandarrowsize}{3.5pt}
\vertex [dot] (v1) at (-0.4,0.8) {};
\vertex [particle] (i2) at (1.3,0.8) {$i^{+}$};
\graph{(i2) --[fer] (v1)};
\end{feynhand}
\end{tikzpicture}= \tilde\lambda_{\dot\alpha} 
&\rightarrow&
\begin{tikzpicture}[baseline=0.7cm]
\begin{feynhand}
\setlength{\feynhandblobsize}{6mm}
\setlength{\feynhandarrowsize}{3.5pt}
\vertex [dot] (v1) at (-0.4,0.8) {};
\vertex [particle] (i2) at (1.3,0.8) {$i^{+}$};
\fer{cyan}{i2}{v1};
\end{feynhand}
\end{tikzpicture}, \\
\begin{tikzpicture}[baseline=0.7cm]
\begin{feynhand}
\setlength{\feynhandblobsize}{6mm}
\setlength{\feynhandarrowsize}{3.5pt}
\vertex [dot] (v1) at (-0.4,0.8) {};
\vertex [particle] (i2) at (1.3,0.8) {$i^{-}$};
\graph{(i2) --[fer] (v1)};
\end{feynhand}
\end{tikzpicture}= \lambda_{\alpha} 
&\rightarrow&
\begin{tikzpicture}[baseline=0.7cm]
\begin{feynhand}
\setlength{\feynhandblobsize}{6mm}
\setlength{\feynhandarrowsize}{3.5pt}
\vertex [dot] (v1) at (-0.4,0.8) {};
\vertex [particle] (i2) at (1.3,0.8) {$i^{-}$};
\fer{red}{i2}{v1};
\end{feynhand}
\end{tikzpicture}.
\end{eqnarray}
where chirality is denoted by different colors, the red color for $t=-\frac{1}{2}$, and cyan for anti-chiral one $t=+\frac{1}{2}$. For anti-fermions, the color assignment is reversed due to their opposite chirality:
\begin{eqnarray}
\begin{tikzpicture}[baseline=0.7cm]
\begin{feynhand}
\setlength{\feynhandblobsize}{6mm}
\setlength{\feynhandarrowsize}{3.5pt}
\vertex [dot] (v1) at (-0.4,0.8) {};
\vertex [particle] (i2) at (1.3,0.8) {$i^{+}$};
\graph{(v1) --[fer] (i2)};
\end{feynhand}
\end{tikzpicture}= \tilde{\lambda}^{\dot\alpha} 
&\rightarrow&
\begin{tikzpicture}[baseline=0.7cm]
\begin{feynhand}
\setlength{\feynhandblobsize}{6mm}
\setlength{\feynhandarrowsize}{3.5pt}
\vertex [dot] (v1) at (-0.4,0.8) {};
\vertex [particle] (i2) at (1.3,0.8) {$i^{+}$};
\fer{red}{v1}{i2};
\end{feynhand}
\end{tikzpicture}, \\
\begin{tikzpicture}[baseline=0.7cm]
\begin{feynhand}
\setlength{\feynhandblobsize}{6mm}
\setlength{\feynhandarrowsize}{3.5pt}
\vertex [dot] (v1) at (-0.4,0.8) {};
\vertex [particle] (i2) at (1.3,0.8) {$i^{-}$};
\graph{(v1) --[fer] (i2)};
\end{feynhand}
\end{tikzpicture}= \lambda^{\alpha} 
&\rightarrow&
\begin{tikzpicture}[baseline=0.7cm]
\begin{feynhand}
\setlength{\feynhandblobsize}{6mm}
\setlength{\feynhandarrowsize}{3.5pt}
\vertex [dot] (v1) at (-0.4,0.8) {};
\vertex [particle] (i2) at (1.3,0.8) {$i^{-}$};
\fer{cyan}{v1}{i2};
\end{feynhand}
\end{tikzpicture}.
\end{eqnarray}
For scalar bosons, the matching reduces to a simple correspondence,
\begin{eqnarray}
\begin{tikzpicture}[baseline=0.7cm]
\begin{feynhand}
\vertex [dot] (v1) at (-0.4,0.8) {};
\vertex [particle] (i2) at (1.3,0.8) {$i^{0}$};
\graph{(i2) --[sca] (v1)};
\end{feynhand}
\end{tikzpicture}= 1 
\rightarrow 
\begin{tikzpicture}[baseline=0.7cm]
\begin{feynhand}
\vertex [dot] (v1) at (-0.4,0.8) {};
\vertex [particle] (i2) at (1.3,0.8) {$i^{0}$};
\sca{i2};
\end{feynhand}
\end{tikzpicture}.
\end{eqnarray}

Using the above connections, it is natural to build the matching for the 3-point massless amplitude and the MHC primary amplitude as follow
\begin{equation} \begin{aligned} \label{eq:FFS_leading}
\begin{tikzpicture}[baseline=0.7cm] \begin{feynhand}
\setlength{\feynhandarrowsize}{3.5pt}
\vertex [particle] (i1) at (0,0.8) {$1^-$}; 
\vertex [particle] (i2) at (1.6,1.6) {$2^-$}; 
\vertex [particle] (i3) at (1.6,0) {$3^0$};  
\vertex (v1) at (0.9,0.8); 
\graph{(i1)--[fer](v1)--[fer](i2)};
\graph{(i3)--[sca] (v1)};  
\end{feynhand} \end{tikzpicture}=Y\langle12\rangle\rightarrow
\Ampthree{1^-}{2^-}{3^0}{\fer{red}{i1}{v1}}{\antfer{cyan}{i2}{v1}}{\sca{i3}},
\end{aligned} \end{equation}
where $Y$ is the coefficient of the massless $FFS$ amplitude. Chirality flips in the MHC amplitude are denoted by color changes. Similarly, the other 3-pt massless amplitude $[12]$ matches to the primary amplitude with opposite helicity and chirality,
\begin{equation} \begin{aligned}
\begin{tikzpicture}[baseline=0.7cm] \begin{feynhand}
\setlength{\feynhandarrowsize}{3.5pt}
\vertex [particle] (i1) at (0,0.8) {$1^+$}; 
\vertex [particle] (i2) at (1.6,1.6) {$2^+$}; 
\vertex [particle] (i3) at (1.6,0) {$3^0$};  
\vertex (v1) at (0.9,0.8); 
\graph{(i1)--[fer](v1)--[fer](i2)};
\graph{(i3)--[sca] (v1)};  
\end{feynhand} \end{tikzpicture}=Y^*[12]\rightarrow
\Ampthree{1^+}{2^+}{3^0}{\fer{cyan}{i1}{v1}}{\antfer{red}{i2}{v1}}{\sca{i3}},
\end{aligned} \end{equation}
where $Y^*$ is the complex conjugate of $Y$. Therefore, the massive coefficients of primary MHC amplitudes inherit the massless coefficient,
\begin{equation} \label{eq:primary_coefficient}
\mathbf{c}_1=Y,\quad \mathbf{c}_5=Y^*.
\end{equation}

The above builds the connection between the 3-pt massless amplitude and the primary MHC amplitude. For descendant MHC amplitudes, additional massless amplitudes are needed. 

\subsection{On-shell Higgs insertion and Higgs splitting}


In the discussion above, we match the 3-pt massless amplitudes to the primary representation of the massive amplitude. We now consider the matching for the higher-order contributions of the massive amplitude, such as the terms with helicity category $(\pm\frac12,\mp\frac12,0)$. As discussed in section~\ref{sec:3pt_massless}, 3-pt massless amplitudes with such helicity category do not exist due to locality constraints. Therefore, these massive terms must be matched form higher-point massless strucutres.

A straightforward method to generate higher-point massless structures is to insert additional massless particles into external lines of the 3-pt massless amplitude. In the present work, we restrict our attention to the case where all such additional particles are Higgs bosons. We refer to this procedure as \textit{on-shell Higgs insertion}, since all particles involved, including the additional Higgs bosons, remain on-shell.

After Higgs insertion, the resulting higher-point massless amplitude generally exhibits pole structures. It is therefore crucial to clarify how these poles should be handled. For simplicity, we focus on the case with a single Higgs boson insertion carrying momentum $p_4$. We now reduce 4-pt massless amplitudes with pole structures to obtain the higher-order terms in the MHC amplitude. In such 4-pt amplitudes, the massless pole structure can always be expressed as $s_{i4}\equiv(p_4+p_i)^2$, where $i=1,2,3$. Taking the on-shell limit $p_4\rightarrow \eta_i$~\footnote{
Let us justify this limit using the Higgs splitting in the anti-collinear limit. Note that a massive amplitude can always be decomposed into two massless light-like momenta $p$ and $\eta$, in the anti-collinear limit. So taking the momentum of the particle 4 to be $\eta$, and combining with the momentum $p$, we recover the massive spinor consisting of the $\lambda$ and $\eta$. }, the pole $s_{i4}$ reduces to the mass squared of particle $i$,
\begin{equation}
\lim_{p_4\rightarrow\eta_i}s_{i4}=(p_i+\eta_i)^2=\mathbf{m}_i^2.
\end{equation}
It shows that the Higgs insertion treats the momentum of the additional Higgs boson as the small component $\eta$ of an massive momentum. In this limit, we can consider the two massless particles $i$ and $4$ as combining to form the massive particle $\mathbf{i}$. Therefore, to match the descendant MHC amplitudes, the on-shell Higgs insertion is always associated with the limit $p_4\rightarrow\eta_i$.

Note that the mass dimension  of an $n$-point amplitude is $4-n$. To match a 4-pt massless amplitude to a 3-pt massive amplitude, we must introduce a quantity of mass dimension 1. A natural choice is the vacuum expectation value (VEV) $v$ of the additional Higgs boson. With this factor, the inserted massless amplitude matches as follows
\begin{equation}
\lim_{p_4\to\eta_i} v\mathcal{A}(1^{h_1}, 2^{h_2}, 3^{h_3};4^0)\rightarrow [\mathcal{M}(\mathbf{1}^{h_1=t_1}, \mathbf{2}^{h_2=t_2}, \mathbf{3}^{h_3=t_3})]_1\sim v.
\end{equation}

Returning to the case of the massive $FFS$ amplitude, we match the subleading terms with helicity $(\pm\frac12,\mp\frac12,0)$ by considering the 4-pt massless amplitudes with single Higgs insertion, which carry helicity $(\pm\frac12,\mp\frac12,0,0)$. We first examine the massless $FFSS$ amplitude with helicity $(-\frac12,+\frac12,0,0)$. It contain terms with different pole structures,
\begin{equation}
\mathcal{A}(1^{-\frac12},2^{+\frac12},3^0,4^0)=G_1\frac{\langle 1|P_{14}|2]}{s_{14}}+G_2\frac{\langle 1|P_{24}|2]}{s_{24}}+G_3\frac{\langle 1|3-4|2]}{s_{34}}.
\end{equation}
where $P_{ij}=p_i+p_j$, and $G_i$ is the coefficient of UV amplitudes. 

Each pole structure $s_{i4}$ correspond to a possible Higgs insertion with the limit $p_4\rightarrow \eta_i$. We first consider the Higgs insertion on particle 1. In this case, only the pole $s_{14}$ is involved in the insertion, while the other poles $s_{24}$ and $s_{34}$ are unrelated and can be rewritten as $s_{13}$ and $s_{12}$. Thus, the massless amplitude reduces to
\begin{equation} \begin{aligned}
\begin{tikzpicture}[baseline=0.7cm] \begin{feynhand}
\setlength{\feynhandarrowsize}{3.5pt}
\vertex [particle] (i1) at (-0.2,0.8) {$1^-$}; 
\vertex [particle] (i2) at (1.6,1.6) {$2^+$}; 
\vertex [particle] (i3) at (1.6,0) {$3^0$};  
\vertex [particle] (i5) at (0.7,0.2) {$4$};
\vertex (v3) at (0.6,0.8);
\vertex (v1) at (0.9,0.8); 
\graph{(i1)--[fer](v3)--[fer](v1)};
\graph{(v1)--[fer](i2)};
\graph{(i3)--[sca] (v1)};  
\graph{(v3)--[sca](i5)};
\end{feynhand} \end{tikzpicture}=
\lim_{p_4\rightarrow \eta_1}v\mathcal{A}&\to\underbrace{v G_1\frac{m_1[\eta_12]}{\mathbf{m}_1^2}}_{\sim v}+\underbrace{v G_2\frac{m_1[\eta_12]}{s_{13}}+v G_3\frac{m_1[\eta_12]}{s_{12}}}_{\sim v^3 E^{-2}}
\end{aligned} \end{equation}
In this expression, the first term gives the leading-order contribution, which corresponds to the MHC amplitude
\begin{equation} \begin{aligned}
vG_1\frac{m_1[\eta_12]}{\mathbf{m}_1^2}
=\Ampthree{1^-}{2^+}{3^0}{\ferflip{1}{180}{red}{cyan}}{\antfer{red}{i2}{v1}}{\sca{i3}},
\end{aligned} \end{equation}
For the Higgs insertion on particle 2, we have
\begin{equation} \begin{aligned} \label{eq:FFSS_FFS}
\begin{tikzpicture}[baseline=0.7cm] \begin{feynhand}
\setlength{\feynhandarrowsize}{3.5pt}
\vertex [particle] (i1) at (0,0.8) {$1^-$}; 
\vertex [particle] (i2) at (1.6,1.6) {$2^+$}; 
\vertex [particle] (i3) at (1.6,0) {$3^0$};  
\vertex [particle] (i4) at (1.6,0.8) {$4$};
\vertex (v2) at (0.9+0.7*0.33,0.8+0.8*0.33);
\vertex (v1) at (0.9,0.8); 
\graph{(i1)--[fer](v1)};
\graph{(v1)--[fer](v2)--[fer](i2)};
\graph{(i3)--[sca] (v1)};  
\graph{(v2)--[sca](i4)};
\end{feynhand} \end{tikzpicture}=
\lim_{p_4\rightarrow \eta_2}v\mathcal{A}&\to vG_2\frac{\tilde{m}_2\langle1\eta_2\rangle}{\mathbf{m}_2^2}
=\Ampthree{1^-}{2^+}{3^0}{\fer{red}{i1}{v1}}{\antferflip{1}{55}{red}{cyan}}{\sca{i3}}.
\end{aligned} \end{equation}
This means that the result of the Higgs insertion is equivalent to the descendant MHC term. Therefore, the term with the $s_{34}$ pole should not contribute to the massive $FFS$ amplitude, because the additional Higgs is inserted into particle 3, which corresponds to a massive scalar that lacks subleading structure.


Similarly, applying limits to massless amplitude with helicity $(+\frac12,-\frac12,0,0)$, we can obtain the other two descendant terms,
\begin{equation} \begin{aligned}
\begin{tikzpicture}[baseline=0.7cm] \begin{feynhand}
\setlength{\feynhandarrowsize}{3.5pt}
\vertex [particle] (i1) at (-0.2,0.8) {$1^+$}; 
\vertex [particle] (i2) at (1.6,1.6) {$2^-$}; 
\vertex [particle] (i3) at (1.6,0) {$3^0$};  
\vertex [particle] (i5) at (0.7,0.2) {$4$};
\vertex (v3) at (0.6,0.8);
\vertex (v1) at (0.9,0.8); 
\graph{(i1)--[fer](v3)--[fer](v1)};
\graph{(v1)--[fer](i2)};
\graph{(i3)--[sca] (v1)};  
\graph{(v3)--[sca](i5)};
\end{feynhand} \end{tikzpicture}
\rightarrow \Ampthree{1^+}{2^-}{3^0}{\ferflip{1}{180}{cyan}{red}}{\antfer{cyan}{i2}{v1}}{\sca{i3}},\quad
\begin{tikzpicture}[baseline=0.7cm] \begin{feynhand}
\setlength{\feynhandarrowsize}{3.5pt}
\vertex [particle] (i1) at (0,0.8) {$1^+$}; 
\vertex [particle] (i2) at (1.6,1.6) {$2^-$}; 
\vertex [particle] (i3) at (1.6,0) {$3^0$};  
\vertex [particle] (i4) at (1.6,0.8) {$4$};
\vertex (v2) at (0.9+0.7*0.33,0.8+0.8*0.33);
\vertex (v1) at (0.9,0.8); 
\graph{(i1)--[fer](v1)};
\graph{(v1)--[fer](v2)--[fer](i2)};
\graph{(i3)--[sca] (v1)};  
\graph{(v2)--[sca](i4)};
\end{feynhand} \end{tikzpicture}
\rightarrow\Ampthree{1^+}{2^-}{3^0}{\fer{cyan}{i1}{v1}}{\antferflip{1}{55}{cyan}{red}}{\sca{i3}}.
\end{aligned} \end{equation}
Together with Eq.~\eqref{eq:FFSS_FFS}, these results give all 1st descendant MHC amplitudes.

Comparing these results with the leading-order matching discussed in the previous subsection, it is obvious that the Higgs insertion only affects the massive state of the inserted line. This allow us to focus on the inserted line and derive general matching rules for on-shell Higgs insertion, similar to the splitting function extracted from the scattering amplitudes. We refere this as the {\it Higgs splitting}. Suppose that the uninserted fermion line carries  helicity $+\frac{1}{2}$. The Higgs insertion flips the helicity to $-\frac12$,
\begin{equation}
\begin{tikzpicture}[baseline=0.7cm]
\begin{feynhand}
\setlength{\feynhandblobsize}{6mm}
\setlength{\feynhandarrowsize}{3.5pt}
\vertex [dot] (v1) at (-0.4,0.8) {};
\vertex [particle] (i2) at (1.3,0.8) {$i^{+}$};
\graph{(i2) --[fer] (v1)};
\end{feynhand}
\end{tikzpicture}
\xrightarrow{\text{Higgs splitting}}
\begin{tikzpicture}[baseline=-0.1cm] \begin{feynhand}
\setlength{\feynhandblobsize}{6mm}
\setlength{\feynhandarrowsize}{3.5pt}
\vertex [particle] (i1) at (2,0) {$i^{-}$};
\vertex [particle] (i2) at (1,0.7) {$4$};
\vertex (v2) at (1,0);
\vertex [dot] (v1) at (0,0) {};
\graph{(i1)--[fer](v2)--[fer](v1)};
\graph{(i2)--[sca](v2)};
\end{feynhand} \end{tikzpicture}.
\end{equation}
Focusing on the non-trivial Lorentz structure, the insertion result can be expressed as the product of a three-point massless $FFS$ amplitude and a single-particle state:
\begin{equation}
\begin{tikzpicture}[baseline=-0.1cm] \begin{feynhand}
\setlength{\feynhandblobsize}{6mm}
\setlength{\feynhandarrowsize}{3.5pt}
\vertex [particle] (i1) at (2,0) {$i^{-}$};
\vertex [particle] (i2) at (1,0.7) {$4$};
\vertex (v2) at (1,0);
\vertex [dot] (v1) at (0,0) {};
\graph{(i1)--[fer](v2)--[fer](v1)};
\graph{(i2)--[sca](v2)};
\end{feynhand} \end{tikzpicture}\rightarrow |\chi]_{\dot\alpha}\times \langle\chi i\rangle=(p_\chi|i\rangle)_{\dot\alpha},
\end{equation}
where the spinors $|\chi]$ and $|\chi\rangle$ correspond to an on-shell momentum $p_\chi$, representing the on-shell part of the internal particle. We define it as
\begin{equation} \label{eq:on_shell_p}
p_\chi\equiv|\chi]\langle\chi|=p_i+p_4-q,
\end{equation}
where $q$ is the off-shell part of the internal momentum $p_i+p_4$ and cannot be expressed in terms of spinors. When $p_\chi\gg q$, we can approximate $p_\chi\sim p_i+p_4$. In this kinematic region, $(p_i+p_4)^2\sim 2 p_\chi\cdot q$ is a small quantity, and we assume it is approximately equal to  $\mathbf m_i^2$. On the other hand, the on-shell limit $p_4\rightarrow \eta_i$ also converts $(p_i+p_4)^2$ to mass squared, so it is consistent with the description in eq.~\eqref{eq:on_shell_p}. Taking this on-shell limit, the Higgs insertion yields
\begin{eqnarray} \label{eq:f_insert1}
\lim_{p_4\rightarrow \eta_i} \left(|\chi]_{\dot\alpha}\times \langle\chi i\rangle\right)
\sim \lim_{p_4\rightarrow \eta_i} |4]\langle4 i\rangle
= |\eta_i]_{\dot\alpha} m_i 
\quad \to \quad 
\Ampone{1.5}{i^-}{\ferflip{1.5}{0}{red}{cyan}}.
\end{eqnarray}
The result of the Higgs insertion is equivalent to a massive state with both an $h$-flip and a $m$-flip, converting  $\lambda_i$ to $m_i \tilde{\eta}_i$. This confirms the condition of helicity-chirality unification.

Similarly, the Higgs insertion can also flip the helicity from $+\frac12$ to $-\frac12$ for a fermion. It gives
\begin{equation} \label{eq:f_insert2}
\begin{tikzpicture}[baseline=-0.1cm] \begin{feynhand}
\setlength{\feynhandblobsize}{6mm}
\setlength{\feynhandarrowsize}{3.5pt}
\vertex [particle] (i1) at (2,0) {$i^{+}$};
\vertex [particle] (i2) at (1,0.7) {$4$};
\vertex (v2) at (1,0);
\vertex [dot] (v1) at (0,0) {};
\graph{(i1)--[fer](v2)--[fer](v1)};
\graph{(i2)--[sca](v2)};
\end{feynhand} \end{tikzpicture}=
\lim_{p_4\rightarrow \eta_i} \left(|\chi\rangle_{\alpha}\times [\chi i]\right)
\sim |\eta_i\rangle_{\alpha} \tilde m_i 
\quad \to \quad 
\Ampone{1.5}{i^+}{\ferflip{1.5}{0}{cyan}{red}}.
\end{equation}
This result also exhibits both an $h$-flip and a $m$-flip. The Higgs boson can be also inserted into the external line of anti-fermion, the result is analogous to eqs.~\eqref{eq:f_insert1} and \eqref{eq:f_insert2} with exchanging color. 

Therefore, for a massive fermion, the results of leading and subleading matching can be summarized as follows 
\begin{equation} \label{eq:f_table}
\begin{tabular}{c|c|c}
\hline
& $h=t=-\frac12$ & $h=t=+\frac12$ \\
\hline
$c=-\frac12$ & $\lambda_\alpha$ & $\tilde m\eta_\alpha$  \\
$c=+\frac12$ & $m\tilde\eta_{\dot\alpha}$ & $\tilde\lambda_{\dot\alpha}$  \\
\hline
\end{tabular}
\end{equation}
where $c$ denotes the chirality of each state. The diagonal entries correspond to the primary MHC amplitude, while the off-diagonal entries correspond to the descendant MHC amplitude. The effect of Higgs insertion is equivalent to a horizontal transition in the table above, mapping primary to descendant amplitudes. The table can be re-expressed using the Feynman diagram
\begin{equation}
\begin{tabular}{c|c|c}
\hline
& $h=t=-\frac12$ & $h=t=+\frac12$ \\
\hline
$c=-\frac12$ & 
\begin{tikzpicture}[baseline=0.7cm]
\begin{feynhand}
\setlength{\feynhandblobsize}{6mm}
\setlength{\feynhandarrowsize}{3.5pt}
\vertex [dot] (v1) at (-0.4,0.8) {};
\vertex [particle] (i2) at (1.3,0.8) {$i^{-}$};
\graph{(i2) --[fer] (v1)};
\end{feynhand}
\end{tikzpicture} & 
\begin{tikzpicture}[baseline=-0.1cm] \begin{feynhand}
\setlength{\feynhandblobsize}{6mm}
\setlength{\feynhandarrowsize}{3.5pt}
\vertex [particle] (i1) at (2,0) {$i^{+}$};
\vertex [particle] (i2) at (1.9,0.7) {$4$};
\vertex (v2) at (1,0);
\vertex [dot] (v1) at (0,0) {};
\graph{(i1)--[fer](v2)--[fer](v1)};
\graph{(i2)--[sca](v2)};
\draw[decorate,decoration=brace] (2.2,0.85)--(2.2,-0.15);
\end{feynhand} \end{tikzpicture}  \\
\hline
$c=+\frac12$ & 
\begin{tikzpicture}[baseline=-0.1cm] \begin{feynhand}
\setlength{\feynhandblobsize}{6mm}
\setlength{\feynhandarrowsize}{3.5pt}
\vertex [particle] (i1) at (2,0) {$i^{-}$};
\vertex [particle] (i2) at (1.9,0.7) {$4$};
\vertex (v2) at (1,0);
\vertex [dot] (v1) at (0,0) {};
\graph{(i1)--[fer](v2)--[fer](v1)};
\graph{(i2)--[sca](v2)};
\draw[decorate,decoration=brace] (2.2,0.85)--(2.2,-0.15);
\end{feynhand} \end{tikzpicture} & 
\begin{tikzpicture}[baseline=0.7cm]
\begin{feynhand}
\setlength{\feynhandblobsize}{6mm}
\setlength{\feynhandarrowsize}{3.5pt}
\vertex [dot] (v1) at (-0.4,0.8) {};
\vertex [particle] (i2) at (1.3,0.8) {$i^{+}$};
\graph{(i2) --[fer] (v1)};
\end{feynhand}
\end{tikzpicture}  \\
\hline
\end{tabular}
\end{equation}

Using this table, one can derive the massive coefficients of descendant MHC amplitudes through the complete subleading matching. To do so, we include the contribution of the pole $s_{i4}$ and the vev $v$
\begin{equation} \begin{aligned}
\lim_{p_4\to \eta_i} \left(\text{primary state}\times \frac{1}{s_{i4}}\times v\mathcal{A}_3\right) = \frac{Yv}{\mathbf{m}_i^2}\times\text{descendant state}
\end{aligned} \end{equation}
where $Y$ is the coefficient of the massless amplitude $\langle12\rangle$. When we use the amplitude $[12]$, the coefficient need to be replaced by $Y^*$. This allows us to establish the relation between the coefficients of the primary and the 1st descendant MHC amplitude:
\begin{equation} \begin{aligned}
\mathbf{c}_2&=\frac{Y^* v}{\mathbf{m}_1^2}\mathbf{c}_1,&
\mathbf{c}_3&=\frac{Y^* v}{\mathbf{m}_2^2}\mathbf{c}_1, \\
\mathbf{c}_6&=\frac{Y v}{\mathbf{m}_1^2}\mathbf{c}_5,&
\mathbf{c}_7&=\frac{Y v}{\mathbf{m}_2^2}\mathbf{c}_5.
\end{aligned} \end{equation}
Substituting eq.~\eqref{eq:primary_coefficient}, we obtain
\begin{equation} \begin{aligned}
\mathbf{c}_2&=\mathbf{c}_6=\frac{Y^* Y v}{\mathbf{m}_1^2},&
\mathbf{c}_3&=\mathbf{c}_7=\frac{Y^* Y v}{\mathbf{m}_2^2}.
\end{aligned} \end{equation}
Therefore, once the coefficient of the primary amplitude is determined, the coefficients of the descendant amplitudes can be systematically obtained through Higgs insertions in the UV theory.

\subsection{Higher order matching}

We now consider higher-order matching from the massless amplitudes with more than one Higgs insertion. For a 5-pt massless amplitude,
\begin{equation}
\mathcal{A}(1^{+\frac12},2^{+\frac12},3^0;4^0,5^0)=G_1\frac{[1|P_{15}P_{24}|2]}{s_{15}s_{24}}+G_2\frac{[1|P_{245}P_{24}|2]}{s_{245}s_{24}}+\cdots.
\end{equation}
where particles 4 and 5 are additional Higgs bosons. We analyze two typcial cases of Higgs insertion. 

In the first case, two Higgs bosons insert into different lines, with their momenta taking different on-shell limits,
\begin{equation} \begin{aligned}
\begin{tikzpicture}[baseline=0.7cm] \begin{feynhand}
\setlength{\feynhandarrowsize}{3.5pt}
\vertex [particle] (i1) at (-0.2,0.8) {$1^+$}; 
\vertex [particle] (i2) at (1.6,1.6) {$2^+$}; 
\vertex [particle] (i3) at (1.6,0) {$3^0$};
\vertex [particle] (i4) at (1.6,0.8) {$4$};  
\vertex [particle] (i5) at (0.7,0.2) {$5$};
\vertex (v3) at (0.6,0.8);
\vertex (v2) at (0.9+0.7*0.33,0.8+0.8*0.33);
\vertex (v1) at (0.9,0.8); 
\graph{(i1)--[fer](v3)--[fer](v1)};
\graph{(v1)--[fer](v2)--[fer](i2)};
\graph{(i3)--[sca] (v1)};  
\graph{(v3)--[sca](i5)};
\graph{(v2)--[sca](i4)};
\end{feynhand} \end{tikzpicture}=
\lim_{\substack{p_4\rightarrow \eta_2\\p_5\rightarrow \eta_1}}
v^2\mathcal{A}
\to v^2 G_1\frac{\tilde m_1 \tilde m_2\langle\eta_1\eta_2\rangle}{\mathbf{m}_1^4}=
\Ampthree{1^+}{2^+}{3^0}{\ferflip{1}{180}{cyan}{red}}{\antferflip{1}{60}{red}{cyan}}{\sca{i3}}.
\end{aligned} \end{equation}
Here, the factor $v^2$ accounts for the insertion of two additional Higgs bosons. This result corresponds to the 2nd descendant MHC amplitude.

In the second case, two Higgs bosons insert into the same line, say particle $2$. Since three massless momenta are not required to form a massive momentum, we can take one Higgs to be soft, $p_5\rightarrow0$, while the other one still take the limit $p_4\rightarrow\eta_2$. We obtain
\begin{equation} \begin{aligned}
\begin{tikzpicture}[baseline=0.7cm] \begin{feynhand}
\setlength{\feynhandarrowsize}{3.5pt}
\vertex [particle] (i1) at (-0.2,0.8) {$1^+$}; 
\vertex [particle] (i2) at (1.6,1.6) {$2^+$}; 
\vertex [particle] (i3) at (1.6,0) {$3^0$};
\vertex [particle] (i4) at (1.7,1.0) {$4$};  
\vertex [particle] (i5) at (1.7,0.5) {$5$};
\vertex (v3) at (0.9+0.7*0.4,0.8+0.8*0.4);
\vertex (v2) at (0.9+0.7*0.2,0.8+0.8*0.2);
\vertex (v1) at (0.9,0.8); 
\graph{(i1)--[fer](v1)};
\graph{(v1)--[fer](v2)--[fer](v3)--[fer](i2)};
\graph{(i3)--[sca] (v1)};  
\graph{(v2)--[sca](i5)};
\graph{(v3)--[sca](i4)};
\end{feynhand} \end{tikzpicture}=
\lim_{\substack{p_4\rightarrow \eta_2\\p_5\rightarrow 0}}
v^2 \mathcal{A}
=v^2 G_2\frac{\mathbf{m}_2^2 [12]}{\mathbf{m}_2^4}\rightarrow
\Ampthree{1^+}{2^+}{3^0}{\fer{cyan}{i1}{v1}}{\antfer{red}{i2}{v1}}{\sca{i3}}.
\end{aligned} \end{equation}
This limit guarantee that two poles $s_{24}$ and $s_{245}$ both convert to physical mass. The result repeat the primary representation, which have already been matched from 3-pt massless amplitude, and therefore does not give new terms for the MHC amplitude.

In the general higher-order matching, we can still focus on the inserted line. Multi-Higgs insertions for a single fermion line produce only two distinct results. Suppose we insert $(n-3)$ Higgs boson into line $i$; only one of their momenta needs to be non-vanishing
\begin{equation} \begin{aligned}
\begin{tikzpicture}[baseline=-0.1cm] \begin{feynhand}
\setlength{\feynhandblobsize}{6mm}
\setlength{\feynhandarrowsize}{3.5pt}
\vertex [particle] (i1) at (3,0) {$i^{-\frac12}$};
\vertex [particle] (i2) at (0.8,0.7) {$4$};
\vertex [particle] (i3) at (1.4,0.7) {$\cdots$};
\vertex [particle] (i4) at (2,0.7) {$n$};
\vertex (v2) at (0.8,0);
\vertex (v3) at (2,0);
\vertex [dot] (v1) at (0,0) {};
\graph{(i1)--[fer](v3)--[fer](v2)--[fer](v1)};
\graph{(i2)--[sca](v2)};
\graph{(i4)--[sca](v3)};
\end{feynhand} \end{tikzpicture} &=
\lim_{\substack{p_n\rightarrow \eta_i\\p_3,\cdots,p_{n-1}\rightarrow 0}}(P_{4 \cdots n i}\cdots P_{ni}|i\rangle)\\
&\rightarrow
\begin{cases}
\Ampone{1.5}{i^-}{\fer{red}{i1}{v1}},\quad \text{$n$ is even number}\\
\Ampone{1.5}{i^-}{\ferflip{1.5}{0}{red}{cyan}},\quad \text{$n$ is odd number}
\end{cases}
\end{aligned} \end{equation}
where $P_{n i}=p_i+p_{n}$ and $P_{4\cdots n i}=p_i+\sum_{j=4}^n p_j$.

Therefore, higher-point massless amplitudes with more Higgs insertion do not give more information for the massive $FFS$ amplitude. All massive coefficients can be derived by repeatedly applying the inserted-line analysis. For the $FFS$ case, we also obtain the relation between the primary and the 2nd descendant MHC amplitude, 
\begin{equation} \begin{aligned}
\mathbf{c}_4=\frac{Y^{*2} v^2}{\mathbf{m}_1^2 \mathbf{m}_2^2}\mathbf{c}_1=\frac{Y^{*2} Y v^2}{\mathbf{m}_1^2 \mathbf{m}_2^2},\\
\mathbf{c}_8=\frac{Y^2 v^2}{\mathbf{m}_1^2 \mathbf{m}_2^2}\mathbf{c}_5=\frac{Y^* Y^2 v^2}{\mathbf{m}_1^2 \mathbf{m}_2^2}.
\end{aligned} \end{equation}

Finally, we obtain the complete MHC amplitude for $FFS$ case,
\begin{equation} \begin{aligned}
\mathcal{M}(FFS)= 
&Y \langle12\rangle  - \frac{Y^* Y v}{\mathbf{m}_1^2} m_1 \langle\eta_12\rangle - \frac{Y^* Y v}{\mathbf{m}_2^2} m_2 \langle1\eta_2\rangle + \frac{Y^{*2} Y v^2}{\mathbf{m}_1^2 \mathbf{m}_2^2} m_1 m_2 \langle\eta_1\eta_2\rangle\\
&+Y^* [12] + \frac{Y^* Y v}{\mathbf{m}_1^2} \tilde{m}_1[\eta_12] + \frac{Y^* Y v}{\mathbf{m}_2^2} \tilde{m}_2 [1\eta_2] + \frac{Y^* Y^2 v^2}{\mathbf{m}_1^2 \mathbf{m}_2^2} \tilde{m}_1 \tilde{m}_2 [\eta_1\eta_2].
\end{aligned} \end{equation}

\section{On-shell Higgs Mechanism for Massive Gauge Boson}
\label{sec:vector}

The massless-massive matching is quite straightforward in the case of fermions and scalars. From above $FFS$ amplitudes, we have seen that the primary MHC amplitudes, with additional Higgs insertion should be enough to achieve the matching. However, if a massive gauge boson is involved, the matching procedure needs additional treatments. These arise from the need to account for gauge fixing in the transverse component and the Goldstone boson equivalence theorem in the longitudinal component. In this section, we would discuss the matching procedure for amplitudes involving a massive spin-1 vector boson, taking the $FFV$ amplitudes as an illustrative example.

\subsection{$FFV$ MHC amplitudes and conserved currents}

In analogy with the analysis of the $FFS$ amplitude, we now study the MHC amplitudes for the $FFV$ case. The general $FFV$ amplitude has four Lorentz structures, but only $ [\mathbf{23}]\langle\mathbf{31}\rangle $ and $ \langle\mathbf{23}\rangle[\mathbf{31}]$ are relevant for the renomralizable thoeries. We first consider $[\mathbf{23}]\langle\mathbf{31}\rangle$ and perform its MHC expansion. The result takes the form
\begin{equation} \begin{aligned}
&\mathcal{M}(FFV)\\
&\sim \mathbf c_{13} [23]\langle31\rangle - \mathbf c_{14} \tilde m_1 [23]\langle3 \eta_1\rangle + \mathbf c_{15} m_2 [\eta_23]\langle31\rangle - \mathbf c_{16}\tilde m_3 [23]\langle\eta_31\rangle \\
&+ \mathbf c_{17} m_3 [2\eta_3]\langle31\rangle - \frac{\mathbf c_{13}}{\mathbf m_3^2}m_3\tilde m_3 [2\eta_3]\langle\eta_31\rangle - \mathbf c_{18} \tilde m_1 m_2 [\eta_23]\langle3\eta_1\rangle + \mathbf c_{19} \tilde m_1 \tilde m_3 [23]\langle\eta_3\eta_1\rangle \\
&- \mathbf c_{20} \tilde m_1 m_3 [2\eta_3]\langle3\eta_1\rangle - \mathbf c_{21} m_2 \tilde m_3 [\eta_23]\langle\eta_31\rangle + \mathbf c_{22} m_2 m_3 [\eta_2\eta_3]\langle31\rangle + \frac{\mathbf c_{14}}{\mathbf m_3^2}\tilde m_1 m_3\tilde m_3  [2\eta_3]\langle\eta_3\eta_1\rangle  \\
& - \frac{\mathbf c_{15}}{\mathbf m_3^2}m_2 m_3\tilde m_3  [\eta_2\eta_3]\langle\eta_31\rangle + \mathbf c_{23} \tilde m_1 m_2\tilde m_3 [\eta_23]\langle\eta_3\eta_1\rangle - \mathbf c_{24} \tilde m_1 m_2 m_3[\eta_2\eta_3]\langle3\eta_1\rangle \\
&+ \frac{\mathbf c_{18}}{\mathbf m_3^2}\tilde m_1 m_2 m_3\tilde m_3 [\eta_2\eta_3]\langle\eta_3\eta_1\rangle \label{eq:FFV_expand}
\end{aligned} \end{equation}
It is important to note that terms within a given helicity category are not independent. For example, both $\mathbf{c}_{13} [\mathbf{23}]\langle\mathbf{31}\rangle$ and $- (\mathbf{c}_{13}/\mathbf{m}_3^2) m_3 \tilde{m}_3 [\mathbf{2}\eta_3]\langle\eta_3\mathbf{1}\rangle$ are related to the same coefficient $\mathbf{c}_{13}$, as they originate from the MHV expansion of a single massive structure. Therefore, for each helicity category, we only need to determine the coefficient of one representative MHV amplitude. The corresponding MHC diagrams for eq.~\eqref{eq:FFV_expand} are shown below,
\begin{equation} \begin{aligned} \label{eq:FFV_expand_diagram}
\includegraphics[width=0.7\linewidth]{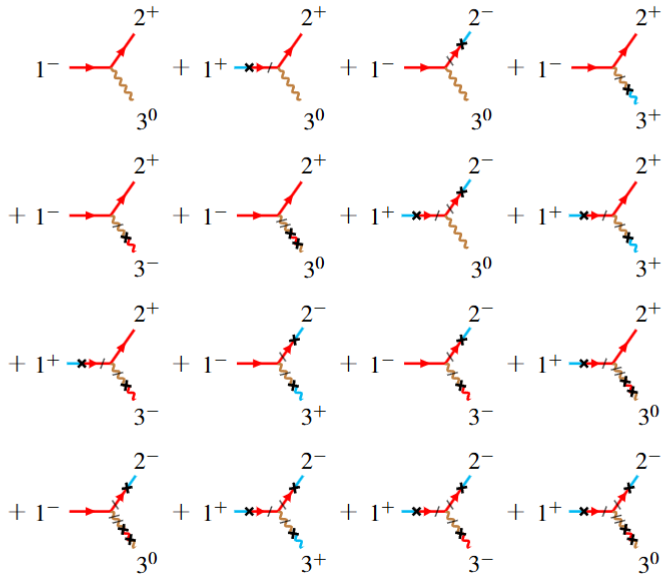}
\end{aligned} \end{equation}

Following the same logic as in the $FFS$ case, we now proceed to match massless amplitudes $\mathcal{A}$ to MHC amplitudes $\mathcal{M}$. At first glance, since massless amplitudes involve only massless spinors $\lambda$ and $\tilde\lambda$, one might assume they naturally match to primary MHC amplitude. However, this intuition fails for amplitudes involving massive vectors. Consider, for example, the massless $FFV$ and $FFS$ amplitudes matching,
\begin{equation}
\begin{tikzpicture}[baseline=0.7cm] \begin{feynhand}
\setlength{\feynhandarrowsize}{3.5pt}
\vertex [particle] (i1) at (0,0.8) {$1^-$}; 
\vertex [particle] (i2) at (1.6,1.6) {$2^+$}; 
\vertex [particle] (i3) at (1.6,0) {$3^+$};  
\vertex (v1) at (0.9,0.8); 
\graph{(i1)--[fer](v1)--[fer](i2)};
\graph{(i3)--[bos] (v1)};  
\end{feynhand} \end{tikzpicture}\xrightarrow{?}
\Ampthree{1^-}{2^+}{3^+}{\fer{red}{i1}{v1}}{\antfer{red}{i2}{v1}}{\bos{i3}{cyan}},\quad
\begin{tikzpicture}[baseline=0.7cm] \begin{feynhand}
\setlength{\feynhandarrowsize}{3.5pt}
\vertex [particle] (i1) at (0,0.8) {$1^-$}; 
\vertex [particle] (i2) at (1.6,1.6) {$2^-$}; 
\vertex [particle] (i3) at (1.6,0) {$3^0$};  
\vertex (v1) at (0.9,0.8); 
\graph{(i1)--[fer](v1)--[fer](i2)};
\graph{(i3)--[sca] (v1)};  
\end{feynhand} \end{tikzpicture}\xrightarrow{?}
\Ampthree{1^-}{2^-}{3^0}{\fer{red}{i1}{v1}}{\antfer{cyan}{i2}{v1}}{\bos{i3}{brown}}.
\end{equation}
We find these two MHC diagrams do not correspond to any $FFV$ primary amplitude. This suggests that the 3-pt massless amplitudes do not map to primary representations in the presence of massive vectors.

Let us understand the $FFV$ primary amplitudes
\begin{equation} \begin{aligned} \label{eq:FFV_primary_example}
\Ampthree{1^-}{2^+}{3^0}{\fer{red}{i1}{v1}}{\antfer{red}{i2}{v1}}{\bos{i3}{brown}}=[23]\langle31\rangle.
\end{aligned} \end{equation}
This expression vanishes when we impose momentum conservation, $p_1+p_2+p_3=0$. The vanishing behavior can be understood by identifying a 2-pt massless structure within the $FFV$ primary amplitudes, where the spinors associated with particle 3 drop out. This 2-pt structure corresponds to a conserved current 
\begin{equation} \begin{aligned} \label{eq:current_1}
J^{{\alpha\dot\alpha}}(1^{-\frac12},2^{+\frac12})=-\langle1|^{\alpha}[2|^{\dot\alpha},
\end{aligned} \end{equation}
where the superscripts $\pm \frac12$ represent the helicity of particles in the current. This current corresponds to the Noether current $\bar{\psi}\gamma^\mu\psi$ in the quantum field theory and therefore must satisfy the conserved current condition,
\begin{equation} \begin{aligned}
\partial\cdot J=-\langle1|1|2]-\langle1|2|2]=0,
\end{aligned} \end{equation}
Using momentum coservation $p_3=-p_1-p_2$, one recovers the primary MHC amplitude in eq.~\eqref{eq:FFV_primary_example}.

In fact, the conserved current is the key to connect the massive vector boson to a well-defined UV theory. This can be seen by examining whether the massless limit of the massive interaction involving the massive vector is continuous. In the massive theory, the interaction can be expressed as the coupling between the massive vector boson $\mathbf{A}$ and a massive current $\mathbf{J}$. The squared S-matrix elements then take the form 
\begin{equation}
|\mathbf J\cdot\mathbf A|^2=\mathbf J_\mu \mathbf J^*_\nu \Pi^{\mu\nu},
\end{equation}
where $\Pi^{\mu\nu}$ is the polarization sum of $\mathbf A^{\mu}\mathbf A^{*\nu}$ and can be viewed as the numerator of the massive vector propagator. In the massless limit, the massive current $\mathbf J$ must reduce to a conserved current $J$. Otherwise, the amplitude would blow up, due to $\mathbf p^\mu \mathbf p^\nu/\mathbf m^2$ divergence in $\Pi^{\mu\nu}$.

It is therefore natural to derive the 3-pt massive structure from the current structure and the polarization vector. Take the massive structure $[\mathbf{23}]\langle\mathbf{31}\rangle$ as example, the massive current and massive polarization vector can be expressed in spinor form as 
\begin{equation} \begin{aligned}
\mathbf{J}^{\alpha\dot\alpha}(\mathbf{1}^{\frac12},\mathbf{2}^{\frac12})&=\langle\mathbf{1}|^{\alpha}[\mathbf{2}|^{\dot\alpha}, \quad
\mathbf{A}^{\alpha\dot\alpha} = |\mathbf{3}\rangle^{\alpha}|\mathbf{3}]^{\dot\alpha}.
\end{aligned}\end{equation}
Here we adopt the normalization $\mathbf A\cdot \mathbf A=\mathbf m^2$ for massive polarization vector. 
In analogous to the MHC state and MHC amplitude, let us define the MHC current, expanded from the above current. The MHC amplitude can be expanded into the MHC current and the MHC vector with the chirality-helicity unification condition at the different order as below
\begin{equation} \begin{aligned}
\relax
[\mathbf{J}^{-+}]_0&=-\langle1|^{\alpha}[2|^{\dot\alpha},& 
[\mathbf{A}^0]_0&=-|3\rangle^{\alpha}|3]^{\dot\alpha},&\\
[\mathbf{J}^{++}]_1&=\tilde m_1 \langle\eta_1|^{\alpha}[2|^{\dot\alpha},&
[\mathbf{A}^-]_1&=-m_3 |3\rangle^{\alpha}|\eta_3]^{\dot\alpha},&\\
[\mathbf{J}^{--}]_1&= -m_2 \langle1|^{\alpha}[\eta_2|^{\dot\alpha},&
[\mathbf{A}^+]_1&= \tilde m_3|\eta_3\rangle^{\alpha}|3]^{\dot\alpha},&\\
[\mathbf{J}^{+-}]_2&=\tilde m_1 m_2 \langle\eta_1|^{\alpha}[\eta_2|^{\dot\alpha},&
[\mathbf{A}^0]_2&=m_3\tilde m_3|\eta_3\rangle^{\alpha}|\eta_3]^{\dot\alpha}.&\\
\end{aligned} \end{equation}
where the superscripts $\pm,0$ denotes the helicity of current $\mathbf J$ or polarization vector $\mathbf A$. Their contraction gives
\begin{equation} \begin{aligned}
\relax
[\mathbf{J}^{-+}]_0\cdot [\mathbf{A}^0]_0&=\langle13\rangle[32]=0,\\
[\mathbf{J}^{++}]_{1}\cdot [\mathbf{A}^0]_0&=-\tilde m_1\langle\eta_1 3\rangle[32],&
[\mathbf{J}^{--}]_{1}\cdot [\mathbf{A}^0]_0&=m_2\langle13\rangle[3\eta_2]\\
[\mathbf{J}^{-+}]_0\cdot [\mathbf{A}^-]_{1}&=-\tilde m_3 \langle1 \eta_3\rangle[32],&
[\mathbf{J}^{-+}]_0\cdot [\mathbf{A}^+]_{1}&=-m_3 \langle1 3\rangle[\eta_32],\\
[\mathbf{J}^{-+}]_0\cdot [\mathbf{A}^0]_2&= -m_3 \tilde m_3\langle1 \eta_3\rangle[\eta_32],\\
&\cdots
\end{aligned} \end{equation}

Thus, the MHC amplitude in eq.~\eqref{eq:FFV_expand} can be expressed in terms of the MHC current and the MHC polarization vector,
\begin{equation} \begin{aligned}
\mathcal{M}(FFV)
&\sim \mathbf c_{13} [\mathbf{J}^{-+}]_0 \cdot[\mathbf{A}^{0}]_0 + \mathbf c_{14} [\mathbf J^{++}]_1 \cdot[\mathbf{A}^{0}]_0 + \mathbf c_{15} [\mathbf J^{--}]_1 \cdot[\mathbf{A}^{0}]_0 + \mathbf c_{16} [\mathbf{J}^{-+}]_0 \cdot[\mathbf A^+]_1 \\
&+ \mathbf c_{17} [\mathbf{J}^{-+}]_0 \cdot[\mathbf A^-]_1 + \frac{\mathbf c_{13}}{\mathbf m_3^2}[\mathbf{J}^{-+}]_0 \cdot[\mathbf{A}^{0}]_0 + \mathbf c_{18} [\mathbf{J}^{+-}]_2 \cdot[\mathbf{A}^{0}]_0 + \mathbf c_{19} [\mathbf J^{++}]_1 \cdot[\mathbf A^+]_1 \\
&+ \mathbf c_{20} [\mathbf J^{++}]_1 \cdot[\mathbf A^-]_1 + \mathbf c_{21} [\mathbf J^{--}]_1 \cdot[\mathbf A^+]_1 + \mathbf c_{22} [\mathbf J^{--}]_1 \cdot[\mathbf A^-]_1 + \frac{\mathbf c_{14}}{\mathbf m_3^2}[\mathbf J^{++}]_1 \cdot[\mathbf{A}^{0}]_0  \\
& + \frac{\mathbf c_{15}}{\mathbf m_3^2}[\mathbf J^{--}]_1 \cdot[\mathbf{A}^{0}]_0 + \mathbf c_{23} [\mathbf{J}^{+-}]_2 \cdot[\mathbf A^+]_1 + \mathbf c_{24} [\mathbf{J}^{+-}]_2 \cdot[\mathbf A^-]_1 + \frac{\mathbf c_{18}}{\mathbf m_3^2}[\mathbf{J}^{+-}]_2 \cdot[\mathbf{A}^{0}]_0
\end{aligned} \end{equation}
The contraction of the MHC current and the MHC polarization vector correspond to the MHC diagram
\begin{align}
[\mathbf{J}^{-+}]_0\cdot [\mathbf{A}^0]_0 &= \Ampthree{1^-}{2^+}{3^0}{\fer{red}{i1}{v1}}{\antfer{red}{i2}{v1}}{\bos{i3}{brown}},& \\
\relax
[\mathbf{J}^{++}]_{1}\cdot [\mathbf{A}^0]_0 &=
\Ampthree{1^+}{2^+}{3^0}{\ferflip{1}{180}{cyan}{red}}{\antfer{red}{i2}{v1}}{\bos{i3}{brown}},&
[\mathbf{J}^{--}]_{1}\cdot [\mathbf{A}^0]_0 &=
\Ampthree{1^-}{2^-}{3^0}{\fer{red}{i1}{v1}}{\antferflip{1}{55}{cyan}{red}}{\bos{i3}{brown}},\\
\relax
[\mathbf{J}^{-+}]_0\cdot [\mathbf{A}^+]_{1} &= \Ampthree{1^-}{2^+}{3^+}{\fer{red}{i1}{v1}}{\antfer{red}{i2}{v1}}{\bosflip{1}{-55}{brown}{cyan}},&
[\mathbf{J}^{-+}]_0\cdot [\mathbf{A}^-]_{1} &= \Ampthree{1^-}{2^+}{3^-}{\fer{red}{i1}{v1}}{\antfer{red}{i2}{v1}}{\bosflip{1}{-55}{brown}{red}},\\
\relax
[\mathbf{J}^{-+}]_0\cdot [\mathbf{A}^0]_2 &= \Ampthree{1^-}{2^+}{3^0}{\fer{red}{i1}{v1}}{\antfer{red}{i2}{v1}}{\bosflipflip{1}{-55}{brown}{red}{brown}},& &\cdots
\end{align}

For another structure $\langle\mathbf{23}\rangle[\mathbf{31}]$, the primary representation is 
\begin{equation} \begin{aligned}
\Ampthree{1^+}{2^-}{3^0}{\fer{cyan}{i1}{v1}}{\antfer{cyan}{i2}{v1}}{\bos{i3}{brown}}=\langle23\rangle[31].
\end{aligned} \end{equation}
It corresponds to the conserved current with the opposite helicity to eq.~\eqref{eq:current_1},
\begin{equation} \begin{aligned}
J^{{\alpha\dot\alpha}}(1^{+\frac12},2^{-\frac12})=-[1|^{\dot\alpha}\langle2|^{\alpha}.
\end{aligned} \end{equation}
Similarly, we can extend it to the massive current. Expand it and the polarization vector with chirality-helicity unification, their contraction corresponds to the other MHC amplitudes in $FFV$ amplitudes.

\subsection{Leading $FFV$ massless-massive matching and amplitude deformation}
\label{sec:FFV_leading}

Given the current and polarization vector decomposition, let us first discuss the connection between the massive vector and massless state. Recall that, the MHC state for a massive vector with helicity $h$ is defined as
\begin{equation}
\mathbf{A} =  
\left\{\begin{aligned}      
h&=+1:& [\mathbf{A}^+]_1&= \tilde m_3|\eta_3\rangle^{\alpha}|3]^{\dot\alpha} \\
h&=0:& [\mathbf{A}^0]_0&=|3\rangle^{\alpha}|3]^{\dot\alpha},\qquad
[\mathbf{A}^0]_2= m_3 \tilde m_3|\eta_3\rangle^{\alpha}|\eta_3]^{\dot\alpha}, \\
h&=-1:& [\mathbf{A}^-]_1&= m_3|3\rangle^{\alpha}|\eta_3]^{\dot\alpha},
\end{aligned}\right.
\end{equation}
The transverse polarizations ($h = \pm 1$) are described by a single term $[\mathbf{A}]_1$, while the longitudinal polarization ($h = 0$) comprises two terms, $[\mathbf{A}]_0$ and $[\mathbf{A}]_2$. In the leading matching of the $FFV$ amplitude, it is sufficient to consider the leading contribution from each polarization, namely $[\mathbf{A}^0]_0$ and $[\mathbf{A}^{\pm}]_1$.

Let us first match $[\mathbf{A}^0]_0$ with zero helicity. For the massless spin-1 particle, the only zero helicity state is the derivative of massless scalar $\partial_\mu \phi$, which can be recognize to be the Goldstone boson. Thus we expect to match the $[\mathbf{A}^0]_0$ with the  zero helicity massless spin-1 state
\begin{eqnarray}
\begin{tikzpicture}[baseline=0.7cm]
\begin{feynhand}
\vertex [dot] (v1) at (-0.4,0.8) {};
\vertex [particle] (i2) at (1.3,0.8) {$i^0$};
\graph{(v1) --[sca] (i2)};
\end{feynhand}
\end{tikzpicture} = 
\partial_{\alpha, \dot{\alpha}} \phi  \equiv \lambda_{\alpha}\tilde{\lambda}_{\dot{\alpha}}
\quad \rightarrow \quad
\begin{tikzpicture}[baseline=0.7cm]
\begin{feynhand}
\setlength{\feynhandblobsize}{6mm}
\vertex [dot] (v1) at (-0.4,0.8) {};
\vertex [particle] (i2) at (1.3,0.8) {$i^{0}$};
\bos{i2}{brown};
\end{feynhand}
\end{tikzpicture},
\label{eq:goldstone-vector-match}
\end{eqnarray}
which denotes a massless Goldstone boson. It couples to the $[\mathbf{J}^{-+}]_0$ as the conserved current, although it also couples to the sub-leading currents $[\mathbf{J}^{\pm\pm}]_1$ and  $[\mathbf{J}^{+-}]_2$, which are non-vanishing.

Then for sub-leading term $[\mathbf{A}^{\pm}]_1$, it should match to the particle state from massless gauge boson. The massless gauge boson has only two helicity states, with the polarization vectors 
\begin{eqnarray} \label{eq:massless_polarization}
A_{\alpha\dot{\alpha}}^+: \frac{\xi_{\alpha}\tilde{\lambda}_{\dot{\alpha}}}{\langle \xi \lambda \rangle }, \quad 
A_{\alpha\dot{\alpha}}^-: -\frac{\lambda_{\alpha}\tilde{\xi}_{\dot{\alpha}}}{[ \tilde{\lambda} \tilde{\xi} ] }.
\end{eqnarray}
where superscript $\pm$ denotes the helicity of the gauge boson. The reference spinor $\xi$ can be interpreted as fixing the light-cone gauge condition $A^\mu \xi_\mu=0$. By taking $\xi$ to be $\eta$, these two massless helicity states are identified as 
\begin{eqnarray}
\begin{tikzpicture}[baseline=0.7cm]
\begin{feynhand}
\vertex [dot] (v1) at (-0.4,0.8) {};
\vertex [particle] (i1) at (-0.4,0.8) {};
\vertex [particle] (i2) at (1.3,0.8) {$i^{+1}$};
\graph{(i1) --[bos] (i2)};
\end{feynhand}
\end{tikzpicture} = 
\frac{\eta_{\alpha}\tilde{\lambda}_{\dot{\alpha}}}{\langle \eta \lambda \rangle },  \quad \quad
\begin{tikzpicture}[baseline=0.7cm]
\begin{feynhand}
\vertex [dot] (v1) at (-0.4,0.8) {};
\vertex [particle] (i1) at (-0.4,0.8) {};
\vertex [particle] (i2) at (1.3,0.8) {$i^{-1}$};
\graph{(i1) --[bos] (i2)};
\end{feynhand}
\end{tikzpicture} = 
 -\frac{\lambda_{\alpha}\tilde{\eta}_{\dot{\alpha}}}{[ \tilde{\lambda} \tilde{\eta} ] }.
\end{eqnarray}
The form is similar to the transverse polarizations of the massive vector, but they have the different mass dimension.
For positive polarization, the transverse massless gauge boson can be deformed as follows
\begin{eqnarray}
\begin{tikzpicture}[baseline=0.7cm]
\begin{feynhand}
\vertex [dot] (v1) at (-0.4,0.8) {};
\vertex [particle] (i1) at (-0.4,0.8) {};
\vertex [particle] (i2) at (1.3,0.8) {$i^{+1}$};
\graph{(i1) --[bos] (i2)};
\end{feynhand}
\end{tikzpicture} 
= 
\frac{\eta_{\alpha}\tilde{\lambda}_{\dot{\alpha}}}{\langle \eta \lambda \rangle }
= \frac{1}{\mathbf m^2}\times\tilde m\eta_{\alpha}\tilde{\lambda}_{\dot{\alpha}} 
\quad \rightarrow \quad
\begin{tikzpicture}[baseline=-0.1cm] \begin{feynhand}
\setlength{\feynhandblobsize}{6mm}
\setlength{\feynhandarrowsize}{5pt}
\vertex [particle] (i1) at (2,0) {$i^{+1}$};
\vertex (v2) at (1,0);
\vertex [dot] (v1) at (0,0) {};
\begin{scope}[]
\clip (0,-0.1) rectangle (1,0.1); 
\draw[brown,thick bos] (v1)--(i1);
\end{scope}
\begin{scope}[]
\clip (1,-0.1) rectangle (1.6,0.1); 
\draw[cyan,thick bos] (v1)--(i1);
\end{scope}
\draw[very thick] plot[mark=x,mark size=2.5] coordinates {(1,0)};
\draw (0.55-0.03,-0.08) -- (0.55+0.03,+0.08);
\end{feynhand} \end{tikzpicture},
\end{eqnarray}
where the MHC amplitude has both chirality and helicity flip, which is sub-leading compared to the Eq.~\eqref{eq:goldstone-vector-match}. This agrees with the power counting $[\mathbf{A}^{\pm}]_1 \ll [\mathbf{A}^0]_0$.  
Another polarization gives
\begin{eqnarray}
    \begin{tikzpicture}[baseline=0.7cm]
\begin{feynhand}
\vertex [dot] (v1) at (-0.4,0.8) {};
\vertex [particle] (i1) at (-0.4,0.8) {};
\vertex [particle] (i2) at (1.3,0.8) {$i^{-1}$};
\graph{(i1) --[bos] (i2)};
\end{feynhand}
\end{tikzpicture} = 
-\frac{\lambda_{\alpha}\tilde{\eta}_{\dot{\alpha}}}{[ \tilde{\lambda} \tilde{\eta} ] } \quad \rightarrow \quad
\begin{tikzpicture}[baseline=-0.1cm] \begin{feynhand}
\setlength{\feynhandblobsize}{6mm}
\setlength{\feynhandarrowsize}{5pt}
\vertex [particle] (i1) at (2,0) {$i^{-1}$};
\vertex (v2) at (1,0);
\vertex [dot] (v1) at (0,0) {};
\begin{scope}[]
\clip (0,-0.1) rectangle (1,0.1); 
\draw[brown,thick bos] (v1)--(i1);
\end{scope}
\begin{scope}[]
\clip (1,-0.1) rectangle (1.6,0.1); 
\draw[red,thick bos] (v1)--(i1);
\end{scope}
\draw[very thick] plot[mark=x,mark size=2.5] coordinates {(1,0)};
\draw (0.55-0.03,-0.08) -- (0.55+0.03,+0.08);
\end{feynhand} \end{tikzpicture}.
\end{eqnarray}

After building the connection between massless and massive vector boson, it is ready to match the 3-point massless amplitudes with the leading non-vanishing MHC amplitudes. Focusing on the case of $\mathbf J\cdot\mathbf A=[\mathbf{23}]\langle\mathbf{31}\rangle$, let us list the primary and 1st descendant MHC amplitudes,
\begin{eqnarray}
\mathbf{J}\cdot \mathbf{A} &\rightarrow &
\mathbf c_{13}[\mathbf{J}^{-+}]_0\cdot [\mathbf{A}^0]_0+
\mathbf c_{14}[\mathbf{J}^{++}]_{1}\cdot [\mathbf{A}^0]_0+
\mathbf c_{15}[\mathbf{J}^{--}]_{1}\cdot [\mathbf{A}^0]_0+
\mathbf c_{16}[\mathbf{J}^{-+}]_0\cdot [\mathbf{A}^+]_{1} \nonumber \\
&&+\mathbf c_{17}[\mathbf{J}^{-+}]_0\cdot [\mathbf{A}^-]_{1}
+ \cdots 
\end{eqnarray}
They correspond to the following MHC diagrams,
\begin{eqnarray}
\Ampthree{1^-}{2^+}{3^0}{\fer{red}{i1}{v1}}{\antfer{red}{i2}{v1}}{\bos{i3}{brown}} +
\Ampthree{1^+}{2^+}{3^0}{\ferflip{1}{180}{cyan}{red}}{\antfer{red}{i2}{v1}}{\bos{i3}{brown}} +
\Ampthree{1^-}{2^-}{3^0}{\fer{red}{i1}{v1}}{\antferflip{1}{55}{cyan}{red}}{\bos{i3}{brown}}+
\Ampthree{1^-}{2^+}{3^+}{\fer{red}{i1}{v1}}{\antfer{red}{i2}{v1}}{\bosflip{1}{-55}{brown}{cyan}}
+\Ampthree{1^-}{2^+}{3^-}{\fer{red}{i1}{v1}}{\antfer{red}{i2}{v1}}{\bosflip{1}{-55}{brown}{red}}
\end{eqnarray}
The first diagram is the primary MHC amplitude with the helicity combination $(-,+,0)$. It corresponds to vanishing conserved current, and thus no corresponding massless amplitude. The other four diagrams are 1st descendant MHC amplitudes with helicity combinations $(\pm,\pm,0)$ and $(-,+,\pm)$. The corresponding massless 3-point amplitudes on the these helicity categories are
\begin{equation}
\begin{tikzpicture}[baseline=0.7cm] \begin{feynhand}
\setlength{\feynhandarrowsize}{3.5pt}
\vertex [particle] (i1) at (0,0.8) {$1^-$}; 
\vertex [particle] (i2) at (1.6,1.6) {$2^+$}; 
\vertex [particle] (i3) at (1.6,0) {$3^\pm$};  
\vertex (v1) at (0.9,0.8); 
\graph{(i1)--[fer](v1)--[fer](i2)};
\graph{(i3)--[bos] (v1)};  
\end{feynhand} \end{tikzpicture},\quad
\begin{tikzpicture}[baseline=0.7cm] \begin{feynhand}
\setlength{\feynhandarrowsize}{3.5pt}
\vertex [particle] (i1) at (0,0.8) {$1^\pm$}; 
\vertex [particle] (i2) at (1.6,1.6) {$2^\pm$}; 
\vertex [particle] (i3) at (1.6,0) {$3^0$};  
\vertex (v1) at (0.9,0.8); 
\graph{(i1)--[fer](v1)--[fer](i2)};
\graph{(i3)--[sca] (v1)};  
\end{feynhand} \end{tikzpicture}.
\end{equation}
In the following, we will take $(-\frac12,+\frac12,+ 1)$ and $(-\frac12,-\frac12,0)$ to show the matching procedure, while the other two take the same routine. 

\paragraph{Transverse Gauge Boson Matching}

First let us match the helicity  $(-\frac12,+\frac12,+1)$ between the massless and MHC amplitude. Given the helicity configuration for both massless and MHC amplitudes, the corresponding diagrams tell
\begin{equation} \begin{aligned} \label{eq:FFV_to_FFV}
\begin{tikzpicture}[baseline=0.7cm] \begin{feynhand}
\setlength{\feynhandarrowsize}{3.5pt}
\vertex [particle] (i1) at (0,0.8) {$1^-$}; 
\vertex [particle] (i2) at (1.6,1.6) {$2^+$}; 
\vertex [particle] (i3) at (1.6,0) {$3^+$};  
\vertex (v1) at (0.9,0.8); 
\graph{(i1)--[fer](v1)--[fer](i2)};
\graph{(i3)--[bos] (v1)};  
\end{feynhand} \end{tikzpicture}
\quad \Longrightarrow \quad [\mathbf{J}^{-+}]_0\cdot [\mathbf{A}^+]_{1}
=
\Ampthree{1^-}{2^+}{3^+}{\fer{red}{i1}{v1}}{\antfer{red}{i2}{v1}}{\bosflip{1}{-55}{brown}{cyan}}
.
\end{aligned} \end{equation}
At the same time, the above diagrams have the corresponding amplitudes
\begin{equation}\label{eq:FFVgaugeleadingmatch}
    \frac{[23]^2}{[12]}
\quad \Longrightarrow \quad  \tilde m_3\langle1\eta_3\rangle[23].
\end{equation}
Therefore, let us write the massless $FFV$ amplitude with the coefficient $T_f$ from the UV physics
\begin{eqnarray}
\begin{tikzpicture}[baseline=0.7cm] \begin{feynhand}
\setlength{\feynhandarrowsize}{3.5pt}
\vertex [particle] (i1) at (0,0.8) {$1^-$}; 
\vertex [particle] (i2) at (1.6,1.6) {$2^+$}; 
\vertex [particle] (i3) at (1.6,0) {$3^+$};  
\vertex (v1) at (0.9,0.8); 
\graph{(i1)--[fer](v1)--[fer](i2)};
\graph{(i3)--[bos] (v1)};  
\end{feynhand} \end{tikzpicture} = 
T_f\frac{[23]^2}{[12]}, \label{eq:FFV1}
\end{eqnarray}
and perform the matching to massive MHC amplitude.
To match the above amplitudes, the following criteria for the massless amplitude deformation should be satisfied:  
\begin{itemize}
    \item The appearance of the $\eta_3$ in the MHC amplitude but not the massless one indicates an amplitude deformation should be taken on particle 3, which can also be seen from the mass insertion on particle 3 in the MHC diagram; 
\item The spurious pole $[12]$ in the massless amplitude, but not existing in the MHC one, indicates that it should be eliminated by either the IBP, the EOM or the Schouten identity.  
\end{itemize}

From the above criteria, let us perform the matching in an intuitive way. The mass insertion on the particle 3 in the MHC amplitude indicates that the amplitude deformation can be taken by multiplying the massless amplitude with the unit quantity $\frac{\langle\eta_3 3\rangle}{m_3}$, and then to eliminate the spurious pole, the IBP $p_3 = |3\rangle [3|= - p_1 - p_2$ can be taken. Thus the amplitude deformation has 
\begin{equation} \label{eq:FFVdeformationmatching}
\frac{[23]^2}{[12]}=\frac{\langle\eta_3 3\rangle}{m_3}\times \frac{[32]^2}{[12]} = \frac{1}{m_3}\times \frac{\langle \eta_3|3|2][32]}{[12]}\overset{\text{IBP}}{=}\frac{1}{m_3}\times[23]\langle\eta_3 1\rangle= -\frac{1}{\mathbf{m}_3^2}\times \tilde m_3\langle1\eta_3\rangle[23].
\end{equation}
This deformation eliminates the spurious pole in the massless amplitudes, yielding an explicitly local massive MHC amplitude. The amplitude matching gives the additional coefficient $1/\mathbf{m}_3^2$. Comparing the eq.~\eqref{eq:FFVgaugeleadingmatch} with eq.~\eqref{eq:FFVdeformationmatching}, the matching tells the coefficients for this helicity amplitudes
\begin{equation}
\label{eq:FFVleading1}
\mathbf{c}_{16}=\frac{T_f}{\mathbf{m}_3^2},\quad
\mathbf{c}_{17}=\frac{T_f}{\mathbf{m}_3^2}.
\end{equation}

Note that the above amplitude deformation does not utilize the correspondence between massless and massive gauge boson. Here we understand the amplitude deformation by connecting the massless gauge boson to the transverse part of massive gauge boson. To this end, in the massless amplitudes, a reference spinor $\xi$ for the massless gauge boson, served as the gauge parameter, is introduced by deforming the spinor structure of the gauge boson as in eq.~\eqref{eq:massless_polarization}. This allows us to rewrite the massless amplitude, by multiplying the unit quantity $\frac{\langle\xi 3\rangle}{\langle\xi 3\rangle}$, as
\begin{eqnarray} 
\frac{[23]^2}{[12]}&=&\frac{[23][2|3|\xi\rangle}{[12]\langle3\xi\rangle}=\frac{[23]\langle1\xi\rangle}{\langle3\xi\rangle}
=-\langle1|^{\alpha}[2|^{\dot\alpha}\times \frac{|\xi\rangle_{\alpha}|3]_{\dot{\alpha}}}{\langle \xi 3 \rangle} ,
\end{eqnarray}
where the current $\langle1|^{\alpha}[2|^{\dot\alpha}$ is also identified along with the gauge boson.
Here, the spurious pole becomes $\langle\xi3\rangle$, which is absorbed into the massless polarization vector. By taking $\xi$ to be $\eta_3$, which indicates that the light-cone gauge-fixing is chosen, the $\langle\xi3\rangle$ becomes a spurion mass $ m_3=\langle\eta_3 3\rangle$, eliminating the spurious pole, so this massless helicity state matches to the transverse polarization part of a massive vector.
Therefore the massless amplitude deforms to the final expression
\begin{equation}
\left.-\frac{[23]\langle1\xi\rangle}{\langle3\xi\rangle}\right|_{\xi=\eta_3}=-\frac{[23]\langle1\eta_3\rangle}{m_3}= -\tilde m_3\times \frac{\langle1\eta_3\rangle[23]}{\mathbf{m}_3^2}.
\end{equation}

In the above treatment, the gauge boson $A$ and the current $J$ are separated in both the massless and massive amplitudes. However, for the massless gauge boson, a spurious pole needs to be specified, which makes the deformation complicated. In the following, we take a systematic approach to separate the gauge boson and the current by identifying the scaling structure of each particles in the massless and massive amplitudes.

The procedure of this systematic approach is as follows. The 3-pt amplitude depends on the spinors $\lambda$ or $\tilde\lambda$, $(m\tilde\eta)$ and $(\tilde{m}\eta)$. Thus its scaling behavior can be read out directly from the analytic expression. For the massless amplitude with the helicity $(-\frac12,+\frac12,+1)$, the spinor scaling reads
\begin{eqnarray} 
\begin{tabular}{ccc}
\begin{tikzpicture}[baseline=0.7cm] \begin{feynhand}
\setlength{\feynhandarrowsize}{3.5pt}
\vertex [particle] (i1) at (0,0.8) {$1^-$}; 
\vertex [particle] (i2) at (1.6,1.6) {$2^+$}; 
\vertex [particle] (i3) at (1.6,0) {$3^+$};  
\vertex (v1) at (0.9,0.8); 
\graph{(i1)--[fer](v1)--[fer](i2)};
\graph{(i3)--[bos] (v1)};  
\end{feynhand} \end{tikzpicture}$\displaystyle=\frac{[23]^2}{[12]}$
& $\Rightarrow$\qquad &
$\left\{\begin{aligned}
\text{particle 1}:\quad &\tilde\lambda_1^{-1},&\\
\text{particle 2}:\quad &\tilde\lambda_2,& \\
\text{particle 3}:\quad &\tilde\lambda_3^2.& \\
\end{aligned}\right.$  \\
\end{tabular}
\end{eqnarray}
while the massive amplitude reads
\begin{eqnarray} 
\begin{tabular}{ccc}
\Ampthree{1^-}{2^+}{3^+}{\fer{red}{i1}{v1}}{\antfer{red}{i2}{v1}}{\bosflip{1}{-55}{brown}{cyan}}$\displaystyle=
\tilde m_2[\eta_23]\langle31\rangle$
& $\Rightarrow$\qquad &
$\left\{\begin{aligned}
\text{particle 1}:\quad &\lambda_1,&\\
\text{particle 2}:\quad &\tilde\lambda_2,& \\
\text{particle 3}:\quad &\tilde\lambda_3(\tilde m_3\eta_3),& \\
\end{aligned}\right.$  \\
\end{tabular}
\end{eqnarray}

When matching the massless amplitude to the MHC amplitude, we should identify the vector and the current,
\begin{eqnarray} 
\begin{tikzpicture}[baseline=0.7cm] \begin{feynhand}
\setlength{\feynhandarrowsize}{3.5pt}
\vertex [particle] (i1) at (0,0.8) {$1^-$}; 
\vertex [particle] (i2) at (1.6,1.6) {$2^+$}; 
\vertex [particle] (i3) at (1.6,0) {$3^+$};  
\vertex (v1) at (0.9,0.8); 
\graph{(i1)--[fer](v1)--[fer](i2)};
\graph{(i3)--[bos] (v1)};  
\end{feynhand} \end{tikzpicture}\to
\Ampthree{1^-}{2^+}{3^+}{\fer{red}{i1}{v1}}{\antfer{red}{i2}{v1}}{\bosflip{1}{-55}{brown}{cyan}}\qquad
\begin{aligned}
&\text{current $\mathbb J$: particles 1 and 2} \\
&\text{vector $\mathbb A$: particle 3}
\end{aligned}
\end{eqnarray}
Since the scaling of the particle 3 is known for both the massless and massive one, we can identify the scaling behavior for the current  and the vector
\begin{eqnarray} 
\begin{tabular}{c|cc}
\hline
 & current scaling & vector scaling  \\
\hline
massless & $\mathbb J(\tilde\lambda^0)$ &  $\mathbb A(\tilde\lambda_3^2)$  \\
MHC & $\mathbb J(\tilde\lambda \lambda)$ & $\mathbb A(\tilde\lambda_3\tilde m_3\eta_3)$  \\
\hline
\end{tabular}
\end{eqnarray}
Note that in the current $\mathbb{J}$, we drop the label of spinor, e.g. $\tilde\lambda^0\sim \tilde\lambda_1^{-1}\tilde\lambda_2$.

The key point is that the deformation needs to be taken from the $\mathbb A(\tilde\lambda_3^2)$ to $\mathbb A(\tilde\lambda_3\tilde m_3\eta_3)$. 
To do this, we multiply the vector $\mathbb{A}$ by a factor $\frac{\langle \lambda \eta \rangle}{m}$ and then apply the IBP to reach the scaling behavior of the massive vector:
\begin{equation} \begin{aligned}
\mathbb J(\tilde\lambda^{0})\cdot\mathbb A^+(\tilde\lambda_3^2) 
&\xrightarrow{\times\frac{\langle\lambda\eta\rangle}{m}}
\mathbb J(\tilde\lambda^{0})\cdot\mathbb A^+(\tilde\lambda_3 \tilde\lambda_3^2 m_3^{-1}\eta_3) \\
&\xrightarrow{\text{IBP}}
\mathbb J(\tilde\lambda \lambda)\cdot\mathbb A^+(\tilde\lambda_3 m_3^{-1}\eta_3) \\
&=
\mathbb J(\tilde\lambda \lambda)\cdot\frac{1}{\mathbf m_3^2}\mathbb A^+(\tilde\lambda_3 \tilde m_3\eta_3) \\
\end{aligned} \end{equation}
Guided by the scaling behavior, the deformation for the amplitudes take the same flow as below
\begin{equation}
\frac{[23]^2}{[12]}
\xrightarrow{\times\frac{\langle\lambda\eta\rangle}{m}}\frac{\langle \eta_3|3|2][32]}{m_3[12]}
\xrightarrow{\text{IBP}^-}\frac{[23]\langle\eta_3 1\rangle}{m_3}=\frac{1}{\mathbf m_3^2}\times\tilde m_3[23]\langle\eta_3 1\rangle.
\end{equation}

\paragraph{Goldstone Boson Matching}

Then for another helicity $(-\frac12,-\frac12, 0)$, picking up the same helicity configuration for both massless and MHC ones,  the matching should be taken as  
\begin{equation} \begin{aligned}
\begin{tikzpicture}[baseline=0.7cm] \begin{feynhand}
\setlength{\feynhandarrowsize}{3.5pt}
\vertex [particle] (i1) at (0,0.8) {$1^-$}; 
\vertex [particle] (i2) at (1.6,1.6) {$2^-$}; 
\vertex [particle] (i3) at (1.6,0) {$3^0$};  
\vertex (v1) at (0.9,0.8); 
\graph{(i1)--[fer](v1)--[fer](i2)};
\graph{(i3)--[sca] (v1)};  
\end{feynhand} \end{tikzpicture} =  \langle12\rangle \rightarrow [\mathbf{J}^{--}]_{1}\cdot [\mathbf{A}^0]_0 =
\Ampthree{1^-}{2^-}{3^0}{\fer{red}{i1}{v1}}{\antferflip{1}{55}{cyan}{red}}{\bos{i3}{brown}}
=m_2\langle13\rangle[3\eta_2],
\end{aligned} \end{equation}
which indicates that the matching should be taken from the scalar boson in the UV to the longitudinal part of the vector boson. Thus we expect the scalar boson should be the Goldstone boson. Similarly the mass insertion on the MHC diagram for the particle 2, and also the $\eta_2$ in the MHC amplitude, tells that the amplitude deformation should be taken by multiplying the $\frac{[\eta_2 2]}{\tilde m_2}$. Since the deformation is taken to match the scalar boson to the Goldstone dof of the vector boson, we will discuss it in detail too.

In the UV, the massless $FFS$ amplitude is
\begin{equation}
\begin{tikzpicture}[baseline=0.7cm] \begin{feynhand}
\setlength{\feynhandarrowsize}{3.5pt}
\vertex [particle] (i1) at (0,0.8) {$1^-$}; 
\vertex [particle] (i2) at (1.6,1.6) {$2^-$}; 
\vertex [particle] (i3) at (1.6,0) {$3^0$};  
\vertex (v1) at (0.9,0.8); 
\graph{(i1)--[fer](v1)--[fer](i2)};
\graph{(i3)--[sca] (v1)};  
\end{feynhand} \end{tikzpicture} = Y \langle12\rangle,
\label{eq:FFV-deformation-Goldstone}
\end{equation}
where $Y$ denotes the Yukawa coefficient in the UV. The good thing is that there is no spurious pole in the UV amplitude, but since the scalar should be matched to the Goldstone boson, we expect that an explicit momentum behavior $[\mathbf A^0]_0=|3 \rangle \langle 3|$ should be kept in the MHC amplitude. Given the mass insertion for particle 2, we perform the following amplitude deformation by multiplying the $\frac{[\eta_2 2]}{\tilde m_2}$
\begin{equation}
\langle12\rangle=\frac{[\eta_2 2]}{\tilde m_2}\times\langle12\rangle=-\frac{1}{\tilde{m}_2}\times[\eta_2|2|1\rangle
\overset{\text{IBP}}{=}\frac{1}{\tilde{m}_2}\times[\eta_23]\langle31\rangle
= -\frac{1}{\mathbf{m}_2^2}\times
m_2[\eta_23]\langle31\rangle.
\end{equation}
From the right-handed side, the momentum $p_3 = |3\rangle [3|$ can be recognized as the Goldstone boson and then the longitudinal gauge boson using the Goldstone equivalence theorem. From the right-handed side to the left, we note that the Goldstone amplitude with the derivative coupling can be rewritten as the scalar Yukawa coupling, valid only for 3-point interactions. The massless amplitude matching determines the coefficients
\begin{equation}
\label{eq:FFVleading2}
\mathbf{c}_{14}=\frac{Y}{\mathbf{m}_2^2},\quad
\mathbf{c}_{15}=\frac{Y}{\mathbf{m}_1^2},
\end{equation}
Thus the matching tells how to relate the coefficients of the MHC amplitudes by the coefficients of the UV massless amplitudes.

A similar systematic approach can be taken to separate the Goldstone boson and the current. The massless amplitude with helicity $(-\frac12,-\frac12,0)$ reads
\begin{eqnarray} 
\begin{tabular}{ccc}
\begin{tikzpicture}[baseline=0.7cm] \begin{feynhand}
\setlength{\feynhandarrowsize}{3.5pt}
\vertex [particle] (i1) at (0,0.8) {$1^-$}; 
\vertex [particle] (i2) at (1.6,1.6) {$2^-$}; 
\vertex [particle] (i3) at (1.6,0) {$3^0$};  
\vertex (v1) at (0.9,0.8); 
\graph{(i1)--[fer](v1)--[fer](i2)};
\graph{(i3)--[sca] (v1)};  
\end{feynhand} \end{tikzpicture}
$\displaystyle=
\langle12\rangle$
& $\Rightarrow$ \qquad &
$\left\{\begin{aligned}
\text{particle 1}:\quad &\lambda_1,&\\
\text{particle 2}:\quad &\lambda_2,& \\
\text{particle 3}:\quad & 1,& \\
\end{aligned}\right.$  \\
\end{tabular}
\end{eqnarray}
while the MHC amplitude read
\begin{eqnarray} 
\begin{tabular}{ccc}
\Ampthree{1^-}{2^-}{3^0}{\fer{red}{i1}{v1}}{\antferflip{1}{55}{cyan}{red}}{\bos{i3}{brown}}
$\displaystyle=
m_2[\eta_23]\langle31\rangle$
& $\Rightarrow$ \qquad &
$\left\{\begin{aligned}
\text{particle 1}:\quad &\lambda_1,&\\
\text{particle 2}:\quad &m_2\tilde\eta_2,& \\
\text{particle 3}:\quad & \tilde\lambda_3\lambda_3,& \\
\end{aligned}\right.$  \\
\end{tabular}
\end{eqnarray}
Identify the gauge boson to be the particle 3, and combine the particles 1 and 2 into a current, and we obtain the scaling 
\begin{eqnarray} 
\begin{tabular}{c|cc}
\hline
 & current scaling & vector scaling  \\
\hline
massless & $\mathbb J(\lambda^2)$ &  $ \mathbb A(1)$  \\
MHC & $\mathbb J(\lambda m\tilde\eta)$ & $\mathbb A(\tilde\lambda_3\lambda_3)$  \\
\hline
\end{tabular}
\end{eqnarray}

For the Goldstone boson, we need to convert the scaling $\mathbb J(\lambda^2)$ to $\mathbb J(\lambda m\tilde\eta)$ for the current. This should be done by multiplying the current $\mathbb{J}$ by a factor $\frac{[\lambda\eta]}{\tilde m}$ and apply then the IBP to convert the scalar to be the longitudinal Goldstone boson
\begin{equation} \begin{aligned}
\mathbb J(\lambda^2)\cdot\mathbb A(1) 
&\xrightarrow{\times\frac{[\lambda\eta]}{\tilde m}}
\mathbb J(\tilde\lambda \lambda^2 \tilde m^{-1}\tilde\eta)\cdot\mathbb A(1) \\
&\xrightarrow{\text{IBP}}
\mathbb J(\lambda \tilde m^{-1}\tilde\eta)\cdot\mathbb A(\tilde \lambda \lambda) \\
&=\frac{1}{\mathbf m^2}\mathbb J(\lambda m\tilde\eta)\cdot\mathbb A(\tilde \lambda_3 \lambda_3) 
\end{aligned} \end{equation}
Then the amplitude deforms under this scaling behavior as
\begin{equation}
\langle12\rangle\xrightarrow{\times\frac{[\lambda\eta]}{\tilde m}}-\frac{[\eta_2|2|1\rangle}{\tilde{m}_2}
\xrightarrow{\text{IBP}^+}\frac{[\eta_23]\langle31\rangle}{\tilde{m}_2}= -\frac{1}{\mathbf{m}_2^2}\times
m_2[\eta_23]\langle31\rangle.
\end{equation}

\paragraph{Relations among Leading Amplitudes}

From the above the leading MHC amplitudes are deformed to the following massless amplitudes: 
the massless $FFV$
\begin{eqnarray}
(-\frac12,+\frac12,+1): \quad \frac{\tilde{m}_3}{{\textbf{m}^2_3}} \langle1\eta_3\rangle [32] \to  T_f \frac{[23]^2}{[12]}, \\
(-\frac12,+\frac12,-1):\quad \frac{{m}_3}{{\textbf{m}_3^2}}\langle13\rangle [\eta_32]
\to  T_f \frac{\langle 23\rangle^2}{\langle 12\rangle},
\end{eqnarray}
and the massless $FFS$ 
\begin{eqnarray}
(+\frac12,+\frac12,0): \quad \frac{\tilde{m}_1}{\textbf{m}_1^2}\langle\eta_13\rangle [32]
\to Y [12], \\
 (-\frac12,-\frac12,0): \quad\frac{m_2}{{\textbf{m}_2^2}} \langle13\rangle [3\eta_2] 
\to Y \langle 12\rangle.
\end{eqnarray}

\begin{figure}[h]
\centering
\includegraphics[width=0.6\linewidth]{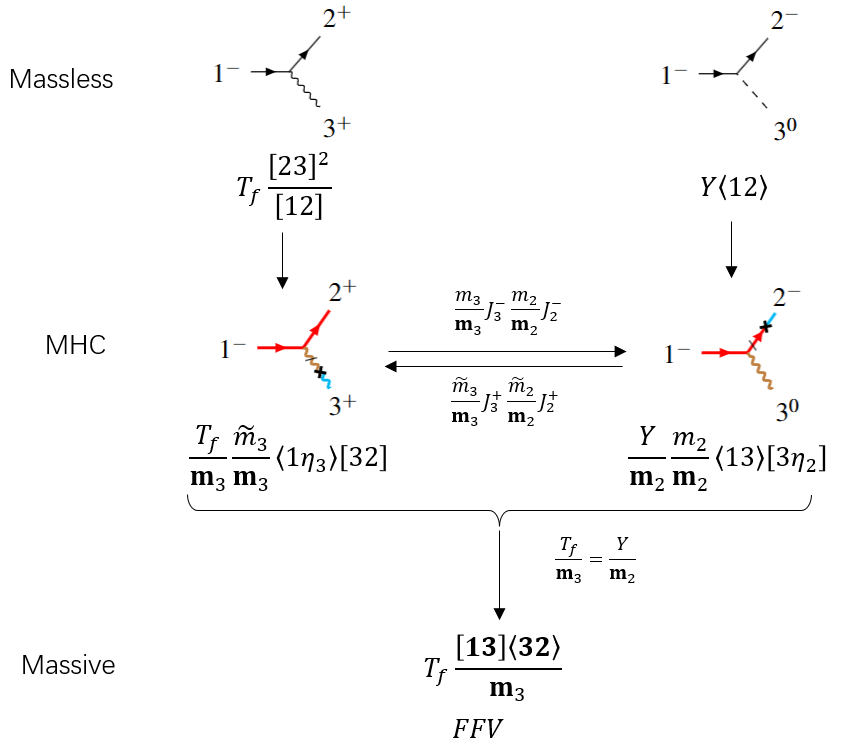}
\caption{Unifying massless $FFS$ and $FFV$ amplitudes into a single massive term, realizing the IR unification, which also validates the $SU(2)$ little group covariance.}
\label{fig:FFS_and_FFV}
\end{figure}

On the other hand, we note that all the leading MHC amplitudes should be related by extended little group symmetry. By restoring the extended little group covariance, we obtain
\begin{eqnarray}
    &&(+\frac12,+\frac12,0): \quad c\frac{\tilde{m}_1}{\textbf{m}_1}\langle\eta_13\rangle [32], \qquad (-\frac12,-\frac12,0): \quad  c\frac{m_2}{{\textbf{m}_2}} \langle13\rangle [3\eta_2], \label{eq:longitudinal_FFV_1}
    \\ 
    &&(-\frac12,+\frac12,+1): \quad c\frac{\tilde{m}_3}{{\textbf{m}_3}} \langle1\eta_3\rangle [32] \qquad (-\frac12,+\frac12,-1):\quad c\frac{{m}_3}{{\textbf{m}_3}}\langle13\rangle [\eta_32], \label{eq:transverse_FFV_1}
\end{eqnarray}
which can be related by the ladder operators $\frac{m_i}{\textbf{m}_i} J_i^\pm \frac{m_j}{\textbf{m}_j} J_j^\pm$. Here $c$ is a coefficient determined by matching from the UV theory
\begin{itemize}
    \item If the UV origin is from gauge interaction, the matching can occur for helicity $(-\frac12,+\frac12,\pm1)$, giving $c=\frac{T_f}{\mathbf m_3}$.
    \item If the UV origin is from Yukawa interaction, the matching can occur for helicity $(\pm\frac12,\pm\frac12,0)$ and we get $c=\frac{Y}{\mathbf m_2}$ and $ c=\frac{Y^*}{\mathbf m_1}$.
\end{itemize}
The extended little group covariance would relate different coefficients
\begin{eqnarray} \label{eq:Yukawa_gauge}
   c =  \frac{Y^*}{\mathbf m_1}=\frac{Y}{\mathbf m_2}=\frac{T_f}{\mathbf m_3},
\end{eqnarray}
which relates the gauge and Yukawa couplings, called the Gauge-Yukawa unification. It also provides the relation on the universal Higgs couplings.
This leads to the following relation among the coefficients of all 1st descendant amplitudes,
\begin{equation}
\mathbf{c}_{14} \mathbf{m}_2=
\mathbf{c}_{15} \mathbf{m}_1=
\mathbf{c}_{16} \mathbf{m}_3=
\mathbf{c}_{17} \mathbf{m}_3.
\end{equation}
As illustrated in Fig.~\ref{fig:FFS_and_FFV}, this reveals the on-shell realization of the universal Higgs mechanism, or saying the IR unification~\cite{Arkani-Hamed:2017jhn}.

The above relation can be derived from the Goldstone equivalence theorem~\cite{Lee:1977eg,Chanowitz:1985hj}. In the light cone gauge, the Goldstone equivalence theorem tells that the massive Goldstone amplitude $\mathbf T_\varphi$ can be related to the massive longitudinal gauge boson amplitude $\mathbf T_V^{\alpha\dot\alpha}$ by replacing the polarization vector with the massive momentum, and applying the Ward identity. In the on-shell language, the Goldstone equivalence gives  
\begin{equation}
(p_{\alpha\dot\alpha}+\eta_{\alpha\dot\alpha}) \mathbf T_V^{\alpha\dot\alpha}= \mathbf{m}_V \mathbf T_{\varphi}.
\end{equation} 
Let us build the connection between two sides:
\begin{itemize}
   
\item Starting from the transverse massive MHC amplitude in eq.~\eqref{eq:transverse_FFV_1} with $c=\frac{T_f}{\mathbf m_3}$, we can use the extended little group covariance to flip the helicity to $(\pm\frac12,\pm\frac12,0)$, which includes the longitudinal polarizations:
\begin{equation}
    \frac{T_f}{\mathbf{m}_3}\left(\frac{m_3}{\mathbf{m}_3} [2\eta_3]\langle31\rangle - \frac{\tilde m_3}{\mathbf{m}_3} [23]\langle\eta_31\rangle\right) 
    \underset{\text{little group}}{\overset{\text{extended}}{\longrightarrow}}
    \frac{T_f}{\mathbf{m}_3}\left(\frac{m_2}{\mathbf{m}_2} [\eta_23]\langle31\rangle - \frac{\tilde m_1}{\mathbf{m}_1} [23]\langle 3\eta_1\rangle\right).
\end{equation}
It is the leading-order part of the term $\frac{p_{\alpha\dot\alpha}}{\textbf m_3} \mathbf T_V^{\alpha\dot\alpha}$.

\item From the Goldstone equivalence theorem, the derivative coupling can be identified as the Goldstone amplitude. Using the relation in eq.~\eqref{eq:Yukawa_gauge}, we obtain
\begin{equation}
    \mathbf T_\varphi=\frac{Y}{\mathbf{m}_2}\frac{m_2}{\mathbf{m}_2} [\eta_23]\langle31\rangle - \frac{Y^*}{\mathbf{m}_1}\frac{\tilde m_1}{\mathbf{m}_1} [23]\langle 3\eta_1\rangle,
    \label{eq:GoldstoneAmp}
\end{equation} 
which agrees with eq.~\eqref{eq:longitudinal_FFV_1} with $c=\frac{Y}{\mathbf{m}_2}$ and $\frac{Y^*}{\mathbf{m}_1}$. 
Thus we start from the transverse MHC amplitude, using the Goldstone equivalence theorem, obtain the longitudinal MHC amplitude.

\item The Goldstone 3-point amplitude with the derivative can be related to the scalar Yukawa 3-point amplitude. From  eq.~\eqref{eq:GoldstoneAmp}, we obtain the scalar 3-point interaction for the Goldstone boson via IBP:
\begin{equation} \begin{aligned}
\mathbf T_\varphi \overset{\text{IBP}}{=}  \frac{Y}{\mathbf{m}_2}\frac{m_2}{\mathbf{m}_2} \tilde{m}_2 \langle21\rangle- \frac{Y^*}{\mathbf{m}_1} \frac{\tilde m_1}{\mathbf{m}_1} \tilde m_1 [21] m_1
\rightarrow Y \langle12\rangle- Y^*[12].
\end{aligned} \end{equation}
This tells how the massive derivative amplitudes recover the UV massless Goldstone amplitudes in the high-energy limit. 
The result is non-vanishing due to the cancellation between the masses in the denominator and numerator.

\end{itemize}

\subsection{Sub-leading matching: Higgs splitting and Higgsing rules}

Having obtained the leading amplitude matching, we now are ready to discuss the sub-leading matching by the Higgs insertion. Similar to the $FFS$ amplitude, the $FFV$ sub-leading amplitudes can be obtained from taking the collinear limit of the higher-point amplitudes $FFVS$ and $FFSS$. In this subsection, we would skip such treatment and directly utilize the Higgs insertion to perform matching, but leave the higher point matching to Appendix~\ref{app:subleading}.

Naively one expect other sub-leading terms from the Higgs insertion, similar to the fermion Higgs insertion, such as
\begin{eqnarray}
\begin{tikzpicture}[baseline=-0.1cm] \begin{feynhand}
\setlength{\feynhandblobsize}{6mm}
\setlength{\feynhandarrowsize}{5pt}
\vertex [particle] (i1) at (2,0) {$i^{0}$};
\vertex [particle] (i2) at (1,0.7) {$4$};
\vertex (v2) at (1,0);
\vertex [dot] (v1) at (0,0) {};
\graph{(i1)--[sca](v2)--[sca](v1)};
\graph{(i2)--[sca](v2)};
\end{feynhand} \end{tikzpicture}, \qquad
\begin{tikzpicture}[baseline=-0.1cm] \begin{feynhand}
\setlength{\feynhandblobsize}{6mm}
\setlength{\feynhandarrowsize}{5pt}
\vertex [particle] (i1) at (2,0) {$i^{\pm}$};
\vertex [particle] (i2) at (1,0.7) {$4$};
\vertex (v2) at (1,0);
\vertex [dot] (v1) at (0,0) {};
\graph{(i1)--[bos](v2)--[bos](v1)};
\graph{(i2)--[sca](v2)};
\end{feynhand} \end{tikzpicture}.
\end{eqnarray}
However, in the SM there is no renormalizable $SSS$ and $VVS$ 3-point amplitudes, therefore, such kinds of Higgs insertion should not exist. On the other hand, there is $VSS$ 3-point amplitude. So we expect there should be a conversion between the Goldstone boson and the transverse gauge boson by the Higgs splitting. 
\begin{equation} \begin{aligned}
\begin{tikzpicture}[baseline=0.7cm]
\begin{feynhand}
\setlength{\feynhandblobsize}{6mm}
\setlength{\feynhandarrowsize}{3.5pt}
\vertex [dot] (v1) at (-0.4,0.8) {};
\vertex [particle] (i2) at (1.3,0.8) {$i^{\pm}$};
\graph{(i2) --[bos] (v1)};
\end{feynhand}
\end{tikzpicture}
&\xrightarrow{\text{Higgs splitting}}
\begin{tikzpicture}[baseline=-0.1cm] \begin{feynhand}
\setlength{\feynhandblobsize}{6mm}
\setlength{\feynhandarrowsize}{5pt}
\vertex [particle] (i1) at (2,0) {$i^{0}$};
\vertex [particle] (i2) at (1,0.7) {$4$};
\vertex (v2) at (1,0);
\vertex [dot] (v1) at (0,0) {};
\graph{(i1)--[sca](v2)--[bos](v1)};
\graph{(i2)--[sca](v2)};
\end{feynhand} \end{tikzpicture}, \\
\begin{tikzpicture}[baseline=0.7cm]
\begin{feynhand}
\setlength{\feynhandblobsize}{6mm}
\setlength{\feynhandarrowsize}{3.5pt}
\vertex [dot] (v1) at (-0.4,0.8) {};
\vertex [particle] (i2) at (1.3,0.8) {$i^{0}$};
\graph{(i2) --[sca] (v1)};
\end{feynhand}
\end{tikzpicture}
&\xrightarrow{\text{Higgs splitting}}
\begin{tikzpicture}[baseline=-0.1cm] \begin{feynhand}
\setlength{\feynhandblobsize}{6mm}
\setlength{\feynhandarrowsize}{5pt}
\vertex [particle] (i1) at (2,0) {$i^{\pm}$};
\vertex [particle] (i2) at (1,0.7) {$4$};
\vertex (v2) at (1,0);
\vertex [dot] (v1) at (0,0) {};
\graph{(i1)--[bos](v2)--[sca](v1)};
\graph{(i2)--[sca](v2)};
\end{feynhand} \end{tikzpicture}.
\end{aligned} \end{equation}
When one scalar obtains VEV, there would be mixing between the Goldstone and the gauge boson.

\paragraph{Symmetric representation}

Following the same gluing technique, we obtain the following matching between the massless amplitude with Higgs insertion and the $[\mathbf{A}]_1$ 
\begin{eqnarray}
\begin{tikzpicture}[baseline=-0.1cm] \begin{feynhand}
\setlength{\feynhandblobsize}{6mm}
\setlength{\feynhandarrowsize}{5pt}
\vertex [particle] (i1) at (2,0) {$i^{-1}$};
\vertex [particle] (i2) at (1,0.7) {$4$};
\vertex (v2) at (1,0);
\vertex [dot] (v1) at (0,0) {};
\graph{(i1)--[bos](v2)--[sca](v1)};
\graph{(i2)--[sca](v2)};
\end{feynhand} \end{tikzpicture}
&=& \lim_{p_4\rightarrow \eta_i} \left(|\chi]_{\dot\alpha}|\chi\rangle_{\beta}\times\frac{\langle\chi i\rangle\langle i4\rangle}{\langle\chi 4\rangle}\right) 
\\ 
&=&  \lim_{p_4\rightarrow \eta_i} \left( (p_\chi|i\rangle)_{\dot\alpha}(p_\chi|4])_{\beta}\frac{\langle i4\rangle}{[4|\chi|4\rangle}\right) \nonumber \\
&\sim&  \lim_{p_4\rightarrow \eta_i} \left( |4]_{\dot\alpha}|i\rangle_{\beta}\langle i4\rangle\right) \nonumber \\
&=&  |\eta_i]_{\dot\alpha}|i\rangle_{\beta}m_i \\
&\rightarrow&
\begin{tikzpicture}[baseline=-0.1cm] \begin{feynhand}
\setlength{\feynhandblobsize}{6mm}
\setlength{\feynhandarrowsize}{5pt}
\vertex [particle] (i1) at (2,0) {$i^{-1}$};
\vertex (v2) at (1,0);
\vertex [dot] (v1) at (0,0) {};
\begin{scope}[]
\clip (0,-0.1) rectangle (1,0.1); 
\draw[brown,thick bos] (v1)--(i1);
\end{scope}
\begin{scope}[]
\clip (1,-0.1) rectangle (1.6,0.1); 
\draw[red,thick bos] (v1)--(i1);
\end{scope}
\draw[very thick] plot[mark=x,mark size=2.5] coordinates {(1,0)};
\draw (0.55-0.03,-0.08) -- (0.55+0.03,+0.08);
\end{feynhand} \end{tikzpicture}. \nonumber
\end{eqnarray}
This indicates that both the transverse massless gauge boson and the massless amplitude with the Higgs insertion, belonged to the same transversality and helicity, match to the same $[\mathbf{A}]_1$. This is a special feature for the spin-1 particle matching: both the Goldstone boson and gauge boson are mixed, following the Higgs mechanism. 

Then, let us match the zero transversality amplitude with Higgs insertion to the $[\mathbf{A}^0]_0$ and $[\mathbf{A}^0]_2$. From the above, $[\mathbf{A}^0]_0$ matches to the Goldstone boson, $[\mathbf{A}^0]_2$ is sub-leading and thus we expect the Higgs insertion
\begin{eqnarray}
\begin{tikzpicture}[baseline=-0.1cm] \begin{feynhand}
\setlength{\feynhandblobsize}{6mm}
\setlength{\feynhandarrowsize}{5pt}
\vertex [particle] (i1) at (2,0) {$i^{0}$};
\vertex [particle] (i2) at (1,0.7) {$4$};
\vertex (v2) at (1,0);
\vertex [dot] (v1) at (0,0) {};
\graph{(i1)--[sca](v2)--[bos](v1)};
\graph{(i2)--[sca](v2)};
\end{feynhand} \end{tikzpicture} 
&=& \lim_{p_4\rightarrow \eta_i} \left( \frac{|\chi]_{\dot\alpha}|\xi\rangle_{\alpha}}{\langle\chi\xi\rangle}\times\frac{\langle4\chi\rangle\langle\chi i\rangle}{\langle4i\rangle}\right)
\label{eq:HiggsingV2} \\
&=& \frac{1}{2}\lim_{p_4\rightarrow \eta_i} \left(\frac{|\chi]_{\dot\alpha}|4\rangle_{\alpha}}{\langle\chi\xi\rangle}\frac{\langle\chi\xi\rangle\langle\chi i\rangle}{\langle4i\rangle}-(i\leftrightarrow 4)\right) \nonumber \\
&=& \frac{1}{2}\lim_{p_4\rightarrow \eta_i} \left(\frac{(p_\chi|i\rangle)_{\dot\alpha}|4\rangle_{\alpha}}{\langle4i\rangle}-(i\leftrightarrow 4)\right) \nonumber \\
&\sim& \frac{1}{2}(\frac{1}{\mathbf m_i^2}\times m_i\tilde m_i|\eta_i]_{\dot\alpha}|\eta_i\rangle_{\alpha}-|i]_{\dot\alpha}|i\rangle_{\alpha}) \nonumber \\
& \rightarrow &
\begin{tikzpicture}[baseline=-0.1cm] \begin{feynhand}
\vertex [particle] (i1) at (2,0) {$i^0$};
\vertex [dot] (v1) at (0,0) {};
\vertex (v2) at (0.95,0);
\vertex (v3) at (1.4,0);
\draw[brown,thick bos] (v3)--(i1);
\draw[red,thick bos] (v2)--(v3);
\draw[brown,thick bos] (v1)--(v2);
\draw[very thick] plot[mark=x,mark size=3.5] coordinates {(v2)};
\draw[very thick] plot[mark=x,mark size=3.5] coordinates {(v3)};
\draw (0.3-0.03,-0.08) -- (0.3+0.03,+0.08);
\draw (0.6-0.03,-0.08) -- (0.6+0.03,+0.08);
\end{feynhand} \end{tikzpicture}
+\begin{tikzpicture}[baseline=-0.1cm] \begin{feynhand}
\setlength{\feynhandblobsize}{6mm}
\setlength{\feynhandarrowsize}{5pt}
\vertex [particle] (i1) at (2,0) {$i^{0}$};
\vertex [dot] (v1) at (0,0) {};
\draw[brown,thick bos] (i1)--(v1);
\end{feynhand} \end{tikzpicture}.
\end{eqnarray}
In the second line we have used the Schouten identity and the conserved current $J^{\alpha\dot\alpha}|\chi]_{\dot\alpha}|\chi\rangle_{\alpha}=0$. Thus the Goldstone and gauge boson are mixed and both woud match to the longitudinal gauge boson.


According to the helicity $h$, transversality $t$ and chirality $c$, the above matching result can be filled into the following table
\begin{equation} \label{eq:vector_table}
\begin{tabular}{c|c|c|c}
\hline
& $h=t=-1$ & $h=t=0$ & $h=t=+1$ \\
\hline
$c=0$ & $m\eta_{\dot\alpha}\lambda_{\alpha}$ & \makecell{$\tilde\lambda_{\dot\alpha}\lambda_{\alpha}$\\$\tilde\lambda_{\dot\alpha}\lambda_{\alpha}-\tilde\eta_{\dot\alpha}\eta_{\alpha}$} & $\tilde m\tilde\lambda_{\dot\alpha}\eta_{\alpha}$ \\
\hline
\end{tabular}
\end{equation}
In the box with $h=\pm 1$, the leading matching and Higgs insertion give the same result. While in the box with $h=0$, they are not the same: $\tilde\lambda\lambda$ for leading matching and $\tilde\lambda \lambda-\tilde\eta \eta$ for Higgs insertion. The effect of Higgs insertion can be viewed as a horizontal movement in the table.

\paragraph{Chiral and anti-chiral representation}

The above discussion focus on the symmetric state for massive vector boson, which has the chirality $c=1$. In more general cases, the vector boson can also have chiral and anti-chiral state, which has other chirality. In the following, let first consider the leading contribution with $c\neq 0$ in each polarization.

For transverse polarization, eliminating spurious pole may not need to explicit introducing reference spinor for all gauge boson. In this case, we can applying derivative to massless gauge boson, the resulting state is nothing but the field strength tensor
\begin{equation}
\mathcal{F}^+=\partial^{\beta}_{(\dot\alpha_1}A^+_{\beta\dot\alpha_2)}=\tilde\lambda_{\dot\alpha_1}\tilde\lambda_{\dot\alpha_2},\quad
\mathcal{F}^-=\partial^{\dot\beta}_{(\alpha_1}A^-_{\alpha_2)\dot\beta}=\lambda_{\alpha_1}\lambda_{\alpha_2},
\end{equation}
which are gauge-invariance state. In the point of view of MHC formalism, such state can be viewed as applying EOM to the gauge boson $A$
\begin{equation}
\text{EOM}(A^\pm)=\mathcal{F}^\pm.
\end{equation}
These states should match to the leading order of massive field-strength tensor $[\mathbf F]_0$
\begin{eqnarray}
\begin{tikzpicture}[baseline=0.7cm]
\begin{feynhand}
\setlength{\feynhandblobsize}{6mm}
\vertex [dot] (v1) at (-0.4,0.8) {};
\vertex [particle] (i2) at (1.3,0.8) {$i^{+1}$};
\graph{(v1) --[bos] (i2)};
\end{feynhand}
\end{tikzpicture}&\overset{\text{EOM}}{=}&\tilde\lambda_{\dot\alpha_1}\tilde\lambda_{\dot\alpha_2}
\rightarrow
\begin{tikzpicture}[baseline=0.7cm]
\begin{feynhand}
\setlength{\feynhandblobsize}{6mm}
\vertex [dot] (v1) at (-0.4,0.8) {};
\vertex [particle] (i2) at (1.3,0.8) {$i^{+1}$};
\bos{i2}{cyan};
\end{feynhand}
\end{tikzpicture}, \\
\begin{tikzpicture}[baseline=0.7cm]
\begin{feynhand}
\setlength{\feynhandblobsize}{6mm}
\vertex [dot] (v1) at (-0.4,0.8) {};
\vertex [particle] (i2) at (1.3,0.8) {$i^{-1}$};
\graph{(v1) --[bos] (i2)};
\end{feynhand}
\end{tikzpicture}&\overset{\text{EOM}}{=}&\lambda_{\alpha_1}\lambda_{\alpha_2}
\rightarrow
\begin{tikzpicture}[baseline=0.7cm]
\begin{feynhand}
\setlength{\feynhandblobsize}{6mm}
\vertex [dot] (v1) at (-0.4,0.8) {};
\vertex [particle] (i2) at (1.3,0.8) {$i^{-1}$};
\bos{i2}{red};
\end{feynhand}
\end{tikzpicture}, 
\end{eqnarray}

For longitudinal polarization, the general leading matching can apply an EOM to the Goldstone boson,
\begin{equation}
\text{EOM}(\partial\phi)=\frac{\tilde\lambda_{\dot\alpha_1}\tilde\eta_{\dot\alpha_2}+\tilde\eta_{\dot\alpha_1}\tilde\lambda_{\dot\alpha_2}}{\tilde m},\quad
\text{EOM}(\partial\phi)=\frac{\lambda_{\alpha_1}\eta_{\alpha_2}+\eta_{\alpha_1}\lambda_{\alpha_2}}{m},
\end{equation}
They match to the longitudinal mode of the massive field-strength tensor $[\mathbf F]_1$,
\begin{eqnarray}
\begin{tikzpicture}[baseline=0.7cm]
\begin{feynhand}
\setlength{\feynhandblobsize}{6mm}
\vertex [dot] (v1) at (-0.4,0.8) {};
\vertex [particle] (i2) at (1.3,0.8) {$i^{0}$};
\graph{(v1) --[sca] (i2)};
\end{feynhand}
\end{tikzpicture} &\overset{\text{EOM}}{=}& \frac{\tilde\lambda_{\dot\alpha_1}\tilde\eta_{\dot\alpha_2}+\tilde\eta_{\dot\alpha_1}\tilde\lambda_{\dot\alpha_2}}{\tilde m}
\rightarrow
\begin{tikzpicture}[baseline=-0.1cm] \begin{feynhand}
\setlength{\feynhandblobsize}{6mm}
\setlength{\feynhandarrowsize}{5pt}
\vertex [particle] (i1) at (2,0) {$i^{0}$};
\vertex (v2) at (1,0);
\vertex [dot] (v1) at (0,0) {};
\begin{scope}[]
\clip (0,-0.1) rectangle (1,0.1); 
\draw[cyan,thick bos] (v1)--(i1);
\end{scope}
\begin{scope}[]
\clip (1,-0.1) rectangle (1.6,0.1); 
\draw[brown,thick bos] (v1)--(i1);
\end{scope}
\draw[very thick] plot[mark=x,mark size=2.5] coordinates {(1,0)};
\draw (0.5-0.03,-0.08) -- (0.5+0.03,+0.08);
\end{feynhand} \end{tikzpicture}. \nonumber
\end{eqnarray}

Inserting a Higgs boson to the transverse polarization, it also match to $[\mathbf F]_1$:
\begin{eqnarray}
\begin{tikzpicture}[baseline=-0.1cm] \begin{feynhand}
\setlength{\feynhandblobsize}{6mm}
\setlength{\feynhandarrowsize}{5pt}
\vertex [particle] (i1) at (2,0) {$i^{0}$};
\vertex [particle] (i2) at (1,0.7) {$4$};
\vertex (v2) at (1,0);
\vertex [dot] (v1) at (0,0) {};
\graph{(i1)--[sca](v2)--[bos](v1)};
\graph{(i2)--[sca](v2)};
\end{feynhand} \end{tikzpicture} &\overset{\text{EOM}}{=}& 
\lim_{p_4\rightarrow \eta_i} \left(|\chi]_{\dot\alpha_1}|\chi]_{\dot\alpha_2}\times\frac{\langle4\chi\rangle\langle\chi i\rangle}{\langle4i\rangle}\right)
\label{eq:HiggsingV3} \\
&\sim& \lim_{p_4\rightarrow \eta_i} \frac{1}{2}\left(\frac{(P_{4i}|i\rangle)_{\dot\alpha_1}(P_{4i}|4\rangle)_{\dot\alpha_2}}{\langle4i\rangle}-(i\leftrightarrow 4)\right) \nonumber \\
&=& \frac{1}{2}(|\eta_i]_{\dot\alpha_1}|i]_{\dot\alpha_2}+|i]_{\dot\alpha_1}|\eta_i]_{\dot\alpha_2}) m_i\nonumber \\
&\rightarrow&
\begin{tikzpicture}[baseline=-0.1cm] \begin{feynhand}
\setlength{\feynhandblobsize}{6mm}
\setlength{\feynhandarrowsize}{5pt}
\vertex [particle] (i1) at (2,0) {$i^{0}$};
\vertex (v2) at (1,0);
\vertex [dot] (v1) at (0,0) {};
\begin{scope}[]
\clip (0,-0.1) rectangle (1,0.1); 
\draw[cyan,thick bos] (v1)--(i1);
\end{scope}
\begin{scope}[]
\clip (1,-0.1) rectangle (1.6,0.1); 
\draw[brown,thick bos] (v1)--(i1);
\end{scope}
\draw[very thick] plot[mark=x,mark size=2.5] coordinates {(1,0)};
\draw (0.5-0.03,-0.08) -- (0.5+0.03,+0.08);
\end{feynhand} \end{tikzpicture} \nonumber
\end{eqnarray}
This result gives the same representation as the scalar boson. Inserting Higgs to the longitudinal polarization, it match to $[\mathbf F]_2$
\begin{eqnarray}
\begin{tikzpicture}[baseline=-0.1cm] \begin{feynhand}
\setlength{\feynhandblobsize}{6mm}
\setlength{\feynhandarrowsize}{5pt}
\vertex [particle] (i1) at (2,0) {$i^{-1}$};
\vertex [particle] (i2) at (1,0.7) {$4$};
\vertex (v2) at (1,0);
\vertex [dot] (v1) at (0,0) {};
\graph{(i1)--[bos](v2)--[sca](v1)};
\graph{(i2)--[sca](v2)};
\end{feynhand} \end{tikzpicture}
&\overset{\text{EOM}}{=}& \lim_{p_4\rightarrow \eta_i} \left(\frac{|\chi]_{\dot\alpha}|\eta_\chi]_{\dot\beta}}{[\eta_\chi\chi]}\times\frac{\langle\chi i\rangle\langle i4\rangle}{\langle\chi 4\rangle}\right) \nonumber \\ 
&=&  \lim_{p_4\rightarrow \eta_i} \left(\frac{|\chi]_{\dot\alpha}|4]_{\dot\beta}}{[\eta_\chi\chi]}\frac{[\chi \eta_\chi]\langle \chi i\rangle\langle i4\rangle}{[4|\chi|4\rangle}\right) \nonumber \\
&=&  \lim_{p_4\rightarrow \eta_i} \left( (p_\chi|i\rangle)_{\dot\alpha}|4]_{\dot\beta}\frac{\langle i4\rangle}{[4|\chi|4\rangle}\right) \nonumber \\
&\sim&  \lim_{p_4\rightarrow \eta_i} \left( |4]_{\dot\alpha}|4]_{\dot\beta}\langle i4\rangle^2\right) \nonumber \\
&=&  |\eta_i]_{\dot\alpha}|\eta_i]_{\dot\beta}m_i^2 \\
&\rightarrow&
\begin{tikzpicture}[baseline=-0.1cm] \begin{feynhand}
\setlength{\feynhandblobsize}{6mm}
\setlength{\feynhandarrowsize}{5pt}
\vertex [particle] (i1) at (2,0) {$i^{-1}$};
\vertex (v2) at (1,0);
\vertex [dot] (v1) at (0,0) {};
\begin{scope}[]
\clip (0,-0.1) rectangle (0.8,0.1); 
\draw[cyan,thick bos] (v1)--(i1);
\end{scope}
\begin{scope}[]
\clip (0.8,-0.1) rectangle (1.2,0.1); 
\draw[brown,thick bos] (v1)--(i1);
\end{scope}
\begin{scope}[]
\clip (1.2,-0.1) rectangle (1.6,0.1); 
\draw[red,thick bos] (v1)--(i1);
\end{scope}
\draw[very thick] plot[mark=x,mark size=2.5] coordinates {(1.2,0)};
\draw[very thick] plot[mark=x,mark size=2.5] coordinates {(0.8,0)};
\draw (0.55-0.03,-0.08) -- (0.55+0.03,+0.08);
\draw (0.3-0.03,-0.08) -- (0.3+0.03,+0.08);
\end{feynhand} \end{tikzpicture} . \nonumber
\end{eqnarray}
In the second line we use the Schouten identity and use an assumption: the term invovling $|\chi]|\chi]$ can be dropped. To make this assumption, we need to know which structure couple to the massive field strength tensor $\mathbf F$. This will be discussed in next section.

\paragraph{Higgsing Rules} 

Similar to the fermion case, we could also summarize the above matching results from the Higgs insertion according to the particle states for the vector boson. Combining all these matching results, we can extend eq.~\eqref{eq:vector_table} to
\begin{equation}\label{eq:vv_table}
\begin{tabular}{c|c|c|c}
\hline
& $h=t=-1$ & $h=t=0$ & $h=t=+1$ \\
\hline
$c=-1$ & $\lambda_{\alpha_1}\lambda_{\alpha_2}$ & $\tilde m\lambda_{(\alpha_1}\eta_{\alpha_2)}$ & $\tilde m^2\eta_{\alpha_1}\eta_{\alpha_2}$ \\
\hline
$c=0$ & $m\tilde\eta_{\dot\alpha}\lambda_{\alpha}$ & \makecell{$\tilde\lambda_{\dot\alpha}\lambda_{\alpha}$\\$m\tilde m\tilde\eta_{\dot\alpha}\eta_{\alpha}$} & $\tilde m\tilde\lambda_{\dot\alpha}\eta_{\alpha}$ \\
\hline
$c=+1$ & $m^2\tilde\eta_{\dot\alpha_1}\tilde{\eta}_{\dot\alpha_2}$ & $m\tilde\lambda_{(\dot\alpha_1}\tilde\eta_{\dot\alpha_2)}$ & $\tilde\lambda_{\dot\alpha_1}\tilde\lambda_{\dot\alpha_2}$ \\
\hline
\end{tabular}
\end{equation}
This table shows that the Higgs insertion correspond to horizontal movements among different particle states. The Higgs insertion can be written diagrammatically as follows
\begin{equation}
\begin{tabular}{c|c|c|c}
\hline
& $h=t=-1$ & $h=t=0$ & $h=t=+1$ \\
\hline
\multirow{2}{*}{$c=-1$} &  & 
\begin{tikzpicture}[baseline=0.7cm]
\begin{feynhand}
\setlength{\feynhandblobsize}{6mm}
\setlength{\feynhandarrowsize}{3.5pt}
\vertex [dot] (v1) at (-0.4,0.8) {};
\vertex [particle] (i2) at (1.3,0.8) {$i^{0}$};
\graph{(i2) --[sca,edge label = $\mathbf D\partial\phi$] (v1)};
\end{feynhand}
\end{tikzpicture} & 
\begin{tikzpicture}[baseline=0.2cm] \begin{feynhand}
\setlength{\feynhandblobsize}{6mm}
\setlength{\feynhandarrowsize}{5pt}
\vertex [particle] (i1) at (2,0) {$i^{+}$};
\vertex [particle] (i2) at (1.9,0.7) {$4$};
\vertex (v2) at (1,0);
\vertex [dot] (v1) at (0,0) {};
\graph{(i1)--[bos,edge label = $\mathcal{F}$](v2)--[sca](v1)};
\graph{(i2)--[sca](v2)};
\draw[decorate,decoration=brace] (2.2,0.85)--(2.2,-0.15);
\node (C) at (2.7,0.3) {$\mathcal{F}\partial^2 h$};
\end{feynhand} \end{tikzpicture} \\
\cline{2-4}
 & \begin{tikzpicture}[baseline=0.7cm]
\begin{feynhand}
\setlength{\feynhandblobsize}{6mm}
\setlength{\feynhandarrowsize}{3.5pt}
\vertex [dot] (v1) at (-0.4,0.8) {};
\vertex [particle] (i2) at (1.3,0.8) {$i^{-}$};
\graph{(i2) --[bos,edge label = $\mathcal{F}$] (v1)};
\end{feynhand}
\end{tikzpicture} & 
\begin{tikzpicture}[baseline=0.2cm] \begin{feynhand}
\setlength{\feynhandblobsize}{6mm}
\setlength{\feynhandarrowsize}{5pt}
\vertex [particle] (i1) at (2,0) {$i^{0}$};
\vertex [particle] (i2) at (1.9,0.7) {$4$};
\vertex (v2) at (1,0);
\vertex [dot] (v1) at (0,0) {};
\graph{(i1)--[sca,edge label = $\partial \phi$](v2)--[bos](v1)};
\graph{(i2)--[sca](v2)};
\draw[decorate,decoration=brace] (2.2,0.85)--(2.2,-0.15);
\node (C) at (2.7,0.3) {$\partial\phi\partial h$};
\end{feynhand} \end{tikzpicture} & \\
\hline
\multirow{2}{*}{$c=0$} & 
\begin{tikzpicture}[baseline=-0.1cm] \begin{feynhand}
\setlength{\feynhandblobsize}{6mm}
\setlength{\feynhandarrowsize}{5pt}
\vertex [particle] (i1) at (2,0) {$i^{-}$};
\vertex [particle] (i2) at (1.9,0.7) {$4$};
\vertex (v2) at (1,0);
\vertex [dot] (v1) at (0,0) {};
\graph{(i1)--[bos,edge label = $A$](v2)--[sca](v1)};
\graph{(i2)--[sca](v2)};
\draw[decorate,decoration=brace] (2.2,0.85)--(2.2,-0.15);
\node (C) at (2.7,0.3) {$A\partial h$};
\end{feynhand} \end{tikzpicture} & 
\begin{tikzpicture}[baseline=0.7cm]
\begin{feynhand}
\setlength{\feynhandblobsize}{6mm}
\setlength{\feynhandarrowsize}{3.5pt}
\vertex [dot] (v1) at (-0.4,0.8) {};
\vertex [particle] (i2) at (1.3,0.8) {$i^{0}$};
\graph{(i2) --[sca,edge label = $\partial\phi$] (v1)};
\end{feynhand}
\end{tikzpicture} & 
\begin{tikzpicture}[baseline=-0.1cm] \begin{feynhand}
\setlength{\feynhandblobsize}{6mm}
\setlength{\feynhandarrowsize}{5pt}
\vertex [particle] (i1) at (2,0) {$i^{+}$};
\vertex [particle] (i2) at (1.9,0.7) {$4$};
\vertex (v2) at (1,0);
\vertex [dot] (v1) at (0,0) {};
\graph{(i1)--[bos,edge label = $A$](v2)--[sca](v1)};
\graph{(i2)--[sca](v2)};
\draw[decorate,decoration=brace] (2.2,0.85)--(2.2,-0.15);
\node (C) at (2.7,0.3) {$A\partial h$};
\end{feynhand} \end{tikzpicture} \\
\cline{2-4}
 & \begin{tikzpicture}[baseline=0.7cm]
\begin{feynhand}
\setlength{\feynhandblobsize}{6mm}
\setlength{\feynhandarrowsize}{3.5pt}
\vertex [dot] (v1) at (-0.4,0.8) {};
\vertex [particle] (i2) at (1.3,0.8) {$i^{-}$};
\graph{(i2) --[bos,edge label = $A$] (v1)};
\end{feynhand}
\end{tikzpicture} & 
\begin{tikzpicture}[baseline=-0.1cm] \begin{feynhand}
\setlength{\feynhandblobsize}{6mm}
\setlength{\feynhandarrowsize}{5pt}
\vertex [particle] (i1) at (2,0) {$i^{0}$};
\vertex [particle] (i2) at (1.9,0.7) {$4$};
\vertex (v2) at (1,0);
\vertex [dot] (v1) at (0,0) {};
\graph{(i1)--[sca,edge label = $\partial\phi$](v2)--[bos](v1)};
\graph{(i2)--[sca](v2)};
\draw[decorate,decoration=brace] (2.2,0.85)--(2.2,-0.15);
\node (C) at (2.9,0.3) {$\partial \phi-\partial h$};
\end{feynhand} \end{tikzpicture} & 
\begin{tikzpicture}[baseline=0.7cm]
\begin{feynhand}
\setlength{\feynhandblobsize}{6mm}
\setlength{\feynhandarrowsize}{3.5pt}
\vertex [dot] (v1) at (-0.4,0.8) {};
\vertex [particle] (i2) at (1.3,0.8) {$i^{+}$};
\graph{(i2) --[bos,edge label = $A$] (v1)};
\end{feynhand}
\end{tikzpicture} \\
\hline
\multirow{2}{*}{$c=+1$} & 
\begin{tikzpicture}[baseline=0.2cm] \begin{feynhand}
\setlength{\feynhandblobsize}{6mm}
\setlength{\feynhandarrowsize}{5pt}
\vertex [particle] (i1) at (2,0) {$i^{-}$};
\vertex [particle] (i2) at (1.9,0.7) {$4$};
\vertex (v2) at (1,0);
\vertex [dot] (v1) at (0,0) {};
\graph{(i1)--[bos,edge label = $\mathcal{F}$](v2)--[sca](v1)};
\graph{(i2)--[sca](v2)};
\draw[decorate,decoration=brace] (2.2,0.85)--(2.2,-0.15);
\node (C) at (2.7,0.3) {$\mathcal{F}\partial^2 h$};
\end{feynhand} \end{tikzpicture} 
& \begin{tikzpicture}[baseline=0.7cm]
\begin{feynhand}
\setlength{\feynhandblobsize}{6mm}
\setlength{\feynhandarrowsize}{3.5pt}
\vertex [dot] (v1) at (-0.4,0.8) {};
\vertex [particle] (i2) at (1.3,0.8) {$i^{0}$};
\graph{(i2) --[sca,edge label = $\mathbf D\partial\phi$] (v1)};
\end{feynhand}
\end{tikzpicture} &  \\
\cline{2-4}
 & & \begin{tikzpicture}[baseline=0.2cm] \begin{feynhand}
\setlength{\feynhandblobsize}{6mm}
\setlength{\feynhandarrowsize}{5pt}
\vertex [particle] (i1) at (2,0) {$i^{0}$};
\vertex [particle] (i2) at (1.9,0.7) {$4$};
\vertex (v2) at (1,0);
\vertex [dot] (v1) at (0,0) {};
\graph{(i1)--[sca,edge label = $\partial \phi$](v2)--[bos](v1)};
\graph{(i2)--[sca](v2)};
\draw[decorate,decoration=brace] (2.2,0.85)--(2.2,-0.15);
\node (C) at (2.7,0.3) {$\partial\phi\partial h$};
\end{feynhand} \end{tikzpicture} & \begin{tikzpicture}[baseline=0.7cm]
\begin{feynhand}
\setlength{\feynhandblobsize}{6mm}
\setlength{\feynhandarrowsize}{3.5pt}
\vertex [dot] (v1) at (-0.4,0.8) {};
\vertex [particle] (i2) at (1.3,0.8) {$i^{+}$};
\graph{(i2) --[bos,edge label = $\mathcal{F}$] (v1)};
\end{feynhand}
\end{tikzpicture} \\
\hline
\end{tabular}
\end{equation}

Of course, it is possible to insert more Higgses. In analogy to the fermion case, we set only the Higgs boson attached with external particle $i$ to be $\eta_i$, other Higgs boson to be soft. Such insertions would match to the same MHC structure,
\begin{equation}
\includegraphics[width=0.8\linewidth]{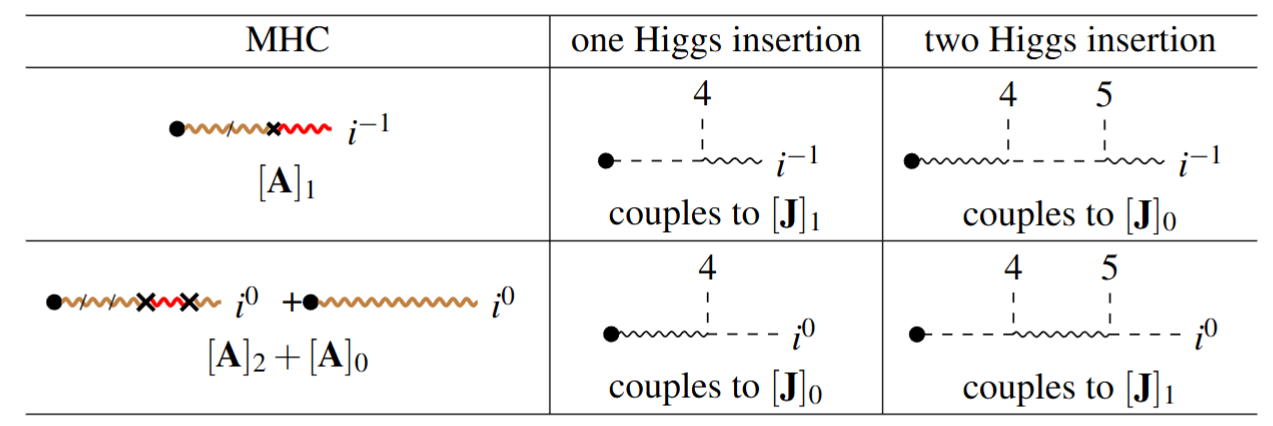}
\end{equation}
Although single Higgs insertion and double Higgs insertion give the same result, they couple to different $[\mathbf{J}]_k$, which depend on the left-most internal particles.
For double insertion, it repeat the $[\mathbf F]_2$
\begin{eqnarray}
\begin{tikzpicture}[baseline=-0.1cm] \begin{feynhand}
\setlength{\feynhandblobsize}{6mm}
\setlength{\feynhandarrowsize}{5pt}
\vertex [particle] (i1) at (3,0) {$i^{-1}$};
\vertex [particle] (i2) at (2,0.7) {$5$};
\vertex [particle] (i3) at (1,0.7) {$4$};
\vertex (v2) at (1,0);
\vertex (v3) at (2,0);
\vertex [dot] (v1) at (0,0) {};
\graph{(i1)--[bos](v3)--[sca](v2)--[bos](v1)};
\graph{(i2)--[sca](v3)};
\graph{(i3)--[sca](v2)};
\end{feynhand} \end{tikzpicture} 
&\rightarrow&
\begin{tikzpicture}[baseline=-0.1cm] \begin{feynhand}
\setlength{\feynhandblobsize}{6mm}
\setlength{\feynhandarrowsize}{5pt}
\vertex [particle] (i1) at (2,0) {$i^{-1}$};
\vertex (v2) at (1,0);
\vertex [dot] (v1) at (0,0) {};
\begin{scope}[]
\clip (0,-0.1) rectangle (0.8,0.1); 
\draw[cyan,thick bos] (v1)--(i1);
\end{scope}
\begin{scope}[]
\clip (0.8,-0.1) rectangle (1.2,0.1); 
\draw[brown,thick bos] (v1)--(i1);
\end{scope}
\begin{scope}[]
\clip (1.2,-0.1) rectangle (1.6,0.1); 
\draw[red,thick bos] (v1)--(i1);
\end{scope}
\draw[very thick] plot[mark=x,mark size=2.5] coordinates {(1.2,0)};
\draw[very thick] plot[mark=x,mark size=2.5] coordinates {(0.8,0)};
\draw (0.55-0.03,-0.08) -- (0.55+0.03,+0.08);
\draw (0.3-0.03,-0.08) -- (0.3+0.03,+0.08);
\end{feynhand} \end{tikzpicture} \label{eq:HiggsingV4}
\end{eqnarray}
Thus, we derive the needed Higgs insertion for the anti-chiral vector.

\subsection{Massless-massive correspondence for massive vector boson}

Since we obtain the leading and sub-leading matching rules for massive boson, now we are able to match any order MHC amplitudes. The 1st descendant amplitudes has been derived through the leading matching in subsection~\ref{sec:FFV_leading}. The higher-order matching can be obtained by insert Higgs to particle line.

According to Higgs insertion technique, the coefficient of higher order MHC amplitude is equal to the 1st descendant coefficient multiplied with a factor, which is written in terms of coupling, particle mass and vev $v$. According to which particle we insert, there are two kinds of factor
\begin{equation} 
\begin{cases}
    \frac{Yv}{\mathbf m_f^2},\frac{Y^*v}{\mathbf m_f^2} &\text{insert to fermion} \\
    \frac{T_s v}{\mathbf m_V^2} & \text{insert to Goldstone/gauge boson}
\end{cases}
\end{equation}
where $\mathbf m_f$ and $\mathbf m_V$ denote the fermion and vector mass.

We first consider the 2nd descendant coefficient, which corresponds to single Higgs insertion. Let us start with the massless FFS amplitude.
For helicity category $(-\frac12,-\frac12,0)$, we can insert a Higgs boson to particle 2 and obtain a 4-pt massless amplitude, which match to a 2nd MHC amplitude
\begin{equation} \begin{aligned}
\begin{tikzpicture}[baseline=0.7cm] \begin{feynhand}
\setlength{\feynhandarrowsize}{3.5pt}
\vertex [particle] (i1) at (0,0.8) {$1^-$}; 
\vertex [particle] (i2) at (1.6,1.6) {$2^-$}; 
\vertex [particle] (i3) at (1.6,0) {$3^0$};  
\vertex (v1) at (0.9,0.8); 
\graph{(i1)--[fer](v1)--[fer](i2)};
\graph{(i3)--[sca] (v1)};
\node (C) at (0.9,-0.8) {$\langle12\rangle$};
\end{feynhand} \end{tikzpicture} 
&\to 
\Ampthree{1^-}{2^-}{3^0}{\fer{red}{i1}{v1}}{\antferflip{1}{55}{cyan}{red}}{\bos{i3}{brown};\node (C) at (-0.2,-1.6) {$\displaystyle\frac{m_2 \langle13\rangle[3\eta_2]}{\mathbf m_2^2}$}}
\xrightarrow{\text{insertion}}
\begin{tikzpicture}[baseline=-0.1cm] \begin{feynhand}
\setlength{\feynhandarrowsize}{3.5pt}
\vertex [particle] (i1) at (-1.1,0) {$1^+$}; 
\vertex [particle] (i2) at (0.579,0.827) {$2^-$}; 
\vertex [particle] (i3) at (0.579,-0.827) {$3^0$};  
\vertex [particle] (i5) at (-0.2,-0.6) {$4$};
\vertex (v3) at (-0.3,0);
\vertex (v1) at (0,0); 
\graph{(v3)--[sca](i5)};
\draw[decoration={markings,mark=at position 0.7 with {\arrow{Triangle[length=4pt,width=4pt]}}},postaction={decorate}] (i1)--(v3);
\draw[red,very thick,decoration={markings,mark=at position 0.8 with {\arrow{Triangle[length=4pt,width=4pt]}}},postaction={decorate}] (v3)--(v1);
\antferflip{1}{55}{cyan}{red};
\bos{i3}{brown};
\node (C) at (-0.2,-1.6) {};
\end{feynhand} \end{tikzpicture}
\rightarrow
\Ampthree{1^+}{2^-}{3^0}{\ferflip{1}{180}{cyan}{red}}{\antferflip{1}{55}{cyan}{red}}{\bos{i3}{brown};\node (C) at (-0.2,-1.6) {$\displaystyle\frac{\tilde m_1 m_2\langle\eta_13\rangle[3\eta_2]}{\mathbf m_1^2 \mathbf m_2^2}$};},
\end{aligned} \end{equation}
This MHC term has a coefficient $\mathbf c_{18}$. Thus, the Higgs insertion build a relation between these two coefficients:
\begin{equation} \begin{aligned}
\mathbf{c}_{18} &=\frac{Yv}{\mathbf m_2^2}\mathbf{c}_{14},\\
\end{aligned} \end{equation}

However, we can also insert to $(+\frac{1}{2},+\frac{1}{2},0)$ and obtain the same result
\begin{equation} \begin{aligned}
\begin{tikzpicture}[baseline=0.7cm] \begin{feynhand}
\setlength{\feynhandarrowsize}{3.5pt}
\vertex [particle] (i1) at (0,0.8) {$1^+$}; 
\vertex [particle] (i2) at (1.6,1.6) {$2^+$}; 
\vertex [particle] (i3) at (1.6,0) {$3^0$};  
\vertex (v1) at (0.9,0.8); 
\graph{(i1)--[fer](v1)--[fer](i2)};
\graph{(i3)--[sca] (v1)};  
\end{feynhand} \end{tikzpicture} 
&\to 
\Ampthree{1^+}{2^+}{3^0}{\ferflip{1}{180}{cyan}{red}}{\antfer{red}{i2}{v1}}{\bos{i3}{brown}}
\xrightarrow{\text{insertion}}
\begin{tikzpicture}[baseline=-0.1cm] \begin{feynhand}
\setlength{\feynhandarrowsize}{3.5pt}
\vertex [particle] (i1) at (-1,0) {$1^+$}; 
\vertex [particle] (i2) at (0.579,0.827) {$2^-$}; 
\vertex [particle] (i3) at (0.579,-0.827) {$3^0$};  
\vertex [particle] (i4) at (0.7,0) {$4$};
\vertex (v2) at (0.579*0.33,0.827*0.33);
\vertex (v1) at (0,0); 
\ferflip{1}{180}{cyan}{red};
\draw[red,very thick,decoration={markings,mark=at position 0.8 with {\arrow{Triangle[length=4pt,width=4pt]}}},postaction={decorate}] (v1)--(v2);
\graph{(v2)--[fer](i2)};
\bos{i3}{brown};
\graph{(v2)--[sca](i4)};
\end{feynhand} \end{tikzpicture}
\rightarrow
\Ampthree{1^+}{2^-}{3^0}{\ferflip{1}{180}{cyan}{red}}{\antferflip{1}{55}{cyan}{red}}{\bos{i3}{brown}},
\end{aligned} \end{equation}
This gives another relation
\begin{equation} \begin{aligned}
\mathbf{c}_{18} &=\frac{Y^*v}{\mathbf m_1^2}\mathbf{c}_{15},
\end{aligned} \end{equation}
These two ways of Higgs insertion can be understood by considering the UV origins of the chirality flips in the MHC diagrams. There are two flips in the diagrams, so one should correspond to Higgs insertion and another should come from restoring the Goldstone behavior for particle 3. Thus we can choose either the flips in particle 1 or particle 2 originate from the Higgs insertion. Note that, since these two Higgs insertion corresponds to taking different on-shell limit for the same massless amplitude $A(1^{-1},2^{-2},3^0,4)$, they will not generate two distinct massive amplitude. We only need to consider one way to perform the Higgs insertion technique.

Then we can consider the massless $FFV$ amplitude. In helicity $(-\frac12,+\frac12,+1)$, the leading matching gives
\begin{equation} \begin{aligned}
\begin{tikzpicture}[baseline=0.7cm] \begin{feynhand}
\setlength{\feynhandarrowsize}{3.5pt}
\vertex [particle] (i1) at (0,0.8) {$1^-$}; 
\vertex [particle] (i2) at (1.6,1.6) {$2^+$}; 
\vertex [particle] (i3) at (1.6,0) {$3^+$};  
\vertex (v1) at (0.9,0.8); 
\graph{(i1)--[fer](v1)--[fer](i2)};
\graph{(i3)--[bos] (v1)};  
\node (C) at (0.9,-0.8) {$\displaystyle\frac{[23]^2}{[13]}$};
\end{feynhand} \end{tikzpicture}
&\to
\Ampthree{1^-}{2^+}{3^+}{\fer{red}{i1}{v1}}{\antfer{red}{i2}{v1}}{\bosflip{1}{-55}{brown}{cyan};\node (C) at (-0.2,-1.6) {$\displaystyle\frac{\tilde m_3[23]\langle\eta_31\rangle}{\mathbf m_3^2}$}}.
\end{aligned} \end{equation}
In this matching result, particle 3 exhibit the spinor structure $\tilde m_3\tilde\lambda_3\eta_3$. We can find this single particle state in the table of eq.~\eqref{eq:vector_table_all}. When we insert Higgs to this gauge boson, the effect of Higgs insertion is equivalent to a horizontal move in that table,
\begin{equation}
    \tilde m_3\tilde\lambda_3\eta_3\xrightarrow{\text{insertion}} \tilde m_3 m_3\eta_3\tilde\eta_3
\end{equation}
Thus the subleading matching can be obtained by using the above repalcement,
\begin{equation} \begin{aligned}
\Ampthree{1^-}{2^+}{3^+}{\fer{red}{i1}{v1}}{\antfer{red}{i2}{v1}}{\bosflip{1}{-55}{brown}{cyan};\node (C) at (-0.2,-1.6) {$\displaystyle\frac{\tilde m_3[23]\langle\eta_31\rangle}{\mathbf m_3^2}$}}
\xrightarrow{\text{insertion}}
\begin{tikzpicture}[baseline=-0.1cm] \begin{feynhand}
\setlength{\feynhandarrowsize}{3.5pt}
\vertex [particle] (i1) at (-1,0) {$1^+$}; 
\vertex [particle] (i2) at (0.579,0.827) {$2^-$}; 
\vertex [particle] (i3) at (0.579,-0.827) {$3^0$};  
\vertex [particle] (i4) at (0.7,0) {$4$};
\vertex (v2) at (0.579*0.4,-0.827*0.4);
\vertex (v1) at (0,0); 
\fer{red}{i1}{v1};
\antfer{red}{i2}{v1};
\begin{scope}[rotate=-55]
\clip (0,-0.1) rectangle (0.2,0.1); 
\draw[brown,thick bos] (0,0)--(0.4,0);
\end{scope}
\begin{scope}[rotate=-55]
\clip (0.2,-0.1) rectangle (0.4,0.1); 
\draw[cyan,thick bos] (0,0)--(0.4,0);
\end{scope}
\draw[very thick,rotate=-55] plot[mark=x,mark size=2.5] coordinates {(0.2,0)};
\draw[rotate=-55] (0.06,-0.08) -- (0.12,+0.08);
\graph{(i3)--[sca](v2)};  
\graph{(v2)--[sca](i4)};
\node (C) at (-0.2,-1.6) {};
\end{feynhand} \end{tikzpicture}\rightarrow
\Ampthree{1^-}{2^+}{3^0}{\fer{red}{i1}{v1}}{\antfer{red}{i2}{v1}}{\bosflipflip{1}{-55}{brown}{cyan}{brown};
\node (C) at (-0.2,-1.6) {$\displaystyle\frac{\tilde m_3 m_3[2\eta_3]\langle\eta_31\rangle}{\mathbf m_3^4}$}}.
\end{aligned} \end{equation}
Similarly, the Higgs insertion for the helicity $(-\frac12,+\frac12,+1)$ yield the same 2nd descendant term,
\begin{equation} \begin{aligned}
\begin{tikzpicture}[baseline=0.7cm] \begin{feynhand}
\setlength{\feynhandarrowsize}{3.5pt}
\vertex [particle] (i1) at (0,0.8) {$1^-$}; 
\vertex [particle] (i2) at (1.6,1.6) {$2^+$}; 
\vertex [particle] (i3) at (1.6,0) {$3^-$};  
\vertex (v1) at (0.9,0.8); 
\graph{(i1)--[fer](v1)--[fer](i2)};
\graph{(i3)--[bos] (v1)};  
\end{feynhand} \end{tikzpicture}
&\to
\Ampthree{1^-}{2^+}{3^-}{\fer{red}{i1}{v1}}{\antfer{red}{i2}{v1}}{\bosflip{1}{-55}{brown}{red}}
\xrightarrow{\text{insertion}}
\begin{tikzpicture}[baseline=-0.1cm] \begin{feynhand}
\setlength{\feynhandarrowsize}{3.5pt}
\vertex [particle] (i1) at (-1,0) {$1^+$}; 
\vertex [particle] (i2) at (0.579,0.827) {$2^-$}; 
\vertex [particle] (i3) at (0.579,-0.827) {$3^0$};  
\vertex [particle] (i4) at (0.7,0) {$4$};
\vertex (v2) at (0.579*0.4,-0.827*0.4);
\vertex (v1) at (0,0); 
\fer{red}{i1}{v1};
\antfer{red}{i2}{v1};
\begin{scope}[rotate=-55]
\clip (0,-0.1) rectangle (0.2,0.1); 
\draw[brown,thick bos] (0,0)--(0.4,0);
\end{scope}
\begin{scope}[rotate=-55]
\clip (0.2,-0.1) rectangle (0.4,0.1); 
\draw[red,thick bos] (0,0)--(0.4,0);
\end{scope}
\draw[very thick,rotate=-55] plot[mark=x,mark size=2.5] coordinates {(0.2,0)};
\draw[rotate=-55] (0.06,-0.08) -- (0.12,+0.08);
\graph{(i3)--[sca](v2)};  
\graph{(v2)--[sca](i4)};
\end{feynhand} \end{tikzpicture}\rightarrow
\Ampthree{1^-}{2^+}{3^0}{\fer{red}{i1}{v1}}{\antfer{red}{i2}{v1}}{\bosflipflip{1}{-55}{brown}{red}{brown}}
\end{aligned} \end{equation}
The corresponding coefficient is $\mathbf c_{13}$. Recall that it is the coefficient of both primary and 2nd descendant MHC amplitudes. Although the primary amplitude vanishes, this coefficient can be determined by the Higgs insertion technique,
\begin{equation} \begin{aligned}
\mathbf{c}_{13} &=\frac{T_s v}{\mathbf m_3^2}\mathbf{c}_{16}=\frac{T_s v}{\mathbf m_3^2}\mathbf{c}_{17},
\end{aligned} \end{equation}

Moreover, the Higgs insertion of massless $FFV$ and $FFS$ amplitude can also match to the same MHC amplitude. We can insert to the fermion and Goldstone in the massless amplitude with helicity category $(-\frac12,+\frac12,+1)$ and $(+\frac12,+\frac12,0)$. They match to
\begin{equation}
\left.\begin{aligned}
\begin{tikzpicture}[baseline=0.7cm] \begin{feynhand}
\setlength{\feynhandarrowsize}{3.5pt}
\vertex [particle] (i1) at (0,0.8) {$1^-$}; 
\vertex [particle] (i2) at (1.6,1.6) {$2^+$}; 
\vertex [particle] (i3) at (1.6,0) {$3^+$};  
\vertex (v1) at (0.9,0.8); 
\graph{(i1)--[fer](v1)--[fer](i2)};
\graph{(i3)--[bos] (v1)};  
\end{feynhand} \end{tikzpicture}\to
\Ampthree{1^-}{2^+}{3^+}{\fer{red}{i1}{v1}}{\antfer{red}{i2}{v1}}{\bosflip{1}{-55}{brown}{cyan}} \xrightarrow{\text{insertion}}
\begin{tikzpicture}[baseline=-0.1cm] \begin{feynhand}
\setlength{\feynhandarrowsize}{3.5pt}
\vertex [particle] (i1) at (-1.1,0) {$1^+$}; 
\vertex [particle] (i2) at (0.7,0.8) {$2^-$}; 
\vertex [particle] (i3) at (0.7,-0.8) {$3^0$};  
\vertex [particle] (i5) at (-0.2,-0.6) {$4$};
\vertex (v3) at (-0.3,0);
\vertex (v1) at (0,0); 
\draw[red,very thick,decoration={markings,mark=at position 0.8 with {\arrow{Triangle[length=4pt,width=4pt]}}},postaction={decorate}] (v3)--(v1);
\graph{(i1)--[fer](v3)};
\antfer{red}{i2}{v1}; 
\bosflip{1}{-55}{brown}{cyan}
\graph{(v3)--[sca](i5)};
\end{feynhand} \end{tikzpicture} \\
\begin{tikzpicture}[baseline=0.7cm] \begin{feynhand}
\setlength{\feynhandarrowsize}{3.5pt}
\vertex [particle] (i1) at (0,0.8) {$1^+$}; 
\vertex [particle] (i2) at (1.6,1.6) {$2^+$}; 
\vertex [particle] (i3) at (1.6,0) {$3^0$};  
\vertex (v1) at (0.9,0.8); 
\graph{(i1)--[fer](v1)--[fer](i2)};
\graph{(i3)--[sca] (v1)};  
\end{feynhand} \end{tikzpicture}\to
\Ampthree{1^+}{2^+}{3^0}{\ferflip{1}{180}{cyan}{red}}{\antfer{red}{i2}{v1}}{\bos{i3}{brown}}\xrightarrow{\text{insertion}}
\begin{tikzpicture}[baseline=-0.1cm] \begin{feynhand}
\setlength{\feynhandarrowsize}{3.5pt}
\vertex [particle] (i1) at (-1,0) {$1^+$}; 
\vertex [particle] (i2) at (0.579,0.827) {$2^-$}; 
\vertex [particle] (i3) at (0.579,-0.827) {$3^0$};  
\vertex [particle] (i4) at (0.7,0) {$4$};
\vertex (v2) at (0.579*0.4,-0.827*0.4);
\vertex (v1) at (0,0); 
\ferflip{1}{180}{cyan}{red};
\antfer{red}{i2}{v1};
\draw[brown,thick bos] (v2)--(v1);
\graph{(i3)--[bos](v2)};  
\graph{(v2)--[sca](i4)};
\end{feynhand} \end{tikzpicture}
\end{aligned}\right.
\to\Ampthree{1^+}{2^+}{3^+}{\ferflip{1}{180}{cyan}{red}}{\antfer{red}{i2}{v1}}{\bosflip{1}{-55}{brown}{cyan}}
\end{equation}

So we have
\begin{equation} \begin{aligned}
\mathbf{c}_{19} &=\frac{T_s v}{\mathbf m_3^2}\mathbf{c}_{14}=\frac{Y^*v}{\mathbf m_1^2}\mathbf{c}_{16},
\end{aligned} \end{equation}
Combing this with the leading matching result $\mathbf c_{16}=\frac{T_f}{\mathbf m_3^2}$ and $\mathbf c_{14}=\frac{Y^*}{\mathbf m_1^2}$, we find that the gauge coupling for fermion and scalar should be equal 
\begin{equation} \begin{aligned}
T_s=T_f.
\end{aligned} \end{equation}
Otherwise, the consistency of amplitude matching would be violated. Therefore, this relation must be satisfied in any consistent UV completion that gives rise to the massive vector boson.

Similarly, we obtain other relations for 2nd descendant coefficients,
\begin{equation} \begin{aligned}
\mathbf{c}_{20} &=\frac{T_s v}{\mathbf m_3^2}\mathbf{c}_{14}=\frac{Y^*v}{\mathbf m_1^2}\mathbf{c}_{17},\\
\mathbf{c}_{21} &=\frac{T_s v}{\mathbf m_3^2}\mathbf{c}_{15}=\frac{Yv}{\mathbf m_2^2}\mathbf{c}_{16},\\
\mathbf{c}_{22} &=\frac{T_s v}{\mathbf m_3^2}\mathbf{c}_{15}=\frac{Yv}{\mathbf m_2^2}\mathbf{c}_{17},
\end{aligned} \end{equation}

The double insertion can derive the 3rd descendant coefficients,
\begin{equation} \begin{aligned}
\mathbf{c}_{23} &=\frac{T_s Y v^2}{\mathbf m_2^2 \mathbf m_3^2}\mathbf{c}_{14}=\frac{T_s Y^* v^2}{\mathbf m_1^2 \mathbf m_3^2}\mathbf{c}_{15}=\frac{Y^*Y v^2}{\mathbf m_1^2 \mathbf m_2^2}\mathbf{c}_{16},\\
\mathbf{c}_{24} &=\frac{T_s Y v^2}{\mathbf m_2^2 \mathbf m_3^2}\mathbf{c}_{14}=\frac{T_s Y^* v^2}{\mathbf m_1^2 \mathbf m_3^2}\mathbf{c}_{15}=\frac{Y^*Y v^2}{\mathbf m_1^2 \mathbf m_2^2}\mathbf{c}_{17},
\end{aligned} \end{equation}
Similarly we obtain the results for the triple insertion, and if more than triple insertion, the MHC amplitudes would repeat as others already obtained.

Finally, we obtain the complete $FFV$ amplitude, as shown in Fig.~\ref{fig:FFV_all}. Using $T_f=T_s$, the amplitude can be expressed as
\begin{equation} \begin{aligned}
&\mathcal{M}(FFV)\\
&= \frac{T_f^2 v}{\mathbf{m}_3^4} [23]\langle31\rangle - \frac{Y^*}{\mathbf{m}_1^2} \tilde m_1 [23]\langle3 \eta_1\rangle + \frac{Y}{\mathbf{m}_2^2} m_2 [\eta_23]\langle31\rangle - \frac{T_f}{\mathbf{m}_3^2} \tilde m_3 [23]\langle\eta_31\rangle \\
&+ \frac{T_f}{\mathbf{m}_3^2} m_3 [2\eta_3]\langle31\rangle - \frac{T_f^2 v}{\mathbf{m}_3^6} m_3\tilde m_3 [2\eta_3]\langle\eta_31\rangle - \frac{Y^* Y v}{\mathbf{m}_1^2 \mathbf{m}_2^2} \tilde m_1 m_2 [\eta_23]\langle3\eta_1\rangle +  \frac{Y^* T_f v}{\mathbf{m}_1^2 \mathbf{m}_3^2} \tilde m_1 \tilde m_3 [23]\langle\eta_3\eta_1\rangle \\
&- \frac{Y^* T_f v}{\mathbf{m}_1^2 \mathbf{m}_3^2} \tilde m_1 m_3 [2\eta_3]\langle3\eta_1\rangle - \frac{Y T_f v}{\mathbf{m}_2^2 \mathbf{m}_3^2} m_2 \tilde m_3 [\eta_23]\langle\eta_31\rangle + \frac{Y T_f v}{\mathbf{m}_2^2 \mathbf{m}_3^2} m_2 m_3 [\eta_2\eta_3]\langle31\rangle + \frac{Y^*}{\mathbf{m}_1^2 \mathbf{m}_3^2} \tilde m_1 m_3\tilde m_3  [2\eta_3]\langle\eta_3\eta_1\rangle  \\
& - \frac{Y}{\mathbf{m}_2^2 \mathbf{m}_3^2} m_2 m_3\tilde m_3  [\eta_2\eta_3]\langle\eta_31\rangle + \frac{Y^* Y T_f v^2}{\mathbf{m}_1^2 \mathbf{m}_2^2 \mathbf{m}_3^2} \tilde m_1 m_2\tilde m_3 [\eta_23]\langle\eta_3\eta_1\rangle - \frac{Y^* Y v}{\mathbf{m}_1^2 \mathbf{m}_2^2 \mathbf{m}_3^2} \tilde m_1 m_2 m_3[\eta_2\eta_3]\langle3\eta_1\rangle \\
&+ \frac{Y^* Y v}{\mathbf{m}_1^2 \mathbf{m}_2^2 \mathbf{m}_3^2} \tilde m_1 m_2 m_3\tilde m_3 [\eta_2\eta_3]\langle\eta_3\eta_1\rangle \label{eq:FFV_expand}
\end{aligned} \end{equation}

\section{Systematic Matching for 3-point Amplitudes}

In previous sections, we have investigated the massless-massive matching for scalar, fermion and vector boson, using the $FFS$ and $FFV$ amplitudes as illustrative examples. For cases involving massive gauge bosons, we have derived matching rules based on the conserved current of the theory. With these matching rules established, we can now systematically derive the massless-massive matching for all three-point scattering amplitudes in the Standard Model.

\subsection{Primary and leading MHC amplitudes}

Based on the Higgs insertion technique, we only need the leading matching to determine all MHC structures. As discussed previously, when a massive gauge boson is present, the primary MHC amplitude may vanish; in such cases, the leading non-vanishing contribution can come from descendant MHC amplitudes. In this subsection, we aim to identify which structures give the leading contributions to the SM amplitude and to explore the physical interpretation of the vanishing amplitudes.

In section 2, the primary and leading MHC amplitudes are identified by expanding the ST amplitudes, and then in section 3 the helicity-chirality unity is utilized to have correspondence to the massless helicity amplitudes.  In this subsection, we would turn this procedure around, and discuss how to obtain the primary and leading MHC amplitude without the need of ST amplitude expansion.

First let us analyze all possible helicity categories for the leading MHC amplitude. For a massive amplitude with spins $(s_1,s_2,s_3)$, all massless amplitudes with $|h_i|\le s_i$ can contribute to the massive structure. The leading MHC amplitude receive the contribution from 3-pt massless amplitude, so the corresponding helicity category must satisfy $h_1+h_2+h_3=\pm1$. The only exception is $SSS$, which corresponds to the 4-point massless amplitude. For convenience, we define the \textit{helicity class} as the set of helicity categories sharing the same number of transverse vector bosons, denoted as $n_T$. In the Standard Model, the helicity classes for all 3-pt massive amplitudes are
\begin{equation}
\begin{tabular}{c|c|c}
\hline
massive particle & helicity class & helicity category \\
\hline
SSS & $n_T=0$ & $(0,0,0)$ \\
\hline
FFS & $n_T=0$ & $(\pm\frac12,\pm\frac12,0)$ \\
\hline
FFV & \makecell{$n_T=0$\\$n_T=1$} & \makecell{$(\pm\frac12,\pm\frac12,0)$\\$(\pm\frac12,\mp\frac12,+1),(\pm\frac12,\mp\frac12,-1)$} \\
\hline
VVS & $n_T=1$ & $(\pm 1,0,0),(0,\pm 1,0)$ \\
\hline
VVV & \makecell{$n_T=1$\\$n_T=3$} & \makecell{$(\pm 1,0,0),(0,\pm 1,0),(0,0,\pm 1)$\\$(\pm 1,\mp 1,\mp 1),(\mp 1,\pm 1,\mp 1),(\mp 1,\mp 1,\pm 1)$} \\
\hline
\end{tabular}
\end{equation}

Then we list the possible helicity categories of primary amplitudes in the Standard Model, and check the total helicity 
\begin{equation}
h=h_1+h_2+h_3
\end{equation}
of these helicity categoriese. If they satisfy $h=\pm 1$ and they do not vanishes, we will say that these primary amplitude is leading. Otherwise, we will use the ladder operator to obtain the descendant MHC amplitudes with total helicity $h+1$ or $h-1$, until we find the non-vanishing amplitudes with total helicity $\pm1$ as the leading order contirbution.

\paragraph{1. FFS \& SSS}

For the $FFS$ amplitude, the primary MHC amplitudes has two helicity categories
\begin{equation} \begin{aligned}
h=-1:\quad &(-\tfrac12,-\tfrac12,0),& \\
h=+1:\quad &(+\tfrac12,+\tfrac12,0),& 
\end{aligned} \end{equation}
They satisfy the condition $h=\pm1$. The corresponding primary amplitude are
\begin{equation} \begin{aligned}
(-\tfrac12,-\tfrac12,0):&\quad\langle12\rangle=\Ampthree{1^-}{2^-}{3^0}{\fer{red}{i1}{v1}}{\antfer{cyan}{i2}{v1}}{\sca{i3}},\\
(+\tfrac12,+\tfrac12,0):&\quad [12]=\Ampthree{1^+}{2^+}{3^0}{\fer{cyan}{i1}{v1}}{\antfer{red}{i2}{v1}}{\sca{i3}}.
\end{aligned}\end{equation}
They do not vanish, they consist of $\lambda$ or $\tilde\lambda$ only. Thus, the priamry $FFS$ amplitude gives the leading contributions.

For the $SSS$ amplitude, it only has one helicity category $(0,0,0)$. The primary MHC amplitude corresponds to the trivial Lorentz structure
\begin{equation} \begin{aligned}
(0,0,0):\quad 1=\Ampthree{1^0}{2^0}{3^0}{\sca{i1}}{\sca{i2}}{\sca{i3}}.
\end{aligned}\end{equation}
It gives the leading contribution obviously.

\paragraph{2. FFV}

For the $FFV$ amplitude, the primary amplitude falls into two helicity categories with total helicity $h=0$,
\begin{equation} \begin{aligned}
h=0:\quad &(-\tfrac12,+\tfrac12,0),& &(+\tfrac12,-\tfrac12,0),& \\
\end{aligned} \end{equation}
The corresponding amplitude are
\begin{equation} \begin{aligned}
\langle13\rangle[32]=\Ampthree{1^-}{2^+}{3^0}{\fer{red}{i1}{v1}}{\antfer{red}{i2}{v1}}{\bos{i3}{brown}},\quad
[13]\langle32\rangle=\Ampthree{1^+}{2^-}{3^0}{\fer{cyan}{i1}{v1}}{\antfer{cyan}{i2}{v1}}{\bos{i3}{brown}}.
\end{aligned} \end{equation}
These are vanishing structures because they contain both $\langle ij\rangle$ and $[ij]$. Due to momentum conservation $\sum p_i = 0$, we have $\lambda_1 \propto \lambda_2 \propto \lambda_3$ or $\tilde\lambda_1 \propto \tilde\lambda_2 \propto \tilde\lambda_3$, which implies that any MHC amplitude involving both $\langle ij\rangle$ and $[ij]$ must vanish.

These vanishing primary amplitudes correspond to the combination between a primary $FF$ current $[\mathbf J]_0$ and a primary vector state $[\mathbf A]_0$. 
\begin{equation} \begin{aligned}
\langle13\rangle[32]&=[\mathbf J(1^{-\frac12},2^{+\frac12})]_0\cdot[\mathbf A(3^0)]_0,\\
[13]\langle32\rangle&=[\mathbf J(1^{+\frac12},2^{-\frac12})]_0\cdot[\mathbf A(3^0)]_0,
\end{aligned} \end{equation}
where the primary currents are
\begin{align}  \label{eq:current_FF_primary}
[\mathbf J(1^{-\frac12},2^{+\frac12})]_0=|2]^{\dot\alpha}\langle1|^{\alpha},\quad
[\mathbf J(1^{+\frac12},2^{-\frac12})]_0=|1]^{\dot\alpha}\langle2|^{\alpha}.
\end{align}
These currents correspond to the Noether current of massless fermion, $\bar\psi\gamma^\mu\psi$. They satisfy the conserved current condition
\begin{equation} \begin{aligned}
\partial\cdot[\mathbf J(1^{-\frac12},2^{+\frac12})]_0=\langle1|1+2|2]=0,\quad
\partial\cdot[\mathbf J(1^{+\frac12},2^{-\frac12})]_0=\langle2|1+2|1]=0.
\end{aligned} \end{equation}
Thus the conserved current condition is closely associated with the vanishing MHC amplitudes.

Then we can apply the ladder operator to the primary amplitude and obtain the 1st descendant amplitudes, which have six helicity categories,
\begin{equation} \begin{aligned}
h=+1:\quad&(+\tfrac12,+\tfrac12,0),(+\tfrac12,-\tfrac12,+1),(-\tfrac12,+\tfrac12,+1)& \\
h=-1:\quad&(-\tfrac12,-\tfrac12,0),(+\tfrac12,-\tfrac12,-1),(-\tfrac12,+\tfrac12,-1)&
\end{aligned} \end{equation}
They satisfy the condition $h=\pm1$, so the 1st descendant amplitude can provide the leading contribution. As examples, we consider two typical helicity categories: $(-\tfrac12,+\tfrac12,+1)$ with one transverse vector, and $(+\tfrac12,+\tfrac12,0)$ with zero transverse vectors. Their corresponding amplitudes are
\begin{equation} \begin{aligned}
(-\tfrac12,+\tfrac12,+1):&\quad\tilde m_3[23]\langle\eta_31\rangle= \Ampthree{1^-}{2^+}{3^+}{\fer{red}{i1}{v1}}{\antfer{red}{i2}{v1}}{\bosflip{1}{-55}{brown}{cyan}},\\
(+\tfrac12,+\tfrac12,0):&\quad\tilde m_1[23]\langle3\eta_1\rangle=
\Ampthree{1^+}{2^+}{3^0}{\ferflip{1}{180}{cyan}{red}}{\antfer{red}{i2}{v1}}{\bos{i3}{brown}},\quad
\tilde m_2\langle\eta_23\rangle[31]=
\Ampthree{1^+}{2^+}{3^0}{\fer{cyan}{i1}{v1}}{\antferflip{1}{55}{red}{cyan}}{\bos{i3}{brown}},
\end{aligned} \end{equation}
where the crosses in the diagrams indicate the action of the ladder operator. We have verified that these expressions are non-vanishing, confirming that the 1st descendant amplitudes indeed contribute at leading order. Amplitudes for other helicity categories can be derived in a similar way, with corresponding diagrams differing in color and the position of the cross.

Recall that we decompose the primary amplitude as the form $[\mathbf J]_0\cdot[\mathbf A]_0$. By applying the ladder operator $mJ^-$ or $\tilde mJ^+$ to the primary current $[\mathbf{J}]_0$, we obtain the 1st descendant current $[\mathbf{J}]_1$. For example, both primary currents with helicity $(\mp\frac12,\pm\frac12)$ can be flipped into $(+\frac12,+\frac12)$ or $(-\frac12,-\frac12)$, giving 
\begin{equation} \begin{aligned}
\relax
[\mathbf J(1^{+},2^{+})]_1&=
\tilde m_2J_2^+\circ [\mathbf J(1^{+},2^{-})]_0+
\tilde m_1J_1^+\circ [\mathbf J(1^{-},2^{+})]_0\\
[\mathbf J(1^{-},2^{-})]_1&=
m_1J^-_1\circ [\mathbf J(1^{+},2^{-})]_0+
m_2J^-_2\circ [\mathbf J(1^{-},2^{+})]_0
\end{aligned}\end{equation}
Substituting from eq.~\eqref{eq:current_FF_primary}, we obtain the 1st descendant currents
\begin{align}
(+\tfrac{1}{2},+\tfrac{1}{2}):\quad [\mathbf J(1^{+},2^{+})]_1&=c_1\tilde m_1|2]^{\dot\alpha}\langle\eta_1|^{\alpha}+c_2\tilde m_2|1]^{\dot\alpha}\langle\eta_2|^{\alpha},\\
(-\tfrac{1}{2},-\tfrac{1}{2}):\quad [\mathbf J(1^{-},2^{-})]_1&=-c_1 m_2|\eta_2]^{\dot\alpha}\langle1|^{\alpha}-c_2 m_1|\eta_1]^{\dot\alpha}\langle2|^{\alpha}.
\end{align}
This current is not conserved, so the product $[\mathbf{J}]_1 \cdot [\mathbf{A}]_0$ does not vanish.

On the other hand, applying the ladder operator to the primary vector $[\mathbf{A}]_0$ gives the 1st descendant vector $[\mathbf{A}]_1$. It couples to the conserved current $[\mathbf{J}]_0$, and both $[\mathbf{J}]_1 \cdot [\mathbf{A}]_0$ and $[\mathbf{J}]_0 \cdot [\mathbf{A}]_1$ yield non-vanishing results.
Therefore, the 1st descendant amplitudes can be expressed in the following two forms
\begin{equation}
\begin{tabular}{c|c|c}
\hline
helicity & current form & diagram \\
\hline
$(-\tfrac12,+\tfrac12,+1)$ & $[\mathbf{J}(1^{-\frac12},2^{+\frac12})]_0\cdot [\mathbf{A}(3^+)]_1$ & $\Ampthree{1^-}{2^+}{3^+}{\fer{red}{i1}{v1}}{\antfer{red}{i2}{v1}}{\bosflip{1}{-55}{brown}{cyan}}$\\
\hline
$(+\tfrac12,+\tfrac12,0)$ & $[\mathbf{J}(1^{+\frac12},2^{+\frac12})]_1\cdot [\mathbf{A}(3^0)]_0$ & 
$\Ampthree{1^+}{2^+}{3^0}{\ferflip{1}{180}{cyan}{red}}{\antfer{red}{i2}{v1}}{\bos{i3}{brown}}+\Ampthree{1^+}{2^+}{3^0}{\fer{cyan}{i1}{v1}}{\antferflip{1}{55}{red}{cyan}}{\bos{i3}{brown}}$\\
\hline 
\end{tabular}
\end{equation}
Note that the last line contains two MHC amplitudes, since the descendant current $[\mathbf{J}(1^{+\frac12}, 2^{+\frac12})]_1$ consists of two terms. Similarly, MHC amplitudes in other helicity categories can also be rewritten as $[\mathbf{J}]_1 \cdot [\mathbf{A}]_0$ or $[\mathbf{J}]_0 \cdot [\mathbf{A}]_1$.

\paragraph{3. VVS}

For the $VVS$ amplitude, the primary amplitude has only one  helicity category,
\begin{equation} \begin{aligned}
h=0:\quad (0,0,0).
\end{aligned}\end{equation}
The corresponding amplitude is given by
\begin{equation} \begin{aligned}
(0,0,0):\quad \langle12\rangle[21]=\Ampthree{1^0}{2^0}{3^0}{\bos{i1}{brown}}{\bos{i2}{brown}}{\sca{i3}}
\end{aligned}\end{equation}
This amplitude vanishes due to the simultaneous presence of $\langle12\rangle$ and $[21]$. The vanishing primary amplitude can also be expressed as the current form $[\mathbf{J}]_0 \cdot [\mathbf{A}]_0$. There are two equivalent ways to perform the decomposition: 
\begin{equation} \begin{aligned}
\langle12\rangle[21]=[\mathbf J(2^{0},3^{0})]_0\cdot[\mathbf A(1^{0})]_0=[\mathbf J(1^{0},3^{0})]_0\cdot[\mathbf A(2^{0})]_0,
\end{aligned}\end{equation}
where the primary $VS$ current $[\mathbf{J}]_0$ is defined as
\begin{equation} \begin{aligned}
\relax
[\mathbf{J}(2^{0},3^{0})]_0&=|2]^{\dot\alpha}\langle2|^{\alpha}, \\
[\mathbf{J}(1^{0},3^{0})]_0&=|1]^{\dot\alpha}\langle1|^{\alpha}.
\end{aligned}\end{equation}
This current is conserved and can be interpreted as the Noether current for a massless scalar boson, $\phi^* (D^\mu \phi) - (D^\mu \phi) \phi$~\footnote{Strictly speaking, the Noether current $\phi^ (D^\mu \phi) - (D^\mu \phi^*) \phi$ corresponds to the spinor structure $\tfrac12(|2]^{\dot\alpha}\langle2|^{\alpha} - |3]^{\dot\alpha}\langle3|^{\alpha})$, which is not identical to the primary MHC current $[\mathbf{J}]_0$. However, when coupled to a vector $[\mathbf{A}]_0$, the two become equivalent: $[\mathbf{J}]_0 \cdot [\mathbf{A}]_0 = \langle12\rangle[21] = \tfrac12(\langle12\rangle[21] - \langle13\rangle[31])$.}.

Applying the ladder operator to the primary $VVS$ amplitude yields the first descendant amplitudes, which fall into four helicity categories,
\begin{equation} \begin{aligned}
h=+1:\quad&(-1,0,0),(0,-1,0)& \\
h=-1:\quad&(+1,0,0),(0,+1,0)&
\end{aligned} \end{equation}
Since these satisfy $h = \pm 1$, the corresponding amplitudes are leading contributions:
\begin{equation} \begin{aligned}
m_1\langle12\rangle[2\eta_1]&=\Ampthree{1^-}{2^0}{3^0}{\bosflip{1}{180}{brown}{red}}{\bos{i2}{brown}}{\sca{i3}},&
-\tilde m_1\langle\eta_12\rangle[21]&=\Ampthree{1^+}{2^0}{3^0}{\bosflip{1}{180}{brown}{cyan}}{\bos{i2}{brown}}{\sca{i3}} \\
m_2\langle12\rangle[\eta_21]&=\Ampthree{1^0}{2^-}{3^0}{\bos{i1}{brown}}{\bosflip{1}{55}{brown}{red}}{\sca{i3}},&
-\tilde m_2\langle1\eta_2\rangle[21]&=\Ampthree{1^0}{2^+}{3^0}{\bos{i1}{brown}}{\bosflip{1}{55}{brown}{cyan}}{\sca{i3}}.\\
\end{aligned}\end{equation}
These can also be rewritten in terms of MHC currents and vector states. For each helicity, two decomposition forms remain available. As an example, consider the helicity category $(-1,0,0)$:
\begin{equation}
\begin{tabular}{c|c|c}
\hline
helicity & current form  & diagram \\
\hline
$(-1,0,0)$ & \makecell{$[\mathbf{J}(2^{0},3^{0})]_0\cdot [\mathbf{A}(1^-)]_1$\\$[\mathbf{J}(1^{-1},3^{0})]_1\cdot [\mathbf{A}(1^0)]_0$} & $\Ampthree{1^-}{2^0}{3^0}{\bosflip{1}{180}{brown}{red}}{\bos{i2}{brown}}{\sca{i3}}$\\
\hline
\end{tabular}
\end{equation}
Amplitudes for other helicities can similarly be rewritten as $[\mathbf{J}]_1 \cdot [\mathbf{A}]_0$ or $[\mathbf{J}]_0 \cdot [\mathbf{A}]_1$. 
By applying ladder operators to the primary current, we can change the helicity to $(\pm1,0)$ and obtain two 1st descendant currents
\begin{align}
[\mathbf J(1^{0},2^{0})]_0
&\xrightarrow{m_1 J_1^-}
[\mathbf J(1^{-},2^{0})]_1=-m_1|\eta_1]^{\dot\alpha}\langle1|^{\alpha},\\
[\mathbf J(1^{0},2^{0})]_0
&\xrightarrow{\tilde m_1 J_1^+}
[\mathbf J(1^{+},2^{0})]_1=\tilde m_1|1]^{\dot\alpha}\langle\eta_1|^{\alpha}
\end{align}
In analogy with the $FFV$ case, the 1st descendant current $[\mathbf{J}]_1$ is not conserved.

\paragraph{4. VVV}

For the $VVV$ amplitude, the primary amplitudes fall into six helicity categories, classified by the total helicity $h$,
\begin{equation} \begin{aligned}
h=-1:\quad &(-1,0,0),(0,-1,0),(0,0,-1),& \\
h=+1:\quad &(+1,0,0),(0,+1,0),(0,0,+1).& \\
\end{aligned} \end{equation}
The corresponding amplitudes are
\begin{equation} \begin{aligned} \label{eq:VVV_primary}
&\langle12\rangle[23]\langle31\rangle,&
&\langle12\rangle\langle23\rangle[31],& 
&[12]\langle23\rangle\langle31\rangle, \\  
&[12]\langle23\rangle[31],&
&[12][23]\langle31\rangle,&
&\langle12\rangle[23][31].
\end{aligned} \end{equation}
Although these primary amplitudes carry total helicity $h = \pm 1$, they vanish because each contains both angle and square brackets $\langle ij\rangle$ and $[ij]$. Diagrammatically, these vanishing primary amplitudes can be represented as
\begin{equation} \begin{aligned}
&\Ampthree{1^-}{2^0}{3^0}{\bos{i1}{red}}{\bos{i2}{brown}}{\bos{i3}{brown}},\quad
\Ampthree{1^0}{2^-}{3^0}{\bos{i1}{brown}}{\bos{i2}{red}}{\bos{i3}{brown}}, \quad
\Ampthree{1^0}{2^0}{3^-}{\bos{i1}{brown}}{\bos{i2}{brown}}{\bos{i3}{red}},\quad \\  
&\Ampthree{1^+}{2^0}{3^0}{\bos{i1}{cyan}}{\bos{i2}{brown}}{\bos{i3}{brown}},\quad
\Ampthree{1^0}{2^+}{3^0}{\bos{i1}{brown}}{\bos{i2}{cyan}}{\bos{i3}{brown}},\quad
\Ampthree{1^0}{2^0}{3^+}{\bos{i1}{brown}}{\bos{i2}{brown}}{\bos{i3}{cyan}}.
\end{aligned} \end{equation}

In these diagrams, only the brown line corresponds to the vector $[\mathbf{A}]_0$. Therefore, each primary $VVV$ amplitude can be expressed in two equivalent ways as the current form $[\mathbf{J}]_0 \cdot [\mathbf{A}]_0$. For the helicity category $(-1,0,0)$: 
\begin{equation} \begin{aligned}
\langle12\rangle[23]\langle31\rangle=[\mathbf J(1^{-1},3^{0})]_0\cdot[\mathbf A(2^{0})]_0=[\mathbf J(1^{-1},2^{0})]_0\cdot[\mathbf A(3^{0})]_0,
\end{aligned}\end{equation}
where the primary $VV$ current $[\mathbf{J}]_0$ is conserved current and defined as
\begin{align}
[\mathbf J(1^+,2^0)]_0&=-c_1[12]|1]^{\dot\alpha}\langle2|^{\alpha},\\
[\mathbf J(1^-,2^0)]_0&=-c_2\langle12\rangle |2]^{\dot\alpha}\langle1|^{\alpha},\\
[\mathbf J(1^0,2^+)]_0&=-c_3[12]|2]^{\dot\alpha}\langle1|^{\alpha},\\
[\mathbf J(1^0,2^-)]_0&=-c_4\langle12\rangle |1]^{\dot\alpha}\langle2|^{\alpha},
\end{align}
These two primary currents share the same 1-pt structure, the field strength tensor of particle 1:
\begin{equation} \begin{aligned}
\mathcal{F}(1^-)&=\langle1|^{\alpha} \langle1|^{\beta}.
\end{aligned}\end{equation}
In this sense, the primary $VVV$ amplitudes correspond to 1-pt massless structures, consistent with the power counting analysis in subsection~\ref{sec:expansion}. Priamry amplitudes for other helicitiy categoreis can likewise be decomposed into $[\mathbf{J}]_0 \cdot [\mathbf{A}]_0$ in two ways, each corresponding to a field strength tensor with different helicity or associated with different legs.

For the 1st descendant MHC amplitudes, there are thirteen helicity categories,
\begin{equation} \begin{aligned}
h&=+2:& &(+1, +1,0),\quad(0,+1,+1),\quad(+1,0,+1),\\
h&=0:& &(\pm1, \mp1,0),\quad(0,\pm1,\mp1),\quad(\pm1,0,\mp1),\quad (0,0,0)\\
h&=-2:& &(-1, -1,0),\quad(0,-1,-1),\quad(-1,0,-1).
\end{aligned} \end{equation}
These do not satisfy the condition $h=\pm1$, so the 1st descendant $VVV$ amplitude do not correspond to the massless amplitude in the SM, and they should either vanish or belong to the EFT category, depending on whether the current is conserved or not. The primary current could induce the 1st descendant current, in five helicity categories $(\pm1,\pm1)$, $(\pm1,\mp1)$ and $(0,0)$
\begin{align}
[\mathbf J(1^+,2^+)]_1=&
c_1\tilde m_2 [12]|1]^{\dot\alpha}\langle\eta_2|^{\alpha}+
c_3\tilde m_1 [12] |2]^{\dot\alpha}\langle\eta_1|^{\alpha} ,\\
[\mathbf J(1^+,2^-)]_1=&
-c_1 m_2[1\eta_2] |1]^{\dot\alpha}\langle2|^{\alpha}+c_4\tilde m_1\langle\eta_12\rangle |1]^{\dot\alpha}\langle2|^{\alpha},\\
[\mathbf J(1^-,2^+)]_1=& 
c_2 \tilde m_2\langle1\eta_2\rangle |2]^{\dot\alpha}\langle1|^{\alpha}
-c_3 m_1[\eta_1 2] |2]^{\dot\alpha}\langle1|^{\alpha},\\
[\mathbf J(1^-,2^-)]_1=& 
-c_2 m_2\langle12\rangle |\eta_2]^{\dot\alpha}\langle1|^{\alpha}+
-c_4 m_1\langle12\rangle |\eta_1]^{\dot\alpha}\langle2|^{\alpha},\\
[\mathbf J(1^0,2^0)]_1=&
-c_1 m_1([\eta_12]|1]^{\dot\alpha}\langle2|^{\alpha}+[12]|\eta_1]^{\dot\alpha}\langle2|^{\alpha}) \nonumber\\
&+c_2 \tilde m_1(\langle\eta_12\rangle |2]^{\dot\alpha}\langle1|^{\alpha}+\langle12\rangle |2]^{\dot\alpha}\langle\eta_1|^{\alpha}) \nonumber\\
&-c_3 m_2([1\eta_2]|2]^{\dot\alpha}\langle1|^{\alpha}+[12]|\eta_2]^{\dot\alpha}\langle1|^{\alpha}) \nonumber\\
&+c_4 \tilde m_2(\langle1\eta_2\rangle |1]^{\dot\alpha}\langle2|^{\alpha}+\langle12\rangle |1]^{\dot\alpha}\langle\eta_2|^{\alpha}),
\end{align}

Based on the helicity categories and the currents, we classify the 1st descendant amplitudes into three cases:
\begin{itemize}
\item First consider the helicity categories with $h=\pm2$, which have two transverse vectors of the same helicity. For example, the helicity $(-1,-1,0)$ includes two terms,
\begin{equation} \begin{aligned}
(-1,-1,0):&\quad 
m_2\langle12\rangle[\eta_2 3]\langle31\rangle
=\Ampthree{1^-}{2^-}{3^0}{\bos{i1}{red}}{\bosflip{1}{55}{brown}{red}}{\bos{i3}{brown}},
m_1\langle12\rangle\langle23\rangle[3\eta_1]
=\Ampthree{1^-}{2^-}{3^0}{\bosflip{1}{180}{brown}{red}}{\bos{i2}{red}}{\bos{i3}{brown}}.
\end{aligned} \end{equation}
They are derived by acting ladder operators on the primary amplitudes with helicities $(-1,0,0)$ and $(0,-1,0)$. Each of the two terms can be identified as $[\mathbf{J}]_1 \cdot [\mathbf{A}]_0$, while together they can be identified as $[\mathbf{J}]_0 \cdot [\mathbf{A}]_1$:
\begin{equation} \begin{aligned}
[\mathbf J(1^{-1},2^{+1})]_1\cdot[\mathbf A(3^0)]_0=
c_1 m_2\langle12\rangle[\eta_2 3]\langle31\rangle
+c_2 m_1\langle12\rangle\langle23\rangle[3\eta_1]
\end{aligned} \end{equation}
where $c_i$ are coefficients. Although the two terms are individually non-zero, their sum can vanish
\begin{equation} \begin{aligned}
\frac{1}{\mathbf m^2_2} m_2\langle12\rangle[\eta_2 3]\langle31\rangle
+\frac{1}{\mathbf m^2_1}m_1\langle12\rangle\langle23\rangle[3\eta_1]=0
\end{aligned} \end{equation}
Hence, the 1st descendant current $[\mathbf{J}(1^{-1},2^{-1})]_1$ is conserved only if takes the form 
\begin{equation} \label{eq:conserved_descendant}
[\mathbf J(1^{-1},2^{-1})]_1= -\frac{1}{\mathbf m_2^2} m_2\langle12\rangle |\eta_2]^{\dot\alpha}\langle1|^{\alpha}
-\frac{1}{\mathbf m_1^2} m_1\langle12\rangle |\eta_1]^{\dot\alpha}\langle2|^{\alpha}
\end{equation}
For the other helicity category with $h=\pm2$, each category corresponds to one such conserved current.

This conserved current corresponds is related to the Yang-Mills current $F^{\mu\nu}A_\nu$. According to the Weinberg-Witten theorem, a gauge-invariant conserved current for Yang-Mills fields does not exist. Thus, in the spinor representation, the conserved current must depend on the reference spinor $\xi_i$ (which serve as a gauge parameter). For helicity $(-1,+1)$, the Yang-Mills current is
\begin{equation} \begin{aligned}
J(1^{-1},2^{-1})&=\frac{\langle12\rangle}{[\xi_2 2]} |\xi_2]^{\dot\alpha}\langle1|^{\alpha}+\frac{\langle12\rangle}{[\xi_1 1]} |\xi_1]^{\dot\alpha}\langle2|^{\alpha},
\end{aligned}\end{equation}
Taking $\xi_i=\eta_i$, this expression matches the 1st descendant current in eq.~\eqref{eq:conserved_descendant}.

\item For helicity categories with $h=0$, there are two cases. We first consider the helicity $(0,0,0)$. It has six terms,
\begin{align}
(0,0,0):\quad&\tilde m_1(\langle \eta_12\rangle[23]\langle31\rangle+\langle 12\rangle[23]\langle3\eta_1\rangle)
=\Ampthree{1^0}{2^0}{3^0}{\bosflip{1}{180}{red}{brown}}{\bos{i2}{brown}}{\bos{i3}{brown}},& \nonumber \\
&\tilde m_2(\langle1\eta_2\rangle\langle23\rangle[31]+\langle12\rangle\langle\eta_23\rangle[31])
=\Ampthree{1^0}{2^0}{3^0}{\bos{i1}{brown}}{\bosflip{1}{55}{red}{brown}}{\bos{i3}{brown}},& \nonumber \\
&\tilde m_3([12]\langle2\eta_3\rangle\langle31\rangle+[12]\langle23\rangle\langle\eta_31\rangle)
=\Ampthree{1^0}{2^0}{3^0}{\bos{i1}{brown}}{\bos{i2}{brown}}{\bosflip{1}{-55}{red}{brown}}, \nonumber \\  
&m_1([\eta_12]\langle23\rangle[31]+[12]\langle23\rangle[3\eta_1])
=\Ampthree{1^0}{2^0}{3^0}{\bosflip{1}{180}{cyan}{brown}}{\bos{i2}{brown}}{\bos{i3}{brown}},& \nonumber \\
&m_2([1\eta_2][23]\langle31\rangle+[12][\eta_23]\langle31\rangle)
=\Ampthree{1^0}{2^0}{3^0}{\bos{i1}{brown}}{\bosflip{1}{55}{cyan}{brown}}{\bos{i3}{brown}},& \nonumber \\
&m_3(\langle12\rangle[2\eta_3][31]+\langle12\rangle[23][\eta_31])
=\Ampthree{1^0}{2^0}{3^0}{\bos{i1}{brown}}{\bos{i2}{brown}}{\bosflip{1}{-55}{cyan}{brown}}. 
\end{align}
If we identify particle 3 as the vector $[\mathbf A]_0$, there are four terms can compose the form $[\mathbf J]_1\cdot [\mathbf A]_0$, 
\begin{equation} \begin{aligned}
[\mathbf J(1^{0},2^{0})]_1\cdot[\mathbf A(3^0)]_0=
&c_1\tilde m_1(\langle \eta_12\rangle[23]\langle31\rangle+\langle 12\rangle[23]\langle3\eta_1\rangle)\\
&+c_2\tilde m_2(\langle1\eta_2\rangle\langle23\rangle[31]+\langle12\rangle\langle\eta_23\rangle[31])\\
&+c_3 m_1([\eta_12]\langle23\rangle[31]+[12]\langle23\rangle[3\eta_1])\\
&+c_4 m_2([1\eta_2][23]\langle31\rangle+[12][\eta_23]\langle31\rangle),
\end{aligned} \end{equation}
These terms contain $[ij]$ and $\langle ij \rangle$ simultaneously, and thus vanish individually. Therefore, the current $[\mathbf{J}(1^{0},2^{0})]_1$ is also conserved, with no constraints on the four coefficients.

\item Finally, we consider the other helicity categories with $h=0$, which has two transverse vector of opposite helicity. For example, the helicity $(-1,+1,0)$ includes two terms,
\begin{equation} \begin{aligned}
(-1,+1,0):&\quad \tilde m_2\langle1\eta_2\rangle[23]\langle31\rangle 
=\Ampthree{1^-}{2^+}{3^0}{\bos{i1}{red}}{\bosflip{1}{55}{brown}{cyan}}{\bos{i3}{brown}},
m_1[\eta_1 2][23]\langle31\rangle
=\Ampthree{1^-}{2^+}{3^0}{\bosflip{1}{180}{brown}{red}}{\bos{i2}{cyan}}{\bos{i3}{brown}},\\
\end{aligned} \end{equation}
which are derived by acting ladder operators on the primary amplitudes with helicities $(-1,0,0)$ and $(0,+1,0)$. We identify these two terms as the current form $[\mathbf J]_1\cdot[\mathbf A]_0$:
\begin{equation} \begin{aligned}
[\mathbf J(1^{-1},2^{-1})]_1\cdot[\mathbf A(3^0)]_0=
c_1\tilde m_2\langle1\eta_2\rangle[23]\langle31\rangle
+c_2 m_1[\eta_1 2][23]\langle31\rangle,
\end{aligned} \end{equation}
Both terms contain $[23]$ and $\langle 31 \rangle$, and thus vanish individually.
\begin{equation} \begin{aligned}
\tilde m_2\langle1\eta_2\rangle[23]\langle31\rangle
=m_1[\eta_1 2][23]\langle31\rangle=0
\end{aligned} \end{equation}
Therefore, the current $[\mathbf{J}(1^{-1},2^{-1})]_1$ is conserved, with no constraints on the coefficients $c_1$ and $c_2$. The same holds for currents in other helicity categories with $h = \pm 2$.

\end{itemize}

In summary, we have shown that all 1st descendant MHC amplitudes can vanish and found all conserved current $[\mathbf J]_1$. These currents and the primary currents $[\mathbf{J}]_0$ will be used next to analyze the 2nd descendant MHC amplitudes.

For 2nd descendant MHC amplitudes, it has twlve helicity categories,
\begin{equation} \begin{aligned}
h&=+1:& &(-1,+1,+1),& &(+1,-1,+1),& &(+1,+1,-1),&\\
&& &(0,0,+1),& &(0,+1,0),& &(+1,0,0),&\\
h&=-1:& &(-1, -1,+1),& &(+1,-1,-1),& &(-1,+1,-1),&\\
&& &(0,0,-1),& &(0,-1,0),& &(-1,0,0),&
\end{aligned} \end{equation}
They satisfy the condition $h=\pm1$, and one can verify that the MHC amplitudes do not vanish. Therefore, the leading $VVV$ contribution arises from 2nd descendant MHC amplitudes. Due to the large number of such amplitudes, we restrict our consideration to two typical helicity categories $(+1,+1,-1)$ and $(-1,0,0)$, which has three and one transverse vector bosons. The other helicity categories can be obtained in a similar way, whose diagram can be derived by change color and location of mass insertion.

The 2nd descendant amplitudes would originate from $[J]_2 \cdot [A]_0$ and others. Here we list the 2nd descendant currents: 
\begin{align}
[\mathbf J(1^+,2^0)]_2=&
c_1\tilde m_2 m_2[1\eta_2]|1]^{\dot\alpha}\langle\eta_2|^{\alpha}
-c_2\tilde m_1^2\langle\eta_12\rangle |2]^{\dot\alpha}\langle\eta_1|^{\alpha} \nonumber\\
&+c_3\tilde m_1 m_2([1\eta_2]|2]^{\dot\alpha}\langle\eta_1|^{\alpha}+[12]|\eta_2]^{\dot\alpha}\langle\eta_1|^{\alpha})\nonumber\\
&-c_4\tilde m_1 \tilde m_2(\langle\eta_1\eta_2\rangle |1]^{\dot\alpha}\langle2|^{\alpha}+\langle\eta_12\rangle |1]^{\dot\alpha}\langle\eta_2|^{\alpha}),\\
[\mathbf J(1^-,2^0)]_2=&
-c_1 m_1^2[\eta_12]|\eta_1]^{\dot\alpha}\langle2|^{\alpha}
+c_2\tilde m_2 m_2\langle1\eta_2\rangle |\eta_2]^{\dot\alpha}\langle1|^{\alpha} \nonumber\\
&-c_3 m_1 m_2([\eta_1\eta_2]|2]^{\dot\alpha}\langle1|^{\alpha}+[\eta_12]|\eta_2]^{\dot\alpha}\langle1|^{\alpha})\nonumber \\
&+c_4 m_1\tilde m_2(\langle1\eta_2\rangle |\eta_1]^{\dot\alpha}\langle2|^{\alpha}+\langle1\eta_2\rangle |\eta_1]^{\dot\alpha}\langle2|^{\alpha}),\\
[\mathbf J(1^0,2^+)]_2=&
c_1 m_1\tilde m_2([\eta_12]|1]^{\dot\alpha}\langle\eta_2|^{\alpha}+[12]|\eta_1]^{\dot\alpha}\langle\eta_2|^{\alpha}) \nonumber\\
&-c_2 m_1m_2(\langle\eta_1\eta_2\rangle |2]^{\dot\alpha}\langle1|^{\alpha}+\langle1\eta_2\rangle |2]^{\dot\alpha}\langle\eta_1|^{\alpha}) \nonumber\\
&+c_3 \tilde m_1 m_1[\eta_12]|2]^{\dot\alpha}\langle\eta_1|^{\alpha}
-c_4 \tilde m_2^2\langle1\eta_2\rangle |1]^{\dot\alpha}\langle\eta_2|^{\alpha},\\
[\mathbf J(1^0,2^-)]_2=&
-c_1 m_1 m_2([\eta_1\eta_2]|1]^{\dot\alpha}\langle2|^{\alpha}+ [1\eta_2]|\eta_1]^{\dot\alpha}\langle2|^{\alpha}) \nonumber\\
&+c_2 \tilde m_1 m_2(\langle\eta_12\rangle |\eta_2]^{\dot\alpha}\langle1|^{\alpha}+\langle12\rangle |\eta_2]^{\dot\alpha}\langle\eta_1|^{\alpha}) \nonumber\\
&-c_3 m_2^2[1\eta_2]|\eta_2]^{\dot\alpha}\langle1|^{\alpha}
+c_4 \tilde m_1^2\langle\eta_12\rangle |\eta_1]^{\dot\alpha}\langle2|^{\alpha},
\end{align}

For the $(+1,+1,-1)$ helicity category, there are three terms,
\begin{equation} \begin{aligned} \label{eq:VVV_leading_1}
\tilde m_1 m_3 [12][2 \eta_3]\langle3\eta_1\rangle
&=\Ampthree{1^+}{2^+}{3^-}{\bosflip{1}{180}{brown}{cyan}}{\bos{i2}{cyan}}{\bosflip{1}{-55}{brown}{red}},&
\tilde m_2 m_3 [12]\langle\eta_23\rangle[\eta_31]
&=\Ampthree{1^+}{2^+}{3^-}{\bos{i1}{cyan}}{\bosflip{1}{55}{brown}{cyan}}{\bosflip{1}{-55}{brown}{red}},& \\
\tilde m_1 \tilde m_2 [12]\langle\eta_23\rangle\langle 3 \eta_1\rangle
&=\Ampthree{1^+}{2^+}{3^-}{\bosflip{1}{180}{brown}{cyan}}{\bosflip{1}{55}{brown}{cyan}}{\bos{i3}{red}},&
\end{aligned}\end{equation}
In this case, any two of the above 2nd descendant amplitudes can be identified as a massive current $[\mathbf{J}]_1$ that couples to a vector $[\mathbf{A}]_1$:
\begin{equation}
\begin{tabular}{c|c}
\hline
current form  & diagram \\
\hline
$[\mathbf{J}(1^+,2^+)]_1\cdot [\mathbf{A}(3^-)]_1$ & 
$\Ampthree{1^+}{2^+}{3^-}{\bosflip{1}{180}{brown}{cyan}}{\bos{i2}{cyan}}{\bosflip{1}{-55}{brown}{red}}+\Ampthree{1^+}{2^+}{3^-}{\bos{i1}{cyan}}{\bosflip{1}{55}{brown}{cyan}}{\bosflip{1}{-55}{brown}{red}}$\\
\hline
$[\mathbf{J}(1^+,3^-)]_1\cdot [\mathbf{A}(2^+)]_1$ & $\Ampthree{1^+}{2^+}{3^-}{\bos{i1}{cyan}}{\bosflip{1}{55}{brown}{cyan}}{\bosflip{1}{-55}{brown}{red}}+\Ampthree{1^+}{2^+}{3^-}{\bosflip{1}{180}{brown}{cyan}}{\bosflip{1}{55}{brown}{cyan}}{\bos{i3}{red}}$\\
\hline
$[\mathbf{J}(2^+,3^-)]_1\cdot [\mathbf{A}(1^+)]_1$ & $\Ampthree{1^+}{2^+}{3^-}{\bosflip{1}{180}{brown}{cyan}}{\bos{i2}{cyan}}{\bosflip{1}{-55}{brown}{red}}+\Ampthree{1^+}{2^+}{3^-}{\bosflip{1}{180}{brown}{cyan}}{\bosflip{1}{55}{brown}{cyan}}{\bos{i3}{red}}$\\
\hline 
\end{tabular}
\end{equation}
where $[\mathbf{J}]_1$ is the conserved current we previously derived from the vanishing 1st descendant MHC amplitudes.

For the $(-1,0,0)$ helicity category, there are seven terms,
\begin{align}
-m_1 \tilde m_3 [\eta_1 2]\langle2\eta_3\rangle\langle31\rangle
=\Ampthree{1^-}{2^0}{3^0}{\bosflip{1}{180}{brown}{red}}{\bos{i2}{brown}}{\bosflip{1}{-55}{red}{brown}},\qquad 
-m_1 \tilde m_2 \langle12\rangle\langle\eta_23\rangle[3\eta_1]
&=\Ampthree{1^-}{2^0}{3^0}{\bosflip{1}{180}{brown}{red}}{\bosflip{1}{55}{red}{brown}}{\bos{i3}{brown}},& \\
-m_3\tilde m_3\langle12\rangle[2\eta_3]\langle\eta_31\rangle
=\Ampthree{1^-}{2^0}{3^0}{\bos{i1}{red}}{\bos{i2}{brown}}{\bosflipflip{1}{-55}{brown}{red}{brown}},\qquad
-m_2\tilde m_2\langle1\eta_2\rangle[\eta_23]\langle31\rangle
&=\Ampthree{1^-}{2^0}{3^0}{\bos{i1}{red}}{\bosflipflip{1}{55}{brown}{red}{brown}}{\bos{i3}{brown}},&  \\
m_1 m_3 (\langle12\rangle[2\eta_3][3\eta_1]+\langle12\rangle[23][\eta_3\eta_1])
&=\Ampthree{1^-}{2^0}{3^0}{\bosflip{1}{180}{brown}{red}}{\bos{i2}{brown}}{\bosflip{1}{-55}{cyan}{brown}},& \\
m_1 m_2 ([\eta_1 2][\eta_2 3]\langle31\rangle+[\eta_1 \eta_2][2 3]\langle31\rangle)
&=\Ampthree{1^-}{2^0}{3^0}{\bosflip{1}{180}{brown}{red}}{\bosflip{1}{55}{cyan}{brown}}{\bos{i3}{brown}},& \\
c_4 m_1^2[\eta_1 2]\langle23\rangle[3 \eta_1]
&=\Ampthree{1^-}{2^0}{3^0}{\bosflipflip{1}{180}{cyan}{brown}{red}}{\bos{i2}{brown}}{\bos{i3}{brown}}.&
\end{align}

In this case, we can identify several 2nd descendant amplitudes as the current form $[\mathbf{J}]_0\cdot[\mathbf{A}]_2$, $[\mathbf{J}]_1\cdot[\mathbf{A}]_1$ or $[\mathbf{J}]_2\cdot[\mathbf{A}]_0$:
\begin{equation}
\begin{tabular}{c|c}
\hline
current form  & diagram \\
\hline
$[\mathbf{J}(1^-,2^0)]_2\cdot [\mathbf{A}(3^0)]_0$ & 
$\Ampthree{1^-}{2^0}{3^0}{\bosflip{1}{180}{brown}{red}}{\bosflip{1}{55}{red}{brown}}{\bos{i3}{brown}}
+\Ampthree{1^-}{2^0}{3^0}{\bos{i1}{red}}{\bosflipflip{1}{55}{brown}{red}{brown}}{\bos{i3}{brown}}
+\Ampthree{1^-}{2^0}{3^0}{\bosflip{1}{180}{brown}{red}}{\bosflip{1}{55}{cyan}{brown}}{\bos{i3}{brown}}
+\Ampthree{1^-}{2^0}{3^0}{\bosflipflip{1}{180}{cyan}{brown}{red}}{\bos{i2}{brown}}{\bos{i3}{brown}}$\\
$+[\mathbf{J}(1^-,2^0)]_0\cdot [\mathbf{A}(3^0)]_2$ & 
$+\Ampthree{1^-}{2^0}{3^0}{\bos{i1}{red}}{\bos{i2}{brown}}{\bosflipflip{1}{-55}{brown}{red}{brown}}$\\
\hline
$[\mathbf{J}(1^-,3^0)]_2\cdot [\mathbf{A}(2^0)]_0$ & $\Ampthree{1^-}{2^0}{3^0}{\bosflip{1}{180}{brown}{red}}{\bos{i2}{brown}}{\bosflip{1}{-55}{red}{brown}}+\Ampthree{1^-}{2^0}{3^0}{\bos{i1}{red}}{\bos{i2}{brown}}{\bosflipflip{1}{-55}{brown}{red}{brown}}+\Ampthree{1^-}{2^0}{3^0}{\bosflip{1}{180}{brown}{red}}{\bos{i2}{brown}}{\bosflip{1}{-55}{cyan}{brown}}
+\Ampthree{1^-}{2^0}{3^0}{\bosflipflip{1}{180}{cyan}{brown}{red}}{\bos{i2}{brown}}{\bos{i3}{brown}}$ \\
\makecell{$+[\mathbf{J}(1^-,3^0)]_0\cdot [\mathbf{A}(2^0)]_2$} & \makecell{
+\Ampthree{1^-}{2^0}{3^0}{\bos{i1}{red}}{\bosflipflip{1}{55}{brown}{red}{brown}}{\bos{i3}{brown}}} \\
\hline
$[\mathbf{J}(2^0,3^0)]_1\cdot [\mathbf{A}(1^-)]_1$ & $\Ampthree{1^-}{2^0}{3^0}{\bosflip{1}{180}{brown}{red}}{\bos{i2}{brown}}{\bosflip{1}{-55}{red}{brown}}
+\Ampthree{1^-}{2^0}{3^0}{\bosflip{1}{180}{brown}{red}}{\bosflip{1}{55}{red}{brown}}{\bos{i3}{brown}}
+\Ampthree{1^-}{2^0}{3^0}{\bosflip{1}{180}{brown}{red}}{\bos{i2}{brown}}{\bosflip{1}{-55}{cyan}{brown}}
+\Ampthree{1^-}{2^0}{3^0}{\bosflip{1}{180}{brown}{red}}{\bosflip{1}{55}{cyan}{brown}}{\bos{i3}{brown}}$\\
\hline 
\end{tabular}
\end{equation}
Here $[\mathbf J]_2$ is not conserved and is absent from the primary and first-descendant $VVV$ amplitudes. Determining the current form of each diagram will help us analyze the leading matching in a later subsection.

\subsection{Systematic leading amplitude matching}

To describe the matching between massless and massive amplitudes, we need to match the ones with the helicity-chirality unity. Note that the spinor structures of massless and massive amplitudes differ significantly: the massive amplitude is explicitly local, while the massless one may be not. Thus we need to deform the massless spinor structures to match the massive form. A useful and systematic approach is to focus on the scaling behavior of spinors rather than their exact form. Furthermore, all the difficulties originate from the matching between the gauge boson and vector boson, we need to separate the amplitudes into conserved current and the gauge boson order by order.

Let us present a systematic procedure for the leading matching between the massless amplitudes and massive MHC 3-point amplitudes:
\begin{itemize}
    \item First, identify all the massless 3-point helicity categories, then select the massive 3-point MHC amplitudes in the same helicity categories. 

    \item Second, for the massive MHC diagram, if there is no gauge boson, perform a direct matching between massless and massive amplitude. If the gauge boson is involved (brown color particle in the primary vertex of the MHC diagram), separate the gauge boson and conserved current structure. 
    To match the amplitude involving a massive vector, we need to decompose the amplitude into vector and current parts, 
\begin{equation} \begin{aligned}
\mathcal{A},\mathcal{M}\to\mathbb J\cdot\mathbb A
\end{aligned} \end{equation}
where $\mathbb{J}$ denotes the current and $\mathbb{A}$ the vector. 
Correspondingly, the massless diagram with the same helicity is separated into the gauge boson and current structure. 
The gauge boson and the current satisfy the following correspondence. 
    \begin{itemize}
\item \textit{Gauge Deformation}: The massless gauge boson matches to the subleading massive gauge boson $[\mathbf A]_1$, while other massless particle (i.e. the two fermions in the $FFV$ case) match to the leading current $[\mathbf{J}]_0$. This gauge boson matching means that there should be a chirality flip must occur in the massive vector state,
\begin{equation}
\begin{tikzpicture}[baseline=(current bounding box.center)]
\node (A1) at (0,2) {
\begin{tikzpicture}[baseline=-0.1cm] \begin{feynhand}
\vertex [particle] (i1) at (1.5,0) {$+1$};
\vertex [dot] (v1) at (0,0) {};
\graph{(i1)--[bos](v1)};
\end{feynhand} \end{tikzpicture}
};
\node (A2) at (6.2,2) {
$[\mathbf{A}]_1=$
\begin{tikzpicture}[baseline=-0.1cm] \begin{feynhand}
\setlength{\feynhandblobsize}{6mm}
\setlength{\feynhandarrowsize}{5pt}
\vertex [particle] (i1) at (2,0) {$+1$};
\vertex (v2) at (1,0);
\vertex [dot] (v1) at (0,0) {};
\begin{scope}[]
\clip (0,-0.1) rectangle (1,0.1); 
\draw[brown,thick bos] (v1)--(i1);
\end{scope}
\begin{scope}[]
\clip (1,-0.1) rectangle (1.6,0.1); 
\draw[cyan,thick bos] (v1)--(i1);
\end{scope}
\draw[very thick] plot[mark=x,mark size=2.5] coordinates {(1,0)};
\draw (0.55-0.03,-0.08) -- (0.55+0.03,+0.08);
\end{feynhand} \end{tikzpicture}
};
\node (B1) at (0,0) {
\begin{tikzpicture}[baseline=-0.1cm] \begin{feynhand}
\setlength{\feynhandarrowsize}{4pt}
\vertex [particle] (i2) at (-0.579,0.827) {$+$};
\vertex [particle] (i3) at (-0.579,-0.827) {$-$};
\vertex [dot] (v1) at (0,0) {};
\graph{(i3)--[double](v1)--[double](i2)};
\end{feynhand} \end{tikzpicture}
};
\node (B2) at (5.5,0) {
$[\mathbf{J}]_0=$
\begin{tikzpicture}[baseline=-0.1cm] \begin{feynhand}
\setlength{\feynhandarrowsize}{4pt} 
\vertex [particle] (i2) at (-0.579,0.827) {$+$}; 
\vertex [particle] (i3) at (-0.579,-0.827) {$-$}; 
\vertex [dot] (v1) at (0,0) {};
\antfer{red}{i2}{v1}
\fer{red}{i3}{v1};
\end{feynhand} \end{tikzpicture}
};
\draw[-Stealth] (1.1,2) -- node[pos=0.5,above]{\small flip in this particle}(4.2,2);
\draw[-Stealth] (1.1,0) -- (4.2,0);
\end{tikzpicture}
\end{equation}

\item \textit{Goldstone Deformation}: When a Goldstone boson is present, it should match to the leading massive gauge boson $[\mathbf A]_0$. This requires the other two massless particles to match to the subleading current $[\mathbf{J}]_1$, meaning the chirality flip occurs in the current,
\begin{equation}
\begin{tikzpicture}[baseline=(current bounding box.center)]
\node (A1) at (0,2) {
\begin{tikzpicture}[baseline=-0.1cm] \begin{feynhand}
\vertex [particle] (i1) at (1.5,0) {\ $0$};
\vertex [dot] (v1) at (0,0) {};
\graph{(i1)--[sca](v1)};
\end{feynhand} \end{tikzpicture}
};
\node (A2) at (6,2) {
$[\mathbf{A}]_0=$
\begin{tikzpicture}[baseline=0.7cm]
\begin{feynhand}
\setlength{\feynhandblobsize}{6mm}
\vertex [dot] (v1) at (-0.4,0.8) {};
\vertex [particle] (i2) at (1.3,0.8) {$0$};
\bos{i2}{brown};
\end{feynhand}
\end{tikzpicture}};
\node (B1) at (0,0) {
\begin{tikzpicture}[baseline=-0.1cm] \begin{feynhand}
\setlength{\feynhandarrowsize}{4pt}
\vertex [particle] (i2) at (-0.579,0.827) {$-$};
\vertex [particle] (i3) at (-0.579,-0.827) {$-$};
\vertex [dot] (v1) at (0,0) {};
\graph{(i3)--[double](v1)--[double](i2)};
\end{feynhand} \end{tikzpicture}
};
\node (B2) at (5.5,0) {
$[\mathbf{J}]_1=$
\begin{tikzpicture}[baseline=-0.1cm] \begin{feynhand}
\setlength{\feynhandarrowsize}{4pt} 
\vertex [particle] (i2) at (-0.579,0.827) {$-$}; 
\vertex [particle] (i3) at (-0.579,-0.827) {$-$}; 
\vertex [dot] (v1) at (0,0) {};
\antferflip{1}{125}{cyan}{red};
\fer{red}{i3}{v1};
\end{feynhand} \end{tikzpicture}};
\draw[-Stealth] (1.1,2) -- (4.2,2);
\draw[-Stealth,double] (2.1,1.9) -- node[pos=0.3,right]{\small flip in the current}(2.1,0.1);
\draw[-Stealth] (1.1,0) -- (4.2,0);
\end{tikzpicture}
\end{equation}

\end{itemize}

    \item Third, for both the massless and massive diagrams, identify the spinor scaling behavior for the gauge boson $\mathbb{A}$ and the current $\mathbb{J}$. Although $\mathbb{A}$ and $\mathbb{J}$ themselves  are not 3-pt amplitudes, they must satisfy the same MHC scaling relation given in eq.~\eqref{eq:MHC_scaling} to ensure the consistency of the amplitude deformation. Perform the amplitude deformation.

    \item Finally, a matching between two amplitudes is obtained in the following form   
\begin{equation}
\mathcal{A}(\lambda,\tilde\lambda) \quad \to  \quad \mathcal{M}(\lambda,\tilde\lambda,\eta,\tilde\eta,m,\tilde{m}). 
\end{equation}

\end{itemize}

\paragraph{Spinor Scaling Behavior}

Let us first consider the 3-pt massless amplitude, which depends only on the spinors $\lambda$ or $\tilde\lambda$. Since the amplitude has mass dimesnion 1 and each spinor carries dimension $\frac12$, the amplitude must scale as:
\begin{equation} \begin{aligned}
\mathcal{A}_3&\sim
\begin{cases}
    \lambda^2,\\
    \tilde\lambda^2.
\end{cases}
\end{aligned} \end{equation}
Ignoring particle labels, this behavior is analogous to that of the primary states of chiral and anti-chiral vectors. 
According to helicity of scalar and gauge boson, we can decompose the spinor scaling behavior of the massless amplitude as follows,
\begin{equation} \label{eq:massless_scaling}
\mathcal{A}(\lambda^2)\to
\left\{\begin{aligned}
\mathbb J(\lambda^0)&\cdot\mathbb A^-(\lambda^2)\\
\mathbb J(\lambda^2)&\cdot\mathbb A^0(\lambda^0)\\
\mathbb J(\lambda^4)&\cdot\mathbb A^+(\lambda^{-2})
\end{aligned}\right.\qquad
\mathcal{A}(\tilde\lambda^2)\to
\left\{\begin{aligned}
\mathbb J(\tilde\lambda^4)&\cdot\mathbb A^-(\tilde\lambda^{-2})\\
\mathbb J(\tilde\lambda^2)&\cdot\mathbb A^0(\tilde\lambda^0)\\
\mathbb J(\tilde\lambda^0)&\cdot\mathbb A^+(\tilde\lambda^2)\\
\end{aligned}\right.
\end{equation}
where the superscript $\pm,0$ denotes the helicity of the vector $\mathbb A$.

On the other hand, let us consider the spinor scaling behavior of the MHC amplitude. In this case, the MHC amplitude depends not only on $\lambda,\tilde\lambda$, but also on $\tilde m\eta,m\tilde\eta$. Thus a general scaling form can be written as
\begin{equation} \begin{aligned} \label{eq:general_scaling}
\mathcal{M} \sim \lambda^x \tilde \lambda^y (\tilde m\eta)^{w}(m\tilde\eta)^{z},\quad x,y,z,w\ge 0
\end{aligned} \end{equation}
Assuming the MHC amplitude has total spin $s = s_1 + s_2 + s_3$ and total helicity $h = h_1 + h_2 + h_3$, the scaling exponents satisfy the constraints
\begin{equation} \begin{aligned}
x+y+w+z &=2s \\
-x+y+w-z &=2h \\
\end{aligned} \end{equation}
Substituting these relations into eq.~\eqref{eq:general_scaling}, we obtain the spinor scaling behavior for the primary and descendant MHC amplitudes,
\begin{equation} \begin{aligned} \label{eq:MHC_scaling}
\text{primary}&:\quad \tilde\lambda^{s+h}\lambda^{s-h},\\
\text{k-th descendant}&:\quad \tilde\lambda^{s+h-w}\lambda^{s-h-z}(\tilde m\eta)^{w}(m\tilde\eta)^{z},\quad w+z=k,
\end{aligned} \end{equation}
In previous subsection, all the leading MHC amplitudes are obtained and they have the following scaling behavior, 
\begin{equation} 
\begin{tabular}{c|c|c|c}
\hline
total spin & massive amplitude & leading term & spinor scaling behavior\\
\hline
0 & $SSS$ & \multirow{2}{*}{primary} & $1$\\
\cline{1-2} \cline{4-4}
1 & $FFS$ & & $\lambda^2,\tilde\lambda^2$\\
\hline
2 & $FFV,VVS$ & 1st descendant & $\tilde\lambda\lambda^2(m \tilde\eta), \tilde\lambda^2\lambda (\tilde m\eta)$ \\
\hline
3 & $VVV$ & 2nd descendant & \makecell{$\tilde\lambda^2\lambda^2(\tilde m\eta)^2,\tilde\lambda^3\lambda(\tilde m\eta)(m \tilde\eta)$\\$\tilde\lambda^2\lambda^2(m \tilde\eta)^2,\tilde\lambda\lambda^3(\tilde m\eta)(m \tilde\eta)$}\\
\hline 
\end{tabular}
\end{equation} 
Similarly, we could obtain the scaling behavior for the vector boson and the current. The spinor scaling of the massive vector $\mathbf A$ can be
\begin{align}
[\mathbf A^+]_1&\sim \lambda (m \tilde \eta),&
[\mathbf A^-]_1&\sim \tilde\lambda (\tilde m\eta),&\\
[\mathbf A^0]_0&\sim \tilde \lambda \lambda,&
[\mathbf A^0]_2&\sim  (m \tilde \eta) (\tilde m\eta),&
\end{align}
The massive current in the 1st descendant MHC amplitude (the current has total spin 1) has the following scaling
\begin{eqnarray}
[\mathbf J]_0&\sim& \tilde \lambda \lambda,\\
{[\mathbf J]_1}&\sim& \tilde\lambda (\tilde m\eta),\lambda (m \tilde \eta)
\end{eqnarray}
The massive current in the 2nd descendant MHC amplitude (the current has total spin 2) has the following scaling
\begin{eqnarray}
[\mathbf J]_0&\sim& \tilde\lambda^3\lambda,\tilde\lambda\lambda^3\\
{[\mathbf J]_1}&\sim& \tilde\lambda^3(\tilde m\eta), \tilde\lambda^2\lambda(m\tilde\eta), \tilde\lambda\lambda^2(\tilde m\eta), \lambda^3(m\tilde \eta),\\
{[\mathbf J]_2}&\sim& \tilde\lambda^2(\tilde m\eta)(m \tilde\eta),\tilde\lambda\lambda(\tilde m\eta)^2,\tilde\lambda\lambda(m \tilde\eta)^2,\lambda^2(\tilde m\eta)(m \tilde\eta),
\end{eqnarray}

\paragraph{Deformation Analysis}

The spinor scaling behavior between massless and massive amplitudes matches only when the massless amplitude corresponds to the primary MHC amplitude. Otherwise, we need to perform the amplitude deformation by multiplying the massless scaling $\lambda^2,\tilde\lambda^2$ by the unit quantities  $\frac{\langle\lambda\eta\rangle}{m},\frac{[\lambda\eta]}{\tilde m}$. 
For the 1st descendant amplitude matching, multiplying once yields
\begin{equation} \begin{aligned}
\lambda^2 &\xrightarrow{\times\frac{[\lambda\eta]}{\tilde m}}\tilde\lambda\lambda^2(\tilde m^{-1} \tilde\eta)=\frac{\tilde\lambda\lambda^2(m \tilde\eta)}{\mathbf m^2},\\
\tilde\lambda^2 &\xrightarrow{\times\frac{\langle\lambda\eta\rangle}{m}}\tilde\lambda^2\lambda (m^{-1}\eta)=\frac{\tilde\lambda^2\lambda (\tilde m\eta)}{\mathbf m^2},
\end{aligned} \end{equation}
which matches to the scaling of the 1st descendant amplitude. 
Depending on the helicity of the vector boson, it is necessary to perform separate scalings for vector boson and current.  
For each helicity of vector $\mathbb A$, we start with the highest scaling behavior in eq.~\eqref{eq:massless_scaling}, and thus the multiplication acts on different particle, obtain gauge boson deformation and Goldstone deformation. 

If the transverse gauge boson is involved, since the $\eta$ is from the gauge boson, a multiplication on the same particle by the unit quantity $\frac{\langle\lambda\eta\rangle}{m},\frac{[\lambda\eta]}{\tilde m}$ is applied. If the longitudinal gauge boson is involved, since the $\eta$ is from the current, a multiplication on the particle in the current is applied by the unit quantity $\frac{\langle\lambda\eta\rangle}{m},\frac{[\lambda\eta]}{\tilde m}$. After applying the multiplication, the gauge boson usually does not have the scaling of the massive vector, and thus a IBP should be applied to transfer the spinor structure from $\mathbb{A}$ to $\mathbb{J}$, or from $\mathbb{J}$ to $\mathbb{A}$. The procedure can be written as
\begin{equation}\begin{aligned}  \label{eq:scaling_deform}
&\text{Gauge deformation}: \\
&\mathbb J(\lambda^0)\cdot\mathbb A^-(\lambda^2)
\overset{\times\frac{[\lambda\eta]}{\vphantom{|}\tilde m}}{\to}
\mathbb J(\lambda^0)\cdot\mathbb A^-(\tilde \lambda \lambda^2 \tilde m^{-1}\tilde \eta)
\overset{\text{IBP}^-}{\to}
\mathbb J(\tilde\lambda\lambda)\cdot\mathbb A^-(\lambda \tilde m^{-1}\tilde \eta)
=\mathbb J(\tilde\lambda\lambda)\cdot\frac{\mathbb A^-(\lambda m\tilde \eta)}{\mathbf m^2} \\
&\mathbb J(\tilde\lambda^0)\cdot\mathbb A^+(\tilde\lambda^2)
\overset{\times\frac{\langle\lambda\eta\rangle}{\vphantom{|}m}}{\to}
\mathbb J(\lambda^0)\cdot\mathbb A^+(\tilde\lambda \tilde\lambda^2 m^{-1}\eta)
\overset{\text{IBP}^-}{\to}
\mathbb J(\tilde\lambda\lambda)\cdot\mathbb A^+(\tilde\lambda m^{-1}\eta)
=\mathbb J(\tilde\lambda\lambda)\cdot\frac{\mathbb A^+(\tilde\lambda \tilde m\eta)}{\mathbf m^2} \\
&\text{Goldstone deformation}: \\
&\mathbb J(\lambda^2)\cdot\mathbb A^0(\lambda^0) 
\overset{\times\frac{[\lambda\eta]}{\vphantom{|}\tilde m}}{\to}
\mathbb J(\tilde\lambda \lambda^2 \tilde m^{-1}\tilde\eta)\cdot\mathbb A^0(\lambda^0)
\overset{\text{IBP}^+}{\to}
\mathbb J(\lambda \tilde m^{-1}\tilde\eta)\cdot\mathbb A^0(\tilde \lambda \lambda)
=\frac{\mathbb J(\lambda m\tilde\eta)}{\mathbf m^2}\cdot\mathbb A^0(\tilde \lambda \lambda) \\
&\mathbb J(\tilde\lambda^2)\cdot\mathbb A^0(\tilde\lambda^0) 
\overset{\times\frac{\langle\lambda\eta\rangle}{\vphantom{|}m}}{\to}
\mathbb J(\lambda \tilde \lambda^2 m^{-1}\eta)\cdot\mathbb A^0(\tilde\lambda^0)
\overset{\text{IBP}^+}{\to}
\mathbb J(\tilde\lambda m^{-1}\eta)\cdot\mathbb A^0(\tilde \lambda \lambda) 
=\frac{\mathbb J(\tilde\lambda \tilde m\eta)}{\mathbf m^2}\cdot\mathbb A^0(\tilde \lambda \lambda) 
\end{aligned} \end{equation}
where the IBP convert the spinor structure between current and vector boson
\begin{align}
\text{IBP}^{+}:\quad \mathbb J(\tilde\lambda^x \lambda^y) \cdot \mathbb A(\tilde\lambda^w \lambda^z)&=\mathbb J(\tilde\lambda^{x-1} \lambda^{y-1})\cdot \mathbb A(\tilde\lambda^{w+1} \lambda^{z+1}),\quad x>0\;\&\;y>0 \\
\text{IBP}^{-}:\quad \mathbb J(\tilde\lambda^x \lambda^y) \cdot \mathbb A(\tilde\lambda^w \lambda^z)&=\mathbb J(\tilde\lambda^{x+1} \lambda^{y+1})\cdot \mathbb A(\tilde\lambda^{w-1} \lambda^{z-1}),\quad w>0\;\&\;z>0
\end{align}

Thus the result can match the 1st descendant scaling. 
The result exhibits the same scaling behavior as the MHC expressions of massive vector and current, allowing for the following matching
\begin{align}
\mathbb A^\pm(\lambda m \tilde \eta),\mathbb A^+( \tilde\lambda \tilde m\eta)&\to[\mathbf A^\pm]_1,& 
\mathbb A(\tilde \lambda \lambda)&\to[\mathbf A^0]_0,&\\
\mathbb J(\lambda m \tilde \eta),\mathbb J( \tilde\lambda\tilde m\eta)&\to[\mathbf J]_1,&
\mathbb J(\tilde\lambda\lambda)&\to [\mathbf J]_0,&
\end{align}
The contraction of these terms collectively yields the 1st descendant scaling
\begin{equation}
    [\mathbf J]_1\cdot[\mathbf A^0]_0+[\mathbf J]_0\cdot[\mathbf A^\pm]_1\sim \tilde\lambda\lambda^2(m\tilde\eta), \tilde\lambda^2\lambda (\tilde m\eta),
\end{equation}
Therefore, a single gauge or Goldstone boson deformation is sufficient to match the 1st descendant MHC amplitude.

For the 2nd descendant MHC amplitude, we need to apply the theses deformations twice, even though the resulting scaling behavior will differ. 
Multiplying $\frac{\langle\lambda\eta\rangle}{m}$ or $\frac{[\lambda\eta]}{\tilde m}$ twice gives 
\begin{equation} \begin{aligned}
\lambda^2 &\xrightarrow{\times\frac{[\lambda\eta]}{\tilde m}}\tilde\lambda\lambda^2(\tilde m^{-1} \tilde\eta)
\xrightarrow{\times\frac{[\lambda\eta]}{\tilde m}}\tilde\lambda^2\lambda^2(\tilde m^{-1} \tilde\eta)^2
=\frac{\tilde\lambda^2\lambda^2(m \tilde\eta)^2}{\mathbf m^4},\\
\tilde\lambda^2 &\xrightarrow{\times\frac{\langle\lambda\eta\rangle}{m}}\tilde\lambda^2\lambda (m^{-1}\eta)
\xrightarrow{\times\frac{\langle\lambda\eta\rangle}{m}}\tilde\lambda^2\lambda^2 (m^{-1}\eta)^2
=\frac{\tilde\lambda^2\lambda^2 (\tilde m\eta)^2}{\mathbf m^4},
\end{aligned} \end{equation}
which corresponds to the 2nd descendant scaling. 
Furthermore, we can also reach the 2nd descendant scaling by multiplying by both $\frac{\langle\lambda\eta\rangle}{m}$ and $\frac{[\lambda\eta]}{\tilde m}$. However, nothing prevents us from applying both factors simultaneously in a single step, which leads to a more complicated scaling behavior
\begin{equation} \begin{aligned}
\lambda^2 &\xrightarrow{\times\frac{[\lambda\eta]}{\tilde m},\frac{\langle\lambda\eta\rangle}{m}}\tilde\lambda\lambda^2(\tilde m^{-1} \tilde\eta)+\lambda^3(m^{-1}\eta)
\xrightarrow{\times\frac{[\lambda\eta]}{\tilde m},\frac{\langle\lambda\eta\rangle}{m}}\tilde\lambda\lambda^3(\tilde m^{-1} \tilde\eta)(m^{-1}\eta)
=\frac{\tilde\lambda\lambda^3(m \tilde\eta)(\tilde m\eta)}{\mathbf m^4},\\
\tilde\lambda^2 &\xrightarrow{\times \frac{[\lambda\eta]}{\tilde m},\frac{\langle\lambda\eta\rangle}{m}}\tilde\lambda^2\lambda (m^{-1}\eta)+\tilde\lambda^3 (\tilde m^{-1} \tilde\eta)\xrightarrow{\times\frac{[\lambda\eta]}{\tilde m},\frac{\langle\lambda\eta\rangle}{m}}\tilde\lambda^3\lambda (\tilde m^{-1} \tilde\eta)(m^{-1}\eta)
=\frac{\tilde\lambda^3\lambda (m \tilde\eta)(\tilde m\eta)}{\mathbf m^4},
\end{aligned} \end{equation}

For example, consider a vector with negative scaling $\mathbb{A}^+(\lambda^{-2})$, which is not covered in eq.~\eqref{eq:scaling_deform}. A single deformation gives
\begin{equation}
\mathbb J(\lambda^4)\cdot\mathbb A^+(\lambda^{-2}) \xrightarrow{\times\frac{\langle\lambda\eta\rangle}{m}}
\mathbb J(\lambda^4)\cdot\mathbb A^+(\lambda^{-1} m^{-1} \eta)
\end{equation}
yielding an amplitude scaling of $\lambda^{3} m^{-1} \eta$.
In this case, IBP cannot be applied due to the absence of a positive exponent in the $\tilde{\lambda}$ scaling. Since our target amplitude is a 2nd descendant involving more than one massive vector, we can decompose it by considering another vector boson. Assuming it has helicity $-1$, its scaling should be $\mathbb{A}^-(\lambda^2)$, as no subleading spinors $\eta, \tilde{\eta}$ have been introduced for this particle. The amplitude scaling can then be decomposed as
\begin{equation}
\lambda^{3} m^{-1} \eta\to \mathbb J(\lambda m^{-1}\eta)\cdot\mathbb A^-(\lambda^2) 
\end{equation}
Here, the previously used vector boson $\mathbb{A}^+$ is now incorporated into the current $\mathbb{J}$. We then apply the gauge deformation again
\begin{equation} \begin{aligned}
\mathbb J(\lambda m^{-1}\eta)\cdot\mathbb A^-(\lambda^2)  
&\xrightarrow{\times\frac{[\lambda\eta]}{\tilde m}} \mathbb J(\lambda m^{-1}\eta)\cdot\mathbb A^-(\tilde \lambda \lambda^2 \tilde m^{-1}\tilde \eta) \\
&\xrightarrow{\text{IBP}^-}
\mathbb J(\tilde\lambda\lambda^2m^{-1}\eta)\cdot\mathbb A^-(\lambda \tilde m^{-1}\tilde \eta)
\end{aligned} \end{equation}
In the last step, IBP changes the scaling of both $\mathbb{A}^-$ and the $\mathbb{A}^+$ contained within the current. The resulting current and vector now exhibit the MHC scaling and can be matched as
\begin{align}
\mathbb A^-(\lambda \tilde m^{-1}\tilde \eta)\to\frac{1}{\mathbf m^2}[\mathbf A^-]_1,\quad
\mathbb J(\tilde\lambda\lambda^2 m^{-1}\eta)\to\frac{1}{\mathbf m^2}[\mathbf J]_1.
\end{align}
Thus, we conclude that applying the deformation twice can correctly reproduces the 2nd descendant MHC amplitude.

\subsection{Leading $VVV$ and $VVS$ matching}

\paragraph{FFV leading matching} 

The $FFV$ leading amplitude matching have been presented in previous section. Let us briefly summarize the procedure
\begin{itemize}

\item For the MHC amplitudes, the primary amplitudes with helicity $(\mp\frac12, \pm\frac12, 0)$ are vanishing, and thus the leading amplitudes should be in the following helicity categories: $(\pm\frac12, \pm\frac12, 0)$, $(-\frac12, +\frac12, \pm1)$, $(+\frac12, -\frac12, \pm1)$. From the helicity-chirality unity, the massless amplitudes should be in the same helicity categories.

\item From the MHC leading amplitudes 
\begin{equation} \begin{aligned}
&(+\tfrac12,+\tfrac12,0):\quad \tilde m_1 [23]\langle3 \eta_1\rangle,\quad \tilde m_2 \langle\eta_23\rangle[31], \\ 
&(-\tfrac12,-\tfrac12,0):\quad  m_2 [\eta_23]\langle31\rangle, \quad m_1 \langle23\rangle[3\eta_1] , \\
&(-\tfrac12,+\tfrac12,+1):\quad  \tilde m_3 [23]\langle\eta_31\rangle, \qquad 
(-\tfrac12,+\tfrac12,-1):\quad  m_3 [2\eta_3]\langle31\rangle \\ 
&(+\tfrac12,-\tfrac12,+1):\quad  \tilde m_3 \langle2\eta_3\rangle[31],\qquad 
(+\tfrac12,-\tfrac12,-1):\quad  m_3 \langle23\rangle[\eta_31],
\end{aligned}
\end{equation}
the deformation should be applied to the particle containing the $\eta$, which indicates the one chirality flip and one helicity flip.

\item \textit{Gauge Deformation}: For the massless $FFV$ amplitude with helicity $(-\frac12,+\frac12,+1)$, we should flip particle 3. Diagrammatically, we have
\begin{equation} \begin{aligned}
\begin{tikzpicture}[baseline=0.7cm] \begin{feynhand}
\setlength{\feynhandarrowsize}{3.5pt}
\vertex [particle] (i1) at (0,0.8) {$1^-$}; 
\vertex [particle] (i2) at (1.6,1.6) {$2^+$}; 
\vertex [particle] (i3) at (1.6,0) {$3^+$};  
\vertex (v1) at (0.9,0.8); 
\graph{(i1)--[fer](v1)--[fer](i2)};
\graph{(i3)--[bos] (v1)};  
\end{feynhand} \end{tikzpicture}\rightarrow
\begin{tikzpicture}[baseline=-0.1cm] \begin{feynhand}
\setlength{\feynhandarrowsize}{3.5pt}
\fill[yellow!50!white, opacity=0.7,rotate=-10] (0.7,-0.3) ellipse (0.9 and 0.6);
\node (C) at (1.1,-0.4) {$\mathbb A^+$};
\vertex [particle] (i1) at (-1.01,0) {$1^-$}; 
\vertex [particle] (i2) at (0.579,0.827) {$2^+$}; 
\vertex [particle] (i3) at (0.579,-0.827) {$3^+$}; 
\vertex (v1) at (0,0);
\fer{red}{i1}{v1};
\antfer{red}{i2}{v1};
\bosflip{1}{-55}{brown}{cyan};
\end{feynhand} \end{tikzpicture}.
\end{aligned} \end{equation}
The subleading spinor $\eta$ will be introduced to particle 3, which is identified as vector $\mathbb A^+$. Applying the gauge deformation, the massless amplitude match to
\begin{equation}
\frac{[23]^2}{[12]}\xrightarrow{\text{Gauge}} \frac{\tilde m_3\langle1\eta_3\rangle[23]}{\mathbf m_3^2}.
\end{equation}

\item \textit{Goldstone Deformation}: When we match the Goldstone boson to a symmetric vector boson, we should flip the chirality of particles in the current. For massless $FFS$ amplitude with helicity $(-\frac12,-\frac12,0)$, particle 3 is a Goldstone boson, so we can only flip particle 1 or 2 in the current $\mathbb J$.  
In this case, one massless diagram corresponds to two MHC diagrams
\begin{equation} \begin{aligned}
\begin{tikzpicture}[baseline=0.7cm] \begin{feynhand}
\setlength{\feynhandarrowsize}{3.5pt}
\vertex [particle] (i1) at (0,0.8) {$1^+$}; 
\vertex [particle] (i2) at (1.6,1.6) {$2^+$}; 
\vertex [particle] (i3) at (1.6,0) {$3^0$};  
\vertex (v1) at (0.9,0.8); 
\graph{(i1)--[fer](v1)--[fer](i2)};
\graph{(i3)--[sca] (v1)};  
\end{feynhand} \end{tikzpicture}&=[12]\to
\begin{tikzpicture}[baseline=-0.1cm] \begin{feynhand}
\setlength{\feynhandarrowsize}{3.5pt}
\fill[yellow!50!white, opacity=0.7,rotate=20] (-0.1,0.4) ellipse (1.2 and 0.6);
\node (C) at (-0.3,0.7) {$\mathbb J$};
\vertex [particle] (i1) at (-1.01,0) {$1^+$}; 
\vertex [particle] (i2) at (0.579,0.827) {$2^+$}; 
\vertex [particle] (i3) at (0.579,-0.827) {$3^0$}; 
\vertex (v1) at (0,0);
\ferflip{1}{180}{cyan}{red};
\antfer{red}{i2}{v1};
\bos{i3}{brown};
\end{feynhand} \end{tikzpicture},
\begin{tikzpicture}[baseline=-0.1cm] \begin{feynhand}
\setlength{\feynhandarrowsize}{3.5pt}
\fill[yellow!50!white, opacity=0.7,rotate=20] (-0.1,0.4) ellipse (1.2 and 0.6);
\node (C) at (-0.3,0.7) {$\mathbb J$};
\vertex [particle] (i1) at (-1.01,0) {$1^+$}; 
\vertex [particle] (i2) at (0.579,0.827) {$2^+$}; 
\vertex [particle] (i3) at (0.579,-0.827) {$3^0$}; 
\vertex (v1) at (0,0);
\fer{cyan}{i1}{v1};
\antferflip{1}{55}{red}{cyan};
\bos{i3}{brown};
\end{feynhand} \end{tikzpicture}\\
\end{aligned} \end{equation}
Thus, applying Goldstone deformation can gives two MHC terms
\begin{equation}
[12]\xrightarrow{\text{Goldstone}} -b_1\frac{\tilde m_1\langle\eta_13\rangle[32]}{\mathbf m_1^2}-b_2\frac{\tilde m_2 [13]\langle3\eta_2\rangle}{\mathbf m_2^2},
\end{equation}
where $b_1+b_2=1$.
    
\end{itemize}

\paragraph{$VVS$ leading matching}

Similar to the $FFV$ case, the $VVS$ leading matching also involves in one gauge boson and the current. Following the similar procedure, we have 
\begin{itemize}

\item The primary amplitude is also vanishing and thus the leading MHC amplitudes are 
\begin{equation} \begin{aligned}
(-1, 0, 0): \quad m_1\langle12\rangle[2\eta_1], \qquad & (+1, 0, 0): \quad 
-\tilde m_1\langle\eta_12\rangle[21] \\
(0, -1, 0): \quad m_2\langle12\rangle[\eta_21], \qquad &
(0, +1, 0): \quad -\tilde m_2\langle1\eta_2\rangle[21].
\end{aligned}\end{equation}
Therefore the massless amplitudes belong to the same helicity, and the diagrammatic matching is straightforward.

\item From the first MHC amplitude, the massless matching has
\begin{equation} \begin{aligned}
\begin{tikzpicture}[baseline=0.7cm] \begin{feynhand}
\setlength{\feynhandarrowsize}{3.5pt}
\vertex [particle] (i1) at (0,0.8) {$1^-$}; 
\vertex [particle] (i2) at (1.6,1.6) {$2^0$}; 
\vertex [particle] (i3) at (1.6,0) {$3^0$};  
\vertex (v1) at (0.9,0.8); 
\graph{(i1)--[bos](v1)--[sca](i2)};
\graph{(i3)--[sca](v1)};  
\end{feynhand} \end{tikzpicture}=\frac{\langle12\rangle\langle31\rangle}{\langle23\rangle}
\quad\rightarrow\quad
\Ampthree{1^-}{2^0}{3^0}{\bosflip{1}{180}{brown}{red}}{\bos{i2}{brown}}{\sca{i3}}=m_1\langle12\rangle[2\eta_1],
\end{aligned} \end{equation}
According to the massless and MHC amplitude, we can  summarize the scaling behavior of each particle in the two cases as follows:
\begin{equation} 
\begin{tabular}{c|c|c|c}
\hline
 & scaling change & \multicolumn{2}{c}{\text{decomposition }$\mathbb J\cdot \mathbb A$} \\
\hline
particle 1 & $\lambda_1^2\to\lambda_1 m_1\tilde\eta_1$ & \multirow{2}{*}{$\mathbb J(\lambda^2)\to\mathbb J(\lambda m\tilde\eta)$} & $\mathbb A^-(\lambda^2)\to\mathbb A^-(\lambda_1 m_1\tilde\eta_1)$ \\
\cline{1-2} \cline{4-4}
particle 3 & $1\to1$ & & \multirow{2}{*}{$\mathbb J(1)\to\mathbb J(\tilde\lambda\lambda)$} \\
\cline{1-3} 
particle 2 & $1\to\tilde\lambda_2\lambda_2$ & $\mathbb A^0(1)\to\mathbb A^0(\tilde\lambda_2\lambda_2)$ & \\
\hline 
\end{tabular}
\end{equation}
Since the $\eta_1$ for the particle 1, we can treat the particle 1 as $\mathbb A$ and use the Gauge Deformation. At the same time, since we identify the particle 2 as the Goldstone boson, the $\eta_1$ could originate from the current $\mathbb J$ in the Goldstone Deformation. These two treatments should be equivalent.

\item \textit{Gauge Deformation}: When the deformation is on the particle 1, the diagram matching is  
\begin{equation} \begin{aligned} 
\begin{tikzpicture}[baseline=0.7cm] \begin{feynhand}
\setlength{\feynhandarrowsize}{3.5pt}
\vertex [particle] (i1) at (0,0.8) {$1^-$}; 
\vertex [particle] (i2) at (1.6,1.6) {$2^0$}; 
\vertex [particle] (i3) at (1.6,0) {$3^0$};  
\vertex (v1) at (0.9,0.8); 
\graph{(i1)--[bos](v1)--[sca](i2)};
\graph{(i3)--[sca](v1)};  
\end{feynhand} \end{tikzpicture}
&=\frac{\langle12\rangle\langle31\rangle}{\langle23\rangle}
\to \begin{tikzpicture}[baseline=-0.1cm] \begin{feynhand}
\fill[yellow!50!white, opacity=0.7,rotate=0] (-0.7,0.2) ellipse (0.9 and 0.6);
\node (C) at (-0.7,0.5) {$\mathbb A^-$};
\vertex [particle] (i1) at (-1.01,0) {$1^-$}; 
\vertex [particle] (i2) at (0.579,0.827) {$2^0$}; 
\vertex [particle] (i3) at (0.579,-0.827) {$3^0$}; 
\vertex (v1) at (0,0);
\bosflip{1}{180}{brown}{red};
\bos{i2}{brown};
\sca{i3};
\end{feynhand} \end{tikzpicture},\\
\end{aligned} \end{equation}
Particle 1 is identified as the vector $\mathbb A^-$, while particles 2 and 3 are identified as the current $\mathbb J$. So we have the scaling change
\begin{equation}\begin{aligned}
\mathbb J(\lambda^0)\cdot\mathbb A^-(\lambda_1^2) \xrightarrow{\times\frac{[\eta\lambda]}{\tilde m}}
\mathbb J(\lambda^0)\cdot\mathbb A^-(\tilde\lambda_1\lambda_1^2 \tilde m_1^{-1} \tilde\eta_1)
&\xrightarrow{\text{IBP}^-}
\mathbb J(\tilde\lambda \lambda)\cdot\mathbb A^-(\lambda_1 \tilde m_1^{-1} \tilde\eta_1)\\
&=\mathbb J(\tilde\lambda \lambda)\cdot\frac{\mathbb A^-(\lambda_1 m_1 \tilde\eta_1)}{\mathbf m_1^2}.
\end{aligned}\end{equation}
The corresponding amplitude deformation is
\begin{equation}\begin{aligned} \label{eq:VVS_deform1}
\frac{\langle12\rangle\langle31\rangle}{\langle23\rangle}
\xrightarrow{\times\frac{[\eta\lambda]}{\tilde m}}
-\frac{\langle12\rangle[\eta_1 13\rangle}{\tilde m_1\langle23\rangle}
\xrightarrow{\text{IBP}^-}
\frac{\langle12\rangle[\eta_1 2]}{\tilde m_1}
=\frac{m_1 \langle12\rangle[\eta_1 2]}{\mathbf m_1^2}.
\end{aligned}\end{equation}
After deformation, the spurious pole $\langle23\rangle$ has been eliminated.

\item \textit{Goldstone Deformation}: Equivalently when the deformation is from the Goldstone boson to the current $\mathbb J$, the diagram matching is
\begin{equation} \begin{aligned}
\begin{tikzpicture}[baseline=0.7cm] \begin{feynhand}
\setlength{\feynhandarrowsize}{3.5pt}
\vertex [particle] (i1) at (0,0.8) {$1^-$}; 
\vertex [particle] (i2) at (1.6,1.6) {$2^0$}; 
\vertex [particle] (i3) at (1.6,0) {$3^0$};  
\vertex (v1) at (0.9,0.8); 
\graph{(i1)--[bos](v1)--[sca](i2)};
\graph{(i3)--[sca](v1)};  
\end{feynhand} \end{tikzpicture}\rightarrow
\begin{tikzpicture}[baseline=-0.1cm] \begin{feynhand}
\fill[yellow!50!white, opacity=0.7,rotate=-20] (-0.1,-0.4) ellipse (1.2 and -0.6);
\node (C) at (-0.3,-0.7) {$\mathbb J$};
\vertex [particle] (i1) at (-1.01,0) {$1^-$}; 
\vertex [particle] (i2) at (0.579,0.827) {$2^0$}; 
\vertex [particle] (i3) at (0.579,-0.827) {$3^0$}; 
\vertex (v1) at (0,0);
\bosflip{1}{180}{brown}{red};
\bos{i2}{brown};
\sca{i3};
\end{feynhand} \end{tikzpicture},
\end{aligned} \end{equation}
In this case, particle 2 is identified as the vector $\mathbb A$, while particles 1 and 3 are identified as the current $\mathbb J$. So the scaling change should be
\begin{equation}\begin{aligned}
\mathbb J(\lambda^2)\cdot\mathbb A^0(\lambda_2^0) \xrightarrow{\times\frac{[\eta\lambda]}{\tilde m}}
\mathbb J(\tilde\lambda\lambda^2\tilde m^{-1}\tilde\eta)\cdot\mathbb A^0(\lambda_2^0)
&\xrightarrow{\text{IBP}^+}
\mathbb J(\lambda\tilde m^{-1}\tilde\eta)\cdot\mathbb A^0(\tilde\lambda_2\lambda_2)\\
&=\frac{\mathbb J(\lambda m\tilde\eta)}{\mathbf m^2}\cdot\mathbb A^0(\tilde\lambda_2\lambda_2).
\end{aligned}\end{equation}
The corresponding amplitude deformation is
\begin{equation}\begin{aligned}
\frac{\langle12\rangle\langle31\rangle}{\langle23\rangle}
\xrightarrow{\times\frac{[\eta\lambda]}{\tilde m}}
-\frac{\langle12\rangle[\eta_1 13\rangle}{\tilde m_1\langle23\rangle}
\xrightarrow{\text{IBP}^+}
\frac{\langle12\rangle[\eta_1 2]}{\tilde m_1}
=\frac{m_1\langle12\rangle[\eta_1 2]}{\mathbf m^2}.
\end{aligned}\end{equation}
It gives the same result as eq.~\eqref{eq:VVS_deform1}.

\end{itemize}

\paragraph{$VVV$ leading matching}

To apply the matching procedure, the first step is analyzing the MHC amplitudes. The primary amplitude is vanishing, and thus the leading amplitudes should be in the following helicity categories: 
\begin{equation} \begin{aligned}
&(\pm, \pm, \mp), 
(\pm, \mp, \pm),
(\mp, \pm, \pm), \\
&(\pm, 0, 0), 
(0, \pm, 0),
(0, 0, \pm).
\end{aligned}\end{equation}
From the helicity-chirality unity, the massless amplitudes should be in the same helicity categories.

Let us separate the leading matching into two different categories. The first category is the matching from massless $VVV$ to massive $VVV$ amplitudes. As shown in eq.~\eqref{eq:VVV_leading_1}, a massless $VVV$ diagram corresponds to three MHC diagrams. Because the massless amplitude has exchange symmetry, it is more convenient to treat the combination of MHC diagrams as the matching object. For helicity $(+1,+1,-1)$, we can consider the exchange symmetry between particles 1 and 2, so the matching yields two cases:  
\begin{equation}
\begin{tikzpicture}[baseline=0.7cm] \begin{feynhand}
\setlength{\feynhandarrowsize}{3.5pt}
\vertex [particle] (i1) at (0,0.8) {$1^+$}; 
\vertex [particle] (i2) at (1.6,1.6) {$2^+$}; 
\vertex [particle] (i3) at (1.6,0) {$3^-$};  
\vertex (v1) at (0.9,0.8); 
\graph{(i1)--[bos](v1)--[bos](i2)};
\graph{(i3)--[bos](v1)};  
\end{feynhand} \end{tikzpicture}
=\frac{[12]^3}{[23][31]}\to
\left\{\begin{aligned}
\Ampthree{1^+}{2^+}{3^-}{\bosflip{1}{180}{brown}{cyan}}{\bosflip{1}{55}{brown}{cyan}}{\bos{i3}{red}} \\
\Ampthree{1^+}{2^+}{3^-}{\bosflip{1}{180}{brown}{cyan}}{\bos{i2}{cyan}}{\bosflip{1}{-55}{brown}{red}}+
\Ampthree{1^+}{2^+}{3^-}{\bos{i1}{cyan}}{\bosflip{1}{55}{brown}{cyan}}{\bosflip{1}{-55}{brown}{red}}
\end{aligned}\right.
\end{equation}
In the first case, we identify partciles 1 and 2 as vector $\mathbb A^+$, so we can apply two Gauge Deformation 
\begin{equation}
\begin{tikzpicture}[baseline=0.7cm] \begin{feynhand}
\setlength{\feynhandarrowsize}{3.5pt}
\vertex [particle] (i1) at (0,0.8) {$1^+$}; 
\vertex [particle] (i2) at (1.6,1.6) {$2^+$}; 
\vertex [particle] (i3) at (1.6,0) {$3^-$};  
\vertex (v1) at (0.9,0.8); 
\graph{(i1)--[bos](v1)--[bos](i2)};
\graph{(i3)--[bos](v1)};  
\end{feynhand} \end{tikzpicture}\to
\begin{tikzpicture}[baseline=-0.1cm] \begin{feynhand}
\fill[yellow!50!white, opacity=0.7,rotate=0] (-0.9,-0.2) ellipse (0.8 and 0.6);
\node (C) at (-0.8,-0.5) {$\mathbb A^+$};
\vertex [particle] (i1) at (-1.01,0) {$1^+$}; 
\vertex [particle] (i2) at (0.579,0.827) {$2^+$}; 
\vertex [particle] (i3) at (0.579,-0.827) {$3^-$}; 
\vertex (v1) at (0,0);
\bosflip{1}{180}{brown}{cyan};
\bosflip{1}{55}{brown}{cyan};
\bos{i3}{red};
\end{feynhand} \end{tikzpicture}
\to\begin{tikzpicture}[baseline=-0.1cm] \begin{feynhand}
\fill[yellow!50!white, opacity=0.7,rotate=0] (-0.9,-0.2) ellipse (0.8 and 0.6);
\fill[green!40!white, opacity=0.7,rotate=-130] (-0.9,0.2) ellipse (0.8 and 0.6);
\node (C1) at (-0.8,-0.5) {$\mathbb A^+$};
\node (C2) at (1.0,0.5) {$\mathbb A^+$};
\vertex [particle] (i1) at (-1.01,0) {$1^+$}; 
\vertex [particle] (i2) at (0.579,0.827) {$2^+$}; 
\vertex [particle] (i3) at (0.579,-0.827) {$3^-$}; 
\vertex (v1) at (0,0);
\bosflip{1}{180}{brown}{cyan};
\bosflip{1}{55}{brown}{cyan};
\bos{i3}{red};
\end{feynhand} \end{tikzpicture}
\end{equation}
where the yellow zone and green zone represent the first and second deformation step. The scaling behavior of each particle in massless and MHC ampliude is
\begin{equation} 
\begin{tabular}{c|c|c|c|c}
\hline
scaling & massless & first step & second step & target \\
\hline
particle 1 & $\tilde\lambda_1^2$ & $\mathbb A^+(\tilde\lambda_1^2)\to\mathbb A^+(\tilde\lambda_1 \tilde m_1\eta_1)$ & \multirow{2}{*}{$\mathbb J(\lambda\tilde m\eta)\to\mathbb J(\tilde\lambda\lambda^2 \tilde m\eta)$} & $\tilde\lambda_1 \tilde m_1\eta_1$ \\
\cline{1-3} \cline{5-5}
particle 3 & $\tilde\lambda_3^{-2}$ & \multirow{2}{*}{$\mathbb J(1)\to\mathbb J(\tilde\lambda \lambda)$} & & $\lambda_3^2$ \\
\cline{1-2} \cline{4-5} 
particle 2 & $\tilde\lambda_2^2$ & & $\mathbb A^+(\tilde\lambda_2^2)\to\mathbb A^+(\tilde\lambda_2 \tilde m_2\eta_2)$ & $\tilde\lambda_2 \tilde m_2\eta_2$ \\
\hline 
\end{tabular}
\end{equation}

Begin with the massless amplitude. In first step, we identify particle 1 as the vector $\mathbb A^+$, particles 2 and 3 as the current $\mathbb J$. So the massless scaling behavior is $\mathbb A^+(\tilde\lambda_1^2)$ and $\mathbb J(1)$. It can be deformed as
\begin{equation} \begin{aligned}
\mathbb J(\tilde\lambda^0)\cdot\mathbb A^+(\tilde\lambda_1^2)
\xrightarrow{\times \frac{\langle\lambda\eta\rangle}{m}}\mathbb J(\tilde\lambda^0)\cdot\mathbb A^+(\tilde\lambda_1^2\lambda_1 m_1^{-1}\eta_1)
&\xrightarrow{\text{IBP}^-}\mathbb J(\tilde\lambda \lambda)\cdot\mathbb A^+(\tilde\lambda_1 m_1^{-1}\eta_1)\\
&=\mathbb J(\tilde\lambda \lambda)\cdot\frac{\mathbb A^+(\tilde\lambda_1\tilde m_1\eta_1)}{\mathbf m^2}
\end{aligned} \end{equation}
The corresponding amplitude deformation is
\begin{equation} \begin{aligned}
\frac{[12]^3}{[23][31]} 
\xrightarrow{\times \frac{\langle\lambda\eta\rangle}{m}}
-\frac{\langle\eta_112][12]^2}{m_1[23][31]}
\xrightarrow{\text{IBP}^-}
-\frac{\langle\eta_1 3\rangle[12]^2}{m_1[31]}
=-\frac{\tilde m_1\langle\eta_1 3\rangle[12]^2}{\mathbf m_1^2[31]}
\end{aligned} \end{equation}

Then we identify particle 2 as the vector $\mathbb A^+$, particles 1 and 3 as the current $\mathbb J$. So we can read the scaling behavior
\begin{equation}
-\frac{\tilde m_1\langle\eta_1 3\rangle[12]^2}{\mathbf m_1^2[31]}\sim
\begin{cases}
\text{current}:\quad {\displaystyle\frac{\mathbb J(\lambda \tilde m\eta)}{\mathbf m^2}\sim\frac{1}{\mathbf m_1^2}}\tilde\lambda_1\tilde m_1\eta_1\tilde\lambda_3^{-1}\lambda_3, \\
\text{vector}:\quad \mathbb A^+(\tilde\lambda_2^2), 
\end{cases}
\end{equation}
This scaling can be deformed as
\begin{equation} \begin{aligned}
\frac{\mathbb J(\lambda \tilde m\eta)}{\mathbf m_1^2}\cdot\mathbb A^+(\tilde\lambda_2^2)
\xrightarrow{\times \frac{\langle\lambda\eta\rangle}{m}}
\frac{\mathbb J(\lambda \tilde m\eta)}{\mathbf m_1^2}\cdot\mathbb A^+(\tilde\lambda_2^2\lambda m_2^{-1}\eta_2)
&\xrightarrow{\text{IBP}^-}\frac{\mathbb J(\tilde\lambda \lambda^2 \tilde m\eta)}{\mathbf m_1^2}\cdot \mathbb A^+(\tilde\lambda_2 m_2^{-1}\eta_2)\\
&=\frac{\mathbb J(\tilde\lambda \lambda^2 \tilde m\eta)}{\mathbf m_1^2}\cdot\frac{\mathbb A^+(\tilde\lambda_2 \tilde m_2\eta_2)}{\mathbf m_2^2}
\end{aligned} \end{equation}
The corresponding amplitude deformation is
\begin{equation} \begin{aligned}
-\frac{\tilde m_1\langle\eta_1 3\rangle[12]^2}{\mathbf m_1^2[31]}
\xrightarrow{\times \frac{\langle\lambda\eta\rangle}{m}}
-\frac{\tilde m_1\langle\eta_1 3\rangle[12\eta_2\rangle[12]}{\mathbf m_1^2 m_2[31]}
&\xrightarrow{\text{IBP}^-}
-\frac{\tilde m_1\langle\eta_1 3\rangle\langle3\eta_2\rangle[12]}{\mathbf m_1^2 m_2}\\
&=
-\frac{\tilde m_1\tilde m_2\langle\eta_1 3\rangle\langle3\eta_2\rangle[12]}{\mathbf m_1^2 \mathbf m_2^2}
\end{aligned} \end{equation}

In summary, the above amplitude deformation can be expressed in a short form as
\begin{equation}
\frac{[12]^3}{[23][31]} \xrightarrow{\text{gauge}}
-\frac{\tilde m_1\langle\eta_1 3\rangle[12]^2}{\mathbf m_1^2[31]}\xrightarrow{\text{gauge}}
-\frac{\tilde m_1\tilde m_2\langle\eta_1 3\rangle\langle3\eta_2\rangle[12]}{\mathbf m_1^2 \mathbf m_2^2}
\end{equation}

For the second case, a massless Feynman diagram corresponds to two MHC diagrams,
\begin{equation} \label{eq:VVVtoVVV2}
\begin{tikzpicture}[baseline=0.7cm] \begin{feynhand}
\setlength{\feynhandarrowsize}{3.5pt}
\vertex [particle] (i1) at (0,0.8) {$1^+$}; 
\vertex [particle] (i2) at (1.6,1.6) {$2^+$}; 
\vertex [particle] (i3) at (1.6,0) {$3^-$};  
\vertex (v1) at (0.9,0.8); 
\graph{(i1)--[bos](v1)--[bos](i2)};
\graph{(i3)--[bos](v1)};  
\end{feynhand} \end{tikzpicture}\to
\makecell{\begin{tikzpicture}[baseline=-0.1cm] \begin{feynhand}
\fill[yellow!50!white, opacity=0.7,rotate=150] (-0.9,0.2) ellipse (0.8 and 0.6);
\node (C2) at (1.0,-0.5) {$\mathbb A^-$};
\vertex [particle] (i1) at (-1.01,0) {$1^+$}; 
\vertex [particle] (i2) at (0.579,0.827) {$2^+$}; 
\vertex [particle] (i3) at (0.579,-0.827) {$3^-$}; 
\vertex (v1) at (0,0);
\bosflip{1}{180}{brown}{cyan};
\bos{i2}{cyan};
\bosflip{1}{-55}{brown}{red};
\end{feynhand} \end{tikzpicture}\\
+\\
\begin{tikzpicture}[baseline=-0.1cm] \begin{feynhand}
\fill[yellow!50!white, opacity=0.7,rotate=150] (-0.9,0.2) ellipse (0.8 and 0.6);
\node (C2) at (1.0,-0.5) {$\mathbb A^-$};
\vertex [particle] (i1) at (-1.01,0) {$1^+$}; 
\vertex [particle] (i2) at (0.579,0.827) {$2^+$}; 
\vertex [particle] (i3) at (0.579,-0.827) {$3^-$}; 
\vertex (v1) at (0,0);
\bos{i1}{cyan};
\bosflip{1}{55}{brown}{cyan};
\bosflip{1}{-55}{brown}{red};
\end{feynhand} \end{tikzpicture}\\}
\to\makecell{\begin{tikzpicture}[baseline=-0.1cm] \begin{feynhand}
\fill[green!40!white, opacity=0.7,rotate=0] (-0.9,-0.2) ellipse (0.8 and 0.6);
\fill[yellow!50!white, opacity=0.7,rotate=150] (-0.9,0.2) ellipse (0.8 and 0.6);
\node (C1) at (-0.8,-0.5) {$\mathbb A^+$};
\node (C2) at (1.0,-0.5) {$\mathbb A^-$};
\vertex [particle] (i1) at (-1.01,0) {$1^+$}; 
\vertex [particle] (i2) at (0.579,0.827) {$2^+$}; 
\vertex [particle] (i3) at (0.579,-0.827) {$3^-$}; 
\vertex (v1) at (0,0);
\bosflip{1}{180}{brown}{cyan};
\bos{i2}{cyan};
\bosflip{1}{-55}{brown}{red};
\end{feynhand} \end{tikzpicture}\\
+\\
\begin{tikzpicture}[baseline=-0.1cm] \begin{feynhand}
\fill[green!40!white, opacity=0.7,rotate=-115] (-0.9,-0.2) ellipse (0.8 and 0.6);
\fill[yellow!50!white, opacity=0.7,rotate=150] (-0.9,0.2) ellipse (0.8 and 0.6);
\node (C1) at (-0.1,0.8) {$\mathbb A^+$};
\node (C2) at (1.0,-0.5) {$\mathbb A^-$};
\vertex [particle] (i1) at (-1.01,0) {$1^+$}; 
\vertex [particle] (i2) at (0.579,0.827) {$2^+$}; 
\vertex [particle] (i3) at (0.579,-0.827) {$3^-$}; 
\vertex (v1) at (0,0);
\bos{i1}{cyan};
\bosflip{1}{55}{brown}{cyan};
\bosflip{1}{-55}{brown}{red};
\end{feynhand} \end{tikzpicture}\\}
\end{equation}
For the MHC diagrams in the first line, we have the following scaling infromation
\begin{equation} 
\begin{tabular}{c|c|c|c|c}
\hline
scaling & massless & first step & second step & target \\
\hline
particle 1 & $\tilde\lambda_1^2$ & \multirow{2}{*}{$\mathbb J(\tilde\lambda^4)\to\mathbb J(\tilde\lambda^4)$} & $\mathbb A^+(\tilde\lambda_1^2)\to\mathbb A^+(\tilde\lambda_1 \tilde m_1\eta_1)$ & $\tilde\lambda_1 \tilde m_1\eta_1$ \\
\cline{1-2} \cline{4-5}
particle 2 & $\tilde\lambda_2^2$ & & \multirow{2}{*}{$\mathbb J(\tilde\lambda m\tilde\eta)\to\mathbb J(\tilde\lambda^2\lambda m\tilde\eta)$} & $\tilde\lambda_2^2$ \\
\cline{1-3} \cline{5-5}
particle 3 & $\tilde\lambda_3^{-2}$ & $\mathbb A^-(\lambda_3^{-2})\to\mathbb A^-(\lambda_3^{-1} m_3\tilde\eta_3)$ & & $\lambda_3 m_3\tilde\eta_3$ \\
\hline 
\end{tabular}
\end{equation}
By exchanging particles 1 and 2, we obtain the scaling information for the MHC diagram in the second line of Eq.~\eqref{eq:VVVtoVVV2}.

Thus, we can first apply the Gauge deformation to match particle 3 to $[\mathbf A]_1$, and then apply the Gauge deformation to particle 1 or 2. The corresponding amplitude deformation takes the form:
\begin{equation} \begin{aligned}
\frac{[12]^3}{[23][31]} 
&\xrightarrow{\text{gauge}} 
\frac{m_3[12]^2[\eta_3 2]}{\mathbf m_3^2[23]}+\frac{m_3[12]^2[\eta_3 1]}{\mathbf m_3^2[31]}\\
&\xrightarrow{\text{gauge}}
\frac{m_3\tilde m_1[12][\eta_3 2]\langle3\eta_1\rangle}{\mathbf m_3^2\mathbf m_1^2}+\frac{m_3\tilde m_2[12]\langle\eta_2 3\rangle[\eta_3 1]}{\mathbf m_3^2\mathbf m_2^2} \\
\end{aligned} \end{equation}
The first term corresponds to the MHC diagram in the first line of Eq.~\eqref{eq:VVVtoVVV2}, while the second term corresponds to the diagram in the second line.

The second category is the matching from the
massless $VSS$ amplitude to massive $VVV$ amplitudes. According to exchange symmetry of two scalar boson, there are four cases
\begin{equation}
\begin{tikzpicture}[baseline=0.7cm] \begin{feynhand}
\setlength{\feynhandarrowsize}{3.5pt}
\vertex [particle] (i1) at (0,0.8) {$1^-$}; 
\vertex [particle] (i2) at (1.6,1.6) {$2^0$}; 
\vertex [particle] (i3) at (1.6,0) {$3^0$};  
\vertex (v1) at (0.9,0.8); 
\graph{(i1)--[bos](v1)--[sca](i2)};
\graph{(i3)--[sca](v1)};  
\end{feynhand} \end{tikzpicture}
=\frac{\langle12\rangle\langle31\rangle}{\langle23\rangle}
\to \left\{\begin{aligned}
\Ampthree{1^-}{2^0}{3^0}{\bosflip{1}{180}{brown}{red}}{\bos{i2}{brown}}{\bosflip{1}{-55}{cyan}{brown}}
+\Ampthree{1^-}{2^0}{3^0}{\bosflip{1}{180}{brown}{red}}{\bosflip{1}{55}{cyan}{brown}}{\bos{i3}{brown}}, \\
\Ampthree{1^-}{2^0}{3^0}{\bosflip{1}{180}{brown}{red}}{\bos{i2}{brown}}{\bosflip{1}{-55}{red}{brown}}
+\Ampthree{1^-}{2^0}{3^0}{\bosflip{1}{180}{brown}{red}}{\bosflip{1}{55}{red}{brown}}{\bos{i3}{brown}}, \\
\Ampthree{1^-}{2^0}{3^0}{\bos{i1}{red}}{\bos{i2}{brown}}{\bosflipflip{1}{-55}{brown}{red}{brown}}+
\Ampthree{1^-}{2^0}{3^0}{\bos{i1}{red}}{\bosflipflip{1}{55}{brown}{red}{brown}}{\bos{i3}{brown}}, \\
\Ampthree{1^-}{2^0}{3^0}{\bosflipflip{1}{180}{cyan}{brown}{red}}{\bos{i2}{brown}}{\bos{i3}{brown}}
\end{aligned}\right.
\end{equation}

Then we consider matching the massless $VSS$ amplitude with helicity $(-1,0,0)$ to the massive $VVV$ amplitude. There are two cases:
\begin{enumerate}
\item \textit{Goldstone + Gauge Deformation:} We can choose a massless gauge boson and a Goldstone boson (e.g. particles 1 and 2) to match non-chiral massive vectors. Since particle 3 is matched to a chiral vector, it should be fliped by the Goldstone rule of particle 2. Additionally, particle 1 undergoes an independent chirality flip as it represents a gauge boson. Thus, the resulting diagram is
\begin{equation} \label{eq:VSStoVVV1}
\begin{tikzpicture}[baseline=0.7cm] \begin{feynhand}
\setlength{\feynhandarrowsize}{3.5pt}
\vertex [particle] (i1) at (0,0.8) {$1^-$}; 
\vertex [particle] (i2) at (1.6,1.6) {$2^0$}; 
\vertex [particle] (i3) at (1.6,0) {$3^0$};  
\vertex (v1) at (0.9,0.8); 
\graph{(i1)--[bos](v1)--[sca](i2)};
\graph{(i3)--[sca](v1)};  
\end{feynhand} \end{tikzpicture}\to
\makecell{
\begin{tikzpicture}[baseline=-0.1cm] \begin{feynhand}
\fill[yellow!50!white, opacity=0.7,rotate=0] (-0.9,-0.2) ellipse (0.8 and 0.6);
\node (C1) at (-0.8,-0.5) {$\mathbb A^-$};
\vertex [particle] (i1) at (-1.01,0) {$1^-$}; 
\vertex [particle] (i2) at (0.579,0.827) {$2^0$}; 
\vertex [particle] (i3) at (0.579,-0.827) {$3^0$}; 
\vertex (v1) at (0,0);
\bosflip{1}{180}{brown}{red};
\bosflip{1}{55}{cyan}{brown};
\bos{i3}{brown};
\end{feynhand} \end{tikzpicture} \\
+\\
\begin{tikzpicture}[baseline=-0.1cm] \begin{feynhand}
\fill[yellow!50!white, opacity=0.7,rotate=0] (-0.9,-0.2) ellipse (0.8 and 0.6);
\node (C1) at (-0.8,-0.5) {$\mathbb A^-$};
\vertex [particle] (i1) at (-1.01,0) {$1^-$}; 
\vertex [particle] (i2) at (0.579,0.827) {$2^0$}; 
\vertex [particle] (i3) at (0.579,-0.827) {$3^0$}; 
\vertex (v1) at (0,0);
\bosflip{1}{180}{brown}{red};
\bos{i2}{brown};
\bosflip{1}{-55}{cyan}{brown};
\end{feynhand} \end{tikzpicture}}
\to
\makecell{
\begin{tikzpicture}[baseline=-0.1cm] \begin{feynhand}
\fill[green!40!white, opacity=0.7,rotate=20] (-0.1,0.4) ellipse (1.2 and 0.6);
\node (C1) at (-0.3,0.6) {$\mathbb J$};
\vertex [particle] (i1) at (-1.01,0) {$1^-$}; 
\vertex [particle] (i2) at (0.579,0.827) {$2^0$}; 
\vertex [particle] (i3) at (0.579,-0.827) {$3^0$}; 
\vertex (v1) at (0,0);
\bosflip{1}{180}{brown}{red};
\bosflip{1}{55}{cyan}{brown};
\bos{i3}{brown};
\end{feynhand} \end{tikzpicture}\\
+\\
\begin{tikzpicture}[baseline=-0.1cm] \begin{feynhand}
\fill[green!40!white, opacity=0.7,rotate=150] (0.1,0.4) ellipse (1.2 and 0.6);
\node (C1) at (-0.3,-0.6) {$\mathbb J$};
\vertex [particle] (i1) at (-1.01,0) {$1^-$}; 
\vertex [particle] (i2) at (0.579,0.827) {$2^0$}; 
\vertex [particle] (i3) at (0.579,-0.827) {$3^0$}; 
\vertex (v1) at (0,0);
\bosflip{1}{180}{brown}{red};
\bos{i2}{brown};
\bosflip{1}{-55}{cyan}{brown};
\end{feynhand} \end{tikzpicture}},
\end{equation}
When we match the massless amplitude to the first term, we have the following scaling information,
\begin{equation} 
\begin{tabular}{c|c|c|c|c}
\hline
scaling & massless & first step & second step & target \\
\hline
particle 1 & $\lambda_1^2$ & $\mathbb A^-(\lambda_1^2)\to\mathbb A^-(\lambda_1 m_1\tilde\eta_1)$ & \multirow{2}{*}{$\mathbb J(\tilde\lambda\lambda^2 m\tilde\eta)\to\mathbb J(\tilde\lambda\lambda m^2\tilde\eta^2)$} & $\lambda_1 m_1\tilde\eta_1$ \\
\cline{1-3} \cline{5-5}
particle 2 & $1$ & \multirow{2}{*}{$\mathbb J(1)\to\mathbb J(\tilde\lambda\lambda)$} & & $\tilde\lambda_2 m_2\tilde\eta_2$ \\
\cline{1-2} \cline{4-5}
particle 3 & $1$ & & $\mathbb A^0(1) \to\mathbb A^0(\tilde\lambda_3\lambda_3)$ & $\tilde\lambda_3\lambda_3$ \\
\hline 
\end{tabular}
\end{equation}
By exchanging particles 2 and 3, we can obtain the scaling information for the second MHC term.

In this case, the gauge boson rule must be applied before the Goldstone rule
\begin{equation} \begin{aligned}
\frac{\langle12\rangle\langle31\rangle}{\langle23\rangle}
&\xrightarrow{\text{Gauge}} b_1\frac{m_1 [\eta_12]\langle21\rangle}{\mathbf m_1^2}-b_2\frac{m_1 [\eta_13]\langle31\rangle}{\mathbf m_1^2} \\
&\xrightarrow{\text{Goldstone}}b_1\frac{m_1 m_2\langle13\rangle[3\eta_2][2\eta_1]}{\mathbf m_1^2 \mathbf m_2^2}+ b_2\frac{m_1 m_3 \langle12\rangle[2\eta_3][3\eta_1]}{\mathbf m_1^2 \mathbf m_3^2},
\end{aligned} \end{equation}
where $b_1+b_2=1$.

There is another case, the corresponding diagrams are 
\begin{equation}
\begin{tikzpicture}[baseline=0.7cm] \begin{feynhand}
\setlength{\feynhandarrowsize}{3.5pt}
\vertex [particle] (i1) at (0,0.8) {$1^-$}; 
\vertex [particle] (i2) at (1.6,1.6) {$2^0$}; 
\vertex [particle] (i3) at (1.6,0) {$3^0$};  
\vertex (v1) at (0.9,0.8); 
\graph{(i1)--[bos](v1)--[sca](i2)};
\graph{(i3)--[sca](v1)};  
\end{feynhand} \end{tikzpicture}\to
\makecell{
\begin{tikzpicture}[baseline=-0.1cm] \begin{feynhand}
\fill[yellow!50!white, opacity=0.7,rotate=0] (-0.9,-0.2) ellipse (0.8 and 0.6);
\node (C1) at (-0.8,-0.5) {$\mathbb A^-$};
\vertex [particle] (i1) at (-1.01,0) {$1^-$}; 
\vertex [particle] (i2) at (0.579,0.827) {$2^0$}; 
\vertex [particle] (i3) at (0.579,-0.827) {$3^0$}; 
\vertex (v1) at (0,0);
\bosflip{1}{180}{brown}{red};
\bosflip{1}{55}{red}{brown};
\bos{i3}{brown};
\end{feynhand} \end{tikzpicture}\\
+\\
\begin{tikzpicture}[baseline=-0.1cm] \begin{feynhand}
\fill[yellow!40!white, opacity=0.7,rotate=20] (-0.1,0.4) ellipse (1.2 and 0.6);
\node (C1) at (-0.3,0.6) {$\mathbb J $};
\vertex [particle] (i1) at (-1.01,0) {$1^-$}; 
\vertex [particle] (i2) at (0.579,0.827) {$2^0$}; 
\vertex [particle] (i3) at (0.579,-0.827) {$3^0$}; 
\vertex (v1) at (0,0);
\bosflip{1}{180}{brown}{red};
\bosflip{1}{55}{red}{brown};
\bos{i3}{brown};
\end{feynhand} \end{tikzpicture}\\
+\\
\begin{tikzpicture}[baseline=-0.1cm] \begin{feynhand}
\fill[yellow!50!white, opacity=0.7,rotate=0] (-0.9,-0.2) ellipse (0.8 and 0.6);
\node (C1) at (-0.8,-0.5) {$\mathbb A^-$};
\vertex [particle] (i1) at (-1.01,0) {$1^-$}; 
\vertex [particle] (i2) at (0.579,0.827) {$2^0$}; 
\vertex [particle] (i3) at (0.579,-0.827) {$3^0$}; 
\vertex (v1) at (0,0);
\bosflip{1}{180}{brown}{red};
\bos{i2}{brown};
\bosflip{1}{-55}{red}{brown};
\end{feynhand} \end{tikzpicture}\\
+\\
\begin{tikzpicture}[baseline=-0.1cm] \begin{feynhand}
\fill[yellow!40!white, opacity=0.7,rotate=150] (0.1,0.4) ellipse (1.2 and 0.6);
\node (C1) at (-0.3,-0.6) {$\mathbb J$};
\vertex [particle] (i1) at (-1.01,0) {$1^-$}; 
\vertex [particle] (i2) at (0.579,0.827) {$2^0$}; 
\vertex [particle] (i3) at (0.579,-0.827) {$3^0$}; 
\vertex (v1) at (0,0);
\bosflip{1}{180}{brown}{red};
\bos{i2}{brown};
\bosflip{1}{-55}{red}{brown};
\end{feynhand} \end{tikzpicture}
}\to
\makecell{
\begin{tikzpicture}[baseline=-0.1cm] \begin{feynhand}
\fill[green!40!white, opacity=0.7,rotate=20] (-0.1,0.4) ellipse (1.2 and 0.6);
\node (C1) at (-0.3,0.6) {$\mathbb J $};
\vertex [particle] (i1) at (-1.01,0) {$1^-$}; 
\vertex [particle] (i2) at (0.579,0.827) {$2^0$}; 
\vertex [particle] (i3) at (0.579,-0.827) {$3^0$}; 
\vertex (v1) at (0,0);
\bosflip{1}{180}{brown}{red};
\bosflip{1}{55}{red}{brown};
\bos{i3}{brown};
\end{feynhand} \end{tikzpicture}\\
+\\
\begin{tikzpicture}[baseline=-0.1cm] \begin{feynhand}
\fill[green!50!white, opacity=0.7,rotate=0] (-0.9,-0.2) ellipse (0.8 and 0.6);
\node (C1) at (-0.8,-0.5) {$\mathbb A^-$};
\vertex [particle] (i1) at (-1.01,0) {$1^-$}; 
\vertex [particle] (i2) at (0.579,0.827) {$2^0$}; 
\vertex [particle] (i3) at (0.579,-0.827) {$3^0$}; 
\vertex (v1) at (0,0);
\bosflip{1}{180}{brown}{red};
\bosflip{1}{55}{red}{brown};
\bos{i3}{brown};
\end{feynhand} \end{tikzpicture}\\
+\\
\begin{tikzpicture}[baseline=-0.1cm] \begin{feynhand}
\fill[green!40!white, opacity=0.7,rotate=150] (0.1,0.4) ellipse (1.2 and 0.6);
\node (C1) at (-0.3,-0.6) {$\mathbb J$};
\vertex [particle] (i1) at (-1.01,0) {$1^-$}; 
\vertex [particle] (i2) at (0.579,0.827) {$2^0$}; 
\vertex [particle] (i3) at (0.579,-0.827) {$3^0$}; 
\vertex (v1) at (0,0);
\bosflip{1}{180}{brown}{red};
\bos{i2}{brown};
\bosflip{1}{-55}{red}{brown};
\end{feynhand} \end{tikzpicture}\\
+\\
\begin{tikzpicture}[baseline=-0.1cm] \begin{feynhand}
\fill[green!50!white, opacity=0.7,rotate=0] (-0.9,-0.2) ellipse (0.8 and 0.6);
\node (C1) at (-0.8,-0.5) {$\mathbb A^-$};
\vertex [particle] (i1) at (-1.01,0) {$1^-$}; 
\vertex [particle] (i2) at (0.579,0.827) {$2^0$}; 
\vertex [particle] (i3) at (0.579,-0.827) {$3^0$}; 
\vertex (v1) at (0,0);
\bosflip{1}{180}{brown}{red};
\bos{i2}{brown};
\bosflip{1}{-55}{red}{brown};
\end{feynhand} \end{tikzpicture}},
\end{equation}
In this case, we need to apply Gauge and Goldstone deformation simultaneously in each step,
\begin{equation} \begin{aligned}
\frac{\langle12\rangle\langle31\rangle}{\langle23\rangle}
\xrightarrow[\text{Gauge}]{\text{Goldstone}}&
-\frac{\mathbf m_3^2}{2\mathbf m_1^2-\mathbf m_2^2-\mathbf m_3^2}\frac{m_1 [\eta_1 3]\langle31\rangle}{\mathbf m_1^2}
+\frac{\mathbf m_2^2}{2\mathbf m_1^2-\mathbf m_2^2-\mathbf m_3^2}\frac{m_1\langle12\rangle[2\eta_1]}{\mathbf m_1^2} \\
&-\frac{2\mathbf m_2^2}{2\mathbf m_1^2-\mathbf m_2^2-\mathbf m_3^2}\frac{\tilde m_2\langle12\rangle\langle1\eta_2\rangle}{\mathbf m_2^2}
-\frac{2\mathbf m_3^2}{2\mathbf m_1^2-\mathbf m_2^2-\mathbf m_3^2}\frac{\tilde m_3\langle31\rangle\langle1\eta_3\rangle}{\mathbf m_3^2} \\
\xrightarrow[\text{Gauge}]{\text{Goldstone}}&
\frac{\mathbf m_3^2}{2\mathbf m_1^2-\mathbf m_2^2-\mathbf m_3^2}\frac{m_1\tilde m_3[\eta_1 2]\langle2\eta_3\rangle\langle31\rangle+[\eta_1 2]\langle23\rangle\langle\eta_31\rangle}{\mathbf m_1^2 \mathbf m_3^2}\\
&+\frac{\mathbf m_2^2}{2\mathbf m_1^2-\mathbf m_2^2-\mathbf m_3^2}\frac{m_1\tilde m_2\langle12\rangle\langle\eta_23\rangle[3\eta_1]+\langle1\eta_2\rangle\langle23\rangle[3\eta_1]}{\mathbf m_1^2 \mathbf m_2^2}
\end{aligned} \end{equation}

\item \textit{Two Goldstone Deformations:} When we choose two Goldstone boson (e.g. particles 2 and 3) to match the non-chiral massive vector, we should not flip these two Goldstone bosons themselves. If there is chirality flip, the only choice is to flip the gauge boson (e.g. particle 1) twice. The corresponding diagram is
\begin{equation} \begin{aligned}
\begin{tikzpicture}[baseline=0.7cm] \begin{feynhand}
\setlength{\feynhandarrowsize}{3.5pt}
\vertex [particle] (i1) at (0,0.8) {$1^-$}; 
\vertex [particle] (i2) at (1.6,1.6) {$2^0$}; 
\vertex [particle] (i3) at (1.6,0) {$3^0$};  
\vertex (v1) at (0.9,0.8); 
\graph{(i1)--[bos](v1)--[sca](i2)};
\graph{(i3)--[sca](v1)};  
\end{feynhand} \end{tikzpicture}\rightarrow
\begin{tikzpicture}[baseline=-0.1cm] \begin{feynhand}
\fill[yellow!50!white, opacity=0.7,rotate=20] (0.5,0.4) ellipse (1.0 and 0.6);
\node (C1) at (-0.1,0.6) {$\mathbb J$};
\vertex [particle] (i1) at (-1.01,0) {$1^-$}; 
\vertex [particle] (i2) at (0.579,0.827) {$2^0$}; 
\vertex [particle] (i3) at (0.579,-0.827) {$3^0$}; 
\vertex (v1) at (0,0);
\bosflipflip{1}{180}{cyan}{brown}{red};
\bos{i2}{brown};
\bos{i3}{brown};
\end{feynhand} \end{tikzpicture}\to
\begin{tikzpicture}[baseline=-0.1cm] \begin{feynhand}
\fill[green!40!white, opacity=0.7,rotate=150] (0,0.4) ellipse (1.2 and 0.6);
\node (C1) at (-0.1,-0.6) {$\mathbb J$};
\vertex [particle] (i1) at (-1.01,0) {$1^-$}; 
\vertex [particle] (i2) at (0.579,0.827) {$2^0$}; 
\vertex [particle] (i3) at (0.579,-0.827) {$3^0$}; 
\vertex (v1) at (0,0);
\bosflipflip{1}{180}{cyan}{brown}{red};
\bos{i2}{brown};
\bos{i3}{brown};
\end{feynhand} \end{tikzpicture},\\
\end{aligned} \end{equation}
When we match the massless amplitude to the MHC term, we have the following scaling information,
\begin{equation} 
\begin{tabular}{c|c|c|c|c}
\hline
scaling & massless & first step & second step & target \\
\hline
particle 2 & $1$ & \multirow{2}{*}{$\mathbb J(\lambda^2)\to\mathbb J(\lambda m\tilde\eta)$} & $\mathbb A^0(1)\to\mathbb A^0(\tilde\lambda_2\lambda_2)$ & $\tilde\lambda_2\lambda_2$ \\
\cline{1-2} \cline{4-5}
particle 1 & $\lambda_1^2$ & & \multirow{2}{*}{$\mathbb J(\tilde\lambda\lambda^2 m\tilde\eta)\to\mathbb J(\tilde\lambda\lambda m^2\tilde\eta^2)$} & $m_1^2\tilde\eta_1^2$ \\
\cline{1-3} \cline{5-5}
particle 3 & $1$ & $\mathbb A^0(1)\to\mathbb A^0(\tilde\lambda_3\lambda_3)$ & & $\tilde\lambda_3\lambda_3$ \\
\hline 
\end{tabular}
\end{equation}

Thus, we should perform two Goldstone Deformations and get
\begin{equation}
\frac{\langle12\rangle\langle31\rangle}{\langle23\rangle}
\xrightarrow{\text{Goldstone}} \frac{m_1[\eta_1 3]\langle31\rangle}{\mathbf m_1^2} \xrightarrow{\text{Goldstone}} \frac{m_1^2[\eta_1 3]\langle32\rangle[2 \eta_1]}{\mathbf m_1^4}
\end{equation}

In the last case, we match the massless amplitude to
\begin{equation} \begin{aligned} \label{eq:VSStoVVV2}
\begin{tikzpicture}[baseline=0.7cm] \begin{feynhand}
\setlength{\feynhandarrowsize}{3.5pt}
\vertex [particle] (i1) at (0,0.8) {$1^-$}; 
\vertex [particle] (i2) at (1.6,1.6) {$2^0$}; 
\vertex [particle] (i3) at (1.6,0) {$3^0$};  
\vertex (v1) at (0.9,0.8); 
\graph{(i1)--[bos](v1)--[sca](i2)};
\graph{(i3)--[sca](v1)};  
\end{feynhand} \end{tikzpicture}\rightarrow
\makecell{\begin{tikzpicture}[baseline=-0.1cm] \begin{feynhand}
\fill[yellow!50!white, opacity=0.7,rotate=160] (0.5,0.35) ellipse (0.9 and 0.6);
\node (C1) at (-0.3,-0.5) {$\mathbb J$};
\vertex [particle] (i1) at (-1.01,0) {$1^-$}; 
\vertex [particle] (i2) at (0.579,0.827) {$2^0$}; 
\vertex [particle] (i3) at (0.579,-0.827) {$3^0$}; 
\vertex (v1) at (0,0);
\bos{i1}{red};
\bos{i2}{brown};
\bosflipflip{1}{-55}{brown}{red}{brown};
\end{feynhand} \end{tikzpicture}\\
+\\
\begin{tikzpicture}[baseline=-0.1cm] \begin{feynhand}
\fill[yellow!50!white, opacity=0.7,rotate=30] (-0.45,0.2) ellipse (0.9 and 0.6);
\node (C1) at (-0.5,-0.4) {$\mathbb J$};
\vertex [particle] (i1) at (-1.01,0) {$1^-$}; 
\vertex [particle] (i2) at (0.579,0.827) {$2^0$}; 
\vertex [particle] (i3) at (0.579,-0.827) {$3^0$}; 
\vertex (v1) at (0,0);
\bos{i1}{red};
\bosflipflip{1}{55}{brown}{red}{brown};
\bos{i3}{brown};
\end{feynhand} \end{tikzpicture}}\to
\makecell{\begin{tikzpicture}[baseline=-0.1cm] \begin{feynhand}
\fill[green!50!white, opacity=0.7,rotate=150] (-0.9,0.2) ellipse (0.8 and 0.6);
\node (C2) at (1.0,-0.5) {$\mathbb A^0$};
\vertex [particle] (i1) at (-1.01,0) {$1^-$}; 
\vertex [particle] (i2) at (0.579,0.827) {$2^0$}; 
\vertex [particle] (i3) at (0.579,-0.827) {$3^0$}; 
\vertex (v1) at (0,0);
\bos{i1}{red};
\bos{i2}{brown};
\bosflipflip{1}{-55}{brown}{red}{brown};
\end{feynhand} \end{tikzpicture}\\
+\\
\begin{tikzpicture}[baseline=-0.1cm] \begin{feynhand}
\fill[green!40!white, opacity=0.7,rotate=-115] (-0.9,-0.2) ellipse (0.8 and 0.6);
\node (C1) at (-0.1,0.8) {$\mathbb A^0$};
\vertex [particle] (i1) at (-1.01,0) {$1^-$}; 
\vertex [particle] (i2) at (0.579,0.827) {$2^0$}; 
\vertex [particle] (i3) at (0.579,-0.827) {$3^0$}; 
\vertex (v1) at (0,0);
\bos{i1}{red};
\bosflipflip{1}{55}{brown}{red}{brown};
\bos{i3}{brown};
\end{feynhand} \end{tikzpicture}}
\end{aligned} \end{equation}
When we match the massless amplitude to the first MHC term, we have the following scaling information,
\begin{equation} 
\begin{tabular}{c|c|c|c|c}
\hline
scaling & massless & first step & second step & target \\
\hline
particle 2 & $1$ & $\mathbb A^0(1)\to\mathbb A^0(\tilde\lambda_2\lambda_2)$ & \multirow{2}{*}{$\mathbb J(\tilde\lambda\lambda^3)\to\mathbb J(\tilde\lambda\lambda^3)$} & $\tilde\lambda_2\lambda_2$ \\
\cline{1-3} \cline{5-5}
particle 1 & $\lambda_1^2$ & \multirow{2}{*}{$\mathbb J(\lambda^2)\to\mathbb J(\lambda m\tilde\eta)$} & & $\lambda_1^2$ \\
\cline{1-2} \cline{4-5} 
particle 3 & $1$ &  & $\mathbb A^0(\lambda_3^{-1} m_3\tilde\eta_3)\to
\mathbb A^0(m_3\tilde\eta_3\tilde m_3\eta_3)$ & $m_3\tilde\eta_3\tilde m_3\eta_3$ \\
\hline 
\end{tabular}
\end{equation}
By exchanging particles 2 and 3, we can obtain the scaling information for the second MHC term.

We can use two Goldstone Deformation and obtain
\begin{equation} \begin{aligned}
\frac{\langle12\rangle\langle31\rangle}{\langle23\rangle}
&\xrightarrow{\text{Goldstone}}
\frac{\mathbf m_3^2}{\mathbf m_2^2+\mathbf m_3^2}\frac{m_3\langle12\rangle^2[\eta_3 2]}{\mathbf m_3^2\langle23\rangle}
+\frac{\mathbf m_2^2}{\mathbf m_2^2+\mathbf m_3^2}\frac{m_2[\eta_2 3]\langle31\rangle^2}{\mathbf m_2^2\langle23\rangle} \\
&\xrightarrow{\text{Goldstone}}
-\frac{\mathbf m_3^2}{\mathbf m_1^2}\frac{\tilde m_3 m_3\langle12\rangle[\eta_3 2]\langle1\eta_3\rangle}{\mathbf m_3^4}
-\frac{\mathbf m_2^2}{\mathbf m_1^2}\frac{\tilde m_2 m_2[\eta_2 3]\langle31\rangle\langle1\eta_2\rangle}{\mathbf m_2^4}.
\end{aligned} \end{equation}

\end{enumerate}

Therefore, all the $VVV$ MHC amplitudes can be derived by using Gauge deformation and Goldstone deformation. 
See the Appendix~\ref{app:VVV_matching} for detailed implementations of these rules for $VVV$ matching. If we match massless amplitude with other helicity to the massive $VVV$ amplitude, the diagrams can be similarly derived by permutate external particles and change the color.

\subsection{Systematic subleading matching with power counting}

After deriving the leading structure for all 3-pt massive amplitude in the standard model, we now turn to the subleading matching. The massless amplitude with additional Higgs bosons can systematically match to all subleading structures. 

For a massless amplitude with $n_h$ Higgs insertions, which corresponds to a $(3+n_h)$-point amplitude, dimensional analysis dictates a mass dimension of $1 - n_h$. This reveals a difference in mass dimensions between higher-point massless amplitudes and 3-point massive amplitudes. To establish a consistent matching between them, a careful power counting in both the massless and massive contexts is essential. The massless amplitude only depends solely on the energy scale $E$, scaling as
\begin{equation}
\mathcal{A}_{3+n_h}\sim E^{1-n_h}.
\end{equation}
In contrast, the massive amplitude depends on both the energy scale $E$ and mass $\mathbf{m}$. In the high-energy regime $E\gg v$, we use MHC formalism to describe the massive structure. The MHC amplitude $\mathcal{M}$ admits an expansion of the form
\begin{equation}
\mathcal{M}=\sum_{l=0}^{2s} [\mathcal{M}]_l\sim \sum_{l=0}^{2s} {\mathbf m}^{l-s+1} E^{s-l},
\end{equation}
where $s$ denotes the total spin of the massive amplitude. Given that the physical mass satisfies $\mathbf{m} \sim v$, the power counting can be re-expressed in terms of $E$ and $v$ as
\begin{equation}
[\mathcal{M}]_l\sim E \left(\frac{v}{E}\right)^{l-s+1}.
\end{equation}

\begin{figure}[htbp]
\centering
\includegraphics[width=0.6\linewidth]{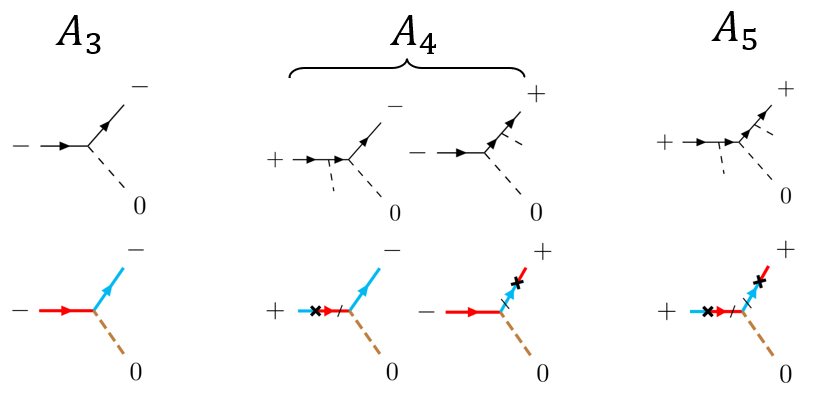}
\caption{The complete top-down matching for the $FFS$ amplitude.}
\label{fig:FFS_all}
\end{figure}

\begin{figure}[htbp]
\centering
\includegraphics[width=0.9\linewidth]{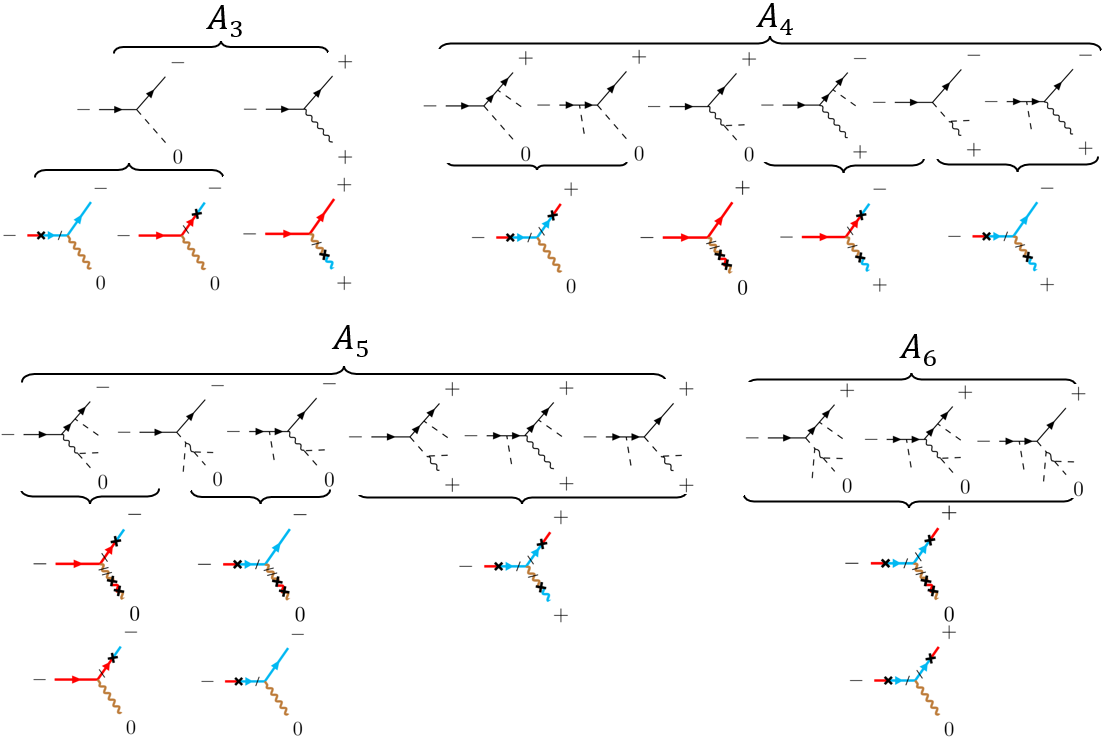}
\caption{The complete top-down matching for the $FFV$ amplitude.}
\label{fig:FFV_all}
\end{figure}

\begin{figure}[htbp]
\centering
\includegraphics[width=\linewidth]{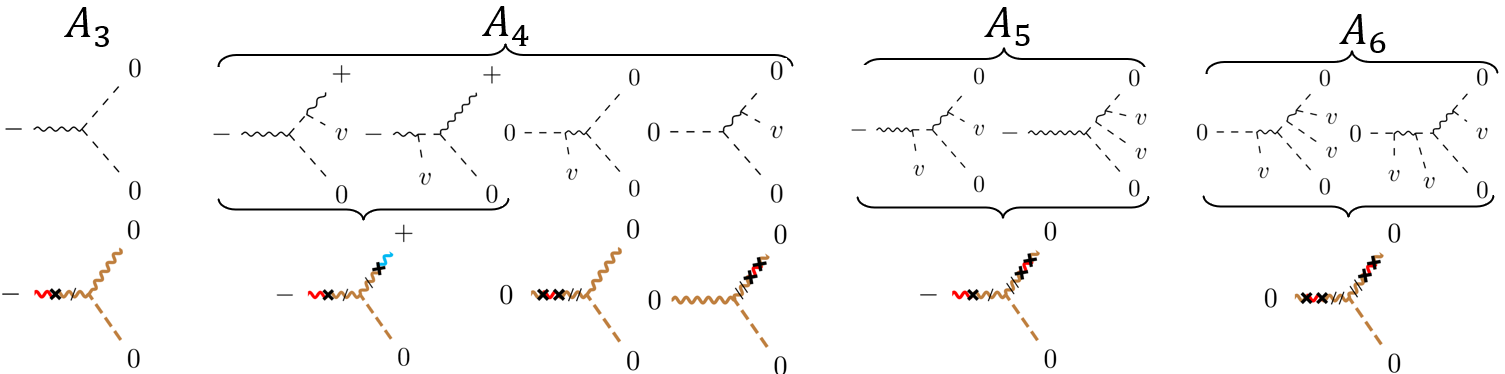}
\caption{The complete top-down matching for the massive $VVS$ amplitude with typical helicity category $(-1,0,0)$ and $(0,0,0)$ .}
\label{fig:VVS_all}
\end{figure}

\begin{figure}[htbp]
\centering
\includegraphics[width=\linewidth]{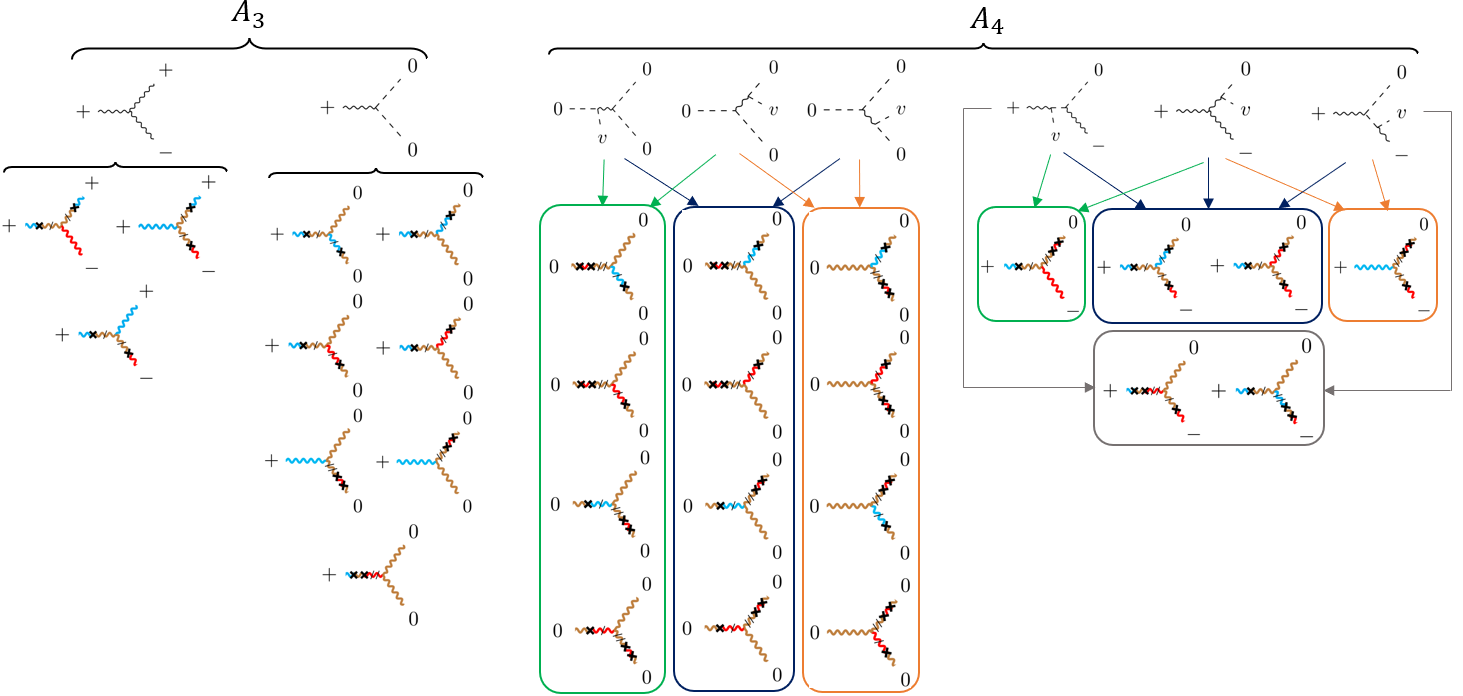}
\caption{The top-down matching from 3-pt and 4-pt massless amplitudes to the massive $VVV$ amplitude.}
\label{fig:VVV_all}
\end{figure}

For $s\ge 1$, the term $[\mathcal{M}]_{s-1}$ has the same scaling behavior as 3-pt massless amplitude $\mathcal{A}_3$. This gives the leading-order matching:
\begin{equation}
\mathcal{A}_3\to [\mathcal{M}]_{s-1}\sim E,
\end{equation}
In subleading matching, introducing factors of $v$ allows the massless amplitude with $n_h$ Higgs insertions to acquires the same scaling behavior as the higher-order MHC term, 
\begin{equation}
v^{n_h} \mathcal{A}_{3+n_h}\rightarrow [\mathcal{M}]_{s-1+n_h}\sim E\left(\frac{v}{E}\right)^{n_h}.
\end{equation}
Combining this with the chirality-helicity unification, we arrive at the {\it Higgs Insertion} introduced earlier, in which the $n$-pt massless amplitude matches as follows
\begin{equation} \label{eq:pc_select}
v^{n_h}\times\mathcal{A}(1^{h_1}, 2^{h_2}, 3^{h_3};\underbrace{4^0,\cdots,n^0}_{n_h\text{ Higgs insertion}})\rightarrow
[\mathcal{M}(\mathbf{1}^{h_1=t_1}, \mathbf{2}^{h_2=t_2}, \mathbf{3}^{h_3=t_3})]_{s-1+n_h},
\end{equation}
where particles $4,\cdots,n$ are all additional Higgs bosons. This means that each additional Higgs boson insertion should be accompanied by a VEV $v$. The power counting analysis establishes a selection rule that systematically match massless amplitudes to MHC amplitudes order-by-order, as shown in Fig.~\ref{fig:FFS_all}, \ref{fig:FFV_all}, \ref{fig:VVS_all} and \ref{fig:VVV_all}. A massless amplitude with $k=n-3$ Higgs insertions matches to the $k$-th order term in the MHC amplitude expansion. Then the {\it Higgs splitting} should be taken by arranging  momenta of the additional Higgs to be along the $\eta$ direction of one particle.

Given the leading-order matching results, we can use the ladder operators to obtain the sub-leading matching. For the primary state with $c=h=+\tfrac12$, acting with $\frac{m}{\mathbf m} J^-$ yields the corresponding descendant state with $h=-\tfrac12$,
\begin{equation} \begin{aligned}
\Ampone{1.5}{+}{\fer{cyan}{i1}{v1}}=\tilde\lambda_{\dot\alpha}
\quad&\xrightarrow{\frac{m}{\mathbf m} J^-}\quad
\Ampone{1.5}{-}{\ferflip{1.5}{0}{red}{cyan}}=\frac{m}{\mathbf m}\tilde\eta_{\dot\alpha} 
\end{aligned} \end{equation}
Further action of $\frac{m}{\mathbf m} J^-$ annihilates this state, consistent with the condition $|h|\le s$. For the primary state with $c=h=-\tfrac12$, we can act with $\frac{\tilde m}{\mathbf m} J^+$ and obtain a similar descendant state with $h=+\tfrac12$. This can be summarized in the Higgsing rules. For a massive fermion, the results of leading and subleading matching can be summarized as follows 
\begin{equation} 
\begin{tabular}{c|c|c}
\hline
& $h=t=-\frac12$ & $h=t=+\frac12$ \\
\hline
$c=-\frac12$ & $\lambda_\alpha$ & $\tilde m\eta_\alpha$  \\
$c=+\frac12$ & $m\tilde\eta_{\dot\alpha}$ & $\tilde\lambda_{\dot\alpha}$  \\
\hline
\end{tabular}
\end{equation}
where $c$ denotes the chirality of each state. The diagonal entries correspond to the primary MHC amplitude, while the off-diagonal entries correspond to the descendant MHC amplitude. The effect of Higgs insertion is equivalent to a horizontal transition in the table above, mapping primary to descendant amplitudes.

For vector states, we first consider the primary state with $c=h=+1$. Acting once and twice with the ladder operator gives the 1st and 2nd descendant states with $|c-h|=1$ and $2$, respectively:
\begin{align}
\Ampone{1.5}{+}{\bos{i1}{cyan}}=\tilde{\lambda}_{\dot{\alpha}_1}\tilde{\lambda}_{\dot{\alpha}_2}
\quad&\xrightarrow{\frac{m}{\mathbf m} J^-}\quad
\Ampone{1.5}{0}{\bosflip{1.5}{0}{cyan}{brown}}=2\frac{m}{\mathbf m}\tilde{\eta}_{(\dot{\alpha}_1}\tilde{\lambda}_{\dot{\alpha}_2)}\\
\Ampone{1.5}{+}{\bos{i1}{cyan}}=\tilde{\lambda}_{\dot{\alpha}_1}\tilde{\lambda}_{\dot{\alpha}_2}
\quad&\xrightarrow{(\frac{m}{\mathbf m} J^-)^2}\quad
\Ampone{1.5}{-}{\bosflipflip{1.5}{0}{cyan}{brown}{red}}=2\frac{m^2}{\mathbf m^2}\tilde{\eta}_{\dot{\alpha}_1}\tilde{\eta}_{\dot{\alpha}_2}
\end{align}
The primary state with $c=h=-1$ gives analogous results with opposite helicity and color. For the primary state with $c=h=0$, the situation is slightly different. Acting with ladder operators yields two distinct 1st descendant states with $|c-h|=1$:
\begin{equation} \begin{aligned}
\Ampone{1.5}{0}{\bos{i1}{brown}}=\lambda_{\alpha}\tilde{\lambda}_{\dot{\alpha}}
\quad&\xrightarrow{\frac{\tilde m}{\mathbf m} J^+}\quad
\Ampone{1.5}{+}{\bosflip{1.5}{0}{brown}{cyan}}=\frac{\tilde m}{\mathbf m}\eta_{\alpha}\tilde{\lambda}_{\dot{\alpha}}\\ 
\Ampone{1.5}{0}{\bos{i1}{brown}}=\lambda_{\alpha}\tilde{\lambda}_{\dot{\alpha}}
\quad&\xrightarrow{\frac{m}{\mathbf m} J^-}\quad
\Ampone{1.5}{-}{\bosflip{1.5}{0}{brown}{red}}=\frac{m}{\mathbf m} \lambda_{\alpha}\tilde{\eta}_{\dot{\alpha}}
\end{aligned} \end{equation}
In this case, the 2nd descendant state has the same helicity as the primary state. TTherefore, we act with both ladder operators $\frac{\tilde m}{\mathbf m} J^+$ and $\frac{m}{\mathbf m} J^-$ to obtain
\begin{equation} \begin{aligned}
\Ampone{1.5}{0}{\bos{i1}{brown}}=\lambda_{\alpha}\tilde{\lambda}_{\dot{\alpha}}
\quad&\xrightarrow{\frac{\tilde m}{\mathbf m} J^+ \frac{m}{\mathbf m} J^-}\quad
\Ampone{1.5}{0}{\bos{i1}{brown}}
+\Ampone{1.5}{0}{\bosflipflip{1.5}{0}{brown}{brown}{brown}}=
\lambda_{\alpha}\tilde{\lambda}_{\dot{\alpha}}+\frac{m\tilde m}{\mathbf m^2}\eta_{\alpha}\tilde{\eta}_{\dot{\alpha}}
\end{aligned} \end{equation}
The result contains both the primary and the 2nd descendant state. Acting the two ladder operators in the opposite order, $\frac{m}{\mathbf m} J^-\frac{\tilde m}{\mathbf m} J^+$, gives the same result.

When $h^{\text{1st}}>h^{\text{2nd}}$, we can act with $\frac{m}{\mathbf m} J^-$ on the 1st descendant state to obtain the 2nd one:
\begin{equation} \begin{aligned}
\Ampone{1.5}{+}{\bosflip{1.5}{0}{brown}{cyan}}=\frac{\tilde m}{\mathbf m}\eta_{\alpha}\tilde{\lambda}_{\dot{\alpha}}
\quad&\overset{\frac{m}{\mathbf m} J^-}{\to}\quad
\left\{\begin{aligned}
\Ampone{1.5}{0}{\bos{i1}{brown}}&=\lambda_{\alpha}\tilde{\lambda}_{\dot{\alpha}}\\
\Ampone{1.5}{0}{\bosflipflip{1.5}{0}{brown}{brown}{brown}}&=\frac{m\tilde m}{\mathbf m^2}\eta_{\alpha}\tilde{\eta}_{\dot{\alpha}}
\end{aligned}\right.\\
\Ampone{1.5}{0}{\bosflip{1.5}{0}{cyan}{brown}}=\frac{m}{\mathbf m}\tilde{\eta}_{(\dot{\alpha}_1}\tilde{\lambda}_{\dot{\alpha}_2)}
\quad&\overset{\frac{\tilde m}{\mathbf m} J^+}{\to}\quad
\Ampone{1.5}{-}{\bosflipflip{1.5}{0}{cyan}{brown}{red}}=\frac{m^2}{\mathbf m^2}\tilde{\eta}_{\dot{\alpha}_1}\tilde{\eta}_{\dot{\alpha}_2}
\end{aligned} \end{equation}
In the first line, the primary and 2nd descendant states for $c=0$ still appear together. Conversely, acting with $\frac{\tilde m}{\mathbf m} J^+$ on the 2nd descendant state yields the 1st one:
\begin{equation} \begin{aligned}
\Ampone{1.5}{0}{\bosflipflip{1.5}{0}{brown}{brown}{brown}}=\frac{m\tilde m}{\mathbf m^2}\eta_{\alpha}\tilde{\eta}_{\dot{\alpha}}
\quad&\overset{\frac{\tilde m}{\mathbf m} J^+}{\to}\quad
\Ampone{1.5}{+}{\bosflip{1.5}{0}{brown}{cyan}}=\frac{\tilde m}{\mathbf m}\eta_{\alpha}\tilde{\lambda}_{\dot{\alpha}}\\
\Ampone{1.5}{-}{\bosflipflip{1.5}{0}{cyan}{brown}{red}}=\frac{m^2}{\mathbf m^2}\tilde{\eta}_{\dot{\alpha}_1}\tilde{\eta}_{\dot{\alpha}_2}
\quad&\overset{\frac{\tilde m}{\mathbf m} J^+}{\to}\quad
\Ampone{1.5}{0}{\bosflip{1.5}{0}{cyan}{brown}}=2\frac{m}{\mathbf m}\tilde{\eta}_{(\dot{\alpha}_1}\tilde{\lambda}_{\dot{\alpha}_2)}
\end{aligned} \end{equation}
For the case $h^{\text{1st}}<h^{\text{2nd}}$, analogous results follow by exchanging the roles of $\frac{\tilde m}{\mathbf m} J^+$ and $\frac{m}{\mathbf m} J^-$. 
Finally similar to the fermion case, we could also summarize the above matching results from the Higgs insertion according to the particle states for the vector boson. Combining all these matching results, we obtain the Higgsing rules
\begin{equation}  \label{eq:rule_table_1}
\begin{tabular}{c|c|c|c}
\hline
& $h=t=-1$ & $h=t=0$ & $h=t=+1$ \\
\hline
$c=-1$ & $\lambda_{\alpha_1}\lambda_{\alpha_2}$ & $\tilde m\lambda_{(\alpha_1}\eta_{\alpha_2)}$ & $\tilde m^2\eta_{\alpha_1}\eta_{\alpha_2}$ \\
\hline
$c=0$ & $m\tilde\eta_{\dot\alpha}\lambda_{\alpha}$ & \makecell{$\tilde\lambda_{\dot\alpha}\lambda_{\alpha}$\\$m\tilde m\tilde\eta_{\dot\alpha}\eta_{\alpha}$} & $\tilde m\tilde\lambda_{\dot\alpha}\eta_{\alpha}$ \\
\hline
$c=+1$ & $m^2\tilde\eta_{\dot\alpha_1}\tilde{\eta}_{\dot\alpha_2}$ & $m\tilde\lambda_{(\dot\alpha_1}\tilde\eta_{\dot\alpha_2)}$ & $\tilde\lambda_{\dot\alpha_1}\tilde\lambda_{\dot\alpha_2}$ \\
\hline
\end{tabular}
\end{equation}
This table shows that the Higgs insertion correspond to horizontal movements among different particle states.

\paragraph{Sub-leading matching examples}

We can now use the Higgsing rules to obtain all subleading matching results from given leading-order matching results for all SM amplitudes. The subleading matching for $FFS$ and $FFV$ amplitudes has already been discussed in previous sections. Here, we focus on the three-boson amplitudes. As an example, we consider the $VVV$ amplitude and apply the Higgsing rules directly to obtain the subleading amplitude. The $VVS$ amplitude can be derived in a similar way.

At leading order, both the massless $VSS$ and $VVV$ amplitudes can match onto the massive $VVV$ amplitude. We first consider the massless $VSS$ amplitude with helicity $(-1,0,0)$. Starting from a chosen leading-order MHC amplitude, the subleading matching is obtained by splitting a Higgs boson from one of the external particles. Suppose we choose particle 1 with $c=0$ and $h=+1$; this particle splits into a Higgs boson and a scalar boson, both with helicity $0$. Using the Higgsing rules, we replace the state of particle 1 in the leading amplitude with the subleading state having $c=h=0$, as given in eq.~\eqref{eq:rule_table_1}. The result is the corresponding 3nd descendant $VVV$ ampltiude:
\begin{equation} 
\begin{aligned}
\begin{tikzpicture}[baseline=0.7cm] \begin{feynhand}
\setlength{\feynhandarrowsize}{3.5pt}
\vertex [particle] (i1) at (0,0.8) {$1^-$}; 
\vertex [particle] (i2) at (1.6,1.6) {$2^0$}; 
\vertex [particle] (i3) at (1.6,0) {$3^0$};  
\vertex (v1) at (0.9,0.8); 
\graph{(i1)--[bos](v1)--[sca](i2)};
\graph{(i3)--[sca] (v1)};
\node (C) at (0.9,-0.8) {$\displaystyle\frac{\langle12\rangle\langle31\rangle}{\langle23\rangle}$};
\end{feynhand} \end{tikzpicture} 
&\to 
\Ampthree{1^-}{2^0}{3^0}{\bosflip{1}{180}{brown}{red}}{\bos{i2}{brown}}{\bosflip{1}{-55}{cyan}{brown};\node (C) at (-0.2,-1.6) {$\displaystyle\frac{\langle12\rangle[2\eta_3][3\eta_1 ]}{\tilde m_1\tilde m_3}$}}
\xrightarrow{\text{insertion}}
\begin{tikzpicture}[baseline=-0.1cm] \begin{feynhand}
\setlength{\feynhandarrowsize}{3.5pt}
\vertex [particle] (i1) at (-1.1,0) {$1^0$}; 
\vertex [particle] (i2) at (0.579,0.827) {$2^0$}; 
\vertex [particle] (i3) at (0.579,-0.827) {$3^0$};  
\vertex [particle] (i5) at (-0.3,-0.6) {$4$};
\vertex (v3) at (-0.4,0);
\vertex (v1) at (0,0); 
\graph{(v3)--[sca](i5)};
\graph{(v3)--[sca](i1)};
\begin{scope}[rotate=180]
\clip (0,-0.1) rectangle (0.2,0.1); 
\draw[brown,thick bos] (0,0)--(1,0);
\end{scope}
\begin{scope}[rotate=180]
\clip (0.2,-0.1) rectangle (0.4,0.1); 
\draw[red,thick bos] (0,0)--(1,0);
\end{scope}
\draw[very thick,rotate=180] plot[mark=x,mark size=2.5] coordinates {(0.2,0)};
\draw[rotate=180] (0.05,-0.08) -- (0.11,+0.08);
\bos{i2}{brown};
\bosflip{1}{-55}{cyan}{brown};
\node (C) at (-0.2,-1.6) {};
\end{feynhand} \end{tikzpicture}
\rightarrow
\Ampthree{1^0}{2^0}{3^0}{\bosflipflip{1}{180}{brown}{red}{brown}}{\bos{i2}{brown}}{\bosflip{1}{-55}{cyan}{brown};\node (C) at (-0.2,-1.6) {$\displaystyle\frac{\langle\eta_1 2\rangle[2\eta_3][3\eta_1 ]}{\tilde m_1 m_1\tilde m_3}$};},
\end{aligned} \end{equation}
Similarly, we can choose to split a Higgs boson from other particles, which yields distinct subleading $VVV$ amplitudes.

Note that a massless $VSS$ amplitude can match more than one leading-order amplitude simultaneously. To obtain the subleading matching in such cases, we must split the Higgs boson from both diagrams, as illustrated below:
\begin{equation} \begin{aligned}
\begin{tikzpicture}[baseline=0.7cm] \begin{feynhand}
\setlength{\feynhandarrowsize}{3.5pt}
\vertex [particle] (i1) at (0,0.8) {$1^-$}; 
\vertex [particle] (i2) at (1.6,1.6) {$2^0$}; 
\vertex [particle] (i3) at (1.6,0) {$3^0$};  
\vertex (v1) at (0.9,0.8); 
\graph{(i1)--[bos](v1)--[sca](i2)};
\graph{(i3)--[sca] (v1)};
\end{feynhand} \end{tikzpicture} 
&\to 
\makecell{\Ampthree{1^-}{2^0}{3^0}{\bos{i1}{red}}{\bosflipflip{1}{55}{brown}{red}{brown}}{\bos{i3}{brown}}\\
\Ampthree{1^-}{2^0}{3^0}{\bos{i1}{red}}{\bos{i2}{brown}}{\bosflipflip{1}{-55}{brown}{red}{brown}}}
\xrightarrow{\text{insertion}}
\makecell{
\begin{tikzpicture}[baseline=-0.1cm] \begin{feynhand}
\setlength{\feynhandarrowsize}{3.5pt}
\vertex [particle] (i1) at (-1.1,0) {$1^0$}; 
\vertex [particle] (i2) at (0.579,0.827) {$2^0$}; 
\vertex [particle] (i3) at (0.579,-0.827) {$3^0$};  
\vertex [particle] (i5) at (-0.3,-0.6) {$4$};
\vertex (v3) at (-0.4,0);
\vertex (v1) at (0,0); 
\graph{(v3)--[sca](i5)};
\graph{(v3)--[sca](i1)};
\draw[red,thick bos] (0,0)--(-0.4,0);
\bosflipflip{1}{55}{brown}{red}{brown};
\bos{i3}{brown};
\end{feynhand} \end{tikzpicture}\\
\begin{tikzpicture}[baseline=-0.1cm] \begin{feynhand}
\setlength{\feynhandarrowsize}{3.5pt}
\vertex [particle] (i1) at (-1.1,0) {$1^0$}; 
\vertex [particle] (i2) at (0.579,0.827) {$2^0$}; 
\vertex [particle] (i3) at (0.579,-0.827) {$3^0$};  
\vertex [particle] (i5) at (-0.3,-0.6) {$4$};
\vertex (v3) at (-0.4,0);
\vertex (v1) at (0,0); 
\graph{(v3)--[sca](i5)};
\graph{(v3)--[sca](i1)};
\draw[red,thick bos] (0,0)--(-0.4,0);
\bos{i2}{brown};
\bosflipflip{1}{-55}{brown}{red}{brown};
\end{feynhand} \end{tikzpicture}}
\rightarrow
\makecell{\Ampthree{1^0}{2^0}{3^0}{\bosflip{1}{180}{red}{brown}}{\bosflipflip{1}{55}{brown}{red}{brown}}{\bos{i3}{brown}}\\
\Ampthree{1^0}{2^0}{3^0}{\bosflip{1}{180}{red}{brown}}{\bos{i2}{brown}}{\bosflipflip{1}{-55}{brown}{red}{brown}}},
\end{aligned} \end{equation}

Then we can consider the massless $VVV$ amplitude. Choosing one of the gauge bosons to split off a Higgs boson and applying the Higgsing rule also gives the subleading result:
\begin{equation} \begin{aligned}
\begin{tikzpicture}[baseline=0.7cm] \begin{feynhand}
\setlength{\feynhandarrowsize}{3.5pt}
\vertex [particle] (i1) at (0,0.8) {$1^+$}; 
\vertex [particle] (i2) at (1.6,1.6) {$2^+$}; 
\vertex [particle] (i3) at (1.6,0) {$3^-$};  
\vertex (v1) at (0.9,0.8); 
\graph{(i1)--[bos](v1)--[bos](i2)};
\graph{(i3)--[bos] (v1)};
\end{feynhand} \end{tikzpicture} 
&\to 
\Ampthree{1^+}{2^+}{3^-}{\bosflip{1}{180}{brown}{cyan}}{\bosflip{1}{55}{brown}{cyan}}{\bos{i3}{red}}
\xrightarrow{\text{insertion}}
\begin{tikzpicture}[baseline=-0.1cm] \begin{feynhand}
\setlength{\feynhandarrowsize}{3.5pt}
\vertex [particle] (i1) at (-1.1,0) {$1^+$}; 
\vertex [particle] (i2) at (0.579,0.827) {$2^0$}; 
\vertex [particle] (i3) at (0.579,-0.827) {$3^-$};  
\vertex [particle] (i5) at (+0.8,0) {$4$};
\vertex (v3) at (0.579*0.4,0.827*0.4);
\vertex (v1) at (0,0); 
\graph{(v3)--[sca](i5)};
\graph{(v3)--[sca](i2)};
\bosflip{1}{180}{brown}{cyan};
\bos{i3}{red};
\begin{scope}[rotate=55]
\clip (0,-0.1) rectangle (0.2,0.1); 
\draw[brown,thick bos] (v1)--(v3);
\end{scope}
\begin{scope}[rotate=55]
\clip (0.2,-0.1) rectangle (0.4,0.1); 
\draw[cyan,thick bos] (v1)--(v3);
\end{scope}
\draw[very thick,rotate=55] plot[mark=x,mark size=2.5] coordinates {(0.2,0)};
\draw[rotate=55] (0.05,-0.08) -- (0.11,+0.08);
\end{feynhand} \end{tikzpicture}
\rightarrow
\Ampthree{1^+}{2^0}{3^-}{\bosflip{1}{180}{brown}{cyan}}{\bosflipflip{1}{55}{brown}{cyan}{brown}}{\bos{i3}{red}},
\end{aligned} \end{equation}
Similarly, by iteratively applying the Higgsing rules to all leading MHC amplitudes, we can obtain the complete subleading $VVV$ amplitude.

\paragraph{Sub-leading matching involving $SSSS$}

Among all massless 4-point amplitudes, the $SSSS$ amplitude is special because it contains a contact term, 
\begin{equation} \label{eq:SSSS_complete}
\mathcal A(1^0,2^0,3^0,4^0)
=T_s^2\frac{s_{12}-s_{14}}{s_{24}}
+T_s^2\frac{s_{12}-s_{24}}{s_{14}}
+\lambda,
\end{equation}
where $T_s$ is the gauge coupling and $\lambda$ corresponds to the contact $\lambda\phi^4$ interaction. Therefore, the subleading matching that involves the $SSSS$ amplitude may receive contributions from the $\lambda\phi^4$ term, which corresponds to a new type of Higgs insertion. In the following, we examine whether such contributions appear in the matching to 3-pt massive bosonic amplitudes: $SSS$, $VVS$ and $VVV$.

We first consider the massive $SSS$ amplitude. Giving a vev to one of the massless scalars, the $\lambda\phi^4$ term naturally matches onto an $SSS$ amplitude,
\begin{equation} \begin{aligned}
\begin{tikzpicture}[baseline=-0.1cm] \begin{feynhand}
\setlength{\feynhandarrowsize}{4pt}
\vertex [particle] (i1) at (-1.01,0) {$1^0$}; 
\vertex [particle] (i2) at (0.579,0.827) {$2^0$}; 
\vertex [particle] (i3) at (0.579,-0.827) {$3^0$}; 
\vertex [particle] (i4) at (1,0) {$v$}; 
\vertex (v1) at (0,0);
\graph{(i1)--[sca](v1)};
\graph{(i2)--[sca](v1)};
\graph{(i3)--[sca](v1)};
\graph{(i4)--[sca](v1)};
\end{feynhand} \end{tikzpicture}=\lambda v
\quad\to\quad
\Ampthree{1^0}{2^0}{3^0}{\sca{i1}}{\sca{i2}}{\sca{i3}}.
\end{aligned}\end{equation}
In this case, the gauge interactions proportional to $T_s^2$ do not contribute to the $SSS$ amplitude. Clearly, this type of Higgs insertion differs significantly from inserting the Higgs boson on an external line, and no special limit on $p_4$ is required.

Next, we consdier the $VVS$ amplitude, in which two types of Higgs insertions can appear simultaneously.  Starting from the massless amplitude in eq.~\eqref{eq:SSSS_complete}, taking the on-shell limits $p_4\to\eta_1$ or $p_4\to\eta_2$ on the first and second terms gives
\begin{align}
\begin{tikzpicture}[baseline=0.7cm] \begin{feynhand}
\setlength{\feynhandarrowsize}{3.5pt}
\vertex [particle] (i1) at (-0.2,0.8) {$1^0$}; 
\vertex [particle] (i2) at (1.6,1.6) {$2^0$}; 
\vertex [particle] (i3) at (1.6,0) {$3^0$};  
\vertex [particle] (i5) at (0.7,0.2) {$v$};
\vertex (v3) at (0.6,0.8);
\vertex (v1) at (0.9,0.8); 
\graph{(i1)--[sca](v3)--[bos](v1)};
\graph{(v1)--[sca](i2)};
\graph{(i3)--[sca] (v1)};  
\graph{(v3)--[sca](i5)};
\end{feynhand} \end{tikzpicture}=g^2 v\frac{s_{12}-s_{24}}{s_{14}}
&\quad\to\quad 
\left\{\begin{aligned}
\Ampthree{1^0}{2^0}{3^0}{\bos{i1}{brown}}{\bos{i2}{brown}}{\sca{i3}}&=g^2 v\frac{\langle12\rangle[21]}{\mathbf m_1^2}, \label{eq:SSSS_match1} \\
\Ampthree{1^0}{2^0}{3^0}{\bosflipflip{1}{180}{brown}{red}{brown}}{\bos{i2}{brown}}{\sca{i3}}&=-g^2 v\frac{\langle\eta_12\rangle[2\eta_1]}{\mathbf m_1^2},
\end{aligned}\right. \\
\begin{tikzpicture}[baseline=0.7cm] \begin{feynhand}
\setlength{\feynhandarrowsize}{3.5pt}
\vertex [particle] (i1) at (-0.2,0.8) {$1^0$}; 
\vertex [particle] (i2) at (1.6,1.6) {$2^0$}; 
\vertex [particle] (i3) at (1.6,0) {$3^0$};
\vertex [particle] (i4) at (1.6,0.8) {$v$};  
\vertex (v2) at (0.9+0.7*0.33,0.8+0.8*0.33);
\vertex (v1) at (0.9,0.8); 
\graph{(i1)--[sca](v1)};
\graph{(v1)--[bos](v2)--[sca](i2)};
\graph{(i3)--[sca] (v1)};  
\graph{(v2)--[sca](i4)};
\end{feynhand} \end{tikzpicture}=g^2 v\frac{s_{12}-s_{14}}{s_{24}}
&\quad\to\quad
\left\{\begin{aligned}
\Ampthree{1^0}{2^0}{3^0}{\bos{i1}{brown}}{\bos{i2}{brown}}{\sca{i3}}&=g^2 v\frac{\langle12\rangle[21]}{\mathbf m_2^2},\\
\Ampthree{1^0}{2^0}{3^0}{\bos{i1}{brown}}{\bosflipflip{1}{55}{brown}{red}{brown}}{\sca{i3}}&=-g^2 v\frac{\langle1\eta_2\rangle[\eta_21]}{\mathbf m_2^2}.
\end{aligned}\right. \label{eq:SSSS_match2}
\end{align}
These yield two different 2nd descendant $VVS$ amplitudes and the same priamry $VVS$ amplitudes, with correct coefficients. The contact $\lambda\phi^4$ term alone does not contribute to massive $VVS$ amplitude.

For a contact-term matching to a massive $VVS$ amplitude, the $\lambda\phi^4$ interaction must mix with gauge interactions. To expose this mixing, we rewrite the massless $SSSS$ amplitude as
\begin{equation}
\mathcal A(1^0,2^0,3^0,4^0)
=-T_s^2\frac{2s_{14}}{s_{24}}
-T_s^2\frac{2s_{24}}{s_{14}}
+(\lambda-2T_s^2).
\end{equation}
In this form, $(\lambda-2T_s^2)$ is identified as the coefficient of the contact term. This new contact term can also contribute to the $VVS$ amplitude when the UV couplings and IR masses satisfy
\begin{equation}
\frac{\lambda}{\mathbf m_S^2}=\frac{T_s^2}{\mathbf m_V^2},
\end{equation}
where $\mathbf m_S$ and $\mathbf m_V$ are the scalar and vector masses. Under this condition, the new contact term $(\lambda-2g)$ with a vev $v$ can be rewritten as
\begin{equation} \begin{aligned}
\begin{tikzpicture}[baseline=-0.1cm] \begin{feynhand}
\setlength{\feynhandarrowsize}{4pt}
\vertex [particle] (i1) at (-1.01,0) {$1^0$}; 
\vertex [particle] (i2) at (0.579,0.827) {$2^0$}; 
\vertex [particle] (i3) at (0.579,-0.827) {$3^0$}; 
\vertex [particle] (i4) at (1,0) {$v$}; 
\vertex (v1) at (0,0);
\graph{(i1)--[sca](v1)};
\graph{(i2)--[sca](v1)};
\graph{(i3)--[sca](v1)};
\graph{(i4)--[sca](v1)};
\end{feynhand} \end{tikzpicture}
&=\lambda v-2T_s^2 v=\frac{T_s^2 v}{\mathbf m_V^2}(\mathbf m_S^2-2 \mathbf m_V^2)=\frac{T_s^2 v}{\mathbf m_V^2}(\mathbf m_3^2-\mathbf m_1^2-\mathbf m_1^2)\\
&=\frac{T_s^2 v}{\mathbf m_V^2}(\langle\eta_12\rangle[2\eta_1]+\langle1\eta_2\rangle[\eta_21])
\end{aligned}\end{equation}
Therefore, in this case the contact amplitude can also match to descendant $VVS$ amplitudes
\begin{equation} \begin{aligned}
\begin{tikzpicture}[baseline=-0.1cm] \begin{feynhand}
\setlength{\feynhandarrowsize}{4pt}
\vertex [particle] (i1) at (-1.01,0) {$1^0$}; 
\vertex [particle] (i2) at (0.579,0.827) {$2^0$}; 
\vertex [particle] (i3) at (0.579,-0.827) {$3^0$}; 
\vertex [particle] (i4) at (1,0) {$v$}; 
\vertex (v1) at (0,0);
\graph{(i1)--[sca](v1)};
\graph{(i2)--[sca](v1)};
\graph{(i3)--[sca](v1)};
\graph{(i4)--[sca](v1)};
\end{feynhand} \end{tikzpicture}
\quad\to\quad
\Ampthree{1^0}{2^0}{3^0}{\bosflipflip{1}{180}{brown}{red}{brown}}{\bos{i2}{brown}}{\sca{i3}}+
\Ampthree{1^0}{2^0}{3^0}{\bos{i1}{brown}}{\bosflipflip{1}{55}{brown}{red}{brown}}{\sca{i3}}=\frac{T_s^2 v}{\mathbf m_V^2}(\langle\eta_12\rangle[2\eta_1]+\langle1\eta_2\rangle[\eta_21]).
\end{aligned}\end{equation}
The two MHC diagrams must appear together. Note that the coefficient here differs from those in eqs.~\eqref{eq:SSSS_match1} and \eqref{eq:SSSS_match2}, because we have not yet included the contributions from the factorized terms,
\begin{equation} \begin{aligned}
\begin{tikzpicture}[baseline=0.7cm] \begin{feynhand}
\setlength{\feynhandarrowsize}{3.5pt}
\vertex [particle] (i1) at (-0.2,0.8) {$1^0$}; 
\vertex [particle] (i2) at (1.6,1.6) {$2^0$}; 
\vertex [particle] (i3) at (1.6,0) {$3^0$};  
\vertex [particle] (i5) at (0.7,0.2) {$v$};
\vertex (v3) at (0.6,0.8);
\vertex (v1) at (0.9,0.8); 
\graph{(i1)--[sca](v3)--[bos](v1)};
\graph{(v1)--[sca](i2)};
\graph{(i3)--[sca] (v1)};  
\graph{(v3)--[sca](i5)};
\end{feynhand} \end{tikzpicture}=-T_s^2 v\frac{2s_{24}}{s_{14}}
&\quad\to\quad 
\Ampthree{1^0}{2^0}{3^0}{\bosflipflip{1}{180}{brown}{red}{brown}}{\bos{i2}{brown}}{\sca{i3}}
=-2T_s^2 v\frac{\langle\eta_12\rangle[2\eta_1]}{\mathbf m_1^2}\\
\begin{tikzpicture}[baseline=0.7cm] \begin{feynhand}
\setlength{\feynhandarrowsize}{3.5pt}
\vertex [particle] (i1) at (-0.2,0.8) {$1^0$}; 
\vertex [particle] (i2) at (1.6,1.6) {$2^0$}; 
\vertex [particle] (i3) at (1.6,0) {$3^0$};
\vertex [particle] (i4) at (1.6,0.8) {$v$};  
\vertex (v2) at (0.9+0.7*0.33,0.8+0.8*0.33);
\vertex (v1) at (0.9,0.8); 
\graph{(i1)--[sca](v1)};
\graph{(v1)--[bos](v2)--[sca](i2)};
\graph{(i3)--[sca] (v1)};  
\graph{(v2)--[sca](i4)};
\end{feynhand} \end{tikzpicture}=-T_s^2 v\frac{2s_{14}}{s_{24}}
&\quad\to\quad
\Ampthree{1^0}{2^0}{3^0}{\bos{i1}{brown}}{\bosflipflip{1}{55}{brown}{red}{brown}}{\sca{i3}}
=-2T_s^2 v\frac{\langle1\eta_2\rangle[\eta_21]}{\mathbf m_2^2}\\
\end{aligned}\end{equation}
In this alternative form, the $SSSS$ amplitude does not match onto the primary $VVS$ amplitude.

Finally, for the massive $VVV$ amplitude, matching from a massless $\phi^4$ interaction is forbidden by the gauge structure of the Standard Model. Only gauge interaction contributes to the $VVV$ amplitude. Therefore, the sub-leading matching involving the $SSSS$ amplitude can be summarized as follows:
\begin{equation}
\begin{tabular}{c|c|c}
\hline
massive amplitude & gauge interaction & $\lambda\phi^4$ interaction \\
\hline
SSS & \mbox{-} & \checkmark \\
\hline
VVS & \checkmark & \checkmark \\
\hline
VVV & \checkmark & \mbox{-} \\
\hline
\end{tabular}
\end{equation}
where \mbox{-} indicates that the matching is forbidden by the gauge structure.

\section{Leading Matching for Standard Model 3-point  Amplitudes}

\subsection{Group structure and Higgs mechanism}

Until now, we classify the massless and massive particles according to their spins. Note that a massless scalar can match to either massive Higgs boson or massive vector boson. In a give spacetime representation, there may exist several particle carrying same helicity and spin. We should distinguish them, and find which one should be match to the massive vector boson. 

For massless particles, the group structure provides a method to organize the particles with same spins. In the following, we will adopt some of the notation from Ref.~\cite{Bachu:2023fjn}. We use the indices $I$, $\hat i$ and $i$ to label gauge boson, fermion and scalar boson. In the Standard Model, these particles are the reducible representation of the gauge group $SU(2)_W\times U(1)_Y$. Among all particle with spin $s\le 1$, we distinguish helicity $\pm\frac12$ particles as left-handed and right-handed fermion. For gauge boson with helicity $\pm 1$, we do not use different notation to distinguish them. Therefore, the possible values of the indices are
\begin{eqnarray}
\begin{array}{c|c|c|c}
\hline
\text{unbroken phase} & \text{gauge boson} & \text{fermion} & \text{scalar boson} \\
\hline
\text{index} & I & \hat{i} & i \\
\hline
\text{upper index value} & W^I=\{W^1,W^2,W^3\};B & \{u_L,d_L\},\{\nu_L,e_L\};u_R,d_R,e_R & \{\phi^+,\phi^0\} \\
\hline
\text{lower index value} & \mbox{-} & \{\bar{u}_L,\bar{d}_L\},\{\bar{\nu}_L,\bar{e}_L\};\bar{u}_R,\bar{d}_R,\bar{e}_R & \{\phi^-,\phi^{0*}\} \\
\hline
\end{array} 
\end{eqnarray}
where $i$ represents the $SU(2)_W$ fundamental indices, denoting the first or second component of $H_i$; $\hat{i}$ stands for the reducible representation for fermion, including both $SU(2)_W$-doublets $\{u_L,d_L\}$, $\{\nu_L, e_L\}$ and $SU(2)_W$-singlets $u_R,d_R,e_R$. The upper index and lower index represent particle and anti-particle separately. For simplicity, here we only consider one generation of fermions.

Thus, the coefficient $G$ of the massless amplitude should be a covariant tensor under $SU(2)_W\times U(1)$. For 3-pt amplitude, coefficient $G$ should have three indices. There are four types of 3-pt massless amplitude, and we can define their coefficient to be
\begin{align} 
FFS&:\quad 7\times7\times2\text{ tensors} \quad \mathbb{Y}_{\hat{i}_2}^{\hat{i}_1 i_3}, \mathbb{Y}_{\hat{i}_2 i_3}^{\hat{i}_1}, \\
FFV&:\quad 7\times7\times4\text{ tensor} \quad (\mathbb{T}_f^{I_3})^{\hat{i}_1}_{\hat{i}_2}, \\
VSS&:\quad 4\times2\times2\text{ tensor} \quad (T_s^{I_1})^{i_2}_{i_3}, \\
VVV&:\quad 4\times4\times4\text{ tensor} \quad f^{I_1 I_2 I_3}.
\end{align}
They are tensors of order 3, so it is not convenient to written their components directly. Here we present them as order-2 matrices by fixing one of the indices. 

1. For $FFS$ amplitude, we fix scalar index $i_3$, so the coefficient can be represented by a $7\times 7$ matrix. According to the chirality of the fermion, these matrices are divided into four blocks. The non-zero component only exist in two off-diagonal blocks, which means that the Yukawa interaction only relate fermion with opposite chirality. 
\begin{align} \label{eq:off_diagonal}
\mathbb{Y}=\left(\begin{array}{c|c}
 & Y \\ \hline
\tilde{Y} & \\
\end{array}\right).
\end{align}
Note that, for massless particle, the chirality is equal to helicity, so the two off-diagonal blocks $Y$ and $\tilde{Y}$ represents the coefficients of $\langle12\rangle$ and $[12]$, respectively. For $\mathbb{Y}_{\hat{i}_2}^{\hat{i}_1 i_3}$, the two off-diagonal blocks are
\begin{align}
Y_{\hat{i}_2}^{\hat{i}_1 i_3} &=-
\begin{pmatrix}
\mathcal{Y}^{(u)} \delta_{\phi^0}^{i_3} & 0 & 0 \\
\mathcal{Y}^{(u)} \delta_{\phi^+}^{i_3} & 0 & 0 \\
0 & 0 & 0 \\
0 & 0 & 0 \\
\end{pmatrix},& &\hat{i}_1=u_L,d_L,\nu_L,e_L\quad \hat{i}_2=\bar u_R,\bar d_R,\bar e_R, \\
\tilde{Y}_{\hat{i}_2}^{\hat{i}_1 i_3} &=-
\begin{pmatrix}
0 & 0 & 0 & 0 \\
\mathcal{Y}^{(d)} \delta^{i_3}_{\phi^+} & \mathcal{Y}^{(d)} \delta^{i_3}_{\phi^0} & 0 & 0 \\
0 & 0 & \mathcal{Y}^{(e)} \delta^{i_3}_{\phi^+} & \mathcal{Y}^{(e)} \delta^{i_3}_{\phi^0} \\
\end{pmatrix},& &\hat{i}_1=u_R,d_R,e_R,\quad \hat{i}_2=\bar u_L,\bar d_L,\bar \nu_L,\bar e_L, 
\end{align}
where $i_3=\phi^+,\phi^0$ and $\mathcal{Y}^{(f)}$ denotes the Yukawa coupling for fermion $f$. There is no right-handed neutrino in the Standard Model, so we do not have a coupling like $\mathcal{Y}^{(\nu)}$ and the above blocks are not square matrices. Similarly, the two off-diagonal blocks of $\mathbb{Y}_{\hat{i}_2 i_3}^{\hat{i}_1}$ are
\begin{align}
Y_{\hat{i}_2 i_3}^{\hat{i}_1} &=-
\begin{pmatrix}
 0 & \mathcal{Y}^{(d)} \delta_{i_3}^{\phi^-} & 0 \\
 0 & \mathcal{Y}^{(d)} \delta_{i_3}^{\phi^0} & 0 \\
 0 & 0 & \mathcal{Y}^{(e)} \delta_{i_3}^{\phi^-} \\
 0 & 0 & \mathcal{Y}^{(e)} \delta_{i_3}^{\phi^0} \\ 
\end{pmatrix},&
\tilde{Y}_{\hat{i}_2 i_3}^{\hat{i}_1} &=-
\begin{pmatrix}
\mathcal{Y}^{(u)} \delta^{\phi^0}_{i_3} & \mathcal{Y}^{(u)} \delta^{\phi^-}_{i_3} & 0 & 0  \\
0 & 0 & 0 & 0  \\
0 & 0 & 0 & 0  \\
\end{pmatrix}.&
\end{align}

2. For $FFV$ amplitude, we fix gauge boson index $I$, so the coefficient is also a $7\times 7$ matrix, which can be also divided into four blocks. since the gauge interaction only relate the fermion with same chirality, the non-zero component only exist in two digaonal blocks,
\begin{align}
\mathbb{T}_f=\left(\begin{array}{c|c}
T_f &  \\ \hline
 & \tilde{T}_f \\
\end{array}\right).
\end{align}
where $T_f$ represents the coefficients of $\frac{\langle23\rangle^2}{\langle12\rangle},\frac{[13]^2}{[12]}$, and $\tilde{T}_f$ represents the ones of $\frac{\langle13\rangle^2}{\langle12\rangle},\frac{[23]^2}{[12]}$.
Since we can fix the gauge boson to be either $SU(2)_W$ gauge boson $W^I$ or $U(1)_Y$ gauge boson $B$, we can write them separately. For $W^I$ boson, the two diagonal blocks are 
\begin{equation} \begin{aligned}
(T_f^{W^I})^{\hat{i}_1}_{\hat{i}_2} &= -\frac{g}{\sqrt{2}} 
\begin{pmatrix}
\sigma^I & \\
 & \sigma^I \\
\end{pmatrix},& 
\hat{i}_1&=u_L,d_L,\nu_L,e_L,\quad \hat{i}_2=\bar u_L,\bar d_L,\bar \nu_L,\bar e_L, \\
(\tilde{T}_f^{W^I})^{\hat{i}_1}_{\hat{i}_2} &=
\begin{pmatrix}
 0 & & \\
 & 0 & \\
 & & 0 \\
\end{pmatrix},&
\hat{i}_1&=u_R,d_R,e_R,\quad \hat{i}_2=\bar u_R,\bar d_R,\bar e_R,  \\
\end{aligned} \end{equation}
where $g$ is the $SU(2)_W$ coupling constant, and $\sigma^I$ are the Pauli matrices acting as $SU(2)W$ generators. For $B$ boson, we have
\begin{equation} \begin{aligned}
(T_f^{B})^{\hat{i}_1}_{\hat{i}_2} =-{\sqrt{2}} {g'}
\begin{pmatrix}
Y_{u_L} & & & \\
 & Y_{d_L} & & \\
 & & Y_{\nu_L} & \\
 &  &  & Y_{e_L} \\ 
\end{pmatrix},\quad
(\tilde{T}_f^{B})^{\hat{i}_1}_{\hat{i}_2} &=-{\sqrt{2}} {g'}
\begin{pmatrix}
 Y_{u_R} & & \\
 & Y_{d_R} & \\
 & & Y_{e_R} \\
\end{pmatrix},
\end{aligned} \end{equation}
where $g^{\prime}$ is $U(1)_Y$ coupling constant. In the Standard Model, the hypercharge $Y$ of $U(1)_Y$ group is set by
\begin{equation} \begin{aligned}
Y_{u_L}&=Y_{d_L}=\frac16,& Y_{u_R}&=\frac23,& Y_{d_R}&=-\frac13, \\
Y_{e_L}&=Y_{\nu_L}=-\frac12,& Y_{e_R}&=-1.& 
\end{aligned}
\end{equation}

3. Similarly, we can fix the gauge boson index of $VSS$ amplitude, the coefficient can be represented by a $2\times 2$ matrix.
\begin{align}
(T_s^{B})^{i_2}_{i_3} = -\sqrt{2} g'  
\begin{pmatrix}
Y_{\phi^+} & 0 \\
0 & Y_{\phi^0}
\end{pmatrix},\quad (T_s^{W^I})^{i_2}_{i_3} = -\frac{g}{\sqrt{2}} 
\begin{pmatrix}
\sigma^I     
\end{pmatrix},
\end{align}
where the hypercharge for scalar boson is $Y_{\phi^+}=Y_{\phi^0}=\frac12$.

4. For massless $VVV$ amplitude, the coefficient is nothing but the Lie algebra of gauge group. In this case, we can fix all three indices and the coefficient become
\begin{align}
f^{W^I W^J W^K} = -i\sqrt{2} g \epsilon^{IJK}.
\end{align}
where $\epsilon^{IJK}$ is the Levi-Civita tensor. The $B$ boson will not participate in the self interaction of gauge particles.

After spontaneously symmetry breaking, we use the indices $\mathbf{I}$, $\hat{\mathbf{i}}$ and $h$ to label vector boson, fermion and scalar boson. Their possible values are
\begin{eqnarray}
\begin{array}{c|c|c|c}
\hline
\text{broken phase} & \text{vector boson} & \text{fermion} & \text{scalar boson} \\
\hline
\text{index} & \mathbf{I} & \hat{\mathbf{i}} & {h} \\
\hline
\text{upper index value} & \{W^+,W^-,Z,A\} & \{u,d;\nu,e\} & \{h\} \\
\hline
\text{lower index value} & \mbox{-} & \{\bar{u},\bar{d};\bar{\nu},\bar{e}\} & \mbox{-} \\
\hline
\end{array}. 
\end{eqnarray}
Since there is only one type of massive scalar, i.e. massive Higgs boson, we use $h$ to represent both particle label and its value. The upper and lower index are still represent particle and anti-particle, except for vector boson.
Though we assign particle indices to massive amplitudes, they do not represent group structure since the broken gauge group is $U(1)_{em}$. 

To match massless particles to massive particles, we introduce a \textit{transformation matrix} to describe the particle conversion. Since the number of massive particle species is fewer than that of massless particles, this transformation matrix need not be square. The elements of the transformation matrix can be determined through symmetry breaking patterns and the counting of degrees of freedom (DOF). It should be noted that the DOF between massless and massive particles may differ, leading to a factor of
\begin{equation}
\sqrt{\frac{\text{massive DOF}}{\text{massless DOF}}},
\end{equation}
where the square root arises from the normalization condition applied to the single-particle state squared (rather than the state itself). We classify four fundamental types of transformation matrices between massless and massive states for scalars ($S$), fermions ($F$) and vectors ($V$):
\begin{itemize}
\item $S\to S$ transitions:

We use $2\times 1$ matrices $\mathcal{U}_{i}^h$ and $\mathcal{U}^{i h}$ to match massless scalar to massive Higgs boson. The non-zero components are
\begin{eqnarray} \label{eq:vevU}
\mathcal{U}^{\phi^{0*}h} = \mathcal{U}^{h}_{\phi^{0}} = \frac{1}{\sqrt{2}}. 
\end{eqnarray}
Here, the complex scalars $\phi^0$ or $\phi^{0*}$ possesses 2 DOF, while the real Higgs boson $h$ has 1 DOF. This leads to a factor of $\frac{1}{\sqrt{2}}$.

\item $S\rightarrow V$ transitions:
 
Massless scalars can transform into massive vectors via $2\times 4$ matrices $\mathcal{U}^{i \mathbf{I}}$ and $\mathcal{U}^{\mathbf{I}}_{i}$, with the non-zero components 
\eq{
\mathcal{U}^{\phi^- W^-} = -\mathcal{U}^{W^+}_{\phi^+} = \sqrt{2}, \quad -\mathcal{U}^{Z}_{\phi^{0}} = \mathcal{U}^{\phi^{0*} Z} = 1. 
}
The factor $\sqrt 2$ reflects the DOF difference between the complex scalar $\phi^{\pm}$ and the longitudinal modes of the massive $W^{\pm}$ boson.~\footnote{In the little group covariant formalism, a massive spin-1 object $\Phi$ is represented by a symmetric tensor $\Phi^{I_1 I_2}$ ($I_1,I_2=1,2$), where $\Phi^{12}$ and $\Phi^{21}$ correspond to longitudinal modes, contributing 2 DOF.} For the neutral sector, matching the complex $\phi^0$ to the real $Z$ boson modifies the factor to $\sqrt{2} \times 1/\sqrt{2} = 1$.

\item $F\rightarrow F$ transitions: 

The conversion between massless and massive fermion is described by $7\times 4$ matrices $\Omega^{\hat{i}\hat{\mathbf{i}}}$ and $\Omega_{\hat{i}\hat{\mathbf{i}}}$. Since there is only one generation, we do not consider the mixing of fermion. For Dirac-type massive fermion, both left- and right-handed chiral components contribute, with the non-zero matrix elements given by:
\begin{eqnarray}
\Omega^{f}_{f_L}= \Omega^{f}_{f_R}= \Omega_{\bar{f}}^{\bar{f}_L} = \Omega_{\bar{f}}^{\bar{f}_R} = 1,\quad f=u,d,e.
\end{eqnarray}
Since both massless and massive fermion are complex particle, we do not have a factor due to DOF. For the neutrino (treated as massless here), only left-handed component contributes,
\begin{eqnarray}
\Omega_{\nu }^{\nu_L} = \Omega_{\bar{\nu}}^{\bar{\nu}_L}  = 1.
\end{eqnarray}

\item $V\rightarrow V$ transitions: 

The $4\times 4$ square matrix $O^{I \mathbf{I}}$ maps massless gauge bosons to massive vectors, incorporating gauge boson mixing. Under the symmetry breaking $SU(2)_W \times U(1)Y \to U(1){\text{em}}$, the non-trivial elements are
\begin{equation} \begin{aligned}
O^{W^1 W^+} &= O^{W^1 W^-} =\frac{1}{\sqrt{2}},&  O^{W^3 Z} &= O^{B A} = \cos\theta_W\equiv\frac{g}{\sqrt{g^2+g^{\prime2}}},& \\ 
O^{W^2 W^+} &= -O^{W^2 W^-} = \frac{i}{\sqrt{2}},& 
-O^{B Z} &= O^{W^3 A} = \sin\theta_W\equiv\frac{g^\prime}{\sqrt{g^2+g^{\prime2}}},& 
\end{aligned} \end{equation}
where $\theta_W$ is the Weinberg angle governing the $W^3$-$B$ mixing.

\end{itemize}

These four types of transformation matrices are summarized as follows
\begin{eqnarray}
\begin{tikzpicture}[baseline=-1.2cm]
 \node (unb) at (0,0) {unbroken};
 \node (b) at (0,-2.5) {broken};
 \node (us) at (2.5,0) {scalar $i$};
 \node (uv) at (6,0) {gauge boson $I$};
 \node (uf) at (9,0) {fermion $\hat{i}$};
 \node (bs) at (2.5,-2.5) {Higgs $h$};
 \node (bv) at (6,-2.5) {massive vector $\mathbf{I}$};
 \node (bf) at (9,-2.5) {fermion $\hat{\mathbf{i}}$};
 \draw [->] (unb) -- (b)
               node[midway] {};
 \draw [->] (us) -- (bs)
               node[midway] {$\mathcal{U}_{i}^{h}, \mathcal{U}^{i h}$};
 \draw [->] (us) -- (bv)
               node[midway] {$\mathcal{U}_{i}^{\mathbf{I}}, \mathcal{U}^{i \mathbf{I}}$};
 \draw [->] (uv) -- (bv)
               node[midway] {$O^{I \mathbf{I}}$};
 \draw [->] (uf) -- (bf)
               node[midway] {$\Omega_{\hat{i}}^{\mathbf{i}}, \Omega^{\hat{i}}_{\mathbf{i}}$};
\end{tikzpicture}
\end{eqnarray}
Note that for $S\to V$ and $S\to S$ cases, there are two kinds of transformation matrices, in which the massive particles are represented by upper index. This means that we should sum over the contribution from particle and anti-particle for massless scalar boson.

For the whole amplitude, we should multiply all transformation matrix for each particle and give the total transformation matrix $\mathcal{T}$. When there is no additional Higgs boson, we contracting it with the 3-pt massless amplitude $\mathcal{A}$ to obtain the 3-pt MHC amplitude in leading order, i.e $[\mathcal{M}]_{0}$. Thus, the complete leading matching is
\begin{equation}
[\mathcal{M}(\mathbf 1,\mathbf 2,\mathbf 3)]_{0}= \mathcal{A}(1,2,3)\times\mathcal{T}+(S\leftrightarrow \bar{S}),
\end{equation}
where $S$ and $\bar{S}$ represent massless scalar and anti-scalar boson. 


For example, when we match massless $VSS$ amplitude with helicity category $(+1,0,0)$ to the massive $VVS$ amplitude, the massless scalar boson $S$ can be either particle and antiparticle. So we should sum these two contribution and the complete matching should be 
\begin{equation} \begin{aligned}
\relax
[\mathcal{M}(\mathbf{1}^{\mathbf I_1},\mathbf{2}^{\mathbf I_1},\mathbf{3}^{h})]_0
&=\mathcal{A}(1^{I_1},2_{i_2},3^{i_3})\times O^{I_1 \mathbf I_1} \mathcal{U}^{i_2 \mathbf I_2} \mathcal{U}^{h}_{i_3}
+\mathcal{A}(1^{I_1},2^{i_2},3_{i_3})\times O^{I_1 \mathbf I_1} \mathcal{U}^{\mathbf I_2}_{i_2 } \mathcal{U}^{i_3 h}\\
&=
\frac{[1|p_2-p_3|\eta_1\rangle}{2m_1}\left((T_s^{I_1})_{i_2}^{i_3}\times O^{I_1 \mathbf I_1} \mathcal{U}^{i_2 \mathbf I_2} \mathcal{U}^{h}_{i_3}
+(T_s^{I_1})_{i_3}^{i_2}\times O^{I_1 \mathbf I_1} \mathcal{U}^{\mathbf I_2}_{i_2 } \mathcal{U}^{i_3 h}\right).
\end{aligned} \end{equation}
The factor in the result can be identified as the massive coefficients. In other cases, there may have a cancelation between two contributions. We can try to match this massless amplitude to the massive $VSS$ amplitude. We obtain
\begin{equation} \begin{aligned}
\relax
[\mathcal{M}(\mathbf{1}^{\mathbf I_1},\mathbf{2}^{h},\mathbf{3}^{h})]_0
&=\mathcal{A}(1^{I_1},2_{i_2},3^{i_3})\times O^{I_1 \mathbf I_1} \mathcal{U}^{i_2 h} \mathcal{U}^{h}_{i_3}
+\mathcal{A}(1^{I_1},2^{i_2},3_{i_3})\times O^{I_1 \mathbf I_1} \mathcal{U}^{h}_{i_2 } \mathcal{U}^{i_3 h}\\
&=
\frac{[1|p_2-p_3|\eta_1\rangle}{2m_1}\left((T_s^{I_1})_{i_2}^{i_3}\times O^{I_1 \mathbf I_1} \mathcal{U}^{i_2 h} \mathcal{U}^{h}_{i_3}
+(T_s^{I_1})_{i_3}^{i_2}\times O^{I_1 \mathbf I_1} \mathcal{U}^{h}_{i_2 } \mathcal{U}^{i_3 h}\right)=0.
\end{aligned} \end{equation}
Under the exchange $2\leftrightarrow 3$, the kinematic structure $[1|p_2-p_3|\eta_1\rangle$ is antisymmetric while the gauge structure is symmetric according to $\mathcal{U}_{h}^{i_2} \mathcal{U}^{i_3 h}=\mathcal{U}^{i_2 h} \mathcal{U}^{h}_{i_3}$. This leads to a vanish result, and it is  consistent with the fact that a vector boson cannot interact with two identical scalar bosons. In Standard Model, there is only one species of massive scalar boson, so there is of course no massive $VSS$ amplitude.

When additional Higgs bosons are present, the complete higher-order matching takes the form
\begin{equation} \begin{aligned} \label{eq:complete_match_3pt}
\frac{v^{n-3}}{(n-3)!}\mathcal{A}(1,2,3;4,\cdots,n)\times\mathcal{T}+(S\leftrightarrow \bar{S})\xRightarrow{\text{Higgs insertion}}[\mathcal{M}(\mathbf 1,\mathbf 2,\mathbf 3)]_{n-3}.
\end{aligned} \end{equation}
These additional Higgs bosons can be interpreted as the Higgs insertions into 3-pt massless amplitude. Exchanging the additional particle line does not generate new massless diagrams for Higgs insertion, resulting in a symmetry factor of $1/(n-3)!$. The Higgs splitting technique must also account for the group structure; the details are provided in Appendix~\ref{app:Higgs_split}.

It is important to note that the higher-point massless amplitude may contain terms without pole structures. Generally, the massless amplitude consists of two distinct contributions,
\begin{equation} \begin{aligned}
\mathcal{A}
=\mathcal{A}^{\text{f}}+\mathcal{A}^{\text{ct}}.
\end{aligned} \end{equation}
where the factorized term $\mathcal{A}^{\text{f}}$ contains pole structure, while the contact term $\mathcal{A}^{\text{ct}}$ does not. In the four-scalar amplitude, the contact term corresponds to the 4-pt $\lambda \phi^4$ interaction: 
\begin{equation} \begin{aligned}
\mathcal{A}^{\text{ct}}(1^{i_1},2_{i_2},3^{i_3},4_{i_4})=-4\lambda \delta^{(i_1}_{i_3} \delta^{i_2)}_{i_4},
\end{aligned} \end{equation}
whereas the factorized term representsthe gauge interaction,
\begin{equation} \begin{aligned}
\mathcal{A}^{\text{f}}(1^{i_1},2_{i_2},3^{i_3},4_{i_4})
=(T^{J}_s)^{i_1}_{i_2} (T^{J}_s)^{i_3}_{i_4}\frac{s_{13}-s_{14}}{2s_{12}}
+(T^{J}_s)^{i_1}_{i_4} (T^{J}_s)^{i_3}_{i_2}\frac{s_{13}-s_{12}}{2s_{14}}.
\end{aligned} \end{equation}
When matching to the 3-pt massive amplitude, particle 4 serves as the additional Higgs boson associated with an VEV, while particles 1,2 and 3 are converted to massive particle with specific spins. The factorized term will match to
\begin{equation} \begin{aligned}
\mathcal{A}^{\text{f}}\mathcal{T}+(S\leftrightarrow \bar{S})
\end{aligned} \end{equation}
Depending on whether this contribution vanishes, we distinguish two matching scenarios:
\begin{itemize}
\item When converting all scalar bosons to massive scalars $h$, the factorized term vanishes. In this case, matching requires only the contact term,
\begin{equation} \begin{aligned}
\mathcal{A}^{\text{ct}}\to \mathcal{M}
\end{aligned} \end{equation}
The $\lambda\phi^4$ interaction then matches to the massive $SSS$ amplitude. This matching does not require taking the limit $p_4\to\eta_i$, as the scalar boson has a unique particle state whose helicity and chirality remain unaffected by the additional Higgs boson.

\item When matching some massless particles to massive vectors $\mathbf I$, we employ the Higgs insertion technique for the factorized term while neglecting the contact term,
\begin{equation} \begin{aligned}
\mathcal{A}^{\text{f}}\to \mathcal{M}
\end{aligned} \end{equation}
This matching applies to massive $VVS$ and $VVV$ amplitudes.

\end{itemize}

In summary, the gauge structure and transformation matrix determine whether matching can occur between particles with given helicities and spins. When combined with the VEV, they define the physical masses of fermions and vector bosons. However, the Higgs mass itself is unrelated to the matching of 3-pt massive amplitudes. We will address in the following subsection.


\subsection{1-point and 2-point SM matching}


We now present the compelte matching results obtained through top-down matching from massless amplitudes. In this work, we focus specifically on the massive contact amplitudes. Our previous discussion established that 3-pt massless amplitudes suffice to determine all non-vanishing leading terms of 3-pt massive amplitudes. Higher-point massless amplitudes with pole structures contribute only to subleading terms, which can alternatively be derived by considering helicity and chirality flips.

Therefore, we conclude that massless contact amplitudes alone are sufficient to determine the massive contact amplitude. In this subsection, we systematically derive as many Standard Model massive contact amplitudes as possible. To achieve this, we generalize the matching equation (Eq.~\eqref{eq:complete_match_3pt}) to a more comprehensive  form that match $n$-point massless amplitudes to $N$-point massive amplitudes:
\begin{equation} \begin{aligned} \label{eq:AntoMn}
\frac{v^{n-N}}{(n-N)!}\mathcal{A}_n\times\mathcal{T}+(S\leftrightarrow \bar{S})\xRightarrow{\text{Higgs insertion}}[\mathcal{M}_N]_{n-N}.
\end{aligned} \end{equation}
Then we bold massless spinors to their massive counterparts  and fix the spurion masses $m,\tilde m$ to the physical mass $\mathbf m$. This gives the $SU(2)_{\text{LG}}$ covariant massive amplitudes,
\begin{equation} \begin{aligned}
\mathcal{M}_N\xRightarrow{\text{Bold}}\mathbf{M}_N.
\end{aligned} \end{equation}
We proceed by considering cases with $N=1,2,3,\cdots$ in ascending order.


We begin with the 1-pt massive amplitude. In quantum field theory, particles are treated as harmonic oscillators, so all 1-pt amplitudes must vanish after symmetry breaking. The top-down matching procedure should reproduce this result. For massless amplitudes with momentum dependence, the matching trivially yields zero due to momentum conservation ($p=0$). Only in the scalar case, where the massless amplitude has no momentum dependence, does an explicit calculation become necessary.

Multiplying the transformation matrix with the four-scalar massless amplitude, the result should includes contributions from scalar and anti-scalar exchange. We obtain
\begin{equation} \begin{aligned} \label{eq:four_scalar}
&\mathcal{A}(1^{i_1},2^{i_2},3_{i_3},4_{i_4}) \mathcal{U}_{i_1}^{h} \mathcal{U}_{i_2}^{h} \mathcal{U}^{i_3 h} \mathcal{U}^{i_4 h}+(S\leftrightarrow \bar{S})\\
=& \mathcal{A}(1_{i_1}, 2_{i_2}, 3^{i_3}, 4^{i_4}) \mathcal{U}_{i_1}^{h} \mathcal{U}_{i_2}^{h} \mathcal{U}^{i_3 h} \mathcal{U}^{i_4 h}
+\mathcal{A}(1_{i_1}, 2^{0 i_2}, 3_{i_3}, 4^{i_4}) \mathcal{U}_{i_1}^{h} \mathcal{U}^{i_2 h} \mathcal{U}_{i_3}^{h} \mathcal{U}^{i_4 h} \\
&+\mathcal{A}(1_{i_1}, 2^{i_2}, 3^{i_3}, 4_{i_4}) \mathcal{U}_{i_1}^{h}  \mathcal{U}^{i_2 h}\mathcal{U}^{i_3 h} \mathcal{U}_{i_4}^{h} 
+\mathcal{A}(1^{i_1}, 2_{i_2}, 3_{i_3}, 4^{i_4}) \mathcal{U}^{i_1 h} \mathcal{U}_{i_2}^{h} \mathcal{U}_{i_3}^{h} \mathcal{U}^{i_4 h} \\
&+\mathcal{A}(1^{i_1}, 2_{i_2}, 3^{i_3}, 4^{i_4}) \mathcal{U}^{i_1 h} \mathcal{U}_{i_2}^{h} \mathcal{U}^{i_3 h} \mathcal{U}_{i_4}^{h}
+\mathcal{A}(1^{i_1}, 2^{i_2}, 3_{i_3}, 4_{i_4}) \mathcal{U}^{i_1 h} \mathcal{U}^{i_2 h} \mathcal{U}_{i_3}^{h} \mathcal{U}_{i_4}^{h} \\
=& -6\lambda,
\end{aligned} \end{equation}
which is non-zero. To cancel this contribution, we introduce a 2-pt massless amplitude
\begin{equation} \begin{aligned}
\mathcal{A}(1^{i_1},2_{i_2})=\lambda v^2 \delta^{i_1}_{i_2},
\end{aligned} \end{equation}
This amplitude is matched to the following massive structure,
\begin{equation} \begin{aligned}
\mathcal{A}(1^{i_1},2_{i_2})\mathcal{U}_{i_1}^{h} \mathcal{U}^{i_2 h}+(S\leftrightarrow \bar{S})
&=\mathcal{A}(1^{i_1},2_{i_2})\mathcal{U}_{i_1}^{h} \mathcal{U}^{i_2 h}+\mathcal{A}(1_{i_1},2_{i_2})\mathcal{U}^{i_1 h}\mathcal{U}_{i_2}^{h} \\
&=\lambda v^2.
\end{aligned} \end{equation}
Using Eq.~\eqref{eq:AntoMn}, these two contributions cancel as follows
\begin{equation} \begin{aligned}
v \mathcal{A}(1^{i_1},2_{i_2})\mathcal{U}_{i_1}^{h} \mathcal{U}_{i_2}^{h} + \frac{v^3}{3!} \mathcal{A}(1^{i_1},2^{i_2},3_{i_3},4_{i_4}) \mathcal{U}_{i_1}^{h} \mathcal{U}_{i_2}^{h} \mathcal{U}^{i_3 h} \mathcal{U}^{i_4 h}+(S\leftrightarrow \bar{S})=0.
\end{aligned} \end{equation}
Diagrammatically, this cancellation can be represented as
\begin{equation}\left(
\begin{tikzpicture}[baseline=0.7cm]
\begin{feynhand}
\vertex [particle] (i1) at (-0.4,0.8) {$1^{0}$};
\vertex [particle] (i2) at (1.3,0.8) {$v$};
\graph{(i1) --[sca] (i2)};
\end{feynhand}
\end{tikzpicture}+
\begin{tikzpicture}[baseline=-0.1cm]
\begin{feynhand}
\vertex [particle] (i1) at (-0.85,0) {$1^0$};
\vertex [particle] (i2) at (0.85,0) {$v$};
\vertex [particle] (i3) at (0,0.6) {$v$};
\vertex [particle] (i4) at (0,-0.6) {$v$};
\vertex (v1) at (0,0);
\graph{(i1) --[sca] (v1)--[sca] (i2)};
\graph{(i3)--[sca](v1) --[sca] (i4)};
\end{feynhand}
\end{tikzpicture}
\right)\times\mathcal{T}+(S\leftrightarrow \bar{S}) \rightarrow 0
\end{equation}

Then we turn to the 2-pt massive amplitude. Since the amplitude is a Lorentz scalar, the two massive particles must have identical spin. We can pick the massless contact amplitudes with at least one scalar boson (i.e. massless $SS$, $FFS$, $VSS$ and $SSSS$ amplitudes), and match them to the 2-pt MHC structures. As shown in Table~\ref{tab:2pt_match}, the non-zero result correspond to massive $SS$, $FF$ and $VV$ amplitude. For the $FF$ and $VV$ cases, the contractions of gauge structures and transformation matrices are absorbed into the fermion and vector masses, as defined in Eqs.~\eqref{eq:fermion_mass} and \eqref{eq:vector_mass}. In analogy to the 3-pt massive amplitude involving a massive vector, the matching for massive $VV$ amplitude still generates the chirality flip.

\begin{table}[htbp]
\centering
\begin{tabular}{c|c|c}
\hline
massive particle & massless diagram & amplitude matching \\
\hline
SS & 
\makecell{\begin{tikzpicture}[baseline=0.7cm]
\begin{feynhand}
\vertex [particle] (i1) at (-0.4,0.8) {$1^{0}$};
\vertex [particle] (i2) at (1.3,0.8) {$2^{0}$};
\graph{(i1) --[sca] (i2)};
\end{feynhand}
\end{tikzpicture}\\+
\begin{tikzpicture}[baseline=-0.1cm]
\begin{feynhand}
\vertex [particle] (i1) at (-0.85,0) {$1^0$};
\vertex [particle] (i2) at (0.85,0) {$2^0$};
\vertex [particle] (i3) at (0,0.6) {$v$};
\vertex [particle] (i4) at (0,-0.6) {$v$};
\vertex (v1) at (0,0);
\graph{(i1) --[sca] (v1)--[sca] (i2)};
\graph{(i3)--[sca](v1) --[sca] (i4)};
\end{feynhand}
\end{tikzpicture}} & $\mathcal{A}_2\times\mathcal{T} + \frac{v^2}{2!} \mathcal{A}_4\times\mathcal{T}+(S\leftrightarrow \bar{S})
=-2\lambda v^2$ \\
\hline
\multirow{2}{*}{FF} & \begin{tikzpicture}[baseline=0.7cm] \begin{feynhand}
\setlength{\feynhandarrowsize}{3.5pt}
\vertex [particle] (i1) at (-0.85,0.8) {$1^-$}; 
\vertex [particle] (i2) at (0.85,0.8) {$2^-$}; 
\vertex [particle] (i3) at (0,1.4) {$v$};  
\vertex (v1) at (0,0.8); 
\graph{(i1)--[fer](v1)--[fer](i2)};
\graph{(i3)--[sca] (v1)};  
\end{feynhand} \end{tikzpicture} & $\lim_{p_3\rightarrow \eta_2} v\mathcal{A}_3\times\mathcal{T}+(S\leftrightarrow \bar{S})=\mathbf m_{\hat{\mathbf i}}\delta^{\hat{\mathbf i}}_{\hat{\mathbf j}}\langle12\rangle$ \\ 
\cline{2-3}
& \begin{tikzpicture}[baseline=0.7cm] \begin{feynhand}
\setlength{\feynhandarrowsize}{3.5pt}
\vertex [particle] (i1) at (-0.85,0.8) {$1^+$}; 
\vertex [particle] (i2) at (0.85,0.8) {$2^+$}; 
\vertex [particle] (i3) at (0,1.4) {$v$};  
\vertex (v1) at (0,0.8); 
\graph{(i1)--[fer](v1)--[fer](i2)};
\graph{(i3)--[sca] (v1)};  
\end{feynhand} \end{tikzpicture} & $\lim_{p_3\rightarrow \eta_2} v\mathcal{A}_3\times\mathcal{T}+(S\leftrightarrow \bar{S})=\mathbf m_{\hat{\mathbf i}}\delta^{\hat{\mathbf i}}_{\hat{\mathbf j}}[12]$ \\
\hline
VV & \begin{tikzpicture}[baseline=0.7cm] \begin{feynhand}
\setlength{\feynhandarrowsize}{3.5pt}
\vertex [particle] (i1) at (-0.85,0.8) {$1^-$}; 
\vertex [particle] (i2) at (0.85,0.8) {$2^0$}; 
\vertex [particle] (i3) at (0,1.4) {$v$};  
\vertex (v1) at (0,0.8); 
\graph{(i1)--[bos](v1)--[sca](i2)};
\graph{(i3)--[sca] (v1)};  
\end{feynhand} \end{tikzpicture} & $\lim_{p_3\rightarrow \eta_2} v\mathcal{A}_3 \times\mathcal{T}+(S\leftrightarrow \bar{S})=\mathbf m_{\mathbf I}\delta^{\mathbf I\mathbf I^*}\frac{\langle1|p_2-\eta_2|\eta_1]}{\tilde{m}_1}$ \\
\hline
\end{tabular}
\caption{The relationship among the 2-pt MHC amplitude and massless contact amplitudes. For $VV$ amplitude, we only list the matching in a typical helicity category $(-1,0,0)$.}
\label{tab:2pt_match}
\end{table}

Bolding these MHC amplitudes, we can obtain the 2-pt little-group covariant massive amplitude,
\begin{equation} \begin{aligned}
\mathbf M(\mathbf 1^0,\mathbf 2^0)&=-\mathbf{m}_h^2,\\
\mathbf M(\mathbf 1^{\frac12},\mathbf 2^{\frac12})&=\mathbf{m}_f(\langle\mathbf{12}\rangle+[\mathbf{12}]),\\
\mathbf M(\mathbf 1^1,\mathbf 2^1)&=[\mathbf{12}]\langle\mathbf{12}\rangle,\\
\end{aligned} \end{equation}
where the Higgs mass $\mathbf{m}_h = \sqrt{2\lambda} v$. These three 2-pt massive structures correspond to the mass term in the Lagrangian, i.e. $m_h\phi^2$, $m_f\bar\psi\psi$ and $m^2_V A^\mu A_\mu$.

\subsection{Complete matching for 3-point SM amplitudes}

We now analyze the 3-pt massive amplitudes. Although we have established the reduction from massless amplitudes to the 3-pt MHC amplitude, the massive coefficients remain undetermined. To fix these coefficients, we need the complete 3-pt and 4-pt massless amplitudes,
\begin{equation}
\mathcal{A}= \mathcal{G}\times \mathcal{K}(\tilde\lambda,\lambda),
\end{equation}
where $\mathcal{G}$ denotes gauge structure and $\mathcal{K}$ represents kinematic structure. Through massless deformation, the 3-pt massless amplitude decomposes into multiple MHC terms, since the massive amplitude can involve more than one type of current. Let $b_i$ parameterize the decomposition factors of the massless amplitude corresponding to the $i$-th term. Therefore, the MHC amplitude can be expressed as
\begin{equation}
\mathcal{M}= \sum_i b_i \mathcal{G}\mathcal{T}\times\mathcal{K}_i(\lambda,\tilde\lambda,\eta,\tilde\eta,m,\tilde{m})+(S\leftrightarrow\bar S),
\end{equation}
with the normalization condition $\sum_i b_i=1$. Finally, we restore $SU(2)_{\text{LG}}$ covariance to obtain the physical massive amplitude,
\begin{equation}
\mathbf{M}=\sum_i \mathbf G_i \times\mathbf{K}_i(\lambda^I,\tilde{\lambda}^I,\mathbf{m}).
\end{equation}
where the massive coefficient $\mathbf G_i$ is given by $\mathbf G_i=b_i \mathcal{G}\mathcal{T}+(S\leftrightarrow\bar S)$.

In this work, we adopt the following strategy to match the gauge structure:
\begin{itemize}
\item \textit{Maxiaml $n_V$ helicity class:} We first determine the massive form using the helicity class with maximal $n_V$. 
Thus, in the helicity categories of this class, the massive coefficient of the term $C_i\times H_i$ is given by
\begin{equation}
\mathbf{G}_i\equiv\frac{1}{\mathbf{n}} \mathcal{G}\mathcal{T}+(S\leftrightarrow\bar S).
\end{equation}
To obtain the full massive structure, we must in principle sum contributions from all helicity categories in this class. However, since multiple massless amplitudes could correspond to identical massive terms $\mathbf{K}_i$, it is sufficient to consider a representative subset of helicity categories to determine all massive structures.

\item \textit{Other helicity classes:} If multiple helicity classes exist, we also consider the matching for the helicity class with smaller $n_V$. In these classes, each helicity category can match to all massive terms. Since the form of $\mathbf{M}$ is already fixed by the maxiaml $n_V$ class, we derive the relation between coefficients,
\begin{equation}
b_i \mathcal{G}\mathcal{T}+(S\leftrightarrow\bar S)=\mathbf{G}_i\times \mathbf{F}_i.
\end{equation}
where $\mathbf{F}_i$ is a dimensionless function of the physical mass $\mathbf{m}$. This relation allow us to determine the decomposition factor $b_i$.

\end{itemize}

Using this strategy, we determine all massive coefficients for 3-pt amplitudes and establish the relation between massless and massive coefficient.

\textbf{1. SSS:} This is the simplest case. We briefly review the $\lambda\phi^4$ interaction
\begin{equation} \begin{aligned}
\mathcal{A}^{\text{ct}}(1^{0},2^{0},3^{0},4^{0})&=-4\lambda \delta^{(i_1}_{i_3} \delta^{i_2)}_{i_4},
\end{aligned} \end{equation}
where the superscript $0$ indicates the helicity of massless scalars. Then match it to the massive $SSS$ amplitude. The result is identical to Eq.~\eqref{eq:four_scalar}, multiplied by a VEV $v$,
\begin{equation} \begin{aligned}
\mathbf{M}({\mathbf{1}}^{0}, \mathbf{2}^{0}, \mathbf{3}^{0}) =-6\lambda v,
\end{aligned} \end{equation}
where the superscript $0$ now denotes the spin of massive scalars.

\textbf{2. FFS:} To match the massless $FFS$ amplitude to the massive case, we consider two massless amplitudes with distinct helicity categories,
\begin{equation} \begin{aligned}
\mathcal{A}(1^{-\frac12},2^{-\frac12},3^{0})&={Y}^{\hat{i}_1 i_3}_{\hat{i}_2}\langle12\rangle,\\
\mathcal{A}(1^{+\frac12},2^{+\frac12},3^{0})&=\tilde{Y}^{\hat{i}_1}_{\hat{i}_2 i_3}[12],
\end{aligned} \end{equation}
where $Y$ and $\tilde Y$ represent the massless couplings. When the transformation matrices $\Omega$ and $\mathcal{U}$ act on the massless couplings $Y$ and $\tilde{Y}$, they become the massive couplings $y$ and $y'$ as follows,
\begin{equation} \begin{aligned}
{Y}^{\hat{i}_1 i_3}_{\hat{i}_2}\times \Omega^{\hat{\mathbf{i}}_1}_{\hat{i}_1} \Omega_{\hat{\mathbf{i}}_2}^{\hat{i}_2} \mathcal{U}_{i_3}^h + {Y}^{\hat{i}_1}_{\hat{i}_2 i_3} \times \Omega^{\hat{\mathbf{i}}_1}_{\hat{i}_1} \Omega_{\hat{\mathbf{i}}_2}^{\hat{i}_2} \mathcal{U}^{i_3 h}
\quad&\Rightarrow\quad {y}^{\hat{\mathbf{i}}_1}_{\hat{\mathbf{i}}_2},\\
\tilde{Y}^{\hat{i}_1 i_3}_{\hat{i}_2}\times \Omega^{\hat{\mathbf{i}}_1}_{\hat{i}_1} \Omega_{\hat{\mathbf{i}}_2}^{\hat{i}_2} \mathcal{U}_{i_3}^h + \tilde{Y}^{\hat{i}_1}_{\hat{i}_2 i_3} \times \Omega^{\hat{\mathbf{i}}_1}_{\hat{i}_1} \Omega_{\hat{\mathbf{i}}_2}^{\hat{i}_2} \mathcal{U}^{i_3 h}
\quad&\Rightarrow\quad {y'}^{\hat{\mathbf{i}}_1}_{\hat{\mathbf{i}}_2} .
\end{aligned} \end{equation}
Here we suppress the index for massive Higgs boson for simplicity. By bolding the massless spinors, we obtain the massive $FFS$ amplitude
\begin{equation} \begin{aligned}
\mathbf{M}({\mathbf{1}}^{1/2}, \mathbf{2}^{1/2}, \mathbf{3}^{0}) = 
{y}^{\hat{\mathbf{i}}_1}_{\hat{\mathbf{i}}_2} \langle\mathbf{12}\rangle +{y'}^{\hat{\mathbf{i}}_1}_{\hat{\mathbf{i}}_2} [\mathbf{12}].
\end{aligned} \end{equation}
The non-zero values of the massive coefficients $y$ and $y'$ are given by
\begin{equation} \begin{aligned}
{y}^{u}_{\bar{u}} = {y'}^{u}_{\bar{u}} &= \frac{1}{\sqrt{2}}\mathcal{Y}^{(u)},\\
{y}^{d}_{\bar{d}} = {y'}^{d}_{\bar{d}} &= \frac{1}{\sqrt{2}}\mathcal{Y}^{(d)},\\
{y}^{e}_{\bar{e}} = {y'}^{e}_{\bar{e}} &= \frac{1}{\sqrt{2}}\mathcal{Y}^{(e)}.\\
\end{aligned} \end{equation}
This implies the absence of CP violation in this case. The result agrees with the Yukawa matrix for massive fermions in the single-flavor case.

\textbf{3. FFV:} First, we match massless $FFV$ amplitudes to the massive one. To avoid overlapping contributions, we  consider only two massless amplitudes with opposite helicity for the fermion and antifermion. Here we choose the massless amplitudes with helicity $(\mp\frac12,\pm\frac12,+1)$ and deform them as follows
\begin{equation} \begin{aligned}
\mathcal{A}(1^{-\frac12},2^{+\frac12},3^{+1})&=(\tilde{T}_f^{I_3})^{\hat{i}_1}_{\hat{i}_2}\frac{[23]^2}{[12]}
\quad \to \quad (\tilde{T}_f^{I_3})^{\hat{i}_1}_{\hat{i}_2}\frac{[23]\langle1\eta_3\rangle}{m_3},\\
\mathcal{A}(1^{+\frac12},2^{-\frac12},3^{+1})&=(T_f^{I_3})^{\hat{i}_1}_{\hat{i}_2}\frac{[13]^2}{[12]}
\quad \to \quad -(T_f^{I_3})^{\hat{i}_1}_{\hat{i}_2}\frac{[13]\langle2\eta_3\rangle}{m_3},
\end{aligned} \end{equation}
where $\tilde T_f$ denotes the coupling for the right-handed current, and $T_f$ for left-handed current. Applying the transformation matrices $O$ and $\Omega$ to convert massless particles to massive ones, we obtain the massive coefficients $X_1$ and $X_2$:
\begin{equation} \begin{aligned} \label{eq:FFV_coef}
(T_f^{I_3})^{\hat{i}_1}_{\hat{i}_2}\times O^{I_3 \mathbf{I}_3} \Omega^{\hat{\mathbf{i}}_1}_{\hat{i}_1} \Omega_{\hat{\mathbf{i}}_2}^{\hat{i}_2}
\quad&\Rightarrow\quad (X_1^{\mathbf{I}_3} )^{\hat{\mathbf{i}}_1}_{\hat{\mathbf{i}}_2},\\
(\tilde{T}_f^{I_3})^{\hat{i}_1}_{\hat{i}_2}\times O^{I_3 \mathbf{I}_3} \Omega^{\hat{\mathbf{i}}_1}_{\hat{i}_1} \Omega_{\hat{\mathbf{i}}_2}^{\hat{i}_2}
\quad&\Rightarrow\quad (X_2^{\mathbf{I}_3} )^{\hat{\mathbf{i}}_1}_{\hat{\mathbf{i}}_2}.
\end{aligned} \end{equation}
By bolding the massless spinors, we arrive at the massive amplitude
\begin{equation} \label{eq:FFV_result}
\mathbf{M}(\mathbf{1}^{1/2}, \mathbf{2}^{1/2}, \mathbf{3}^{1}) = (X_1^{\mathbf{I}_3} )^{\hat{\mathbf{i}}_1}_{\hat{\mathbf{i}}_2} \frac{\langle\mathbf{13}\rangle [\mathbf{23}]}{\mathbf{m}_3} +(X_2^{\mathbf{I}_3} )^{\hat{\mathbf{i}}_1}_{\hat{\mathbf{i}}_2} \frac{\langle\mathbf{23}\rangle [\mathbf{13}]}{\mathbf{m}_3}.
\end{equation}

The coefficients $X_1$ and $X_2$ are defined by Eq.~\eqref{eq:FFV_coef}. Since this relation also applies to massless particles, the coefficient components will subsequently include contributions from neutrinos and photons. We now examine the non-zero components of the massive FFV coefficients based on the vector boson type $\mathbf{I}_3$. For the charged vector boson $\mathbf I_3=W^\pm$, only $X_1$ exists,
\begin{equation} \begin{aligned}
(X_1^{W^{+}} )^{d}_{\bar{u}}&=(X_1^{W^{+}} )^{e}_{\bar{\nu}}=g,\\
(X_1^{W^{-}} )^{u}_{\bar{d}}&=(X_1^{W^{-}} )^{\nu}_{\bar{e}}=g,\\
(X_2^{W^{\pm}} )^{\hat{\mathbf{i}}_1}_{\hat{\mathbf{i}}_2} &=0.
\end{aligned} \end{equation}
This describes the coupling between the $W$ boson and the charged current. For the neutral vector boson $\mathbf{I}_3 = Z$, both coefficients contribute, with $X_1$ and $X_2$ taking different values:
\begin{equation} \begin{aligned}
(X_1^{Z} )^{u}_{\bar{u}}&=-\frac{-\frac13 g^{\prime 2}+g^2}{\sqrt{2(g^2+g^{\prime 2})}},&
(X_1^{Z} )^{d}_{\bar{d}}&=-\frac{-\frac13 g^{\prime 2}-g^2}{\sqrt{2(g^2+g^{\prime 2})}},&\\
(X_1^{Z} )^{\nu}_{\bar{\nu}}&=-\frac{g^{\prime 2}+g^2}{\sqrt{2(g^2+g^{\prime 2})}},&
(X_1^{Z} )^{e}_{\bar{e}}&=-\frac{g^{\prime 2}-g^2}{\sqrt{2(g^2+g^{\prime 2})}},&\\
(X_2^{Z} )^{u}_{\bar{u}}&=\frac{\frac43 g^{\prime 2}}{\sqrt{2(g^2+g^{\prime 2})}},&
(X_2^{Z} )^{d}_{\bar{d}}&=\frac{-\frac23 g^{\prime 2}}{\sqrt{2(g^2+g^{\prime 2})}},& \\
(X_2^{Z} )^{e}_{\bar{e}}&=\frac{-2g^{\prime 2}}{\sqrt{2(g^2+g^{\prime 2})}}.&\\
\end{aligned} \end{equation}
For the photon $\mathbf{I}_3 = A$, the two coefficients $X_1$ and $X_2$ are identical:
\begin{equation} \begin{aligned}
(X_1^{A} )^{u}_{\bar{u}}&=-\frac{\frac43 g g'}{\sqrt{2(g^2+g^{\prime 2})}},\quad
(X_1^{A} )^{d}_{\bar{d}}=-\frac{-\frac23 g g'}{\sqrt{2(g^2+g^{\prime 2})}},\quad
(X_1^{A} )^{e}_{\bar{e}}=-\frac{-2 g g'}{\sqrt{2(g^2+g^{\prime 2})}},\\
(X_2^{A} )^{\hat{\mathbf{i}}_1}_{\hat{\mathbf{i}}_2}&=(X_1^{A} )^{\hat{\mathbf{i}}_1}_{\hat{\mathbf{i}}_2}.\\
\end{aligned} \end{equation}
These correspond to interactions with between the $Z$ boson or photon and the neutral current.

Then we consider the massless $FFS$ amplitude with helicity category $(-\frac12,-\frac12,0)$. It can be matched to two MHC terms,
\begin{equation} \begin{aligned}
\mathcal{A}(1^{-\frac12},2^{-\frac12},3^{0})=
{Y}^{\hat{i}_1 i_3}_{\hat{i}_2}\langle12\rangle&
\quad\to\quad
b_1{Y}^{\hat{i}_1 i_3}_{\hat{i}_2}\frac{\langle23\rangle[3 \eta_1]}{\tilde{m}_1} +b_2{Y}^{\hat{i}_1 i_3}_{\hat{i}_2}\frac{\langle13\rangle[3 \eta_2]}{\tilde{m}_2},\\
\end{aligned} \end{equation}
where $b_1+b_2=1$.
After applying the transformation matrices, the massless coefficient $Y$ becomes
\begin{equation}
{Y}^{\hat{i}_1 i_3}_{\hat{i}_2} \mathcal{U}_{i_3 }^{\mathbf{I}_3} \Omega^{\hat{\mathbf{i}}_1}_{\hat{i}_1} \Omega_{\hat{\mathbf{i}}_2}^{\hat{i}_2} +{Y}^{\hat{i}_1}_{\hat{i}_2 i_3} \mathcal{U}^{i_3 \mathbf{I}_3} \Omega^{\hat{\mathbf{i}}_1}_{\hat{i}_1} \Omega_{\hat{\mathbf{i}}_2}^{\hat{i}_2}
\quad\Rightarrow\quad G
\end{equation}
Bolding the massless spinors, we obtain an alternative expression for the massive amplitude
\begin{equation}
\mathbf{M}(\mathbf{1}^{1/2}, \mathbf{2}^{1/2}, \mathbf{3}^{1}) = b_1 G \frac{\langle\mathbf{13}\rangle [\mathbf{23}]}{\mathbf{m}_2} +b_2 G \frac{\langle\mathbf{23}\rangle [\mathbf{13}]}{\mathbf{m}_1}.
\end{equation}

This amplitude is consistency with eq.~\eqref{eq:FFV_result}, if the decomposition factors $b_1$ and $b_2$ satisfy
\begin{equation} \begin{aligned}
\frac{1}{\mathbf m_2}b_1 G &= -\frac{1}{\mathbf m_3} (X_1^{\mathbf{I}_3})^{\hat{\mathbf{i}}_1}_{\hat{\mathbf{i}}_2},\\
\frac{1}{\mathbf m_1}b_2 G &=
\frac{1}{\mathbf m_3} (X_2^{\mathbf{I}_3})^{\hat{\mathbf{i}}_1}_{\hat{\mathbf{i}}_2}.
\end{aligned} \end{equation}
Substituting the particle species, we determine the  decomposition factor $b_i$
\begin{equation}
\begin{tabular}{c|c|c}
\hline
$\hat{\mathbf{i}}_1 \hat{\mathbf{i}}_2 \mathbf{I}_3$ & $b_1$ & $b_2$ \\
\hline
$f_1\bar f_2 W^\pm$ & $1$ & $0$ \\
\hline
$f_1\bar f_2 A$ & $\frac{1}{2}$ & $\frac{1}{2}$ \\
\hline
$u \bar u Z$ & $\cos^2\theta_W-\frac13\sin^2\theta_W$ & $\frac43\sin^2\theta_W$ \\
$d \bar d Z$ & $\cos^2\theta_W+\frac13\sin^2\theta_W$ & $\frac23\sin^2\theta_W$ \\
$e \bar e Z$ & $\cos^2\theta_W-\sin^2\theta_W$ & $2\sin^2\theta_W$ \\
\hline
\end{tabular}
\end{equation}
where $\theta_W=\arctan\frac{g'}{g}$ is the Weinberg angle. 
The $W$ boson exclusively couples to the left-handed current ($X_2=0$), so the massless $FFS$ amplitude only match to the massive structure $\langle\mathbf{23}\rangle [\mathbf{13}]$. For neutral boson, both $X_1$ and $X_2$ are non-vanishing, so the massless $FFS$ amplitude contributes to both $\langle\mathbf{23}\rangle [\mathbf{13}]$ and $\langle\mathbf{13}\rangle [\mathbf{23}]$. In the case of the $Z$ boson, $b_1\neq b_2$ reflects the CP violation in the neutral current.

\textbf{4. VVS:} In this case, a single massless $VSS$ amplitude with helicity category $(-1,0,0)$ is sufficient, 
\begin{equation} \begin{aligned} \label{eq:massless_VSS}
\mathcal{A}(1^{-1},2^{0},3^{0})&=(T_s^{I_1})^{i_2}_{i_3}\frac{\langle12\rangle\langle31\rangle}{\langle23\rangle}
\quad\to\quad 
\frac{[\eta_1 2]\langle12\rangle}{\tilde m_1},
\end{aligned} \end{equation}
The massless coefficient $T_s$ maps to the massive coupling $\mathbf{g}$ via
\begin{equation}
{({T_s}^{I_1})}_{i_3}^{i_2} O^{I_1 \mathbf{I}_1} \mathcal{U}_{i_2}^{\mathbf{I}_2} \mathcal{U}^{i_3 h} -{({T_s}^{I_1})}_{i_2}^{i_3} O^{I_1 \mathbf{I}_1} \mathcal{U}^{i_2 \mathbf{I}_2} \mathcal{U}_{i_3}^{h}
\quad\Rightarrow\quad \mathbf{g}^{\mathbf{I}_1 \mathbf{I}_2}
\end{equation}
Bolding the massless spinors, we obtain the massive amplitude
\begin{equation}
\mathbf{M}(\mathbf{1}^{1}, \mathbf{2}^{1}, \mathbf{3}^{0}) = 
\mathbf{g}^{\mathbf{I}_1 \mathbf{I}_2} \frac{\langle\mathbf{12}\rangle [\mathbf{12}]}{\mathbf{m}_1},
\end{equation}
where we enforce $\mathbf{m}_1 = \mathbf{m}_2$ for consistency. 
The non-vanishing components of the massive coupling $\mathbf{g}$ are
\begin{equation} \begin{aligned}
\mathbf{g}^{W^+ W^-} = \mathbf{g}^{W^- W^+} = g,
\quad \mathbf{g}^{Z Z} = \frac{g}{\cos\theta_W}.
\end{aligned} \end{equation}

\textbf{5. VVV:} We begin by matching massless $VVV$ amplitudes to their massive counterparts. Note that, one massless $VVV$ amplitude cannot account for all massive terms. To minimize the use of massless amplitudes, we select two massless amplitudes with opposite helicities for the matching procedure. Specifically, we choose:
\begin{equation} \begin{aligned}
\mathcal{A}(1^{+1},2^{+1},3^{-1})&=f^{I_1 I_2 I_3}\frac{[12]^3}{[23][31]},\\
\mathcal{A}(1^{-1},2^{-1},3^{+1})&=f^{I_1 I_2 I_3} \frac{\langle12\rangle^3}{\langle23\rangle\langle31\rangle}.
\end{aligned} \end{equation}
Each massless amplitude admits two distinct deformations to the MHC amplitude. We employ both possibilities equally, assigning the decomposition factors as $\frac12$. The massless deformation thus yields
\begin{equation} \begin{aligned}
f^{I_1 I_2 I_3}\frac{[12]^3}{[23][31]} &\rightarrow 
\frac{1}{2}f^{I_1 I_2 I_3}\frac{[12]\langle\eta_23\rangle\langle 3 \eta_1\rangle}{m_1 m_2}+\frac{1}{2}f^{I_1 I_2 I_3}\left(\frac{[12][2 \eta_3]\langle3\eta_1\rangle}{m_1 \tilde{m}_3}+\frac{[12]\langle\eta_23\rangle[\eta_31]}{m_2 \tilde{m}_3}\right), \\
f^{I_1 I_2 I_3}\frac{\langle12\rangle^3}{\langle23\rangle\langle31\rangle} 
&\rightarrow \frac{1}{2}f^{I_1 I_2 I_3}\frac{\langle 12\rangle[\eta_23][3 \eta_1]}{\tilde m_1 \tilde m_2}+\frac{1}{2}f^{I_1 I_2 I_3}\left(\frac{\langle12\rangle\langle2 \eta_3\rangle[3\eta_1]}{\tilde m_1 m_3}+\frac{\langle12\rangle[\eta_23]\langle\eta_31\rangle}{\tilde m_2 m_3}\right).
\end{aligned} \end{equation}
The massless coefficient $f$ maps to the massive coupling $\mathbf{f}$ via
\begin{equation}
\frac{1}{2}f^{I_1 I_2 I_3} O^{I_1 \mathbf{I}_1} O^{I_2 \mathbf{I}_2} O^{I_3 \mathbf{I}_3}
\quad\Rightarrow\quad \mathbf{f}^{\mathbf{I}_1 \mathbf{I}_2 \mathbf{I}_3} .
\end{equation}
Summing both contributions and restoring $SU(2)_{\text{LG}}$ covariance yields the complete massive $VVV$ amplitude
\begin{equation} \begin{aligned} \label{eq:VVV_result}
\mathbf{M}(\mathbf{1}^{1}, \mathbf{2}^{1}, \mathbf{3}^{1}) 
=& \mathbf{f}^{\mathbf{I}_1 \mathbf{I}_2 \mathbf{I}_3} \left(\frac{[\mathbf{12}]\langle\mathbf{23}\rangle\langle\mathbf{31}\rangle}{\mathbf{m}_1 \mathbf{m}_2}+\frac{[\mathbf{12}][\mathbf{23}]\langle\mathbf{31}\rangle }{\mathbf{m}_1 \mathbf{m}_3}+\frac{[\mathbf{12}]\langle\mathbf{23}\rangle[\mathbf{31}]}{\mathbf{m}_2 \mathbf{m}_3} \right)\\
&+ \mathbf{f}^{\mathbf{I}_1 \mathbf{I}_2 \mathbf{I}_3} \left(\frac{\langle\mathbf{12}\rangle[\mathbf{23}][\mathbf{31}]}{\mathbf{m}_1 \mathbf{m}_2}+\frac{\langle\mathbf{12}\rangle\langle\mathbf{23}\rangle[\mathbf{31}]}{\mathbf{m}_1 \mathbf{m}_3}+\frac{\langle\mathbf{12}\rangle[\mathbf{23}]\langle\mathbf{31}\rangle}{\mathbf{m}_2 \mathbf{m}_3} \right),
\end{aligned} \end{equation}
where all terms share the same massive coefficient.

In analogy to the massless case, $\mathbf{f}^{\mathbf{I}_1 \mathbf{I}_2 \mathbf{I}_3}$ is totally antisymmetric. Substituting the relevant particle types, we obtain the non-zero values,
\begin{equation}
\begin{tabular}{c|c}
\hline
$\mathbf{I}_1\mathbf{I}_2\mathbf{I}_3$ & $\mathbf{f}^{\mathbf{I}_1 \mathbf{I}_2 \mathbf{I}_3}$ \\
\hline
\makecell{$W^+ W^- Z$\\ $W^- Z W^+$\\ $Z W^+ W^-$} & $\frac12 g \cos\theta_W$ \\
\hline
\makecell{$W^- W^+ Z$\\ $W^+ Z W^-$\\ $Z W^- W^+$} & $-\frac12 g \cos\theta_W$  \\
\hline
\end{tabular}
\end{equation}

Then we match massless $VSS$ amplitude to the massive $VVV$ amplitude. For the helicity category $(-1,0,0)$, the massless amplitude matches more MHC terms,
\begin{equation} \begin{aligned}
\mathcal{A}(1^{-1},2^{0},3^{0})=&(T_s^{I_1})^{i_2}_{i_3}\frac{\langle12\rangle\langle31\rangle}{\langle23\rangle}\\
\rightarrow 
&b_1(T_s^{I_1})^{i_2}_{i_3}\left(\frac{\mathbf m_3^2}{2\mathbf m_1^2-\mathbf m_2^2-\mathbf m_3^2}\frac{[\eta_1 2]\langle23\rangle\langle\eta_3 1\rangle+[\eta_1 2]\langle2\eta_3\rangle\langle3 1\rangle}{\tilde{m}_1 m_3}\right.\\
&+\left.\frac{\mathbf m_2^2}{2\mathbf m_1^2-\mathbf m_2^2-\mathbf m_3^2}\frac{\langle12\rangle\langle\eta_23\rangle[3\eta_1]+\langle1\eta_2\rangle\langle23\rangle[3\eta_1]}{\tilde{m}_1 m_2}\right) \\
&+b_2(T_s^{I_1})^{i_2}_{i_3}\frac{[\eta_1 2][\eta_2 3]\langle31\rangle}{\tilde{m}_2 \tilde{m}_1}
+b_3(T_s^{I_1})^{i_2}_{i_3}\frac{\langle12\rangle[2\eta_3][3\eta_1]}{\tilde{m}_1 \tilde{m}_3}\\
&-b_4(T_s^{I_1})^{i_2}_{i_3}\frac{[\eta_1 2]\langle23\rangle[3 \eta_1]}{\tilde{m}_1^2}\\
&-b_5(T_s^{I_1})^{i_2}_{i_3}\left(\frac{\langle12\rangle[2\eta_3]\langle\eta_31\rangle}{\mathbf{m}_1^2}+\frac{\langle1\eta_2\rangle[\eta_23]\langle31\rangle}{\mathbf{m}_1^2}\right).\\
\end{aligned} \end{equation}
After applying the transformation matrices, the gauge tensor $T_s$ becomes
\begin{equation} 
({T_s}^{I_1})^{i_2}_{i_3} O^{I_1 \mathbf{I}_1} \mathcal{U}^{\mathbf{I}_2}_{i_2} \mathcal{U}^{i_3 \mathbf{I}_3} -({T_s}^{I_1})^{i_3}_{i_2} O^{I_1 \mathbf{I}_1} \mathcal{U}^{i_2 \mathbf{I}_2} \mathcal{U}^{\mathbf{I}_3}_{i_3}
\quad\Rightarrow\quad\mathbf G.
\end{equation}
Bolding the massless spinors, we obtain another expression of the massive $VVV$ amplitude
\begin{equation} \begin{aligned}
\mathbf{M}(\mathbf{1}^{1}, \mathbf{2}^{1}, \mathbf{3}^{1}) 
=&b_1\mathbf G\left(\frac{\mathbf{m}_3[\mathbf{12}]\langle\mathbf{23}\rangle\langle\mathbf{31}\rangle}{\mathbf{m}_1(2\mathbf m_1^2-\mathbf m_2^2-\mathbf m_3^2)}+\frac{\mathbf{m}_2\langle\mathbf{12}\rangle\langle\mathbf{23}\rangle[\mathbf{31}]}{\mathbf{m}_1(2\mathbf m_1^2-\mathbf m_2^2-\mathbf m_3^2)}\right)
+b_2\mathbf G\frac{[\mathbf{12}][\mathbf{23}]\langle\mathbf{31}\rangle }{\mathbf{m}_1 \mathbf{m}_2}\\
&+b_3\mathbf G\frac{\langle\mathbf{12}\rangle[\mathbf{23}][\mathbf{31}]}{\mathbf{m}_1 \mathbf{m}_3}+b_4\mathbf G\frac{[\mathbf{12}]\langle\mathbf{23}\rangle[\mathbf{31}]}{\mathbf{m}_1^2}+b_5\mathbf G\frac{\langle\mathbf{12}\rangle[\mathbf{23}]\langle\mathbf{31}\rangle}{\mathbf{m}_1^2}.
\end{aligned} \end{equation}

This amplitude is consistency with Eq.~\eqref{eq:VVV_result}, if the decomposition factors satisfy
\begin{equation} \begin{aligned}
&b_1 \mathbf G = -\frac{2\mathbf m_1^2-\mathbf m_2^2-\mathbf m_3^2}{\mathbf m_2 \mathbf m_3}\mathbf{f}^{\mathbf{I}_1 \mathbf{I}_2 \mathbf{I}_3}\\
&b_2 \mathbf G = \frac{\mathbf m_2}{\mathbf m_3}\mathbf{f}^{\mathbf{I}_1 \mathbf{I}_2 \mathbf{I}_3}, \\
&b_3 \mathbf G = \frac{\mathbf m_3}{\mathbf m_2}\mathbf{f}^{\mathbf{I}_1 \mathbf{I}_2 \mathbf{I}_3}, \\
&b_4 \mathbf G =-b_5 \mathbf G =\frac{\mathbf m_1^2}{\mathbf m_2 \mathbf m_3}\mathbf{f}^{\mathbf{I}_1 \mathbf{I}_2 \mathbf{I}_3}. \\
\end{aligned} \end{equation}
From these relations, we can obtain the decomposition factor $b_i$ for different particle types:
\begin{equation}
\begin{tabular}{c|c|c|c|c}
\hline
$\mathbf{I}_1\mathbf{I}_2\mathbf{I}_3$ & $b_1$ & $b_2$ & $b_3$ & $b_4,-b_5$ \\
\hline
$W^\pm W^\mp Z$ & $\frac{1-\cos^2\theta_W}{2}$ & $\frac12\cos^2\theta_W$ & $\frac12$ & $\frac12\cos^2\theta_W$ \\
\hline
$W^\pm Z W^\mp$ & $\frac{1-\cos^2\theta_W}{2}$  & $\frac12$ & $\frac12\cos^2\theta_W$ & $\frac12\cos^2\theta_W$ \\
\hline
$Z W^\pm W^\mp $ & $\frac{2\cos^2\theta_W-2}{2(2\cos^2\theta_W-1)}$ & $\frac{\cos^2\theta_W}{2(2\cos^2\theta_W-1)}$ & $\frac{\cos^2\theta_W}{2(2\cos^2\theta_W-1)}$ & $\frac{1}{2(2\cos^2\theta_W-1)}$ \\
\hline
\end{tabular}
\end{equation}

In the Standard Model, for massive amplitudes involving more than three particles, only the 4-pt amplitude contains contact terms. In addition to the contact term, the 4-pt massive amplitude also includes factorized contributions. In the leading matching at 4-pt, the contact term and the factorized term cannot be treated separately. This lies beyond the scope of the present work, as our primary focus is on establishing the matching for massive amplitude without pole structures. We leave a detailed study of this case for future research.


\section{Conclusion}

In this work, we investigate the correspondence between three-point massive amplitudes and their high-energy massless counterparts at different orders in the mass/energy expansion. The correspondence is constructed using Minimal Helicity-Chirality (MHC) amplitudes, which are organized according to the number of chirality flips and which embody the principle of helicity--transversality unification. The chirality-flip structure dictates the matching procedure: at leading order, the correspondence is realized either directly for primary amplitude, or through amplitude deformation if conserved current is involved, while at sub-leading orders, it is systematically implemented by the on-shell Higgsing.

The starting point of this correspondence is the massive MHC leading amplitudes and the massless helicity amplitudes. For a massive amplitude with spins $(s_1,s_2,s_3)$, in principle all the MHC amplitudes with the $|h_i|\le s_i$ and helicity-chirality unity would contribute to the massive structures. Only the ones satisfying the massless helicity selection rule would survive as the leading MHC amplitudes. Thus both the massless helicity and massive leading MHC amplitudes are determined. This constraint simultaneously determines both the massless helicity amplitudes and the massive leading MHC components. Given a leading MHC amplitude, its helicity configuration directly determines the number of chirality flips associated with each external particle. In the amplitude's expression, each such chirality flip carries an explicit factor of the particle mass.

Once the massive leading MHC amplitudes and their corresponding massless ultraviolet counterparts are identified, we proceed to implement the leading-order matching through amplitude deformation. This deformation is guided by the chirality-flip structure present in the MHC amplitude. In processes without external vector bosons, there is no need to perform deformation. When a vector boson is involved, however, additional care is required. Specifically, even though the leading MHC amplitudes in this sector contain chirality flips (and thus explicit mass factors), they do not vanish in the high-energy massless limit. This non-vanishing behavior occurs because the mass deformation itself introduces compensating factors that cancel the mass dependence in the numerator of the amplitude.

We have proposed a systematic amplitude deformation procedure based on the scaling properties of the $m\eta$ factor. Three-point MHC amplitudes are first decomposed into a product of a conserved current and a vector-boson state. For the case of a transverse vector boson, the deformation applies a chirality flip to the corresponding massless gauge boson; integration by parts (IBP) then removes the associated spurious pole from the massless amplitude. For a longitudinal vector boson, the chirality flip is applied instead to the conserved current, and subsequent IBP transforms the scalar component into a Goldstone boson, in accordance with the Goldstone equivalence theorem. Following this systematic procedure, we present explicit matching results for the fundamental three-point amplitudes $FFS$, $FFV$, $VVS$, and $VVV$.

Massless helicity amplitudes with additional on-shell Higgs insertions correspond to sub-leading MHC amplitudes. These sub-leading contributions can be obtained by computing the relevant four-point massless amplitude involving a Higgs boson, then taking the collinear limit and factoring out the vacuum expectation value to recover the three-point structure. We identify a more efficient route to this correspondence at sub-leading order: analogous to the use of splitting functions, one can apply {\it Higgsing rules} directly to the leading-order matching result to systematically generate the sub-leading matchings. This procedure is summarized in Eqs.~\eqref{eq:f_table} and~\eqref{eq:vv_table}. A key consequence is that once the leading-order matching is established, all sub-leading matchings are derived—and therefore not independent—through the repeated application of these Higgsing rules.

In the Standard Model, on-shell scattering amplitudes can be constructed recursively from fundamental three-point amplitudes. In this work we establishe the matching between massless and massive three-point amplitudes in the Standard Model. In subsequent work, we will utilize these matched massless three-point amplitudes as building blocks to systematically construct four-point massless amplitudes, thereby extending the constructive framework to higher multiplicity. Furthermore, the massless--massive correspondence developed here can be generalized to effective field theories, establishing a matching procedure between a massless UV-complete EFT and its massive counterpart in the symmetry-broken phase. These promising extensions will be pursued in future studies.

\acknowledgments

We would like to thank Csaba Csaki and Yi-Ning Wang for valuable discussions. This work is supported by the National Science Foundation of China under Grants No. 12347105, No. 12375099 and No. 12447101, and the National Key Research and Development Program of China Grant No. 2020YFC2201501, No. 2021YFA0718304.

\begin{appendix}

\section{The Complete Matching for massive $VVV$ amplitude}
\label{app:VVV_matching}

In the following discussion, we present a detailed matching procedure for the massive $VVV$ amplitude at leading order. The general steps are outlined below:
\begin{itemize}
    \item The massive $VVV$ amplitude has spins $s_1=s_2=s_3=1$. Running over all helicity $h=-1,0,+1$, and picking up the ones with helicity-transversality equality, we can use the massless helicity condition $h_1+h_2+h_3=\pm1$ to select the helicity category of the leading order:
    \begin{equation} \begin{aligned}
    &(\pm1, \pm1, \mp1),& 
    &(\pm1, \mp1, \pm1),&
    &(\mp1, \pm1, \pm1),& \\
    &(\pm1, 0, 0),& 
    &(0, \pm1, 0),&
    &(0, 0, \pm1).&
    \end{aligned}\end{equation} 
    \item Choosing a specific helicity $(h_1,h_2,h_3)$, we can write down the amplitude and its scaling behavior of each massless particle.
    \item Find the corresponding MHC amplitude in this helicity, we can write down the scaling of massive particle. From the scaling behavior, we can identify which corresponds to the symmetric vector $\mathbb A$, while the other two are treated as the current $\mathbb J$.
    \item Both the massless and MHC amplitudes can be expressed in the form $\mathbb{J} \cdot \mathbb{A}$. Starting from the massless scaling, we deform the scaling behaviors of $\mathbb{A}$ and $\mathbb{J}$ so that they match the MHC scaling. Since there are two vector $\mathbb{A}$ in the massive $VVV$ amplitude, this scaling deformation must be applied twice.
    \item  The spinor scaling deformation then guides the corresponding amplitude deformation. The goal of each deformation is to remove potential spurious poles. If any terms containing spurious poles remain after deformation, we convert them back into expressions in terms of the original massless amplitude.
\end{itemize}

\subsection{Massless $VVV$ $\to$ masssive $VVV$}

Let us begin with the massless amplitude with helicity $(+1,+1,-1)$. We can read the scaling behavior of each particle:
\begin{equation} \label{eq:massless_VVV_scaling}
\begin{tikzpicture}[baseline=0.7cm] \begin{feynhand}
\setlength{\feynhandarrowsize}{3.5pt}
\vertex [particle] (i1) at (0,0.8) {$1^+$}; 
\vertex [particle] (i2) at (1.6,1.6) {$2^+$}; 
\vertex [particle] (i3) at (1.6,0) {$3^-$};  
\vertex (v1) at (0.9,0.8); 
\graph{(i1)--[bos](v1)--[bos](i2)};
\graph{(i3)--[bos](v1)};  
\end{feynhand} \end{tikzpicture}=
\frac{[12]^3}{[23][31]}\sim 
\begin{cases}
\text{particle 1}:\quad \tilde\lambda_1^2, \\
\text{particle 2}:\quad \tilde\lambda_2^2, \\
\text{particle 3}:\quad \tilde\lambda_3^{-2}. \\
\end{cases}
\end{equation}

1. Suppose that we want to match the massless amplitude to  the following MHC amplitude with spinor scaling
\begin{equation} \label{eq:VVV_1}
\Ampthree{1^+}{2^+}{3^-}{\bosflip{1}{180}{brown}{cyan}}{\bosflip{1}{55}{brown}{cyan}}{\bos{i3}{red}}
=\tilde m_1\tilde m_2\langle\eta_1 3\rangle\langle3\eta_2\rangle[12]\sim
\begin{cases}
\text{particle 1}:\quad \tilde\lambda_1\tilde m_1\eta_1, \\
\text{particle 2}:\quad \tilde\lambda_2\tilde m_2\eta_2, \\
\text{particle 3}:\quad \lambda_3^2. \\
\end{cases}
\end{equation}
Both particles 1 and 2 can be identified as the transverse vector $\mathbb A^+$, so we need to apply two gauge boson deformation. 

In first step, we identify particle 1 as the vector $\mathbb A^+$, particles 2 and 3 as the current $\mathbb J$. Begin with the massless amplitude, we can identify the scaling behavior
\begin{equation}
\frac{[12]^3}{[23][31]}\sim 
\begin{cases}
\text{current}:\quad \mathbb J(\tilde\lambda^0)\sim\tilde\lambda_2^2\tilde\lambda_3^{-2}, \\
\text{vector}:\quad \mathbb A^+(\tilde\lambda_1^2), \\
\end{cases}
\end{equation}
The gauge boson deformation then yields,
\begin{equation} \begin{aligned}
\mathbb J(\tilde\lambda^0)\cdot\mathbb A^+(\tilde\lambda_1^2)
\xrightarrow{\times \frac{\langle\lambda\eta\rangle}{m}}\mathbb J(\tilde\lambda^0)\cdot\mathbb A^+(\tilde\lambda_1^2\lambda_1 m_1^{-1}\eta_1)
\xrightarrow{\text{IBP}^-}\mathbb J(\tilde\lambda \lambda)\cdot\frac{\mathbb A^+(\tilde\lambda_1 \tilde m_1\eta_1)}{\mathbf m_1^2}.
\end{aligned} \end{equation}
The vector scaling $\mathbb A^+(\tilde\lambda_1 \tilde m_1\eta_1)$ now match to eq.~\eqref{eq:VVV_1}. The corresponding amplitude deformation is
\begin{equation} \begin{aligned}
\frac{[12]^3}{[23][31]} 
\xrightarrow{\times \frac{\langle\lambda\eta\rangle}{m}}
-\frac{\langle\eta_112][12]^2}{m_1[23][31]}
\xrightarrow{\text{IBP}^-}
-\frac{\tilde m_1\langle\eta_1 3\rangle[12]^2}{\mathbf m_1^2[31]}.
\end{aligned} \end{equation}

Then we identify particle 2 as the vector $\mathbb A^+$, particles 1 and 3 as the current $\mathbb J$. So we can read the scaling behavior
\begin{equation}
-\frac{\tilde m_1\langle\eta_1 3\rangle[12]^2}{\mathbf m_1^2[31]}\sim
\begin{cases}
\text{current}:\quad {\displaystyle\frac{\mathbb J(\lambda \tilde m\eta)}{\mathbf m^2}\sim\frac{1}{\mathbf m_1^2}}\tilde\lambda_1\tilde m_1\eta_1\tilde\lambda_3^{-1}\lambda_3, \\
\text{vector}:\quad \mathbb A^+(\tilde\lambda_2^2), 
\end{cases}
\end{equation}
The scaling behavior of the above result can be deformed as
\begin{equation} \begin{aligned}
\frac{\mathbb J(\lambda \tilde m\eta)}{\mathbf m_1^2}\cdot\mathbb A^+(\tilde\lambda_2^2)
\xrightarrow{\times \frac{\langle\lambda\eta\rangle}{m}}
\frac{\mathbb J(\lambda \tilde m\eta)}{\mathbf m_1^2}\cdot\mathbb A^+(\tilde\lambda_2^2\lambda m_2^{-1}\eta_2)
\xrightarrow{\text{IBP}^-}\frac{\mathbb J(\tilde\lambda \lambda^2 \tilde m\eta)}{\mathbf m_1^2}\cdot\frac{\mathbb A^+(\tilde\lambda_2 \tilde m_2\eta_2)}{\mathbf m_2^2}.
\end{aligned} \end{equation}
The corresponding amplitude deformation is
\begin{equation} \begin{aligned}
-\frac{\tilde m_1\langle\eta_1 3\rangle[12]^2}{\mathbf m_1^2[31]}
\xrightarrow{\times \frac{\langle\lambda\eta\rangle}{m}}
-\frac{\tilde m_1\langle\eta_1 3\rangle[12\eta_2\rangle[12]}{\mathbf m_1^2 m_2[31]}
\xrightarrow{\text{IBP}^-}
-\frac{\tilde m_1\tilde m_2\langle\eta_1 3\rangle\langle3\eta_2\rangle[12]}{\mathbf m_1^2 \mathbf m_2^2}.
\end{aligned} \end{equation}

2. The massless amplitude can also be matched to two MHC amplitudes. Their scaling behaviors are given by
\begin{align}
\Ampthree{1^+}{2^+}{3^-}{\bosflip{1}{180}{brown}{cyan}}{\bos{i2}{cyan}}{\bosflip{1}{-55}{brown}{red}}
=m_3\tilde m_1[\eta_3 2][12]\langle\eta_1 3\rangle&\sim 
\begin{cases}
\text{particle 1}:\quad \tilde\lambda_1 (\tilde m_1\eta_1), \\
\text{particle 2}:\quad \tilde\lambda_2^2, \\
\text{particle 3}:\quad \lambda_3 (m_3\tilde\eta_3), 
\end{cases}  \label{eq:VVV_2a}\\
\Ampthree{1^+}{2^+}{3^-}{\bos{i1}{cyan}}{\bosflip{1}{55}{brown}{cyan}}{\bosflip{1}{-55}{brown}{red}}
=-m_3\tilde m_2[\eta_3 1][12]\langle3\eta_2\rangle&\sim
\begin{cases}
\text{particle 1}:\quad  \tilde\lambda_1^2,\\
\text{particle 2}:\quad \tilde\lambda_2(\tilde m_2\eta_2),\\
\text{particle 3}:\quad \lambda_3 (m_3\tilde\eta_3).
\end{cases}  \label{eq:VVV_2b}
\end{align}
These correspond to two transverse vectors: $\mathbb{A}^-$ (particle 3) and $\mathbb{A}^+$ (particle 1 in the first amplitude, particle 2 in the second). 

In the first step, we identify particle 3 as $\mathbb{A}^+$, while particles 1 and 2 are treated as the current $\mathbb{J}$. Begin with the massless amplitude, we can identify the scaling behavior
\begin{equation}
\frac{[12]^3}{[23][31]}\sim 
\begin{cases}
\text{current}:\quad \mathbb J(\tilde\lambda^4)\sim\tilde\lambda_1^2\tilde\lambda_2^2, \\
\text{vector}:\quad \mathbb A^-(\tilde\lambda_3^{-2}). \\
\end{cases}
\end{equation}
The scaling deformation is then performed as
\begin{equation} \begin{aligned}
\mathbb J(\tilde\lambda^4)\cdot\mathbb A^-(\tilde\lambda_3^{-2})
\xrightarrow{\times \frac{[\eta\lambda]}{\tilde m}}\mathbb J(\tilde\lambda^4)\cdot\mathbb A^-(\tilde\lambda_3^{-1} \tilde m_3^{-1}\tilde\eta_3)
=\mathbb J(\tilde\lambda^4)\cdot\frac{\mathbb A^-(\tilde\lambda_3^{-1} m_3\tilde\eta_3)}{\mathbf m_3^2}.
\end{aligned} \end{equation}
In this step, we cannot apply the IBP, because there is no positive scaling of $\tilde\lambda$. Therefore particle 2 will not match to the correct MHC scaling in this step. The  amplitude deformation in this step is 
\begin{equation} \begin{aligned}
\frac{[12]^3}{[23][31]} 
\xrightarrow{\times \frac{[\eta\lambda]}{m}}
\frac{[\eta_3 3][12]^3}{\tilde m_3[23][31]}
\xrightarrow{\text{Schouten}}
-\frac{m_3[\eta_3 1][12]^2}{\mathbf m_3^2[31]}-\frac{m_3[\eta_3 2][12]^2}{\mathbf m_3^2[23]}.
\end{aligned} \end{equation}
Here we use the Schouten identity to eliminate the spurious pole, which produces two terms. 

In the second step, we match the first term to eq.~\eqref{eq:VVV_2a} and the second term to eq.~\eqref{eq:VVV_2b}. Thus, for the first term we identify particle 1 as $\mathbb A^+$, while for the second term we identify particle 2 as $\mathbb A^+$. This leads to the following scaling behavior,
\begin{equation} \begin{aligned}
-\frac{m_3[\eta_3 1][12]^2}{\mathbf m_3^2[31]}&\sim
\begin{cases}
\text{current}:\quad {\displaystyle\frac{\mathbb J(\tilde\lambda m\tilde\eta)}{\mathbf m^2}\sim\frac{1}{\mathbf m_3^2}}\tilde\lambda_2^2\tilde\lambda_3^{-1}m_3\tilde\eta_3, \\
\text{vector}:\quad \mathbb A^+(\tilde\lambda_1^2), 
\end{cases} \\
-\frac{m_3[\eta_3 2][12]^2}{\mathbf m_3^2[23]}&\sim
\begin{cases}
\text{current}:\quad {\displaystyle\frac{\mathbb J(\tilde\lambda m\tilde\eta)}{\mathbf m^2}\sim\frac{1}{\mathbf m_3^2}}\tilde\lambda_1^2\tilde\lambda_3^{-1}m_3\tilde\eta_3, \\
\text{vector}:\quad \mathbb A^+(\tilde\lambda_2^2). 
\end{cases}
\end{aligned} \end{equation}
The scaling behavior of the first term can then be deformed as
\begin{equation} \begin{aligned}
\frac{\mathbb J(\tilde\lambda m\tilde\eta)}{\mathbf m^2}\cdot\mathbb A^+(\tilde\lambda_1^2)
&\xrightarrow{\times \frac{\langle\lambda\eta\rangle}{m}}\frac{\mathbb J(\tilde\lambda m\tilde\eta)}{\mathbf m^2}\cdot\mathbb A^+(\tilde\lambda_1^2\lambda_1 m_1^{-1}\eta_1)\\
&\xrightarrow{\text{IBP}^-}\frac{\mathbb J(\tilde\lambda^2\lambda m\tilde\eta)}{\mathbf m^2}\cdot\frac{\mathbb A^+(\tilde\lambda_1 m_1^{-1}\eta_1)}{\mathbf m_1},
\end{aligned} \end{equation}
For the second term, we change the spinor in $\mathbb A^+$ from particle 1 to particle 2. Therefore, the corresponding amplitude deformation of the two terms gives 
\begin{equation} \begin{aligned}
-\frac{m_3[\eta_3 1][12]^2}{\mathbf m_3^2[31]}-\frac{m_3[\eta_3 2][12]^2}{\mathbf m_3^2[23]}
&\xrightarrow{\times \frac{\langle\lambda\eta\rangle}{m}}
\frac{m_3[\eta_3 2][12]\langle\eta_1 12]}{\mathbf m_3^2 m_1[23]}-\frac{m_3[\eta_3 1][12][12\eta_2\rangle}{\mathbf m_3^2 m_2[31]}\\
&\xrightarrow{\text{IBP}^-}
\frac{m_3 \tilde m_1[\eta_3 2][12]\langle\eta_1 3\rangle}{\mathbf m_3^2 \mathbf m_1^2}
-\frac{m_3 \tilde m_2[\eta_3 1][12]\langle3\eta_2\rangle}{\mathbf m_3^2 \mathbf m_2^2}.
\end{aligned} \end{equation}
In the last step, the IBP give the correct scaling behaivor for both particles 2 and 3. This result completely match eqs.~\eqref{eq:VVV_2a} and ~\eqref{eq:VVV_2b}. 


The above matching result is summarized in the following table:
\begin{equation}
\includegraphics[width=0.9\linewidth]{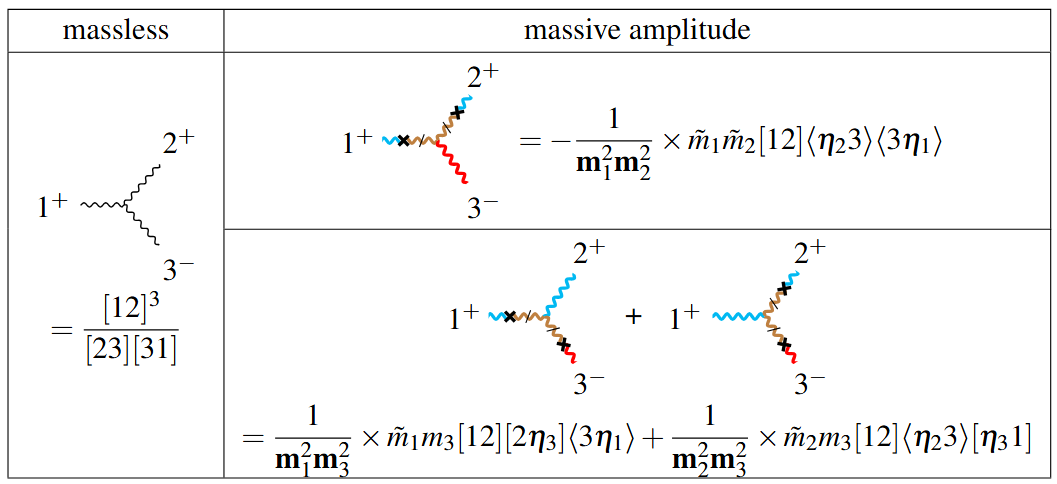}
\end{equation}

\subsection{Massless $VSS$ $\to$ masssive $VVV$}

Then we consider the massless amplitude with helicity $(-1,0,0)$. It has the following amplitude expression and scaling
\begin{equation}
\begin{tikzpicture}[baseline=0.7cm] \begin{feynhand}
\setlength{\feynhandarrowsize}{3.5pt}
\vertex [particle] (i1) at (0,0.8) {$1^-$}; 
\vertex [particle] (i2) at (1.6,1.6) {$2^0$}; 
\vertex [particle] (i3) at (1.6,0) {$3^0$};  
\vertex (v1) at (0.9,0.8); 
\graph{(i1)--[bos](v1)--[sca](i2)};
\graph{(i3)--[sca](v1)};  
\end{feynhand} \end{tikzpicture}=
\frac{\langle12\rangle\langle31\rangle}{\langle23\rangle}\sim  
\begin{cases}
\text{particle 1}:\quad \lambda_1^2, \\
\text{particle 2}:\quad \lambda_2^0, \\
\text{particle 3}:\quad \lambda_3^0. 
\end{cases}
\end{equation}

1. We match this massless $VSS$ amplitude to the following MHC amplitude:
\begin{equation}
\Ampthree{1^-}{2^0}{3^0}{\bosflipflip{1}{180}{cyan}{brown}{red}}{\bos{i2}{brown}}{\bos{i3}{brown}}
=m_1^2[\eta_1 3]\langle32\rangle[2\eta_1]\sim  
\begin{cases}
\text{particle 1}:\quad m_1\tilde\eta_1 \tilde m_1\eta_1, \\
\text{particle 2}:\quad \tilde\lambda_2\lambda_2, \\
\text{particle 3}:\quad \tilde\lambda_3\lambda_3. 
\end{cases}
\end{equation}
Both particles 2 and 3 can be identified as the longitudinal vector $\mathbb A^0$, so we need to apply two Goldstone boson deformation. 

In first step, we identify particle 3 as the vector $\mathbb A^0$, particles 1 and 3 as the current $\mathbb J$. Begin with the massless amplitude, we can identify the scaling behavior
\begin{equation}
\frac{\langle12\rangle\langle31\rangle}{\langle23\rangle}\sim 
\begin{cases}
\text{current}:\quad \mathbb J(\lambda^2)\sim\lambda_1^2\lambda_2^0, \\
\text{vector}:\quad \mathbb A^0(\lambda_3^{0}), \\
\end{cases}
\end{equation}
So the scaling deformation is 
\begin{equation}\begin{aligned}
\mathbb J(\lambda^2)\cdot\mathbb A^0(\lambda_3^0) \xrightarrow{\times\frac{[\eta\lambda]}{\tilde m}}
\mathbb J(\tilde\lambda \lambda^2 \tilde m^{-1}\tilde\eta)\cdot\mathbb A^0(\lambda_3^0)
\xrightarrow{\text{IBP}^+}
\frac{\mathbb J(\lambda m\tilde\eta)}{\mathbf m^2}\cdot\mathbb A^0(\tilde \lambda_3 \lambda_3).
\end{aligned}\end{equation}
The corresponding amplitude deformation is
\begin{equation}\begin{aligned}
\frac{\langle12\rangle\langle31\rangle}{\langle23\rangle}
\xrightarrow{\times\frac{[\eta\lambda]}{\tilde m}}
\frac{[\eta_1 12\rangle\langle31\rangle}{\tilde m_1\langle23\rangle}
\xrightarrow{\text{IBP}^+}
\frac{m_1[\eta_1 3]\langle31\rangle}{\mathbf m_1^2}.
\end{aligned}\end{equation}

Then we identify particle 2 as the vector $\mathbb A^0$, particles 1 and 3 as the current $\mathbb J$. So we can read the scaling behavior
\begin{equation}
\frac{m_1[\eta_1 3]\langle31\rangle}{\mathbf m_1^2}\sim
\begin{cases}
\text{current}:\quad {\displaystyle\frac{\mathbb J(\tilde\lambda\lambda^2 m\tilde\eta)}{\mathbf m^2}\sim\frac{1}{\mathbf m_1^2}}\lambda_1 m_1\tilde\eta_1\tilde\lambda_3\lambda_3, \\
\text{vector}:\quad \mathbb A^+(\lambda_0^2), 
\end{cases}
\end{equation}
The scaling behavior of the above result can be deformed as
\begin{equation}\begin{aligned}
\frac{\mathbb J(\tilde\lambda\lambda^2 m\tilde\eta)}{\mathbf m^2}\cdot\mathbb A^0(\lambda_2^0) 
&\xrightarrow{\times\frac{[\lambda\eta]}{\tilde m}}
\frac{\mathbb J(\tilde\lambda^2 \lambda^2 \tilde m^{-1}m\tilde\eta^2)}{\mathbf m^2}\cdot\mathbb A^0(\lambda_2^0)\\
&\xrightarrow{\text{IBP}^+}
\frac{\mathbb J(\tilde\lambda\lambda m^2\tilde\eta^2)}{\mathbf m^4}\cdot\mathbb A^0(\tilde \lambda_2 \lambda_2).
\end{aligned}\end{equation}
The corresponding amplitude deformation is
\begin{equation}\begin{aligned}
\frac{m_1[\eta_1 3]\langle31\rangle}{\mathbf m_1^2}
\xrightarrow{\times\frac{[\eta\lambda]}{\tilde m}}
-\frac{m_1[\eta_1 3]\langle31\eta_1]}{\mathbf m_1^2\tilde m_1}
\xrightarrow{\text{IBP}^+}
\frac{m_1^2[\eta_1 3]\langle32\rangle[2\eta_1]}{\mathbf m_1^4}.
\end{aligned}\end{equation}

2. The massless amplitude can also be matched to two MHC amplitudes. Their scaling behaviors are given by 
\begin{align}
\Ampthree{1^-}{2^0}{3^0}{\bosflip{1}{180}{brown}{red}}{\bos{i2}{brown}}{\bosflip{1}{-55}{cyan}{brown}}
&=-m_1 m_3[\eta_1 3][\eta_32]\langle21\rangle\sim
\begin{cases}
\text{particle 1}:\quad \lambda_1 m_1\tilde\eta_1, \\
\text{particle 2}:\quad \tilde\lambda_2\lambda_2, \\
\text{particle 3}:\quad \tilde\lambda_3 m_3\tilde\eta_3, 
\end{cases} \label{eq:VVV_3a}\\
\Ampthree{1^-}{2^0}{3^0}{\bosflip{1}{180}{brown}{red}}{\bosflip{1}{55}{cyan}{brown}}{\bos{i3}{brown}}
&=-m_1 m_2\langle13\rangle[3\eta_2][2\eta_1]\sim
\begin{cases}
\text{particle 1}:\quad \lambda_1 m_1\tilde\eta_1, \\
\text{particle 2}:\quad \tilde\lambda_2 m_2\tilde\eta_2, \\
\text{particle 3}:\quad \tilde\lambda_3\lambda_3.
\end{cases} \label{eq:VVV_3b}
\end{align}
These correspond to a transverse vector $\mathbb{A}^-$ (particle 1) and longitudinal vector $\mathbb{A}^0$ (particle 2 in the first amplitude, particle 3 in the second). 

In the first step, we identify particle 1 as $\mathbb{A}^-$, while particles 2 and 3 are treated as the current $\mathbb{J}$. Starting with the massless amplitude, this identification gives the scaling behavior
\begin{equation}
\frac{\langle12\rangle\langle31\rangle}{\langle23\rangle}\sim 
\begin{cases}
\text{current}:\quad \mathbb J(\lambda^0)\sim\lambda_2^0\lambda_3^0, \\
\text{vector}:\quad \mathbb A^-(\lambda_1^{2}).  \\
\end{cases}
\end{equation}
The scaling deformation is then performed as
\begin{equation}\begin{aligned}
\mathbb J(\lambda^0)\cdot\mathbb A^-(\lambda^2_1) \xrightarrow{\times\frac{[\eta\lambda]}{\tilde m}}
\mathbb J(\lambda^0)\cdot\mathbb A^-(\tilde\lambda_1\lambda_1^2\tilde m_1^{-1}\tilde\eta_1)
\xrightarrow{\text{IBP}^-}
\mathbb J(\tilde\lambda\lambda)\cdot\frac{\mathbb A^-(\lambda_1 m_1\tilde\eta_1)}{\mathbf m_1^2}.
\end{aligned}\end{equation}
The corresponding amplitude deformation is
\begin{equation}\begin{aligned}
\frac{\langle12\rangle\langle31\rangle}{\langle23\rangle}
\xrightarrow{\times\frac{[\eta\lambda]}{\tilde m}}
b_1\frac{[\eta_1 12\rangle\langle31\rangle}{\tilde m_1\langle23\rangle}-b_2\frac{\langle12\rangle\langle31\eta_1]}{\tilde m_1\langle23\rangle}
\xrightarrow{\text{IBP}^-}
b_1\frac{m_1[\eta_1 3]\langle31\rangle}{\mathbf m_1^2}-b_2\frac{m_1\langle12\rangle[2\eta_1]}{\mathbf m_1^2}.
\end{aligned}\end{equation}
where $b_1+b_2=1$. The IBP eliminate the spurious pole $\langle23\rangle$ and it generates two terms. They
both exhibiting the scaling behavior of $\mathbb A^-(\lambda_1\tilde m_1^{-1}\tilde\eta_1)$. 

In the second step, we match the first term to eq.~\eqref{eq:VVV_3a} and the second term to eq.~\eqref{eq:VVV_3b}. Thus, for the first term we identify particle 2 as $\mathbb A^0$, while for the second term we identify particle 3 as $\mathbb A^0$. This leads to the following scaling behavior,
\begin{equation} \begin{aligned}
\frac{m_1[\eta_1 3]\langle31\rangle}{\mathbf m_1^2}&\sim
\begin{cases}
\text{current}:\quad {\displaystyle\frac{\mathbb J(\tilde\lambda\lambda^2 m\tilde\eta)}{\mathbf m^2}\sim\frac{1}{\mathbf m_1^2}}\lambda_1m_1\tilde\eta_1\tilde\lambda_3\lambda_3, \\
\text{vector}:\quad \mathbb A^0(\lambda_2^0), 
\end{cases} \\
\frac{m_1\langle12\rangle[2\eta_1]}{\mathbf m_1^2}&\sim
\begin{cases}
\text{current}:\quad {\displaystyle\frac{\mathbb J(\tilde\lambda\lambda^2 m\tilde\eta)}{\mathbf m^2}\sim\frac{1}{\mathbf m_1^2}}\lambda_1m_1\tilde\eta_1\tilde\lambda_2\lambda_2, \\
\text{vector}:\quad \mathbb A^0(\lambda_3^0). 
\end{cases}
\end{aligned} \end{equation}
The scaling behavior of the first term can then be deformed as
\begin{equation}\begin{aligned}
\frac{\mathbb J(\tilde\lambda\lambda^2 m\tilde\eta)}{\mathbf m^2}
\cdot\mathbb A^0(\lambda^0_2) 
&\xrightarrow{\times\frac{[\eta\lambda]}{\tilde m}}
\frac{\mathbb J(\tilde\lambda\lambda^3 \tilde m^{-1}m\tilde\eta^2)}{\mathbf m^2}
\cdot\mathbb A^0(\lambda^0_2) \\
&\xrightarrow{\text{IBP}^+}
\frac{\mathbb J(\lambda^2 m^2\tilde\eta^2)}{\mathbf m^4}\cdot\mathbb A^0(\tilde \lambda_2 \lambda_2).
\end{aligned}\end{equation}
For the second term, we change the spinor in $\mathbb A^0$ from particle 1 to particle 2. Therefore, the corresponding amplitude deformation of the two terms gives 
\begin{equation}\begin{aligned}
b_1\frac{m_1[\eta_1 3]\langle31\rangle}{\mathbf m_1^2}-b_2\frac{m_1\langle12\rangle[2\eta_1]}{\mathbf m_1^2}
&\xrightarrow{\times\frac{[\eta\lambda]}{\tilde m}}
b_1\frac{m_1[\eta_1 3][\eta_331\rangle}{\mathbf m_1^2\tilde m_3}+b_2\frac{m_1\langle12\eta_2][2\eta_1]}{\mathbf m_1^2\tilde m_2}\\
&\xrightarrow{\text{IBP}^+}
-b_1\frac{m_1m_3[\eta_1 3][\eta_32]\langle21\rangle}{\mathbf m_1^2\mathbf m_3^2}-b_2\frac{m_1m_2\langle13\rangle[3\eta_2][2\eta_1]}{\mathbf m_1^2\mathbf m_2^2}.
\end{aligned}\end{equation}
Therefore, we match the massless amplitude to two MHC amplitudes, but the coefficients $b_1$ and $b_2$ do not determined.

3. The deformation described above is achieved by multiplying the amplitude solely by $[\lambda\eta]/\tilde m$. However, some deformations require both $\langle\lambda\eta\rangle/m$ and $[\lambda\eta]/\tilde m$. Consider the following two MHC amplitudes,
\begin{align}
\Ampthree{1^-}{2^0}{3^0}{\bos{i1}{red}}{\bosflipflip{1}{55}{brown}{red}{brown}}{\bos{i3}{brown}}=
-\tilde m_2 m_2[\eta_2 3]\langle31\rangle\langle1\eta_2\rangle\sim 
\begin{cases}
\text{particle 1}:\quad \lambda_1^2, \\
\text{particle 2}:\quad m_2\tilde\eta_2\tilde m_2\eta_2, \\
\text{particle 3}:\quad \tilde\lambda_3\lambda_3, 
\end{cases} \label{eq:VVV_4a}\\
\Ampthree{1^-}{2^0}{3^0}{\bos{i1}{red}}{\bos{i2}{brown}}{\bosflipflip{1}{-55}{brown}{red}{brown}}=
-\tilde m_3 m_3\langle12\rangle[\eta_3 2]\langle1\eta_3\rangle \sim 
\begin{cases}
\text{particle 1}:\quad \lambda_1^2, \\
\text{particle 2}:\quad \tilde\lambda_2\lambda_2, \\
\text{particle 3}:\quad m_3\tilde\eta_3\tilde m_3\eta_3. 
\end{cases} \label{eq:VVV_4b}
\end{align}

In the first step, we identify particle 3 or 2 as $\mathbb{A}^0$, while the other two particles are treated as the current $\mathbf J$. Starting with the massless amplitude, these two identifications give the similar scaling behavior
\begin{equation}
\frac{\langle12\rangle\langle31\rangle}{\langle23\rangle}\sim 
\begin{cases}
\text{current}:\quad \mathbb J(\lambda^2), \\
\text{vector}:\quad \mathbb A^0(\lambda_3^{0})\text{ or }\mathbb A^0(\lambda_2^{0}).  \\
\end{cases}
\end{equation}
When we identify particle 3 as $\mathbb{A}^0$, the scaling behavior can be deformed as
\begin{equation}\begin{aligned}
\mathbb J(\lambda^2)\cdot\mathbb A^0(\lambda_3^0) \xrightarrow{\times\frac{[\eta\lambda]}{\tilde m}}
\mathbb J(\tilde\lambda \lambda^2 \tilde m^{-1}\tilde\eta)\cdot\mathbb A^0(\lambda_3^0)
\xrightarrow{\text{IBP}^+}
\frac{\mathbb J(\lambda m\tilde\eta)}{\mathbf m^2}\cdot\mathbb A^0(\tilde \lambda_3 \lambda_3).
\end{aligned}\end{equation}
For the other identification, we change the spinor in $\mathbb A^0$ from particle 3 to particle 2. Thefore, the corresponding amplitude deformation yields two terms,
\begin{equation}\begin{aligned}
\frac{\langle12\rangle\langle31\rangle}{\langle23\rangle}
\xrightarrow{\times\frac{[\eta\lambda]}{\tilde m}}
-b_1\frac{[\eta_2 21\rangle\langle31\rangle}{\tilde m_2\langle23\rangle}+b_2\frac{\langle12\rangle[\eta_3 31\rangle}{\tilde m_3\langle23\rangle}
\xrightarrow{\text{IBP}^+}
b_1\frac{m_2[\eta_2 3]\langle31\rangle^2}{\mathbf m_2^2\langle23\rangle}+b_2\frac{m_3\langle12\rangle^2[\eta_3 2]}{\mathbf m_3^2\langle23\rangle}.
\end{aligned}\end{equation}
where $b_1+b_2=1$. The IBP generates two terms, corresponding to Goldstone behavior $\tilde \lambda_3 \lambda_3$ and $\tilde \lambda_2 \lambda_2$ respectively.

In the second step, we match the first term to eq.~\eqref{eq:VVV_4a} and the second term to eq.~\eqref{eq:VVV_4b}. Thus, for the first term we identify particle 2 as $\mathbb A^0$, while for the second term we identify particle 3 as $\mathbb A^0$. This leads to the following scaling behavior,
\begin{equation} \begin{aligned}
\frac{m_2[\eta_2 3]\langle31\rangle^2}{\mathbf m_2^2\langle23\rangle}&\sim
\begin{cases}
\text{current}:\quad \mathbb J(\tilde\lambda\lambda^3)\sim\lambda_1^2 \tilde\lambda_3\lambda_3, \\
\text{vector}:\quad {\displaystyle\frac{\mathbb A^0(\lambda_2^{-1}m_2\tilde\eta_2)}{\mathbf m_2^2}}, 
\end{cases} \\
\frac{m_3\langle12\rangle^2[\eta_3 2]}{\mathbf m_3^2\langle23\rangle}&\sim
\begin{cases}
\text{current}:\quad {\mathbb J(\tilde\lambda\lambda^3)\sim}\lambda_1^2 \tilde\lambda_2\lambda_2, \\
\text{vector}:\quad {\displaystyle\frac{\mathbb A^0(\lambda_3^{-1}m_3\tilde\eta_3)}{\mathbf m_3^2}}, 
\end{cases}
\end{aligned} \end{equation}
The scaling behavior of the first term can then be deformed as
\begin{equation}\begin{aligned}
\mathbb J(\tilde\lambda\lambda^3)\cdot\frac{\mathbb A^0(\lambda_2^{-1}m_2\tilde\eta_2)}{\mathbf m_2^2} \xrightarrow{\times\frac{\langle\lambda\eta\rangle}{m}}
\mathbb J(\tilde\lambda\lambda^3)\cdot\frac{\mathbb A^0(\tilde\eta_2 \eta_2)}{\mathbf m_2^2}
=\mathbb J(\tilde\lambda\lambda^3)\cdot\frac{\mathbb A^0(m_2\tilde\eta_2 \tilde m_2\eta_2)}{\mathbf m_2^4}.
\end{aligned}\end{equation}
For the second term, we change the spinor in $\mathbb A^0$ and square mass $\mathbf m^2_2$ from particle 2 to particle 3.  Therefore, the corresponding amplitude deformation of the two terms gives 
\begin{equation}\begin{aligned} \label{eq:deform1}
&b_1\frac{m_2[\eta_2 3]\langle31\rangle^2}{\mathbf m_2^2\langle23\rangle}+b_2\frac{m_3\langle12\rangle^2[\eta_3 2]}{\mathbf m_3^2\langle23\rangle}\\
&\xrightarrow{\times\frac{\langle\lambda\eta\rangle}{m}}
b_1\frac{[\eta_2 3]\langle31\rangle^2\langle2\eta_2\rangle}{\mathbf m_2^2\langle23\rangle}+b_2\frac{\langle12\rangle^2[\eta_3 2]\langle3\eta_3\rangle}{\mathbf m_3^2\langle23\rangle}\\
&\xrightarrow{\text{Schouten}}
-b_1\frac{m_2\tilde m_2[\eta_2 3]\langle31\rangle\langle1\eta_2\rangle}{\mathbf m_2^4}-b_2\frac{m_3\tilde m_3\langle12\rangle[\eta_3 2]\langle1\eta_3\rangle}{\mathbf m_3^4}+\Delta.
\end{aligned}\end{equation}
In the last line, we use the IBP to eliminate the spurious pole. The last term $\Delta$ denotes the term in which the spurious pole is not eliminated. It can be further deformed as
\begin{equation}\begin{aligned}
\Delta 
&=-\left(\frac{b_1}{\mathbf m_2^2}[\eta_2 3]\langle3\eta_2\rangle+\frac{b_2}{\mathbf m_3^2}[\eta_3 2]\langle2\eta_3\rangle\right)\frac{\langle12\rangle\langle31\rangle}{\langle23\rangle}.
\end{aligned}\end{equation}
This expression is proportional to the initial massless amplitude $\frac{\langle12\rangle\langle31\rangle}{\langle23\rangle}$, so we can rewrite the factor in terms of mass,
\begin{equation}\begin{aligned}
\frac{b_1}{\mathbf m_2^2}[\eta_2 3]\langle3\eta_2\rangle+\frac{b_2}{\mathbf m_3^2}[\eta_3 2]\langle2\eta_3\rangle
&\to\frac{b_1}{\mathbf m_2^2}([\eta_2 3]\langle3\eta_2\rangle+[\eta_3 2]\langle2\eta_3\rangle)\\
&=\frac{b_1}{\mathbf m_2^2}(\mathbf m_1^2-\mathbf m_2^2-\mathbf m_3^2).
\end{aligned}\end{equation}
In the last line, we use the identity $[\eta_2 3]\langle3\eta_2\rangle+[\eta_3 2]\langle2\eta_3\rangle=2p_2\cdot p_3=\mathbf m_1^2-\mathbf m_2^2-\mathbf m_3^2$. This rewriting requires the relation
\begin{equation}
\frac{b_1}{\mathbf m_2^2}=\frac{b_2}{\mathbf m_3^2}.
\end{equation}
Combining this with the normalization condition $b_1 + b_2 = 1$, we solve for the coefficients
\begin{equation}
b_1=\frac{\mathbf m_2^2}{\mathbf m_2^2+\mathbf m_3^2},\quad b_2=\frac{\mathbf m_3^2}{\mathbf m_2^2+\mathbf m_3^2}.
\end{equation}
Substituting into eq.~\eqref{eq:deform1} and simplifying, we finally obtain the amplitude matching in this case:
\begin{equation}
\frac{\langle12\rangle\langle31\rangle}{\langle23\rangle}\to
-\frac{\mathbf m_2^2}{\mathbf m_1^2}\frac{m_2\tilde m_2[\eta_2 3]\langle31\rangle\langle1\eta_2\rangle}{\mathbf m_2^4}-\frac{\mathbf m_3^2}{\mathbf m_1^2}\frac{m_3\tilde m_3\langle12\rangle[\eta_3 2]\langle1\eta_3\rangle}{\mathbf m_3^4}.
\end{equation}


4. The deformation using both  $\langle\lambda\eta\rangle/m$ and $[\lambda\eta]/\tilde m$ can also generate the MHC amplitude without descendant longitudinal vector scaling. Consider the following two MHC amplitudes,
\begin{align}
\Ampthree{1^-}{2^0}{3^0}{\bosflip{1}{180}{brown}{red}}{\bos{i2}{brown}}{\bosflip{1}{-55}{red}{brown}}=
m_1\tilde m_3([\eta_1 2]\langle2\eta_3\rangle\langle31\rangle+[\eta_1 2]\langle23\rangle\langle\eta_31\rangle)\sim
\begin{cases}
\text{particle 1}:\quad \lambda_1 m_1\tilde\eta_1, \\
\text{particle 2}:\quad \tilde\lambda_2\lambda_2, \\
\text{particle 3}:\quad \lambda_3\tilde m_3\eta_3, 
\end{cases}\\
\Ampthree{1^-}{2^0}{3^0}{\bosflip{1}{180}{brown}{red}}{\bosflip{1}{55}{red}{brown}}{\bos{i3}{brown}}=
m_1\tilde m_2(\langle12\rangle\langle\eta_23\rangle[3\eta_1]+\langle1\eta_2\rangle\langle23\rangle[3\eta_1])\sim
\begin{cases}
\text{particle 1}:\quad \lambda_1 m_1\tilde\eta_1, \\
\text{particle 2}:\quad \lambda_2\tilde m_2\eta_2, \\
\text{particle 3}:\quad \tilde\lambda_3\lambda_3.
\end{cases}
\end{align}
These correspond to a transverse vector $\mathbb{A}^-$ (particle 1) and a longitudinal vector $\mathbb{A}^0$ (particle 2 in the first amplitude, particle 3 in the second). In this case, we apply the gauge boson deformation and Goldstone boson deformation simultaneously. 

In the first step, we identify each leg as the vector $\mathbb A$, which leads to three types of identifications. Starting from the massless amplitude, these identifications give the scaling behavior
\begin{equation}
\frac{\langle12\rangle\langle31\rangle}{\langle23\rangle}\sim 
\begin{tabular}{c|c|c|c}
\hline
current & $\mathbb J(\lambda^0)$ & \multicolumn{2}{|c}{$\mathbb J(\lambda^2)$} \\
\hline
vector & $\mathbb A^-(\lambda_1^2)$ & $\mathbb A^0(\lambda_2^{0})$ & $\mathbb A^0(\lambda_3^{0})$ \\
\hline
\end{tabular}
\end{equation}
For the first identification, the scaling can be deformed as
\begin{equation}\begin{aligned}
\mathbb J(\lambda^0)\cdot\mathbb A^-(\lambda_1^2)
&\xrightarrow{\times\frac{[\eta\lambda]}{\tilde m}}
\mathbb J(\lambda^0)\cdot\mathbb A^-(\tilde\lambda_1\lambda_1^2\tilde m_1^{-1}\tilde\eta_1)
\xrightarrow{\text{IBP}^-}
\mathbb J(\tilde\lambda\lambda)\cdot\frac{\mathbb A^-(\lambda_1 m_1\tilde\eta_1)}{\mathbf m_1^2}.
\end{aligned}\end{equation}
The corresponding amplitude deformation is
\begin{equation}\begin{aligned} \label{eq:deform4}
\frac{\langle12\rangle\langle31\rangle}{\langle23\rangle}
&\xrightarrow{\times\frac{[\eta\lambda]}{\tilde m}}
b_1\frac{[\eta_1 12\rangle\langle31\rangle}{\tilde m_1\langle23\rangle}-b_2\frac{\langle12\rangle\langle31\eta_1]}{\tilde m_1\langle23\rangle}\\
&\xrightarrow{\text{IBP}^-}
b_1\frac{m_1[\eta_1 3]\langle31\rangle}{\mathbf m_1^2}-b_2\frac{m_1\langle12\rangle[2\eta_1]}{\mathbf m_1^2}.
\end{aligned}\end{equation}
For the second identification, the scaling deforms as
\begin{equation}\begin{aligned}
\mathbb J(\lambda^2)\cdot\mathbb A^0(\lambda_2^0) 
&\xrightarrow{\times\frac{\langle\lambda\eta\rangle}{m}}
\mathbb J(\lambda^3m^{-1}\eta)\cdot\mathbb A^0(\lambda_2^0)
=\frac{\mathbb J(\lambda^3\tilde m\eta)}{\mathbf m^2}\cdot\mathbb A^0(\lambda_2^0).
\end{aligned}\end{equation}
It does not have positive scaling for $\tilde\lambda$, so we cannot apply IBP. For the third identification, we change the spinor in $\mathbb A^0$  from particle 2 to particle 3. Therefore, the corresponding amplitude deformation for the second and the third identifications gives
\begin{equation}\begin{aligned} \label{eq:deform5}
\frac{\langle12\rangle\langle31\rangle}{\langle23\rangle}
&\xrightarrow{\times\frac{\langle\lambda\eta\rangle}{m}}
b_3\frac{\langle12\rangle\langle31\rangle\langle2\eta_2\rangle}{m_2\langle23\rangle}+b_4\frac{\langle12\rangle\langle31\rangle\langle3\eta_3\rangle}{m_3\langle23\rangle}\\
&\xrightarrow{\text{Schouten}}
-b_3\frac{\tilde m_2\langle12\rangle\langle1\eta_2\rangle}{\mathbf m_2^2}-b_4\frac{\tilde m_3\langle31\rangle\langle1\eta_3\rangle}{\mathbf m_3^2}+\Delta.
\end{aligned}\end{equation}
Here $\Delta$ denotes the term with spurious pole,
\begin{equation}\begin{aligned}
\Delta &=-b_3\frac{\tilde m_2\langle12\rangle^2\langle3\eta_2\rangle}{\mathbf m_2^2\langle23\rangle}-b_4\frac{\tilde m_3\langle31\rangle^2\langle2\eta_3\rangle}{\mathbf m_3^2\langle23\rangle}.
\end{aligned}\end{equation}
If $\frac{b_3}{\mathbf m_2^2}=\frac{b_4}{\mathbf m_3^2}$, $\Delta$ can be reduced to a form proportional to the initial massless amplitude,
\begin{equation}\begin{aligned}
\Delta &\to-\frac{b_3}{\mathbf m_2^2}\frac{\tilde m_2\langle12\rangle^2\langle3\eta_2\rangle+\tilde m_3\langle31\rangle^2\langle2\eta_3\rangle}{\langle23\rangle}\\
&=\frac{b_3}{\mathbf m_2^2}([\eta_2 3]\langle3\eta_2\rangle+[\eta_3 2]\langle2\eta_3\rangle)\frac{\langle12\rangle\langle13\rangle}{\langle23\rangle}\\
&=-\frac{b_3}{\mathbf m_2^2}(\mathbf m_1^2-\mathbf m_2^2-\mathbf m_3^2)\frac{\langle12\rangle\langle31\rangle}{\langle23\rangle}.
\end{aligned}\end{equation}

Summing the amplitude deformations from eqs.~\eqref{eq:deform4} and \eqref{eq:deform5}, we obtain:
\begin{equation}\begin{aligned}
\frac{\langle12\rangle\langle31\rangle}{\langle23\rangle}
\to& b_1\frac{m_1[\eta_1 3]\langle31\rangle}{\mathbf m_1^2}-b_2\frac{m_1\langle12\rangle[2\eta_1]}{\mathbf m_1^2}-b_3\frac{\tilde m_2\langle12\rangle\langle1\eta_2\rangle}{\mathbf m_2^2}\\
&-b_4\frac{\tilde m_3\langle31\rangle\langle1\eta_3\rangle}{\mathbf m_3^2}+b_3\frac{\mathbf m_1^2-\mathbf m_2^2-\mathbf m_3^2}{\mathbf m_2^2}\frac{\langle12\rangle\langle31\rangle}{\langle23\rangle}.
\end{aligned}\end{equation}
Rewriting this, the first-step deformation can be expressed as
\begin{equation}\begin{aligned}
\left(1+b_3\frac{\mathbf m_1^2-\mathbf m_2^2-\mathbf m_3^2}{\mathbf m_2^2}\right)\frac{\langle12\rangle\langle31\rangle}{\langle23\rangle}\to&
b_1\frac{m_1[\eta_1 3]\langle31\rangle}{\mathbf m_1^2}-b_2\frac{m_1\langle12\rangle[2\eta_1]}{\mathbf m_1^2}\\
&-b_3\frac{\tilde m_2\langle12\rangle\langle1\eta_2\rangle}{\mathbf m_2^2}-b_4\frac{\tilde m_3\langle31\rangle\langle1\eta_3\rangle}{\mathbf m_3^2}.
\end{aligned}\end{equation}

In the second step, we begin with the Goldstone boson deformation. We identify particle 2 in the $b_1$ term as the vector $\mathbb A^0$, and particle 3 in the $b_2$ term as the vector $\mathbb A^0$. This leads to the following scaling behavior,
\begin{equation} \begin{aligned}
b_1\frac{m_1[\eta_1 3]\langle31\rangle}{\mathbf m_1^2}&\sim
\begin{cases}
\text{current}:\quad {\displaystyle\frac{\mathbb J(\tilde\lambda\lambda^2 m\tilde\eta)}{\mathbf m^2}\sim\frac{1}{\mathbf m_1^2}\lambda_1 m_1\tilde\eta_1 \tilde\lambda_3\lambda_3}, \\
\text{vector}:\quad \mathbb A^0(\lambda_2^0), 
\end{cases} \\
-b_2\frac{m_1\langle12\rangle[2\eta_1]}{\mathbf m_1^2}&\sim
\begin{cases}
\text{current}:\quad {\displaystyle\frac{\mathbb J(\tilde\lambda\lambda^2 m\tilde\eta)}{\mathbf m^2}\sim\frac{1}{\mathbf m_1^2}\lambda_1 m_1\tilde\eta_1 \tilde\lambda_2\lambda_2}, \\
\text{vector}:\quad \mathbb A^0(\lambda_3^0), 
\end{cases}
\end{aligned} \end{equation}
The scaling behavior of the $b_1$ term can then be deformed as
\begin{equation}\begin{aligned}
\frac{\mathbb J(\tilde\lambda\lambda^2 m\tilde\eta)}{\mathbf m^2}\cdot\mathbb A^0(\lambda_2^0) 
&\xrightarrow{\times\frac{\langle\lambda\eta\rangle}{m}}
\frac{\mathbb J(\tilde\lambda\lambda^3 \tilde\eta\eta)}{\mathbf m^2}\cdot\mathbb A^0(\lambda_2^0)\xrightarrow{\text{IBP}^+}
\frac{\mathbb J(\lambda^2 m\tilde\eta \tilde m\eta)}{\mathbf m^4}\cdot\mathbb A^0(\tilde \lambda_2 \lambda_2).
\end{aligned}\end{equation}
For the $b_2$ term, we change the spinor in $\mathbb A^0$ from particle 2 to particle 3. The corresponding amplitude deformation for the $b_1$ and $b_2$ terms is then
\begin{equation}\begin{aligned} \label{eq:Goldstone_result}
&b_1\frac{m_1[\eta_1 3]\langle31\rangle}{\mathbf m_1^2}-b_2\frac{m_1\langle12\rangle[2\eta_1]}{\mathbf m_1^2}\\
&\xrightarrow{\times\frac{\langle\lambda\eta\rangle}{m}}
b_1\frac{m_1[\eta_1 3\eta_3\rangle\langle31\rangle}{\mathbf m_1^2 m_3}+b_2\frac{m_1\langle12\rangle\langle\eta_22\eta_1]}{\mathbf m_1^2 m_2}\\
&\xrightarrow{\text{IBP}^+}
-b_1\frac{m_1\tilde m_3[\eta_1 2]\langle2\eta_3\rangle\langle31\rangle}{\mathbf m_1^2\mathbf m_3^2}-b_2\frac{m_1\tilde m_2\langle12\rangle\langle\eta_23\rangle[3\eta_1]}{\mathbf m_1^2\mathbf m_3^2}\\
&\qquad\quad -b_2\frac{\tilde m_2\langle12\rangle\langle1\eta_2\rangle}{\mathbf m_2^2}-b_1\frac{\tilde m_3\langle31\rangle\langle1\eta_3\rangle}{\mathbf m_3^2}.
\end{aligned}\end{equation}
Only the first two terms exhibit Goldstone behavior $\tilde\lambda_2\lambda_2$ or $\tilde\lambda_3\lambda_3$. The last two terms share the same spinor structure as the $b_3$ and $b_4$ terms, so the total contribution should have the coefficients $(b_2+b_3)$ and $(b_1+b_4)$.

For these two terms, we apply the Gauge boson deformation. Identifying particle 1 as the transverse vector $\mathbb A^-(\lambda_1^2)$, we obtain
\begin{equation} \begin{aligned}
-(b_2+b_3)\frac{\tilde m_2\langle12\rangle\langle1\eta_2\rangle}{\mathbf m_2^2}&\sim
\begin{cases}
\text{current}:\quad {\displaystyle\frac{\mathbb J(\lambda\tilde m\eta)}{\mathbf m^2}\sim\frac{1}{\mathbf m_2^2}\lambda_2\tilde m_2\eta_2}, \\
\text{vector}:\quad \mathbb A^-(\lambda_1^2), 
\end{cases} \\
-(b_1+b_4)\frac{\tilde m_3\langle31\rangle\langle1\eta_3\rangle}{\mathbf m_3^2}&\sim
\begin{cases}
\text{current}:\quad {\displaystyle\frac{\mathbb J(\lambda\tilde m\eta)}{\mathbf m^2}\sim\frac{1}{\mathbf m_3^2}\lambda_3\tilde m_3\eta_3}, \\
\text{vector}:\quad {\displaystyle\mathbb A^-(\lambda_1^2)}, 
\end{cases}
\end{aligned} \end{equation}
The scaling deformation is then given by
\begin{equation}\begin{aligned}
\frac{\mathbb J(\lambda\tilde m\eta)}{\mathbf m^2}\cdot\mathbb A^-(\lambda_1^2)
&\xrightarrow{\frac{[\eta\lambda]}{\tilde m}}
\frac{\mathbb J(\lambda\tilde m\eta)}{\mathbf m^2}\cdot\mathbb A^-(\tilde\lambda_1\lambda_1^2\tilde m_1^{-1}\tilde\eta_1)\xrightarrow{\text{IBP}^-}
\frac{\mathbb J(\tilde\lambda\lambda^2\tilde m\eta)}{\mathbf m^2}\cdot\frac{\mathbb A^-(\lambda_1 m_1\tilde\eta_1)}{\mathbf m_1^2}.
\end{aligned}\end{equation}
The corresponding amplitude deformation becomes
\begin{equation}\begin{aligned} \label{eq:deform2}
&-(b_2+b_3)\frac{\tilde m_2\langle12\rangle\langle1\eta_2\rangle}{\mathbf m_2^2}-(b_1+b_4)\frac{\tilde m_3\langle31\rangle\langle1\eta_3\rangle}{\mathbf m_3^2}\\
&\xrightarrow{\times\frac{[\eta\lambda]}{\tilde m}}
-(b_2+b_3)\frac{\tilde m_2[\eta_112\rangle\langle1\eta_2\rangle}{\mathbf m_2^2\tilde m_1}+(b_1+b_4)\frac{\tilde m_3\langle31\eta_1]\langle1\eta_3\rangle}{\mathbf m_3^2\tilde m_1}\\
&\xrightarrow{\text{IBP}^-}
(b_2+b_3)\frac{\tilde m_2 m_1[\eta_13]\langle32\rangle\langle1\eta_2\rangle}{\mathbf m_2^2 \mathbf m_1^2}-(b_1+b_4)\frac{\tilde m_3 m_1\langle32\rangle[2\eta_1]\langle1\eta_3\rangle}{\mathbf m_3^2 \mathbf m_1^2}.\\
\end{aligned}\end{equation}
We require these two terms to combine with the first two terms in eq.~\eqref{eq:Goldstone_result} as follows
\begin{equation}\begin{aligned} \label{eq:deform3}
\left(1+b_3\frac{\mathbf m_1^2-\mathbf m_2^2-\mathbf m_3^2}{\mathbf m_2^2}\right)\frac{\langle12\rangle\langle31\rangle}{\langle23\rangle}\to
&-b_1\frac{m_1 \tilde m_3[\eta_1 2]\langle2\eta_3\rangle\langle31\rangle+[\eta_1 2]\langle23\rangle\langle\eta_31\rangle}{\mathbf m_1^2 \mathbf m_3^2}\\
&-b_2\frac{m_1\tilde m_2\langle12\rangle\langle\eta_23\rangle[3\eta_1]+\langle1\eta_2\rangle\langle23\rangle[3\eta_1]}{\mathbf m_1^2 \mathbf m_2^2}.
\end{aligned}\end{equation}
This require the relations,
\begin{equation}\begin{aligned}
b_1+b_4&=-b_1,\\
b_2+b_3&=-b_2.\\
\end{aligned}\end{equation}
Combining these with the relation from the first step
\begin{equation}\begin{aligned}
\frac{b_3}{\mathbf m_2^2}=\frac{b_4}{\mathbf m_3^2},
\end{aligned}\end{equation}
and the normalization condition $b_1+b_2+b_3+b_4=1$, we can solve for the coefficients
\begin{equation}\begin{aligned}
b_1&=-\frac{\mathbf m_3^2}{\mathbf m_2^2+\mathbf m_3^2},&
b_2&=-\frac{\mathbf m_2^2}{\mathbf m_2^2+\mathbf m_3^2},&\\
b_3&=\frac{2\mathbf m_2^2}{\mathbf m_2^2+\mathbf m_3^2},&
b_4&=\frac{2\mathbf m_3^2}{\mathbf m_2^2+\mathbf m_3^2}.&
\end{aligned}\end{equation}
Substituting into eq.~\eqref{eq:deform3} and rescaling, we finally obtain
\begin{equation} \begin{aligned}
\frac{\langle12\rangle\langle31\rangle}{\langle23\rangle}
\to&
\frac{\mathbf m_3^2}{2\mathbf m_1^2-\mathbf m_2^2-\mathbf m_3^2}\frac{m_1 \tilde m_3[\eta_1 2]\langle2\eta_3\rangle\langle31\rangle+[\eta_1 2]\langle23\rangle\langle\eta_31\rangle}{\mathbf m_1^2 \mathbf m_3^2}\\
&+\frac{\mathbf m_2^2}{2\mathbf m_1^2-\mathbf m_2^2-\mathbf m_3^2}\frac{m_1\tilde m_2\langle12\rangle\langle\eta_23\rangle[3\eta_1]+\langle1\eta_2\rangle\langle23\rangle[3\eta_1]}{\mathbf m_1^2 \mathbf m_2^2}.
\end{aligned} \end{equation}

The above matching result is summarized in the following table,
\begin{equation}
\includegraphics[width=0.9\linewidth]{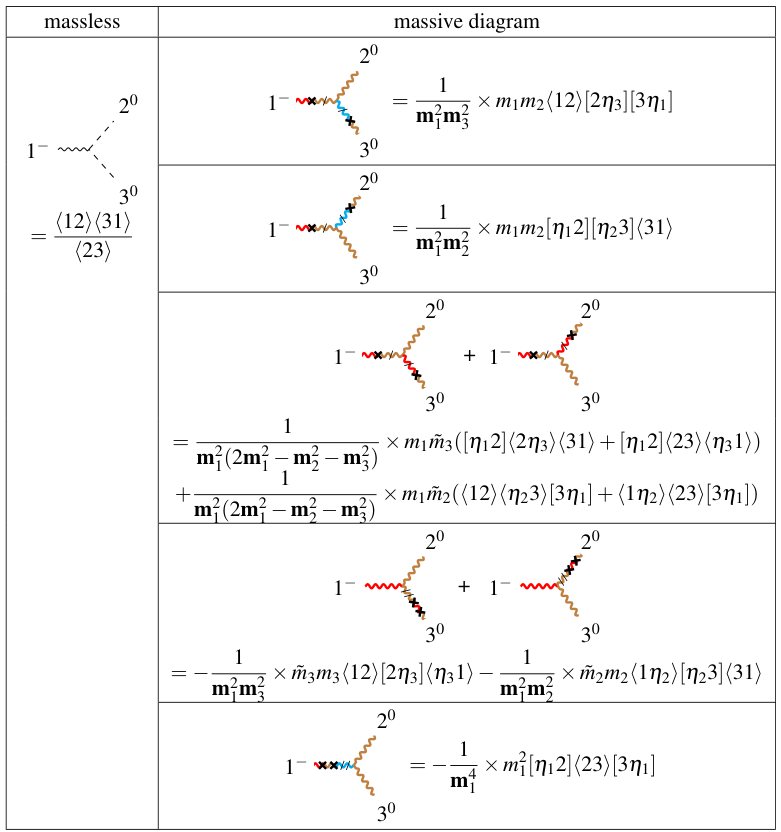}
\end{equation}

\section{Sub-leading matching from higher point massless amplitudes}
\label{app:subleading}

In the main text, we demonstrated that the sub-leading matching can be obtained by combining the leading-order matching with the Higgs insertion technique. However, it is also possible to derive it directly from higher-point massless amplitudes, without relying on the leading-order result. In section~\ref{sec:FFS}, we already considered matching higher-point massless amplitudes to the massive $FFS$ amplitude. In the following, we will use the 4-pt massless amplitude as an example to illustrate how this direct sub-leading matching can be implemented for amplitudes involving massive vector bosons.

\subsection{FFV subleading matching}



First let us consider to match the massless $FFSS$ amplitude to the massive $FFV$ amplitude. The 4-pt massless amplitude can be constructed using methods such as BCFW recursion or bootstrap techniques. These approaches rely on taking the on-shell limit of propagators, which requires knowledge of all possible topologies of the massless amplitude. For the $FFSS$ amplitude, the possible topologies are as follows,
\begin{eqnarray}
&FFSS:&\quad 
\begin{tikzpicture}[baseline=0.7cm] \begin{feynhand}
\setlength{\feynhandarrowsize}{3.5pt}
\vertex [particle] (i1) at (-0.2,0.8) {$1^-$}; 
\vertex [particle] (i2) at (1.6,1.6) {$2^+$}; 
\vertex [particle] (i3) at (1.6,0) {$3^0$};  
\vertex [particle] (i5) at (0.7,0.2) {$4$};
\vertex (v3) at (0.6,0.8);
\vertex (v1) at (0.9,0.8); 
\graph{(i1)--[fer](v3)--[fer](v1)};
\graph{(v1)--[fer](i2)};
\graph{(i3)--[sca] (v1)};  
\graph{(v3)--[sca](i5)};
\end{feynhand} \end{tikzpicture},\quad
\begin{tikzpicture}[baseline=0.7cm] \begin{feynhand}
\setlength{\feynhandarrowsize}{3.5pt}
\vertex [particle] (i1) at (0,0.8) {$1^-$}; 
\vertex [particle] (i2) at (1.6,1.6) {$2^+$}; 
\vertex [particle] (i3) at (1.6,0) {$3^0$};  
\vertex [particle] (i4) at (1.6,0.8) {$4$};
\vertex (v2) at (0.9+0.7*0.33,0.8+0.8*0.33);
\vertex (v1) at (0.9,0.8); 
\graph{(i1)--[fer](v1)};
\graph{(v1)--[fer](v2)--[fer](i2)};
\graph{(i3)--[sca] (v1)};  
\graph{(v2)--[sca](i4)};
\end{feynhand} \end{tikzpicture}, \quad
\begin{tikzpicture}[baseline=0.7cm] \begin{feynhand}
\setlength{\feynhandarrowsize}{3.5pt}
\vertex [particle] (i1) at (0,0.8) {$1^-$}; 
\vertex [particle] (i2) at (1.6,1.6) {$2^+$}; 
\vertex [particle] (i3) at (1.6,0) {$3^0$};  
\vertex [particle] (i4) at (1.6,0.8) {$4$};
\vertex (v2) at (0.9+0.7*0.33,0.8-0.8*0.33);
\vertex (v1) at (0.9,0.8); 
\graph{(i1)--[fer](v1)};
\graph{(v1)--[fer](i2)};
\graph{(i3)--[sca](v2)--[bos] (v1)};  
\graph{(v2)--[sca](i4)};
\end{feynhand} \end{tikzpicture}, \label{eq:FFSS_diagram}
\end{eqnarray}
where particle 4 is the additional Higgs boson. Using these topologies, we obtain the massless amplitude in the helicity $(-\frac12,+\frac12,0,0)$,
\begin{equation} \label{eq:FFSS}
\mathcal{A}^{(-\frac12,+\frac12,0,0)}=
G_1 \frac{\langle 1|P_{14}|2]}{s_{14}}+
G_2 \frac{\langle 1|P_{14}|2]}{s_{24}}+
G_3 \frac{\langle 1|3-4|2]}{s_{34}},
\end{equation}
where $G_i$ are the massless coefficients. The three terms correspond one-to-one with the three diagrams in eq.~\eqref{eq:FFSS_diagram}. This massless amplitude can be matched to the MHC amplitudes with helicity $(-\frac12,+\frac12,0)$,  represented as
\begin{equation} \begin{aligned} \label{eq:MHC_subleading_1}
\Ampthree{1^-}{2^+}{3^0}{\ferflip{1}{180}{red}{cyan}}{\antferflip{1}{55}{red}{cyan}}{\bos{i3}{brown}},
\Ampthree{1^-}{2^+}{3^0}{\fer{red}{i1}{v1}}{\antfer{red}{i2}{v1}}{\bosflipflip{1}{-55}{brown}{red}{brown}}.
\end{aligned} \end{equation}
Each MHC diagram has two chirality flips, but only one comes from the Higgs insertion. 

When the additional Higgs is inserted on particle 1 or 2, the massless amplitude matches the first MHC diagram. Here we choose to insert the Higgs on particle 1, so we need to multiply a unit quantity  $\frac{\langle2\eta_2\rangle}{m_2}$ to expalin
the chirality flip on particle 2,
\begin{equation} \label{eq:multiply_f}
|2]\times \frac{\langle2\eta_2\rangle}{m_2}=\frac{p_2|\eta_2\rangle}{m_2}.
\end{equation}
Applying this to eq.~\eqref{eq:FFSS}, the massless amplitude can be deformed as
\begin{equation} \begin{aligned}
\mathcal{A}^{(-\frac12,+\frac12,0,0)}
&\overset{\times \frac{\langle2\eta_2\rangle}{m_2}}{=}
\frac{1}{m_2}G_1 \frac{\langle 1|P_{14}2|\eta_2\rangle}{s_{14}}+\left(G_2 \frac{\langle 1|P_{14}|2]}{s_{24}}+G_3 \frac{\langle 1|3-4|2]}{s_{34}}\right)\\
&\overset{\text{IBP}}{=}
-\frac{1}{m_2}G_1 \frac{\langle 1|P_{14}|3]\langle3\eta_2\rangle}{s_{14}}+
\left(-\frac{1}{m_2}G_1 \langle 1\eta_2\rangle +G_2 \frac{\langle 1|P_{14}|2]}{s_{13}}+G_3 \frac{\langle 1|3-4|2]}{s_{12}}\right).
\end{aligned} \end{equation}
Only the pole $s_{14}$ is related to the Higgs insertion on particle 1, so we only multiply $\frac{\langle2\eta_2\rangle}{m_2}$ to the term with this pole. For the other poles, we express them in a form that does not involve the Higgs boson momentum $p_4$.
Applying the on-shell limit $p_4\to \eta_1$ to these new form, we obtain the chirality flip for particle 1,
\begin{equation} \begin{aligned}
\begin{tikzpicture}[baseline=0.7cm] \begin{feynhand}
\setlength{\feynhandarrowsize}{3.5pt}
\vertex [particle] (i1) at (-0.2,0.8) {$1^-$}; 
\vertex [particle] (i2) at (1.6,1.6) {$2^+$}; 
\vertex [particle] (i3) at (1.6,0) {$3^0$};  
\vertex [particle] (i5) at (0.7,0.2) {$4$};
\vertex (v3) at (0.6,0.8);
\vertex (v1) at (0.9,0.8); 
\graph{(i1)--[fer](v3)--[fer](v1)};
\graph{(v1)--[fer](i2)};
\graph{(i3)--[sca] (v1)};  
\graph{(v3)--[sca](i5)};
\end{feynhand} \end{tikzpicture}
=\lim_{p_4\rightarrow \eta_1}\mathcal{A}^{(-\frac12,+\frac12,0,0)}\to G_1\frac{1}{m_2}\frac{[\eta_1 3]\langle3\eta_2\rangle}{\tilde{m}_1}+\mathcal{O}(E^{-2}).\\
\end{aligned} \end{equation}
The leading term, which scales as $E^0$, matches the first MHC diagram. Similarly, when we choose the Higgs insertion for particle 2, the chirality flip of the particle 1 should correspond to the multiplication. They match to the same diagrams.
\begin{equation} \begin{aligned} \label{eq:FFSS_FFV_diagram}
\begin{tikzpicture}[baseline=0.7cm] \begin{feynhand}
\setlength{\feynhandarrowsize}{3.5pt}
\vertex [particle] (i1) at (-0.2,0.8) {$1^-$}; 
\vertex [particle] (i2) at (1.6,1.6) {$2^+$}; 
\vertex [particle] (i3) at (1.6,0) {$3^0$};  
\vertex [particle] (i5) at (0.7,0.2) {$4$};
\vertex (v3) at (0.6,0.8);
\vertex (v1) at (0.9,0.8); 
\graph{(i1)--[fer](v3)--[fer](v1)};
\graph{(v1)--[fer](i2)};
\graph{(i3)--[sca] (v1)};  
\graph{(v3)--[sca](i5)};
\end{feynhand} \end{tikzpicture},
\begin{tikzpicture}[baseline=0.7cm] \begin{feynhand}
\setlength{\feynhandarrowsize}{3.5pt}
\vertex [particle] (i1) at (0,0.8) {$1^-$}; 
\vertex [particle] (i2) at (1.6,1.6) {$2^+$}; 
\vertex [particle] (i3) at (1.6,0) {$3^0$};  
\vertex [particle] (i4) at (1.6,0.8) {$4$};
\vertex (v2) at (0.9+0.7*0.33,0.8+0.8*0.33);
\vertex (v1) at (0.9,0.8); 
\graph{(i1)--[fer](v1)};
\graph{(v1)--[fer](v2)--[fer](i2)};
\graph{(i3)--[sca] (v1)};  
\graph{(v2)--[sca](i4)};
\end{feynhand} \end{tikzpicture}\rightarrow
\Ampthree{1^-}{2^+}{3^0}{\ferflip{1}{180}{red}{cyan}}{\antferflip{1}{55}{red}{cyan}}{\bos{i3}{brown}}.
\end{aligned} \end{equation}

Then we consider the Higgs insertion for particle $3$, it only match to the second MHC diagram in eq.~\eqref{eq:MHC_subleading_1}. In that diagram, both chirality flips happen in the particle 3. In this case, we do not need to mutliply any unity quantity, the matching is
\begin{equation} \begin{aligned}
\begin{tikzpicture}[baseline=0.7cm] \begin{feynhand}
\setlength{\feynhandarrowsize}{3.5pt}
\vertex [particle] (i1) at (0,0.8) {$1^-$}; 
\vertex [particle] (i2) at (1.6,1.6) {$2^+$}; 
\vertex [particle] (i3) at (1.6,0) {$3^0$};  
\vertex [particle] (i4) at (1.6,0.8) {$4$};
\vertex (v2) at (0.9+0.7*0.33,0.8-0.8*0.33);
\vertex (v1) at (0.9,0.8); 
\graph{(i1)--[fer](v1)};
\graph{(v1)--[fer](i2)};
\graph{(i3)--[sca](v2)--[bos] (v1)};  
\graph{(v2)--[sca](i4)};
\end{feynhand} \end{tikzpicture}
&=\lim_{p_4\rightarrow \eta_3}\mathcal{A}^{(-\frac12,+\frac12,0,0)} \\
&=\lim_{p_4\rightarrow \eta_3}G_3\frac{\langle 1|3-4|2]}{s_{34}}+\left(G_1 \frac{\langle 1|P_{14}|2]}{s_{23}}+
G_2 \frac{\langle 1|P_{14}|2]}{s_{13}}\right)\\
&\to G_3\frac{\langle 13\rangle[32]-\langle 1\eta_3\rangle[\eta_3 2]}{\mathbf{m}_3^2}+\mathcal{O}(E^{-2}).  
\end{aligned} \end{equation}
The first term corresponds to both the primary and descendant MHC diagrams,
\begin{equation} \begin{aligned}
\begin{tikzpicture}[baseline=0.7cm] \begin{feynhand}
\setlength{\feynhandarrowsize}{3.5pt}
\vertex [particle] (i1) at (0,0.8) {$1^-$}; 
\vertex [particle] (i2) at (1.6,1.6) {$2^+$}; 
\vertex [particle] (i3) at (1.6,0) {$3^0$};  
\vertex [particle] (i4) at (1.6,0.8) {$4$};
\vertex (v2) at (0.9+0.7*0.33,0.8-0.8*0.33);
\vertex (v1) at (0.9,0.8); 
\graph{(i1)--[fer](v1)};
\graph{(v1)--[fer](i2)};
\graph{(i3)--[sca](v2)--[bos] (v1)};  
\graph{(v2)--[sca](i4)};
\end{feynhand} \end{tikzpicture}\rightarrow
\Ampthree{1^-}{2^+}{3^0}{\fer{red}{i1}{v1}}{\antfer{red}{i2}{v1}}{\bos{i3}{brown}}+
\Ampthree{1^-}{2^+}{3^0}{\fer{red}{i1}{v1}}{\antfer{red}{i2}{v1}}{\bosflipflip{1}{-55}{brown}{red}{brown}}.
\end{aligned} \end{equation}
The second MHC diagram gives the subleading result we are interested in.

Next, we consider the massless amplitude with gauge boson. We examine the massless $FFVS$ amplitude with helicity $(-\frac12,-\frac12,+1,0)$ and match it to the massive $FFV$ amplitude. This massless amplitude has three topologies,
\begin{eqnarray}
&FFVS:&\quad 
\begin{tikzpicture}[baseline=0.7cm] \begin{feynhand}
\setlength{\feynhandarrowsize}{3.5pt}
\vertex [particle] (i1) at (0,0.8) {$1^-$}; 
\vertex [particle] (i2) at (1.6,1.6) {$2^-$}; 
\vertex [particle] (i3) at (1.6,0) {$3^+$};  
\vertex [particle] (i5) at (0.7,0.2) {$4$};
\vertex (v3) at (0.6,0.8);
\vertex (v1) at (0.9,0.8); 
\graph{(i1)--[fer](v3)--[fer](v1)};
\graph{(v1)--[fer](i2)};
\graph{(i3)--[bos] (v1)};  
\graph{(v3)--[sca](i5)};
\end{feynhand} \end{tikzpicture}, \quad
\begin{tikzpicture}[baseline=0.7cm] \begin{feynhand}
\setlength{\feynhandarrowsize}{3.5pt}
\vertex [particle] (i1) at (0,0.8) {$1^-$}; 
\vertex [particle] (i2) at (1.6,1.6) {$2^-$}; 
\vertex [particle] (i3) at (1.6,0) {$3^+$};  
\vertex [particle] (i4) at (1.6,0.8) {$4$};
\vertex (v2) at (0.9+0.7*0.33,0.8+0.8*0.33);
\vertex (v1) at (0.9,0.8); 
\graph{(i1)--[fer](v1)};
\graph{(v1)--[fer](v2)--[fer](i2)};
\graph{(i3)--[bos] (v1)};  
\graph{(v2)--[sca](i4)};
\end{feynhand} \end{tikzpicture},\quad
\begin{tikzpicture}[baseline=0.7cm] \begin{feynhand}
\setlength{\feynhandarrowsize}{3.5pt}
\vertex [particle] (i1) at (0,0.8) {$1^-$}; 
\vertex [particle] (i2) at (1.6,1.6) {$2^-$}; 
\vertex [particle] (i3) at (1.6,0) {$3^+$};  
\vertex [particle] (i4) at (1.6,0.8) {$4$};
\vertex (v2) at (0.9+0.7*0.33,0.8-0.8*0.33);
\vertex (v1) at (0.9,0.8); 
\graph{(i1)--[fer](v1)};
\graph{(v1)--[fer](i2)};
\graph{(i3)--[bos](v2)--[sca] (v1)};  
\graph{(v2)--[sca](i4)};
\end{feynhand} \end{tikzpicture}. \label{eq:FFVS_diagram}
\end{eqnarray}
Using either BCFW recursion or bootstrap technique, we obtain 
\begin{equation} \label{eq:FFVS}
\mathcal{A}^{(-\frac12,-\frac12,+1,0)}=G_1 \frac{[13][43]\langle12\rangle\langle14\rangle}{s_{14}s_{34}}+
G_2 \frac{[23][43]\langle21\rangle\langle24\rangle}{s_{24}s_{34}}.
\end{equation}
Due to the multipole structure of this amplitude, the two terms do not correspond one-to-one with the diagrams in eq.~\eqref{eq:FFVS_diagram}. This massless amplitude match to the following MHC structures:
\begin{equation} \begin{aligned}
\Ampthree{1^-}{2^-}{3^+}{\ferflip{1}{180}{red}{cyan}}{\antfer{cyan}{i2}{v1}}{\bosflip{1}{-55}{brown}{cyan}},
\Ampthree{1^-}{2^-}{3^+}{\fer{red}{i1}{v1}}{\antferflip{1}{55}{cyan}{red}}{\bosflip{1}{-55}{brown}{cyan}}.
\end{aligned} \end{equation}

We now match the massless amplitude to the first MHC diagram. The Higgs can be inserted on particle 1 or 3. Choosing to insert it on particle 1, we account for the chirality flip on particle 3 by multiplying with
\begin{equation} \label{eq:multiply_v}
|3]\times\frac{\langle3\eta_3\rangle}{m_3}=\frac{p_3|\eta_3\rangle}{m_3}.
\end{equation}
Applying this to eq.~\eqref{eq:FFVS}, we can isolate the multipole $s_{14}s_{12}$,
\begin{equation} \begin{aligned}
\mathcal{A}^{(-\frac12,-\frac12,+1,0)}
&\overset{\times \frac{\langle3\eta_3\rangle}{m_3}}{=}\frac{1}{m_3} G_1 \frac{[1|3|\eta_3\rangle[43]\langle12\rangle\langle14\rangle}{s_{14}s_{12}}+G_2 \frac{[23][43]\langle21\rangle\langle24\rangle}{s_{13}s_{12}}\\
&\overset{\text{IBP}}{=}\frac{1}{m_3} G_1 \frac{\langle2 \eta_3\rangle[43]\langle14\rangle}{s_{14}}+ \left(\frac{1}{m_3} G_1 \frac{\langle4 \eta_3\rangle[43]\langle12\rangle}{s_{12}}+G_2 \frac{[23][43]\langle21\rangle\langle24\rangle}{s_{13}s_{12}}\right).
\end{aligned} \end{equation}
Here, we do not isolate the other multipole $s_{13}s_{12}$, as it will contribute to subleading terms $\mathcal{O}(E^{-2})$, as shown below.
In the on-shell limit, the corresponding amplitudes reduce to
\begin{equation} \begin{aligned} \label{eq:s14FFVS}
\begin{tikzpicture}[baseline=0.7cm] \begin{feynhand}
\setlength{\feynhandarrowsize}{3.5pt}
\vertex [particle] (i1) at (0,0.8) {$1^-$}; 
\vertex [particle] (i2) at (1.6,1.6) {$2^-$}; 
\vertex [particle] (i3) at (1.6,0) {$3^+$};  
\vertex [particle] (i5) at (0.7,0.2) {$4$};
\vertex (v3) at (0.6,0.8);
\vertex (v1) at (0.9,0.8); 
\graph{(i1)--[fer](v3)--[fer](v1)};
\graph{(v1)--[fer](i2)};
\graph{(i3)--[bos] (v1)};  
\graph{(v3)--[sca](i5)};
\end{feynhand} \end{tikzpicture}=\lim_{p_4\rightarrow \eta_1}\mathcal{A}^{(-\frac12,-\frac12,+1,0)}
&=G_1\frac{1}{m_3}\frac{\langle2 \eta_3\rangle[\eta_1 3]}{\tilde{m}_1}+\mathcal{O}(E^{-2})
\rightarrow
\Ampthree{1^-}{2^-}{3^+}{\ferflip{1}{180}{red}{cyan}}{\antfer{cyan}{i2}{v1}}{\bosflip{1}{-55}{brown}{cyan}}.\\
\end{aligned} \end{equation}
Similarly, inserting the Higgs on particle 2 leads to a matching with the other MHC diagram,
\begin{equation} \begin{aligned}
\begin{tikzpicture}[baseline=0.7cm] \begin{feynhand}
\setlength{\feynhandarrowsize}{3.5pt}
\vertex [particle] (i1) at (0,0.8) {$1^-$}; 
\vertex [particle] (i2) at (1.6,1.6) {$2^-$}; 
\vertex [particle] (i3) at (1.6,0) {$3^+$};  
\vertex [particle] (i4) at (1.6,0.8) {$4$};
\vertex (v2) at (0.9+0.7*0.33,0.8+0.8*0.33);
\vertex (v1) at (0.9,0.8); 
\graph{(i1)--[fer](v1)};
\graph{(v1)--[fer](v2)--[fer](i2)};
\graph{(i3)--[bos] (v1)};  
\graph{(v2)--[sca](i4)};
\end{feynhand} \end{tikzpicture}&\rightarrow
\Ampthree{1^-}{2^-}{3^+}{\fer{red}{i1}{v1}}{\antferflip{1}{55}{cyan}{red}}{\bosflip{1}{-55}{brown}{cyan}}.
\end{aligned} \end{equation}

If the Higgs is inserted on particle 3, the massless amplitude can match both of the two MHC diagrams. In this case, the chirality flip on the two fermion line corresponds to multiplying with $\frac{[\eta\lambda]}{\tilde m}$. For convenience, we rewrite the massless amplitude as 
\begin{equation} \begin{aligned}
\mathcal{A}^{(-\frac12,-\frac12,+1,0)}=-G_1 \frac{[23][43]\langle12\rangle\langle24\rangle}{s_{14}s_{34}}-
G_2 \frac{[13][43]\langle21\rangle\langle14\rangle}{s_{24}s_{34}}.
\end{aligned} \end{equation}
Multiplying with $\frac{[\eta_1 1]}{\tilde m_1}$ and $\frac{[\eta_2 2]}{\tilde m_2}$ to the first and second term respectively, we obtain
\begin{equation} \begin{aligned}
\mathcal{A}^{(-\frac12,-\frac12,+1,0)}
\overset{\times \frac{[\eta\lambda]}{\tilde m}}{=}&-\frac{1}{\tilde m_1}G_1 \frac{[23][43][\eta_1|1|2\rangle\langle24\rangle}{s_{23}s_{34}}
-\frac{1}{\tilde m_2}G_2\frac{[13][43][\eta_2|2|1\rangle\langle14\rangle}{s_{13}s_{34}}\\
\overset{\text{IBP}}{=}&\frac{1}{\tilde m_1}G_1 \frac{[43][\eta_13]\langle24\rangle}{s_{34}}
+\frac{1}{\tilde m_2}G_2\frac{[43][\eta_23]\langle14\rangle}{s_{34}}\\
&+ \left(G_1 \frac{1}{\tilde m_1}\frac{[23][43][\eta_1 4]\langle42\rangle\langle24\rangle}{s_{23}s_{34}}+G_2 \frac{[13][43][\eta_2|4|1\rangle\langle14\rangle}{s_{13}s_{34}}\right).
\end{aligned} \end{equation}
Applying the on-shell limit then gives
\begin{equation}
\begin{tikzpicture}[baseline=0.7cm] \begin{feynhand}
\setlength{\feynhandarrowsize}{3.5pt}
\vertex [particle] (i1) at (0,0.8) {$1^-$}; 
\vertex [particle] (i2) at (1.6,1.6) {$2^-$}; 
\vertex [particle] (i3) at (1.6,0) {$3^+$};  
\vertex [particle] (i4) at (1.6,0.8) {$4$};
\vertex (v2) at (0.9+0.7*0.33,0.8-0.8*0.33);
\vertex (v1) at (0.9,0.8); 
\graph{(i1)--[fer](v1)};
\graph{(v1)--[fer](i2)};
\graph{(i3)--[bos](v2)--[sca] (v1)};  
\graph{(v2)--[sca](i4)};
\end{feynhand} \end{tikzpicture}
=\lim_{p_4\rightarrow \eta_3}\mathcal{A}
=G_1\frac{1}{\tilde{m}_1 m_3}[\eta_1 3]\langle\eta_3 2\rangle+G_2\frac{1}{\tilde{m}_2 m_3}[\eta_2 3]\langle\eta_3 1\rangle+\mathcal{O}(E^{-2}).
\end{equation}
The first term matches eq.~\eqref{eq:s14FFVS}, confirming that the amplitude corresponds to both MHC diagrams
\begin{equation} \begin{aligned}
\begin{tikzpicture}[baseline=0.7cm] \begin{feynhand}
\setlength{\feynhandarrowsize}{3.5pt}
\vertex [particle] (i1) at (0,0.8) {$1^-$}; 
\vertex [particle] (i2) at (1.6,1.6) {$2^-$}; 
\vertex [particle] (i3) at (1.6,0) {$3^+$};  
\vertex [particle] (i4) at (1.6,0.8) {$4$};
\vertex (v2) at (0.9+0.7*0.33,0.8-0.8*0.33);
\vertex (v1) at (0.9,0.8); 
\graph{(i1)--[fer](v1)};
\graph{(v1)--[fer](i2)};
\graph{(i3)--[bos](v2)--[sca] (v1)};  
\graph{(v2)--[sca](i4)};
\end{feynhand} \end{tikzpicture}\rightarrow
\Ampthree{1^-}{2^-}{3^+}{\ferflip{1}{180}{red}{cyan}}{\antfer{cyan}{i2}{v1}}{\bosflip{1}{-55}{brown}{cyan}},
\Ampthree{1^-}{2^-}{3^+}{\fer{red}{i1}{v1}}{\antferflip{1}{55}{cyan}{red}}{\bosflip{1}{-55}{brown}{cyan}}.
\end{aligned} \end{equation}

\subsection{General Sub-Leading Matching}

From the previous example, we see that multiplying the amplitude by the unity factors $\frac{\langle\lambda\eta\rangle}{m}$ and $\frac{[\eta\lambda]}{\tilde m}$ plays a crucial role in subleading matching. This operation can be understood through the EOM for particle states. We will use the example from the last subsection to illustrate this idea.

For a fermion, the multiplication in eq.~\eqref{eq:multiply_f} can be interpreted as
\begin{equation}
|2]^{\dot\alpha}=\frac{p_2^{\dot \alpha\alpha}}{\mathbf m_2^2}\times \tilde m_2|\eta_2\rangle_{\alpha}.
\end{equation}
Thus, applying the EOM converts a primary state $\tilde\lambda$ into a descendant state $\tilde m\eta$.
Recall that we list all particle states for a massive fermion in eq.~\eqref{eq:f_table}. The effect of the EOM can be viewed as vertical transitions among different particle states. For a fermion, this transition can be illustrated as
\begin{equation} \label{eq:f_EOM}
\includegraphics[width=0.5\linewidth]{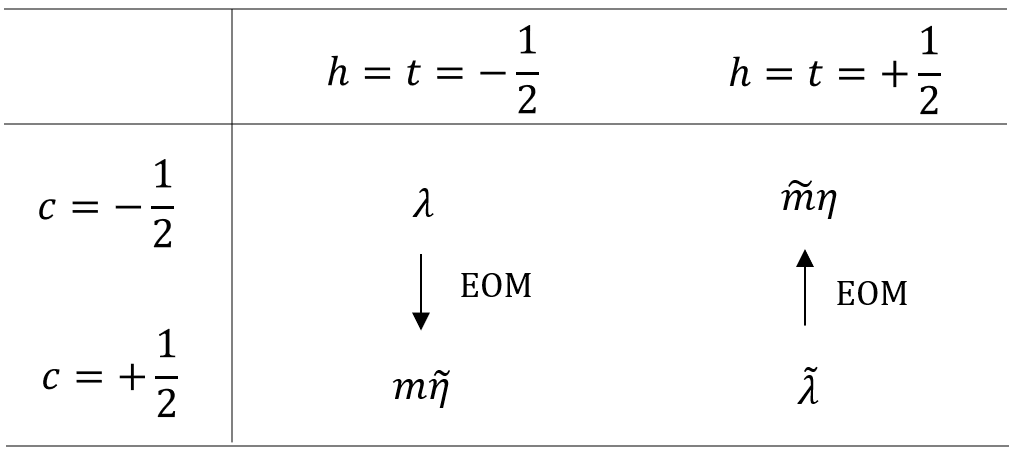}
\end{equation}
Therefore, the EOM transition $\tilde\lambda\to m\tilde\eta$ is in the colomn of $h=t=-\frac12$.

Similarly, for a vector boson, the multiplication in eq.~\eqref{eq:multiply_v} can be explained by
\begin{equation}
|3]^{\dot\alpha}|3]^{\dot\beta}=\frac{p_3^{\dot \alpha\alpha}}{\mathbf m_3^2}\times  \tilde m_3|\eta_3\rangle_{\alpha}|3]^{\dot\beta}.
\end{equation}
This EOM converts primary state $\tilde\lambda^2$ into the descendant state $\tilde m\tilde\lambda\eta$. This effect can also be viewed as vertical transitions among the states listed in eq.~\eqref{eq:vector_table_all}. The following table illustrates the transition,
\begin{equation} \label{eq:V_EOM}
\includegraphics[width=0.8\linewidth]{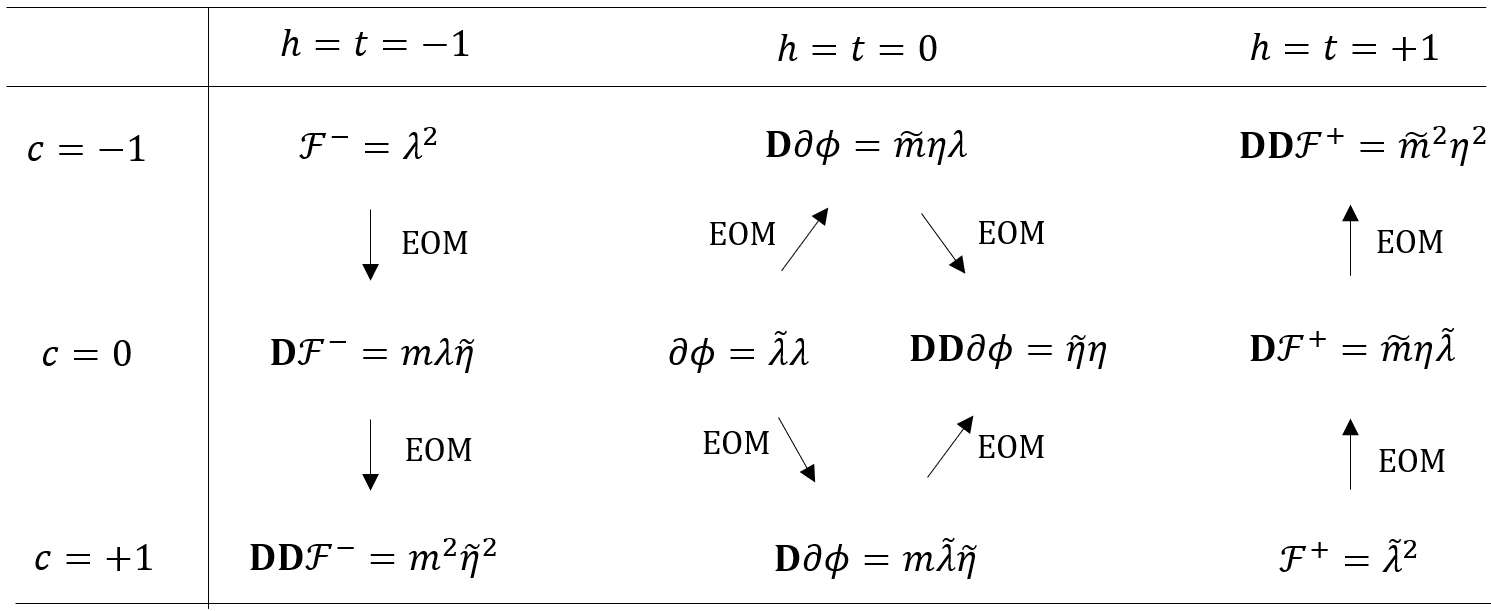}
\end{equation}
where $\mathcal F$ denotes field strength tensor, $\partial \phi$ the Goldstone boson, and $\mathbf D$ the derivative of the massive particle.

When matching a higher-point massless amplitude $\mathcal A$ to a massive amplitude, these two EOM tables can always be used to peform amplitude deformation. This is because the states listed in the EOM tables also appear in $\mathcal{A}$.
In general, a higher-point massless amplitude can be written in the form
\begin{equation} \label{eq:massless_form}
    \mathcal{A}=\sum_i G_i\frac{N_i}{D_i},
\end{equation}
where $D$ denotes the pole structure that carries no helicity information, and $N$ is the Lorentz structure that depends on the helicity of all particles. 
A general Lorentz structure can be decomposed as
\begin{equation}
    N=\text{momentum}\times\text{particle state}.
\end{equation}
For a particle with helicity $h$, the state takes the form
\begin{equation}
\left\{\begin{aligned}
&\lambda^{-h},& h&<0,\\
&\tilde\lambda^{h},& h&\ge 0.\\        
\end{aligned}\right.
\end{equation}
These states are included in eqs.~\eqref{eq:f_EOM} and \eqref{eq:V_EOM}, so we can always use them to deform the higher-point massless amplitude. 

As an example, consider the FFSS amplitude. It takes the form given in eq.~\eqref{eq:massless_form}:
\begin{equation} \label{eq:FFSS_2}
\mathcal{A}^{(-\frac12,+\frac12,0,0)}
G_1 \frac{\langle 1|P_{14}|2]}{s_{14}}+
G_2 \frac{\langle 1|P_{14}|2]}{s_{24}}+
G_3 \frac{\langle 1|3-4|2]}{s_{34}},
\end{equation}
where $s_{ij}$ belongs to the denominator $D$ and each  numerator $N\sim\langle1|p|2]$ carries all helicity information. We can identify particle states $\lambda_1$, $\tilde\lambda_2$ for the two fermions, which are included in the EOM table.

Note that, a general massless amplitude may not have Goldstone state $\tilde\lambda\lambda$, but in gauge theory it is possible to use IBP to convert the momentum of other particle to this state. For the case in eq.~\eqref{eq:FFSS_2}, IBP can be applied to convert another momentum into $\tilde{\lambda}_3\lambda_3$. Therefore, in spontaneous symmetry breaking theories, all primary state can be found in EOM table.

By combining the EOM table with Higgs insertion, we can systematically match a higher-point massless amplitude to a 3-point massive amplitude. We again use the matching from massless $FFSS$ to $FFV$ as an example to illustrate the subleading matching procedure:
\begin{itemize}
    \item For a higher-point massless amplitude with specified helicity $(h_1,h_2,h_3;0,\cdots,0)$, we find the corresponding MHC amplitude. For the $FFSS$ amplitude with heilicity $(-\frac12,+\frac12,0,0)$, we choose the following MHC diagram
    \begin{equation} \begin{aligned} \label{eq:MHC_subleading_2}
    \Ampthree{1^-}{2^+}{3^0}{\ferflip{1}{180}{red}{cyan}}{\antferflip{1}{55}{red}{cyan}}{\bos{i3}{brown}}.
    \end{aligned} \end{equation}
    \item Based on the MHC diagram, we identify which particle lines undergo chirality flips, and decide which of these should be explained by Higgs insertion. This also tells us which pole structure of the amplitude to focus on. For eq.~\eqref{eq:MHC_subleading_2}, we choose particle 1 to be explianed by the Higgs insertion. Thus, we focus on the term with the pole $s_{14}$,
    \begin{equation}
    G_1 \frac{\langle 1|P_{14}|2]}{s_{14}}.
    \end{equation}
    \item If chirality flips occur on lines without Higgs insertion, use the EOM table to convert the corresponding particle states. In eq.~\eqref{eq:MHC_subleading_2}, particle 2 has a chirality but the Higgs insert on particle 1. Therefore, particle 2 requires an EOM. In above massless $FFSS$ amplitude, the spinor structure of particle 2 is $|2]$. Referring to the EOM table in eq.~\eqref{eq:V_EOM}, this can be converted as
    \begin{equation}
    |2]^{\dot\alpha}=\frac{p_2^{\dot \alpha\alpha}}{\mathbf m_2^2}\times \tilde m_2|\eta_2\rangle_{\alpha},
    \end{equation}
     so we deform the massless amplitude as
    \begin{equation}
    G_1 \frac{\langle 1|P_{14}|2]}{s_{14}}
    \overset{\text{EOM}}{=}
    \frac{\tilde m_2}{\mathbf m_2^2}G_1 \frac{\langle 1|P_{14} 2|\eta_2\rangle}{s_{14}}.
    \end{equation}
    \item Apply IBP to recover Goldstone behavior or isolate the multipoles,
    \begin{equation}
    \frac{\tilde m_2}{\mathbf m_2^2}G_1 \frac{\langle 1|P_{14} 2|\eta_2\rangle}{s_{14}}
    \overset{\text{IBP}}{=}
    -\frac{\tilde m_2}{\mathbf m_2^2}G_1 \frac{\langle 1|P_{14}|3]\langle3\eta_2\rangle}{s_{14}}-\frac{\tilde m_2}{\mathbf m_2^2}G_1 \langle 1\eta_2\rangle.
    \end{equation}
    The first term restores Goldstone behavior for particle 2. The second term lacks the $s_{14}$ pole and is therefore irrelevant for the Higgs insertion on particle 1. In this case, no multipole structure appears.
    \item For particle lines with Higgs insertion, take the on-shell limit and extract the leading contribution, which matches the desired MHC amplitude. For the FFSS amplitude, taking the limit $p_4 \to \eta_1$ gives
    \begin{equation}
    \lim_{p_4\rightarrow \eta_1}\left(-\frac{\tilde m_2}{\mathbf m_2^2}G_1 \frac{\langle 1|P_{14}|3]\langle3\eta_2\rangle}{s_{14}}-\frac{\tilde m_2}{\mathbf m_2^2}G_1 \langle 1\eta_2\rangle\right)
    =G_1\frac{\tilde m_2 m_1[\eta_1 3]\langle3\eta_2\rangle}{\mathbf m_2^2 \mathbf m_1^2}+\mathcal{O}(E^{-2}).
    \end{equation}
\end{itemize}

\subsection{VVS subleading matching}
\label{sec:VVS_subleading}

We now apply the EOM and Higgs insertion method to perform subleading matching for the $VVS$ amplitude.

First consider the massless amplitude with helicity $(0,0,0,0)$:
\begin{equation}
\mathcal{A}^{(0,0,0,0)}=G_1 \frac{(p_1-p_4)\cdot(p_2-p_3)}{s_{14}}+
G_2 \frac{(p_1-p_3)\cdot(p_2-p_4)}{s_{13}}.
\end{equation}
It matches the MHC amplitude with helicity $(0,0,0)$, 
\begin{equation} \begin{aligned}
\Ampthree{1^0}{2^0}{3^0}{\bosflipflip{1}{180}{brown}{red}{brown}}{\bos{i2}{brown}}{\sca{i3}}.
\end{aligned} \end{equation}
We see that two chirality flips occur on the same particle, namely particle 1. Therefore, the UV origin of this diagram must come from Higgs insertion on particle 1, and no EOM rule is needed in this case.

We can deform the massless amplitude to restore the Goldstone behavior $\tilde\lambda\lambda$ for particle 2,
\begin{equation} \begin{aligned}
\mathcal{A}^{(0,0,0,0)}
&\overset{\text{IBP}}{=}
2G_1\frac{\langle2|p_1-p_4|2]}{s_{14}}+G_2 \frac{(p_1-p_3)\cdot(p_2-p_4)}{s_{13}}.
\end{aligned} \end{equation}
Taking the on-shell limit then gives,
\begin{equation} \begin{aligned}
\begin{tikzpicture}[baseline=0.7cm] \begin{feynhand}
\setlength{\feynhandarrowsize}{3.5pt}
\vertex [particle] (i1) at (-0.2,0.8) {$1^0$}; 
\vertex [particle] (i2) at (1.6,1.6) {$2^0$}; 
\vertex [particle] (i3) at (1.6,0) {$3^0$};  
\vertex [particle] (i5) at (0.7,0.2) {$4$};
\vertex (v3) at (0.6,0.8);
\vertex (v1) at (0.9,0.8); 
\graph{(i1)--[sca](v3)--[bos](v1)};
\graph{(v1)--[sca](i2)};
\graph{(i3)--[sca] (v1)};  
\graph{(v3)--[sca](i5)};
\end{feynhand} \end{tikzpicture}
&=\lim_{p_4\rightarrow \eta_1} 2G_1\frac{\langle2|p_1-p_4|2]}{s_{14}}+G_2 \frac{(p_1-p_3)\cdot(p_2-p_4)}{s_{13}} \\
&= 2G_1\frac{(\langle2 1\rangle[1 2]-\langle2\eta_1\rangle[\eta_1 2])}{\tilde m_1 m_1}+\mathcal{O}(E^{-2}).
\end{aligned} \end{equation}
Therefore, the result matches both the primary and descendant MHC amplitudes,
\begin{equation} \begin{aligned} \label{eq:VVS_subleading}
\begin{tikzpicture}[baseline=0.7cm] \begin{feynhand}
\setlength{\feynhandarrowsize}{3.5pt}
\vertex [particle] (i1) at (-0.2,0.8) {$1^0$}; 
\vertex [particle] (i2) at (1.6,1.6) {$2^0$}; 
\vertex [particle] (i3) at (1.6,0) {$3^0$};  
\vertex [particle] (i5) at (0.7,0.2) {$v$};
\vertex (v3) at (0.6,0.8);
\vertex (v1) at (0.9,0.8); 
\graph{(i1)--[sca](v3)--[bos](v1)};
\graph{(v1)--[sca](i2)};
\graph{(i3)--[sca] (v1)};  
\graph{(v3)--[sca](i5)};
\end{feynhand} \end{tikzpicture}&\rightarrow
\Ampthree{1^0}{2^0}{3^0}{\bosflipflip{1}{180}{brown}{red}{brown}}{\bos{i2}{brown}}{\sca{i3}}+
\Ampthree{1^0}{2^0}{3^0}{\bos{i1}{brown}}{\bos{i2}{brown}}{\sca{i3}}.
\end{aligned} \end{equation}





Then we consider the massless amplitude with helicity $(-1,+1,0,0)$, 
\begin{equation}
\mathcal{A}^{(-1,+1,0,0)}=G_1 \frac{\langle1|3-4|2]^2}{s_{14}s_{34}}+
G_2 \frac{\langle1|3-4|2]^2}{s_{24}s_{34}}.
\end{equation}
It matches the MHC amplitude:
\begin{equation} \begin{aligned}
\Ampthree{1^-}{2^+}{3^0}{\bosflip{1}{180}{brown}{red}}{\bosflip{1}{55}{brown}{cyan}}{\sca{i3}}.
\end{aligned} \end{equation}
Here, chirality flips occur on both particles 1 and 2. We choose to use the EOM to account for the chirality flip on particle 1. In the numerator, the spinor structure of particle 1 is $|1\rangle^2$. Referring to the EOM table in eq.~\eqref{eq:V_EOM}, this can be converted as
\begin{equation}
|1\rangle_{\alpha}|1\rangle_{\beta}=\frac{p_{1\alpha\dot \alpha}}{\mathbf m_1^2}\times  m_1|\eta_1]^{\dot\alpha}|1\rangle_{\dot\beta}.
\end{equation}
Therefore massless amplitude can then be deformed as follows,
\begin{equation} \begin{aligned}
\mathcal{A}^{(-1,+1,0,0)}
&\overset{\text{EOM}}{=}G_2 \frac{m_1}{\mathbf m_1^2}\frac{[\eta_1 1(3-4)2]\langle1|3-4|2]}{s_{24}s_{12}}+
G_1 \frac{\langle1|3-4|2]^2}{s_{23}s_{12}}\\
&\overset{\text{IBP}}{=}
-2G_2 \frac{m_1}{\mathbf m_1^2}\frac{[\eta_1 2]\langle1|4|2]}{s_{24}}+
\left(-4 G_2 \frac{m_1}{\mathbf m_1^2}\frac{[\eta_1 2]\langle1|4|2]}{s_{12}}+
G_1 \frac{\langle1|3-4|2]^2}{s_{23}s_{12}}\right).
\end{aligned} \end{equation}
We only isolate the multipole $s_{12}s_{24}$, as the term with the other multipole $s_{23}s_{12}$ will be suppressed. The chirality flip on particle 1 is accounted for by Higgs insertion. Applying the on-shell limit, we obtain
\begin{equation} \begin{aligned}
\begin{tikzpicture}[baseline=0.7cm] \begin{feynhand}
\setlength{\feynhandarrowsize}{3.5pt}
\vertex [particle] (i1) at (0,0.8) {$1^-$}; 
\vertex [particle] (i2) at (1.6,1.6) {$2^+$}; 
\vertex [particle] (i3) at (1.6,0) {$3^0$};  
\vertex [particle] (i4) at (1.6,0.8) {$4$};
\vertex (v2) at (0.9+0.7*0.33,0.8+0.8*0.33);
\vertex (v1) at (0.9,0.8); 
\graph{(i1)--[bos](v1)};
\graph{(v1)--[sca](v2)--[bos](i2)};
\graph{(i3)--[sca] (v1)};  
\graph{(v2)--[sca](i4)};
\end{feynhand} \end{tikzpicture}
&=\lim_{p_4\rightarrow \eta_2}-2G_2 \frac{m_1}{\mathbf m_1^2}\frac{[\eta_1 2]\langle1|4|2]}{s_{24}}-
\left(4 G_2 \frac{m_1}{\mathbf m_1^2}\frac{[\eta_1 2]\langle1|4|2]}{s_{12}}+
G_1 \frac{\langle1|3-4|2]^2}{s_{23}s_{12}}\right)\\
&= -2G_2\frac{m_1 \tilde m_2 \langle1\eta_2\rangle[2\eta_1]}{\mathbf m_1^2\mathbf m_2^2}+\mathcal{O}(E^{-2})\\
&\to \Ampthree{1^-}{2^+}{3^0}{\bosflip{1}{180}{brown}{red}}{\bosflip{1}{55}{brown}{cyan}}{\sca{i3}}.
\end{aligned} \end{equation}

\section{Higgs splitting with group structure}
\label{app:Higgs_split}

In sections~\ref{sec:FFS} and~\ref{sec:vector}, we presented a simplified analysis of Higgs splitting without considering group structure. For the Standard Model case, however, we must take these structures into account and verify that the 3-pt gauge structure and vev can be consistently incorporated into the physical mass $\mathbf{m}$.

As a specific example, consider Higgs insertion on a fermion line. The $h=-\frac{1}{2}$ state with gauge tensor has the form 
\begin{equation} \begin{aligned}
\Ampone{1.5}{-\frac12}{\graph{(i1)--[fer] (v1)}} &= \mathcal{G}^{\hat i}|p\rangle,\\
\end{aligned} \end{equation}
where $\hat i$ denotes the fermion species, and $\mathcal{G}^{\hat i}$ represents the group structure associated with this massless fermion.  

For a single Higgs insertion, the helicity flips to $h=+\frac12$. In the on-shell limit $p_h\to\eta$, we obtain
\begin{equation} \begin{aligned} \label{eq:insert_f_complete}
\begin{tikzpicture}[baseline=-0.1cm] \begin{feynhand}
\setlength{\feynhandblobsize}{6mm}
\setlength{\feynhandarrowsize}{5pt}
\vertex [particle] (i1) at (2,0) {$+\frac12$};
\vertex [particle] (i2) at (1,0.7) {$v$};
\vertex (v2) at (1,0);
\vertex [dot] (v1) at (0,0) {};
\graph{(i1)--[fer](v2)--[fer](v1)};
\graph{(i2)--[sca](v2)};
\end{feynhand} \end{tikzpicture} &= \mathcal{G}^{\hat j}|\chi\rangle\times v Y_{\hat j}^{\hat i i_h}[\chi p],\\
&\to v\mathcal{G}^{\hat j}\delta^{\hat k}_{\hat j}Y_{\hat k}^{\hat i i_h}\times \tilde m |\eta\rangle.\\
\end{aligned} \end{equation}
Here, $i_h$ indicates the species of the additional scalar boson, while $\hat i$ still refers to the external fermion. We introduce a Dirac delta $\delta^{\hat k}_{\hat j}$ to represent the group structure of the internal fermion. The normalization condition of the transformation matrix $\Omega$ also gives this internal group structure, 
\begin{equation} \begin{aligned}
\delta^{\hat k}_{\hat j} &= \Omega^{\hat k}_{\hat{\mathbf j}} \Omega_{\hat j}^{\hat{\mathbf j}}.\\
\end{aligned} \end{equation}
Using these conditions, we can convert internal fermions from massless to massive. For external particles, we must still multiply by the transformation matrices $\mathcal U$ and $\Omega$. After converting all particles to massive states, the group tensor yields
\begin{equation} \begin{aligned}
&v\mathcal{G}^{\hat j}\delta^{\hat k}_{\hat j}{Y}_{\hat{k}}^{\hat{i} i_h}\times\mathcal{U}^{h}_{i_h} \Omega^{\hat{\mathbf{i}}}_{\hat{i}}
+(S\leftrightarrow \bar{S}) \\
=&\mathcal{G}^{\hat j}\Omega^{\hat{j} \hat{\mathbf j}}\times \left(v{Y}_{\hat{j}}^{\hat{i} i_h}\mathcal{U}^{h}_{i_h}\Omega^{\hat{\mathbf{i}}}_{\hat{i}} \Omega_{\hat{\mathbf j}}^{\hat{j}}
+v{Y}_{\hat{j} i_h}^{\hat{i}}\mathcal{U}^{i_h h}\Omega^{\hat{\mathbf{i}}}_{\hat{i}} \Omega_{\hat{\mathbf j}}^{\hat{j}}\right).
\end{aligned} \end{equation}
The bracketed structures exists only when $\hat{\mathbf i}$ and $\hat{\mathbf j}$ correspond to the same particle species. This quantity effectively defines the fermion mass
\begin{equation} \label{eq:fermion_mass}
\mathbf m_{\hat{\mathbf i}}
\equiv v\left({Y}_{\hat{j}}^{\hat{i} i_h}\mathcal{U}^{h}_{i_h}
+{Y}_{\hat{j} i_h}^{\hat{i}}\mathcal{U}^{i_h h}\right)\Omega^{\hat{\mathbf{i}}}_{\hat{i}} \Omega_{\hat{\mathbf j}}^{\hat{j}}. \\
\end{equation}
Thus the single insertion result reduces to
\begin{equation} \begin{aligned}
\begin{tikzpicture}[baseline=-0.1cm] \begin{feynhand}
\setlength{\feynhandblobsize}{6mm}
\setlength{\feynhandarrowsize}{5pt}
\vertex [particle] (i1) at (2,0) {$+\frac12$};
\vertex [particle] (i2) at (1,0.7) {$v$};
\vertex (v2) at (1,0);
\vertex [dot] (v1) at (0,0) {};
\graph{(i1)--[fer](v2)--[fer](v1)};
\graph{(i2)--[sca](v2)};
\end{feynhand} \end{tikzpicture}
&\Rightarrow \mathbf{G}^{\hat{\mathbf i}} \mathbf{m}_{\hat{\mathbf i}} \times \tilde m|\eta\rangle, \\
\end{aligned} \end{equation}
where $\mathbf{G}^{\hat{\mathbf j}}=\mathcal{G}^{\hat j}\Omega^{\hat{j} \hat{\mathbf j}}$ represents the coefficient including the massive fermion $\hat j$. When we restore the $SU(2)_{LG}$ convariance, the spurion mass $\tilde m$ should be identified with the physical mass, which in this case is precisely $\mathbf m_{\hat{\mathbf i}}$ . The insertion for the opposite helicity fermion will give the same definition of fermion mass.

Similarly, we can analyze the group structure in the subleading matching for massive vector bosons. Consider the particle states with helicity $h=0$ and $-1$,
\begin{equation} \begin{aligned}
\Ampone{1.5}{-1}{\graph{(i1)--[bos](v1)}} &= \mathcal{G}^{I} \frac{|p\rangle|\eta]}{[p\eta]}, \\
\Ampone{1.5}{0}{\graph{(i1)--[sca](v1)}} &= \mathcal{G}^i |p]|p\rangle,\\ 
\end{aligned} \end{equation}
where $\mathcal{G}^{i}$ and $\mathcal{G}^{I}$ represent the group structure associated with the massless scalar and gauge boson, respectively.  

By inserting an additional Higgs boson, we can increase the helicity
\begin{eqnarray}
\begin{tikzpicture}[baseline=-0.1cm] \begin{feynhand}
\setlength{\feynhandblobsize}{6mm}
\setlength{\feynhandarrowsize}{5pt}
\vertex [particle] (i1) at (2,0) {$0$};
\vertex [particle] (i2) at (1,0.7) {$v$};
\vertex (v2) at (1,0);
\vertex [dot] (v1) at (0,0) {};
\graph{(i1)--[sca](v2)--[bos](v1)};
\graph{(i2)--[sca](v2)};
\end{feynhand} \end{tikzpicture} 
&\to&
\frac12 v\mathcal{G}^{J}\delta^{JK}(T_s^{K})^{i}_{i_4}\times (|p]|p\rangle-|\eta]|\eta\rangle),  \label{eq:insert2} \\
\begin{tikzpicture}[baseline=-0.1cm] \begin{feynhand}
\setlength{\feynhandblobsize}{6mm}
\setlength{\feynhandarrowsize}{5pt}
\vertex [particle] (i1) at (2,0) {$+1$};
\vertex [particle] (i2) at (1,0.7) {$v$};
\vertex (v2) at (1,0);
\vertex [dot] (v1) at (0,0) {};
\graph{(i1)--[bos](v2)--[sca](v1)};
\graph{(i2)--[sca](v2)};
\end{feynhand} \end{tikzpicture} &\to&
\mathcal{G}^{j} v\delta^k_j(T_s^{I_i})^{i_4}_{k}  \times \tilde m|p]|\eta\rangle.  \label{eq:insert3}
\end{eqnarray}
Here $\delta_{j}^{k}$ and $\delta^{JK}$ denote the gauge structures of the internal scalar and gauge boson. These can be expressed in terms of the bosonic transformation matrices $\mathcal{U}$ and $\mathcal O$ through their normalization conditions:
\begin{eqnarray}
\delta_{j}^{k} &=& -\frac{1}{2}\mathcal{U}^{k \mathbf I} \mathcal{U}_{j}^{\mathbf I^*}+\mathcal{U}^{k h} \mathcal{U}_{j}^{h}, \label{eq:gauge2}\\
\delta^{JK} &=& \mathcal{O}^{J \mathbf J} \mathcal{O}^{K \mathbf J^*}. 
\end{eqnarray}
The term $\mathcal{U}^{kh}\mathcal{U}^{h}_j$ does not contribute to Eq.~\eqref{eq:insert3}, since the massive $VSS$ amplitude is forbiden. Therefore, the insertion result simplify to
\begin{eqnarray}
\begin{tikzpicture}[baseline=-0.1cm] \begin{feynhand}
\setlength{\feynhandblobsize}{6mm}
\setlength{\feynhandarrowsize}{5pt}
\vertex [particle] (i1) at (2,0) {$0$};
\vertex [particle] (i2) at (1,0.7) {$v$};
\vertex (v2) at (1,0);
\vertex [dot] (v1) at (0,0) {};
\graph{(i1)--[sca](v2)--[bos](v1)};
\graph{(i2)--[sca](v2)};
\end{feynhand} \end{tikzpicture} &\Rightarrow&
\mathbf{G}^{\mathbf I} \mathbf{m}_{\mathbf I}\times(|p]|p\rangle-|\eta]|\eta\rangle), \\
\begin{tikzpicture}[baseline=-0.1cm] \begin{feynhand}
\setlength{\feynhandblobsize}{6mm}
\setlength{\feynhandarrowsize}{5pt}
\vertex [particle] (i1) at (2,0) {$+1$};
\vertex [particle] (i2) at (1,0.7) {$v$};
\vertex (v2) at (1,0);
\vertex [dot] (v1) at (0,0) {};
\graph{(i1)--[bos](v2)--[sca](v1)};
\graph{(i2)--[sca](v2)};
\end{feynhand} \end{tikzpicture} &\Rightarrow&
\mathbf{G}^{\mathbf I} \mathbf{m}_{\mathbf I}\times\tilde m|p]|\eta\rangle,
\end{eqnarray}
where the vector boson mass is defined as
\begin{equation} \label{eq:vector_mass}
\mathbf m_{\mathbf I}
\equiv\frac12 v\left((T_s^{K})^{i}_{i_h}\mathcal{U}_{i}^{\mathbf{I}} \mathcal{U}^{i_h h}+ (T_s^{K})^{i_h}_{i} \mathcal{U}^{i \mathbf{I}}\mathcal{U}_{i_h}^{h}\right)O^{K \mathbf I}.
\end{equation}
The factor of $\frac12$ comes from Eqs.~\eqref{eq:insert2} and \eqref{eq:gauge2}. This mass definition applies to both $h=\pm 1$ and $0$ states. 

For specific particle species, we can verify the mass value for both vector bosons and fermions,
\begin{equation} \begin{aligned}
&\text{fermion}:& \mathbf{m}_{f} &= \mathcal{Y}^{(f)}\frac{v}{\sqrt 2},\quad f=u,d,e, \\
&\text{vector}:& \mathbf{m}_{W^+} &=\mathbf{m}_{W^-}= \frac{1}{2} gv,\quad\mathbf{m}_Z = \frac{\mathbf{m}_W}{\cos\theta_W}. 
\end{aligned} \end{equation}
These results agree with the tree-level predictions obtained from the Lagrangian with spontaneous symmetry breaking.

%
%
%
%

\end{appendix}

\bibliographystyle{JHEP}
\bibliography{ref}

\end{document}